\documentclass[12pt,preprint]{aastex}

\newcommand{\logg}{\ensuremath{\log g}}
\newcommand{\teff}{\ensuremath{T_\mathrm{eff}}}

\newcommand{\massy}{\ensuremath{{\log \beta_\mathrm{He}}}}
\newcommand{\massc}{\ensuremath{{\log \beta_\mathrm{C}}}}
\newcommand{\massn}{\ensuremath{{\log \beta_\mathrm{N}}}}

\usepackage{natbib,twoopt}
\usepackage{lscape}
\usepackage[breaklinks=true]{hyperref} 
\bibpunct{(}{)}{;}{a}{}{,}             
\makeatletter
  \newcommandtwoopt{\citeads}[3][][]{\href{http://adsabs.harvard.edu/abs/#3}%
    {\def\hyper@linkstart##1##2{}%
     \let\hyper@linkend\@empty\citealp[#1][#2]{#3}}}
  \newcommandtwoopt{\citepads}[3][][]{\href{http://adsabs.harvard.edu/abs/#3}%
    {\def\hyper@linkstart##1##2{}%
     \let\hyper@linkend\@empty\citep[#1][#2]{#3}}}
  \newcommandtwoopt{\citetads}[3][][]{\href{http://adsabs.harvard.edu/abs/#3}%
    {\def\hyper@linkstart##1##2{}%
     \let\hyper@linkend\@empty\citet[#1][#2]{#3}}}
  \newcommandtwoopt{\citeyearads}[3][][]%
    {\href{http://adsabs.harvard.edu/abs/#3}
    {\def\hyper@linkstart##1##2{}%
     \let\hyper@linkend\@empty\citeyear[#1][#2]{#3}}}
\makeatother

\shorttitle{Hot Subluminous Stars}
\shortauthors{Heber}

\begin{document}


\title{Hot Subluminous Stars
}


\author{U. Heber\altaffilmark{1}}
\affil{Dr. Remeis-Sternwarte \& ECAP, Astronomical Institute, University of Erlangen-N\"urnberg, Germany}




\begin{abstract}
Hot subluminous stars of spectral type B and O are core helium-burning stars at the blue end of the horizontal branch or have evolved even beyond that stage. Most hot subdwarf stars are chemically highly peculiar and provide a laboratory to study diffusion processes that cause these anomalies. 
The most obvious anomaly lies with helium, which may be a trace element in the atmosphere of some stars (sdB, sdO) while it may be the dominant species in others (He-sdB, He-sdO). Strikingly, the distribution in the Hertzsprung-Russell diagram of He-rich vs. He-poor hot subdwarf stars of the globular clusters $\omega$ Cen and NGC~2808  differ from that of their field counterparts. 
The metal-abundance patterns of hot subdwarfs are typically characterized by strong deficiencies of some lighter elements as well as large enrichments of heavy elements. 

A large fraction of sdB stars are found in close binaries with white dwarf or very low-mass main sequence companions, which must have gone through a common-envelope phase of evolution. Because the binaries are detached they 
provide a clean-cut laboratory to study this important but yet purely understood phase of stellar evolution. Hot subdwarf binaries with sufficiently massive white dwarf companions are viable candidate progenitors of type Ia supernovae both in the double degenerate as well as in the single degenerate scenario as helium donors for double detonation supernovae. The hyper-velocity He-sdO star  US~708 may be the surviving donor of such a double detonation supernova. 

Substellar companions to sdB stars have also been found. For HW~Vir systems the companion mass distribution extends from the stellar into the brown dwarf regime.
A giant planet to the acoustic-mode pulsator V391 Peg was the first discovery of a planet that survived the red giant evolution of its host star.
Evidence for Earth-size planets to two pulsating sdB stars have been reported  and circumbinary giant planets or brown dwarfs have been found around HW~Vir systems from eclipse timings. The high incidence of circumbinary substellar objects suggests that most of the planets are formed from the remaining common-envelope material (second generation planets). 
  
Several types of pulsating star have been discovered among hot subdwarf stars, the most common are the gravity-mode sdB pulsators (V1093 Her) and their hotter siblings, the p-mode pulsating V361 Hya stars. Another class of multi-periodic pulsating hot subdwarfs has been found in the globular cluster $\omega$ Cen that is unmatched by any field star.  
Asteroseismology has advanced enormously thanks to the high-precision {\it Kepler} photometry and allowed stellar rotation rates to be determined, the interior structure of  gravity-mode pulsators to be probed and stellar ages to be estimated. Rotation rates turned out to be unexpectedly slow calling for very efficient angular momentum loss on the red giant branch or during the helium core flash. The convective cores were found to be larger than predicted by standard stellar evolution models requiring very efficient angular momentum transport on the red giant branch. 

The masses of hot subdwarf stars, both single or in binaries, are the key to understand the stars' evolution. A few pulsating sdB stars in eclipsing binaries have been found that allow both techniques to be applied for mass determination. The results, though few, are in good agreement with predictions from binary population synthesis calculations.

New classes of binaries, hosting so-called extremely low mass (ELM) white dwarfs (M$<$0.3 M$_\odot$), have recently been discovered, filling a gap in the mosaic of binary stellar evolution. Like most sdB stars the ELM white dwarfs are the stripped cores of red giants, the known companions are either white dwarfs, neutron stars (pulsars) or F- or A-type main sequence stars (''EL CVn'' stars).

In the near future, the Gaia mission will provide high-precision astrometry for a large sample of subdwarf stars to disentangle the different stellar populations in the field and to compare the field subdwarf population with the globular clusters' hot subdwarfs. New fast-moving subdwarfs will allow  the mass of the Galactic dark matter halo to be constrained and additional unbound hyper-velocity stars may be discovered. 

\end{abstract}





\newpage

\section{Introduction}\label{sect:intro}

\begin{figure}
\epsscale{1.0}
\plotone{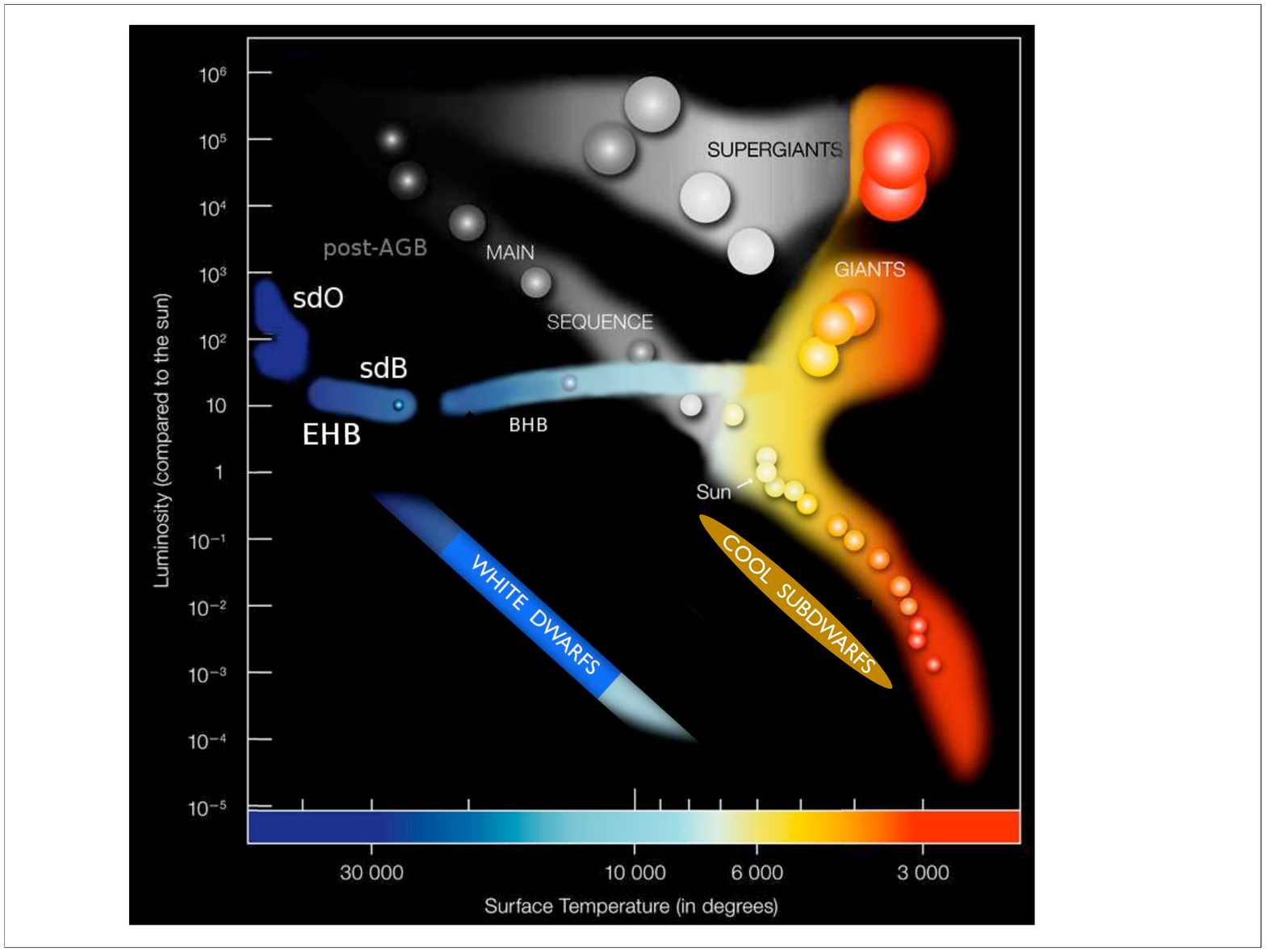}
\caption{Hertzsprung-Russell diagram highlighting the position of the extreme horizontal branch (EHB) populated by subluminous B stars and located between he upper main sequence and the white dwarf sequence. The EHB is separated from the blue horizontal branch (BHB). The location of subluminous O stars as well as stars having evolved from the asymptotic giant branch is shown for comparison. From \citetads{2009ARA&A..47..211H}; copyright Annual Review of Astronomy \& Astrophysics; reproduced with permission.}\label{fig:hrd_sketch}
\end{figure}

The hot subluminous stars of B and O-type  (sdB, sdO) represent several late stages in 
the evolution of low-mass stars.
In the Hertzsprung-Russell diagram they can be found between the main sequence and the white-dwarf sequence
(see Fig. \ref{fig:hrd_sketch}).  
The discovery of subluminous blue stars at high Galactic latitudes dates back to the 1950s after exploitation of the \citetads{1947ApJ...105...85H} 
photometric survey of the North Galactic
Pole and Hyades regions \citepads[e.g.][]{1953AJ.....58...75L,1956bsms.conf...11G,1958ApJ...127..642M}. 
The number of known objects
remained small until the 1980s when the Palomar-Green survey \citepads[PG,][]{1986ApJS...61..305G} of the northern
Galactic hemisphere was published \citepads[see ][ for overviews of early developments]{1987fbs..conf....3G,2004Ap&SS.291..197L}. 
Several other photometric surveys followed, e.g. the Kitt Peak-Downes survey of the Galactic plane \citepads[KPD,][]{1986ApJS...61..569D} 
and the Edinburgh-Cape Survey \citepads[EC,][]{1997MNRAS.287..848S} for the southern sky, as well as objective prism surveys, e.g. Byurakan surveys \citepads[FBS, SBS,][]{2007A&A...464.1177M}, 
the Hamburg Quasar Survey \citepads[HS,][]{1995A&AS..111..195H} for the northern and the Hamburg ESO survey \citepads[HE,][]{1996A&AS..115..227W} for the southern sky.
The Sloan Digital Sky survey has doubled the number of known hot subdwarfs within a few years and the Galaxy Evolution Explorer ({\it GALEX}) all-sky survey extended the list further.

Most of the B-type subdwarfs were identified as helium burning stars of about half a solar mass
at the blue end of the horizontal branch, the so-called Extreme Horizontal Branch \citepads[EHB,][]{1986A&A...155...33H}. 
 In contrast to the normal horizontal branch (HB) stars, the hydrogen envelope is too thin  (M$_\mathrm{env} < 0.01$ M$_\odot$) to sustain hydrogen shell burning. Hence, subluminous B stars are the stripped cores of red giants, which managed to ignite helium while retaining just a little bit of hydrogen as their envelope.  
 
\citetads{2001MNRAS.326.1391M} 
found two-thirds of all sdBs to be in detached binary systems with short periods from hours to days and their companion stars to be mostly white dwarfs. More recent investigations indicate that half of the sdBs reside in such close binaries \citepads{2004Ap&SS.291..321N,2011MNRAS.415.1381C}, still a very high fraction. 
 Their orbital separations are only a few solar radii. Since the sdB stars have evolved from red giants, much larger in size than the present binary orbit, the progenitor system must have undergone a common-envelope phase, during which the orbit shrank when the companion was engulfed in the red giant's envelope. Finally, the common envelope is ejected and a tight, detached binary system remains. The evolution through common-envelope ejection is crucial to an understanding of various types of binaries, including X-ray binaries, double neutron stars, double white dwarfs and sdB binaries. However, common-envelope formation and evolution is poorly understood because of the complexity of the physics involved \citepads[for a review see][]{2013A&ARv..21...59I}. 
Amongst the various types of post-common-envelope systems, the sdB binaries provide an important laboratory for studying common-envelope evolution. Because the systems are non-interacting, the properties of the components can be derived in more detail, with higher precision, and selection effects are less severe than for interacting systems such as cataclysmic variables \citepads{2007MNRAS.374.1495P}. 
Hence, sdB stars form an important piece in the puzzle of common-envelope evolution.

Many sdB binaries host unseen white dwarf companions and sdBs with neutron star or black hole partners are predicted to exist. Such systems join the zoo of compact binary systems and play a role as progenitors of the cosmologically-important type Ia supernovae and other transient phenomena via mergers driven by gravitational wave radiation. Because of their short periods the orbital gravitational wave emission may be strong enough to be detectable by future space-based missions. An extensive review of the formation and evolution of compact binary star systems has been published recently by \citetads{2014LRR....17....3P}. 

However, besides degenerate companions, main-sequence stars and brown dwarfs have been discovered to orbit sdB stars. 
Binary population synthesis has identified several possible channels for the formation of hot subdwarfs, both in binaries as well as single stars 
that involve Roche lobe overflow, common-envelope ejection and gravitational wave-driven mergers of helium white dwarfs. A common-envelope phase does not necessarily lead to an envelope ejection but may also result in a core merger, e.g. for a sub-stellar companion.
Single hot subdwarf stars have been suggested to result from mergers of two helium white dwarfs \citepads{2002MNRAS.336..449H,2003MNRAS.341..669H} 
but may also form via internal instabilities and mixing processes in the envelope of the progenitor.   
 
Though sdBs and sdOs occupy neighboring regions in the Hertzsprung-Russell diagram, they are quite different both with respect to their chemical compositions and evolutionary status. 
The atmospheres of sdBs are mostly helium poor, their helium abundances might be as low as a thousandth of the solar value or less. 
Subluminous O stars, on the other hand, show a variety of helium abundances, ranging from a hundredth of the solar content to pure helium atmospheres (He-sdO).
Therefore, a direct evolutionary link between these two types of stars has always been considered questionable.
It is not clear how diffusion, which is thought to be responsible for the helium deficiency of sdBs, should turn them into helium rich objects when an EHB star evolves to higher effective temperatures after core helium burning has ceased. Because of their low binary frequency He-sdO stars are thought to have formed by a merger of two helium white dwarfs or in isolation via internal mixing in the hot flasher scenario \citepads{2008A&A...491..253M}.
 
Asteroseismology has become an important tool to study the internal structure of hot subluminous stars, because several classes of pulsating star have been discovered among them. Their light variations are multi-periodic, many showing a multitude of acoustic and/or gravity modes. The analysis of their light curves using sophisticated models allowed the internal structure, mass, and rotation of several sdB pulsators to be probed and ages to be estimated. 
Although amplitude variations are frequently encountered, some stars show modes that are sufficiently stable for years and allow a search for periodic light travel time variations caused by planetary companions. Indeed, a giant planet companion of the pulsator V391~Pegasi has been discovered \citepads{2007Natur.449..189S} and several circumbinary planets followed in recent years.   

Because they are UV-bright and relatively luminous, hot subdwarf stars have been used as UV-light sources to probe the interstellar medium (ISM) out to $\approx$1\,kpc or more from the Galactic plane for all components of the ISM \citepads[e.g.][]{1999A&A...352..287B, 2004PASP..116..895L,2006ApJ...647.1106L,2006ASPC..348..401T,2013ApJ...764...25J} as well as to constrain the distances and chemical composition to high velocity interstellar clouds \citepads[e.g.][]{2001ApJS..136..463W,2004MNRAS.352.1279S,2013ApJ...777...19H}. 
The ultraviolet upturn phenomenon in elliptical galaxies is suspected to be caused by extreme horizontal branch stars. Binary population models have recently been found to be a promising explanation of this this phenomenon as well. First results  
suggest that the UV upturn may not be an age indicator as previously assumed \citepads{2010Ap&SS.329...41H} 

A broad review of hot subluminous stars has been given by \citetads{2009ARA&A..47..211H}. The field is flourishing thanks to many new observational opportunities provided by the {\it Kepler} and {\it GALEX} satellites, and the Sloan Digital Sky Survey (SDSS) to name a few and it is timely to report on recent progress. In this paper I shall 
address the atmospheric properties, abundance patterns, rotation, and magnetic fields of hot subdwarf stars in Sect. \ref{sect:atmos}. 
 Section \ref{sect:evolution} deals with the formation and evolution of hot subdwarfs, followed in Sect. \ref{sect:cluster} by a brief discussion of hot subdwarfs in stellar clusters. The progress made in understanding binary stars from observations are reported in Sect. \ref{sect:binaries}. Substellar companions to sdB binaries are introduced in Sect. \ref{sect:planets} and a brief account of the recent developments in asteroseismology follows in Sect. \ref{sect:asteroseismology}. A tight relation between binary subdwarf stars and the new classes of binaries, hosting low-mass helium-core objects (also called extremely low mass (ELM) white dwarfs) is suggested. Such objects are also found as companions to early-type main-sequence stars (EL CVn variables), reported in Sect. \ref{sect:lmwd_elm}. 
 The kinematics of hot subdwarf stars and a unique hyper-velocity star are discussed in Sect. \ref{sect:kinematic}.
 
A summary and some conclusions are presented at the end of the paper.    
A comprehensive overview of the progress in this field of research can be found in the proceedings of the fourth, fifth, and sixth meetings on hot subdwarf stars and related objects \citepads{2010Ap&SS.329....1H,2012ASPC..452.....K,2014ASPC..481.....V}.

\newpage
\section{Atmospheric properties and chemical composition of hot subdwarf stars}
\label{sect:atmos}

Because of their peculiar spectra the hot subluminous stars cannot be classified in the MK spectral classification scheme.
Therefore, \citetads{2013A&A...551A..31D} 
introduced three additional luminosity classes and added helium classes (0 to 40) as a third dimension to the MK scheme. Additional sequences were identified with respect to the carbon and nitrogen line spectra.

\begin{figure}
\begin{center}
\includegraphics[width=0.8\textwidth]{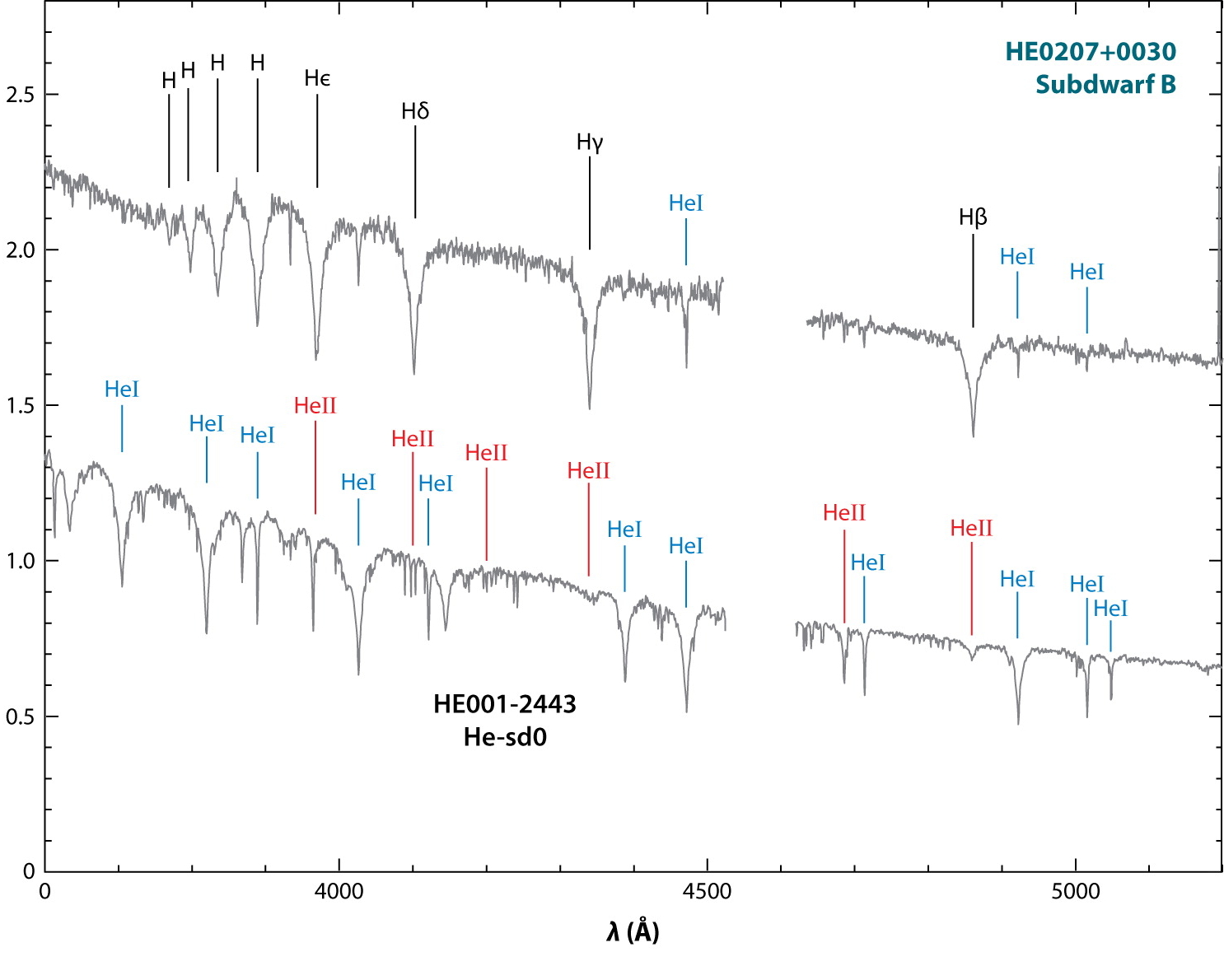}
\end{center}
\caption{Spectra of typical sdB and He-sdO stars \citepads{2008ASPC..392..139N}.  Important
lines of hydrogen and helium are indicated. Helium lines are weak in sdB stars.
Balmer lines are absent in He-sdOs, but note that every hydrogen Balmer line is
blended with a He {\sc ii} Pickering line. From \citetads{2009ARA&A..47..211H}, copyright ARAA, reproduced with permission.}
\label{fig:spectrum_sd}
\end{figure}

A less detailed classification has become common practice, though. According to their helium line spectra hot subdwarf stars are grouped into sdB, sdOB, sdO, He-sdB and He-sdO. This scheme is not well defined. The classes sdOB and sdB are often merged into sdB, although sdOB stars are characterized by the presence of He {\sc ii} 4686\AA, which becomes strong in sdO stars (see Fig. \ref{fig:spectrum_sd}). An extension of this nomenclature is proposed by \citetads{2010IBVS.5927....1K} 
for pulsating hot subdwarf stars.

About 10\% of the hot subluminous stars have helium-dominated spectra \citepads{1986ApJS...61..305G} 
and come in two flavors, the He-sdBs and the He-sdOs.
The term He-sdB is inconsistent as the He {\sc ii} 4686\AA\ line is weakly present, but He {\sc i} lines dominate. Nevertheless the term is useful to distinguish them from the He-sdO stars, in  which He {\sc ii} lines prevail.   
Spectral classification is based on medium-resolution ($\approx$ 2\AA) spectra and, therefore, of limited value to study the chemical composition of stars beyond the helium-to-hydrogen ratio. 

Hence quantitative spectral analyses are necessary to provide us with detailed information on atmospheric parameters (
effective temperature, gravity, and helium abundances, see Sects. \ref{sect:atmos_param} and \ref{sect:abu_helium}).
The chemical composition of sdB and sdO stars are reviewed in 
in Sect. \ref{sect:abu_metal} and \ref{sect:sdo}.

It is a general consensus that the helium-deficiency of the atmospheres of most sdB stars as well as of many (mostly helium poor) sdO stars is caused by diffusion processes as first proposed by \citetads{1967Natur.213..871G}. 
Therefore, we briefly revisit the diffusion theory in Sect. \ref{sect:diffusion}. 
In addition, the surface rotation can be studied from high-resolution spectra (Sect. \ref{sect:rotation}), while
spectropolarimetry or Zeeman splittings of spectral lines provide information on the magnetic fields of hot subdwarf stars
(Sect. \ref{sect:mag}).  

\subsection{Model atmospheres}

The first attempt to model the spectrum of a hot subdwarf (an He-sdO) star was carried out by \citetads{1958ApJ...127..642M}. 
Pioneering work by \citetads{1965ApJS....9..321M} 
and the Uns\"old school 
followed thereafter by \citetads[][three sdO stars]{1970ApJ...162..239T} 
and \citetads{1971A&A....14..415R} for the sdO \object{HD~49798}. 
Models for sdB stars were constructed by \citetads{1966ApJ...145..652S}, and 
\citetads{1970ApJS...19..327B} demonstrating that sdB stars are helium deficient. 

For the O-type subdwarfs is became clear that departures from local thermodynamic equilibrium (LTE) are significant \citepads{1979LIACo..22..295K}. 
The complete linearisation method pioneered by \citetads{1969ApJ...158..641A}
marked a milestone and allowed \citetads{1976A&A....52...11K}, \citetads{1978A&A....70..653K}
and \citetads{1981PhDT.......113G} to carry out the first non-LTE analyses of sdO stars.
A breakthrough was achieved by the invention of the accelerated lambda iteration technique using approximate lambda operators  \citepads{1985A&A...148..417W,1986A&A...161..177W} 
which allowed more detailed model atoms to be implemented in computations of non-LTE model atmospheres.  
This technique was applied to He-sdO star e.g. by 
 \citetads{1990A&A...235..234D} 
and \citetads{1993A&A...273..212D}. 

\subsection{Atmospheric parameters} \label{sect:atmos_param}

Thanks to the efforts of several research teams,
atmospheric parameters have become available for many hundreds of  hot subdwarfs, %
which allow the distribution of stars in the T$_{\rm eff}$ - $\log$ g - parameter space to be investigated. It has to be taken into account that results may differ systematically because of different model atmospheres used (LTE vs. NLTE; 
treatment of line blanketing, see \citetads{2012ASPC..452...41J} for a detailed discussion), 
different observational material (spectral resolution, wavelength coverage, normalization procedure
\footnote{\citetads{2012MNRAS.427.2180N}
extensively discuss the effect arising from different instrumentation and spectral resolution.}) and analysis strategy (choice of spectral lines, fitting procedure). Hence it would be  timely to compare 
the results of different published analyses. Few such attempts have been made
\citepads[e.g.][]{2014ApJ...788...65L}, though. 

\subsubsection{Results from major surveys}\label{sect:results_maforjor}

We shall consider five major projects which studied samples of hot subdwarf stars recently. The only study that makes use of high-resolution spectra (spectral resolution better than 0.4\AA) is the sample drawn from the ESO/SPY-survey \citepads{2001AN....322..411N}. We regard this sample as a benchmark because the spectra also allowed the abundances of many chemical elements to be determined (see Sect: \ref{sect:abu_metal}). Furthermore, the results of three studies at intermediate (1--5\AA) and one at low resolution ($\approx$ 8.7\AA) will be discussed and compared. Altogether, effective temperatures, surface gravities and helium abundances  
have become available for many hundreds of hot subdwarf stars in the field. 
The largest project is the  Arizona-Montr\'eal Spectroscopic Program \citepads{2008ASPC..392...75G,2014ASPC..481...83F}.
Because the stars are drawn from different surveys they might represent different stellar populations. The {\it GALEX} sample and those drawn mostly from the Palomar Green \citepads[PG,][]{1986ApJS...61..305G} 
and Edinburgh-Cape  \citepads[EC,][]{1997MNRAS.287..848S} 
surveys
have relatively bright average magnitudes, whereas the ESO/SPY sample includes fainter and, therefore, on average more distant {{stars. The}} Hamburg quasar survey extends {{the sample of hot subdwarfs deeper by another magnitude.}} The SDSS survey will extend the distribution in depth reaching out into the halo.    

\paragraph{The ESO/SPY survey -- the benchmark sample} 
The ESO Supernova Ia Progenitor Survey \citepads{2001AN....322..411N} obtained high-resolution optical spectra of more than 1000 white dwarf candidates, containing some 140 previously misclassified hot subdwarfs of various types.
The targets were largely drawn from the Hamburg-ESO survey \citepads{1996A&AS..115..227W} 
which was a wide-angle survey for bright quasars (12.5$<$B$<$~17.5 mag) in the southern hemisphere, based on objective prism plates taken with the ESO Schmidt telescope over an effective area of $\approx 1000$ square degrees. 
Observations were obtained at the ESO Very Large Telescope with the UV-Visual Echelle Spectrograph (UVES) at a spectral resolution of 0.36~\AA\ or better. Wavelength coverage of 3300--6650~\AA\ was achieved, with gaps at 4500--4600~\AA\ and 5600--5700~\AA.

The sample of sdB stars was analyzed by
\citetads{2005A&A...430..223L} 
using metal-line blanketed LTE models of solar composition  and the LINFOR program for spectrum synthesis \citepads{1999A&A...348L..25H,2000A&A...363..198H}. 
The spectra of the subluminous O stars from that sample were originally analyzed by  \citetads{2007A&A...462..269S} 
using NLTE models of hydrogen/helium composition calculated with the TMAP 
package \citepads{1999JCoAM.109...65W,2003ASPC..288...31W,2003ASPC..288..103R}\footnote{\url{http://www.uni-tuebingen.de/de/41621}}. A more sophisticated analysis of the same data using NLTE models of H/He/C or H/He/N composition also calculated with the TMAP package was carried out by \citet{phd_hirsch2009},
see also \citetads{2010AIPC.1314...79H} for a brief account.

\paragraph{The Arizona-Montr\'eal Spectroscopic Program}  The largest sample of hot subdwarfs presently under investigation is the Arizona-Montr\'eal Spectroscopic Program \citepads{2014ASPC..481...83F}, which targets mostly bright stars including many from the Palomar-Green survey. It is based on low-resolution spectra ($\approx$8.7\AA) covering the spectral range from the Balmer jump to 6800\AA, taken over more than a decade with the 
Bok 2.3m telescope of the Steward Observatory Kitt Peak Station. The data set is characterized by its high S/N (typically above 200) and its homogeneity, surpassing any other sample.  \citetads{2014ASPC..481...83F} give a brief account of effective temperatures, gravities and helium abundances for 320 of the 421 stars that make up the sample (see Fig. \ref{fig:arizona}). The team uses TLUSTY and SYNSPEC  \citepads{1988CoPhC..52..103H,1995ApJ...439..875H} to construct grids of non-LTE model atmospheres and to synthesize optical spectra \citepads{2010AIPC.1273..259B}, which are then matched to the observed spectra to extract effective temperatures, surface gravities and helium abundances.  

\paragraph{The Palomar-Green \& Edinburgh-Cape Surveys} 
This study was restricted to PG and EC stars classified as B-type subdwarfs; O-types were excluded. 
Observations of the northern targets were obtained with
the IDS at the 2.5-m Isaac Newton
Telescope on La Palma at a resolution of $\approx$ 1.5\AA.  
Southern targets were observed with
the grating spectrograph mounted on the 1.9-m
telescope at the South African Astronomical
Observatory  (SAAO) at a resolution of better than 1 \AA.
This sample of sdB stars has been analyzed 
using the same metal-line blanketed LTE models used for the samples from the ESO/SPY and HQS surveys \citepads{2001MNRAS.326.1391M,2003MNRAS.338..752M,2011MNRAS.415.1381C}. The sample comprises about 150 
subluminous B stars. 

\paragraph{The Hamburg Quasar Survey -- the deepest sample} 
The Hamburg Quasar Survey (HQS), an objective prism survey (spectral resolution of 45\,$\AA$ FWHM at H$_\gamma$ covering the magnitude range 13 $.\!\!^{\rm m}$5 $\le B \le$ 18 $.\!\!^{\rm m}$5) was carried out, starting in 1980, with the 80 cm Schmidt telescope at the German-Spanish Astronomical Center (DSAZ) on Calar Alto, Spain \citepads{1995A&AS..111..195H}. 
Although it aimed primarily at finding quasars, it is also a very rich source of faint blue stars. Spectroscopic follow-up of visually selected candidates of hot stars at the Calar Alto observatory mostly with the TWIN Spectrograph at the 3.5m telescope resulted in a sample of 400 faint blue stars \citepads{1991ASIC..336..109H}. Half of them turned out to be hot white dwarfs \citepads[e.g.][]{1995A&A...303L..53D,1996A&A...311L..17H,1998A&A...338..563H} 
and 
PG~1159 stars \citepads[e.g.][]{1994A&A...286..463D,1996A&A...309..820D}, 
whereas the other half were classified as hot subdwarfs of B- and O-type. A sample of 107 sdB stars has been analyzed by \citetads{2003A&A...400..939E} 
using the same metal line-blanketed LTE atmospheres as in \citetads{2000A&A...363..198H} 
and 
another 58 O-type subdwarfs  were analyzed using TMAP non-LTE model atmospheres by \citetads{1997fbs..conf..375L} 
and \citet{diploma_stroeer2004}. 

\paragraph{The {\it GALEX} sample - the brightest one} 

The Galaxy Evolution Explorer ({\it GALEX}) all-sky survey provides ultraviolet photometry in two bands, the FUV ($\approx$154\,nm)  and NUV ($\approx$ 232\,nm), as described by \citetads{2007ApJS..173..682M}.
Because the hot subdwarfs are the dominant population of UV-bright sources at high Galactic latitudes they should be easy to find in large numbers in the {\it GALEX} data base, provided the white dwarf contamination can be eliminated, e.g. by making use of reduced proper motions. Most previous surveys were limited to high Galactic latitudes, whereas the {\it GALEX} survey covers all Galactic latitudes, making it an important tool to study the hot subdwarf population in or near the Galactic plane, a region yet poorly explored with respect to hot subdwarf stars and white dwarfs \citepads[but see ][]{2013MNRAS.434.2727V,2014MNRAS.438....2V}. 
   
\citetads{2011MNRAS.410.2095V} 
identified 280 hot subluminous stars amongst the bright {\it GALEX} sources (NUV$<$14 mag) based on UV, optical, and IR color indices, with about 120 hot subdwarfs previously uncovered. Even amongst the brightest stars (V $<$ 12 mag) two were uncatalogued and nine previously unstudied.   
\citetads{2012MNRAS.427.2180N} 
presented a homogeneously modelled sample of 124 sdB and 42 sdO stars from the {\it GALEX} survey and
determined non-LTE atmospheric parameters 
by considering H, He and CNO opacities in their computations using TLUSTY/SYNSPEC \citepads{1995ApJ...439..875H}.

\subsubsection{Effective temperatures and surface gravities from major samples in comparison}

Besides systematic differences in the quantitative spectral analyses, the samples suffer from different selection biases. Therefore, it is worthwhile to compare the distribution of parameters derived by the different projects. 
In section 
\ref{sect:cluster} we shall also compare with the sample of hot subdwarfs in the globular cluster $\omega$\ Cen. 

\begin{figure*}
\begin{center}
\includegraphics[width=0.65\textwidth]{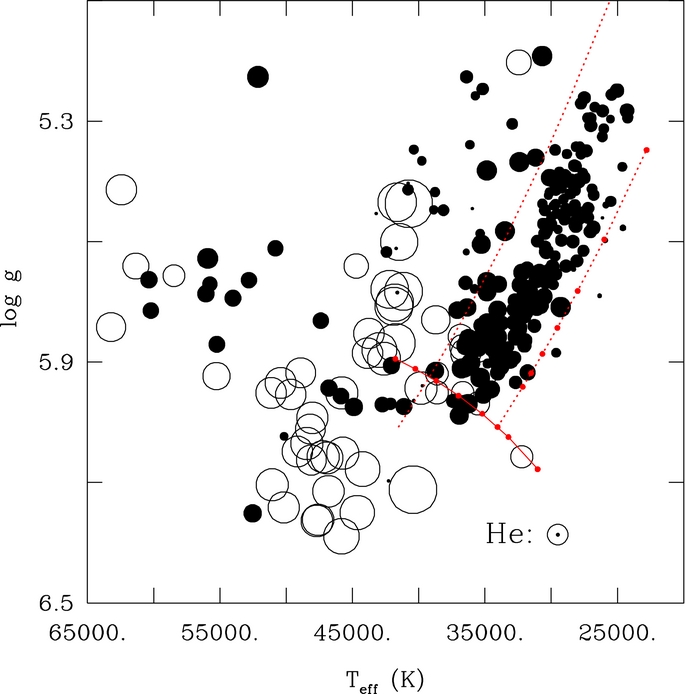}
\end{center}
\caption{The distribution of hot subdwarfs stars from the Arizona-Montr\'eal Spectroscopic Program \citepads{2008ASPC..392...75G,2014ASPC..481...83F} is
depicted in the log g -- T$_{\rm eff}$  plane. The size of a given circle is a logarithmic measure of the helium abundance relative to
that of hydrogen {{(the solar He abundance is indicated by the symbol in the lower right corner).}} He-poor and He-rich stars are represented by filled and open circles, respectively. Also shown are the zero-age helium main-sequence (ZAHEMS, full drawn), as well as zero-age extreme horizontal branch (ZAEHB) and the  terminal-age extreme horizontal branch (TAEHB, dotted lines) for a core mass of 0.47 M$_\odot$ and a metallicity of Z = 0.02. 
From \citetads{2014ApJ...795..106L}, copyright ApJ, reproduced  with permission. 
} 
\label{fig:arizona}
\end{figure*}

\begin{figure*}
\begin{center}
\includegraphics[width=0.65\textwidth]{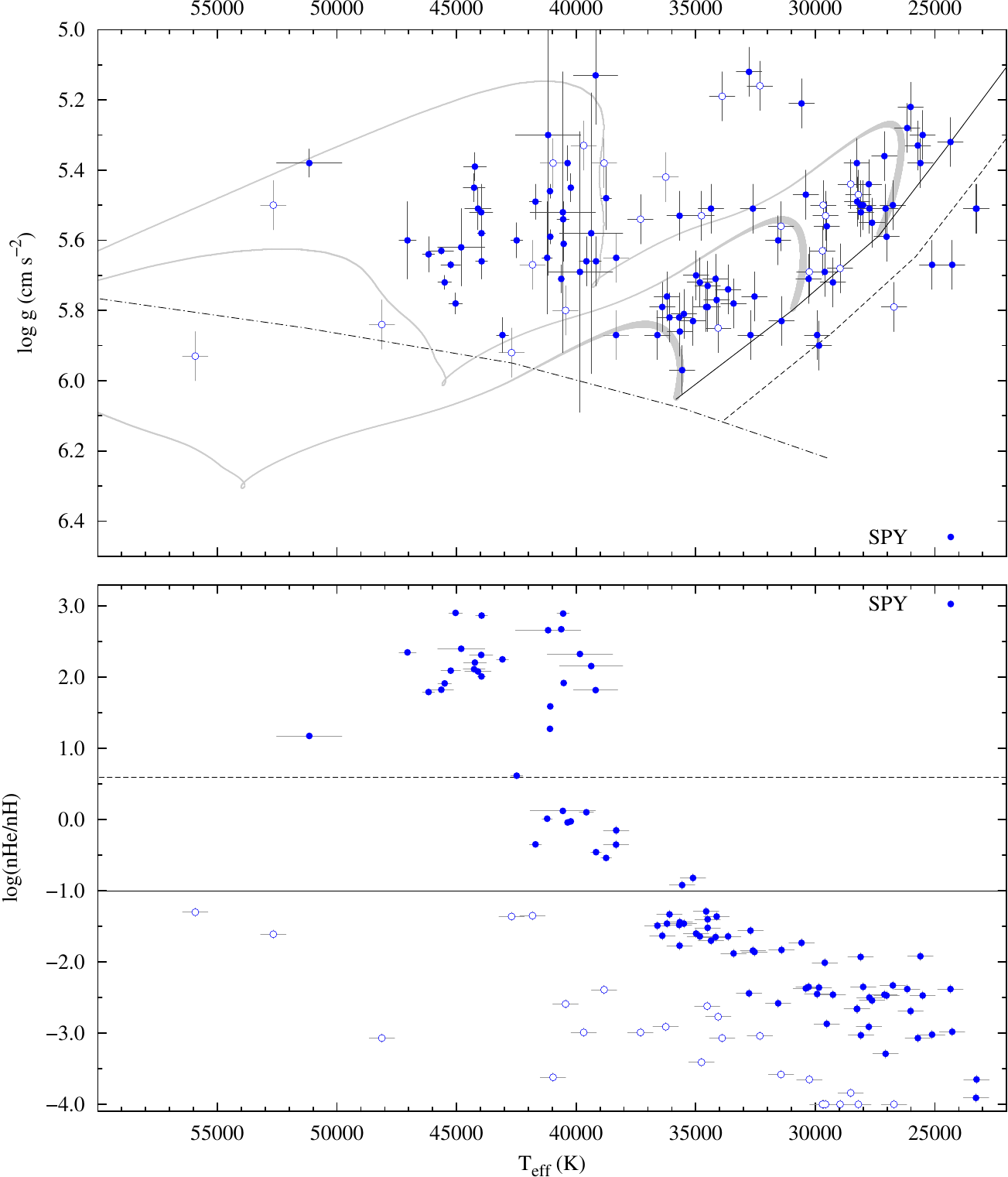}
\end{center}
\caption{
{\bf Upper panel:} 
Distribution of hot subdwarf stars drawn from the ESO/SPY project \citepads{2005A&A...430..223L,phd_hirsch2009,2010AIPC.1314...79H} in the T$_{\rm eff}$ - $\log g$- plane. The location of the ZAEHB is shown for two masses, (0.45 M$_\odot$, dotted, and 0.5 M$_\odot$, full drawn). Evolutionary tracks for three envelope masses \citepads[0.000 M$_\odot$, 0.001 M$_\odot$, and 0.005 M$_\odot$, from left to right by][]{2002MNRAS.336..449H} 
 are shown as dark grey lines.
Linewidths are proportional to evolutionary time-scales. 
  The helium main sequence is shown as a dashed-dotted line.
{\bf Lower panel:} distribution of the helium abundances vs. T$_{\rm eff}$. The solar helium abundance is shown as a dotted horizontal line.
From N{\'e}meth (priv. comm.).
} 
\label{fig:spy}
\end{figure*}

\begin{figure*}
\begin{center}
\includegraphics[width=0.65\textwidth]{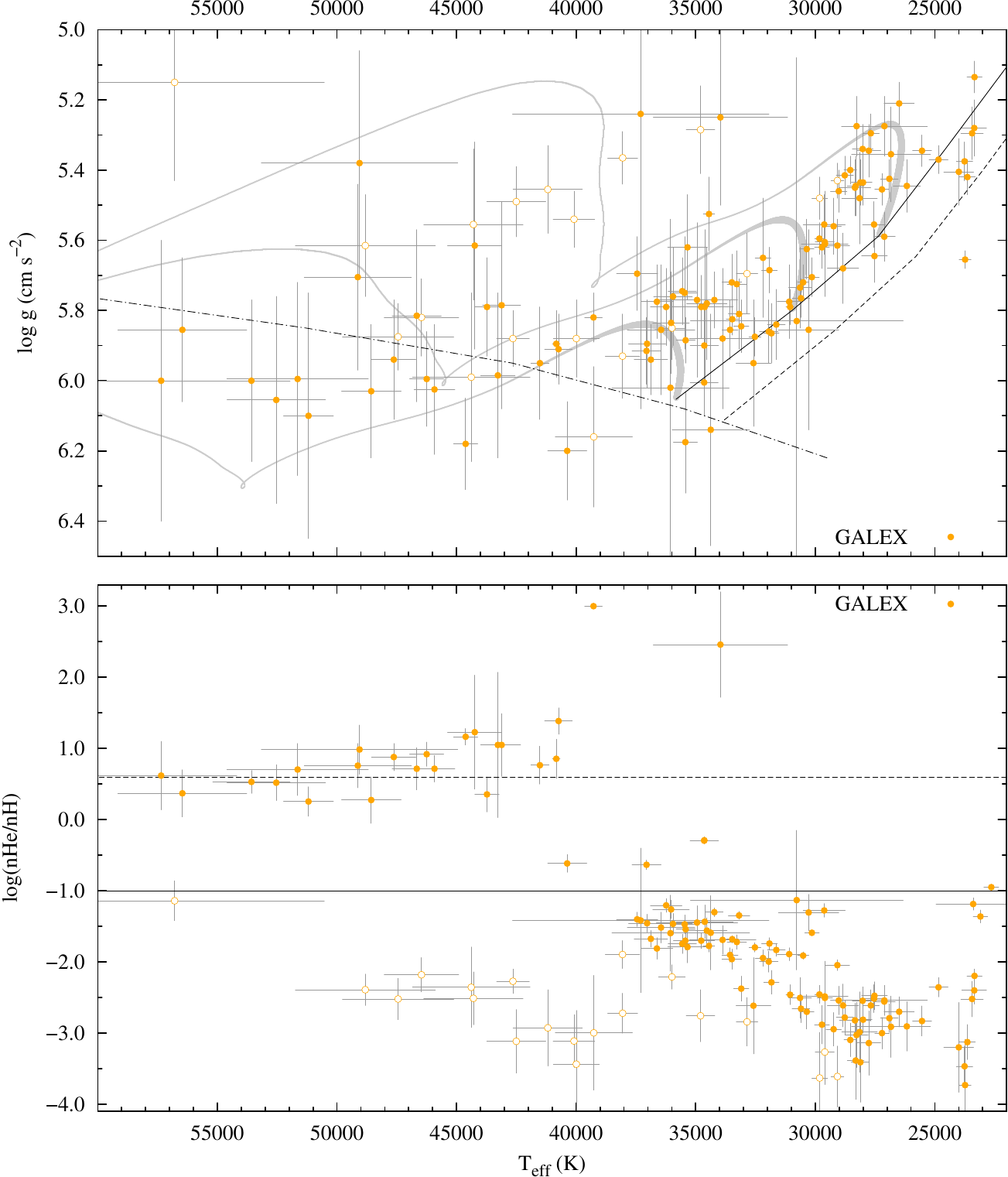}
\end{center}
\caption{
Same as Fig. \ref{fig:spy} but for the {\it GALEX} sample \citepads{2012MNRAS.427.2180N}.
From N{\'e}meth (priv. comm.).
} 
\label{fig:galex}
\end{figure*}

\begin{figure*}
\begin{center}
\includegraphics[width=0.65\textwidth]{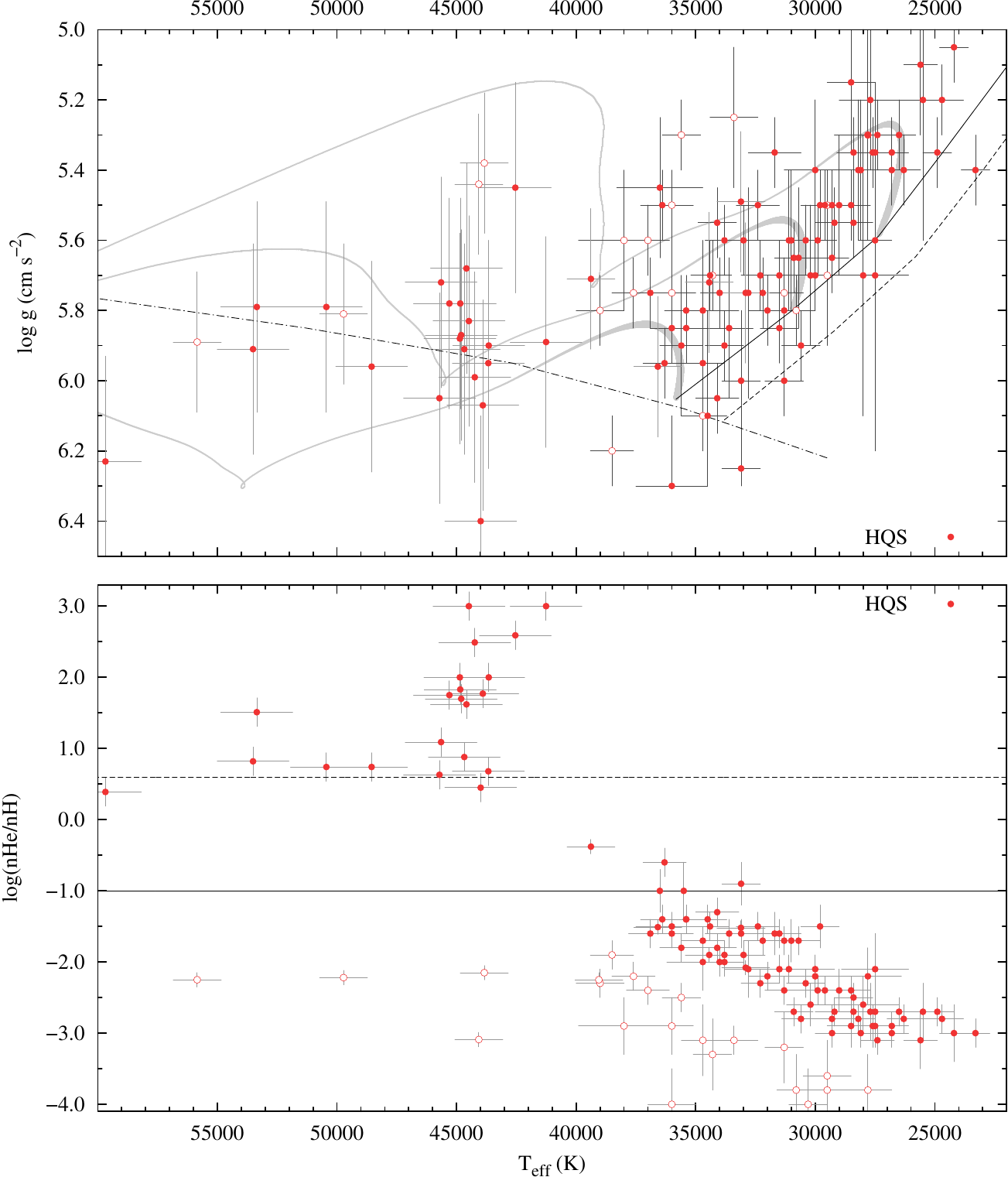}
\end{center}
\caption{
Same as Fig. \ref{fig:spy} but for the hot subdwarf sample drawn from the HQS project.
From N{\'e}meth (priv. comm.).
} 
\label{fig:hqs}
\end{figure*}

\begin{figure*}
\begin{center}
\includegraphics[width=0.65\textwidth]{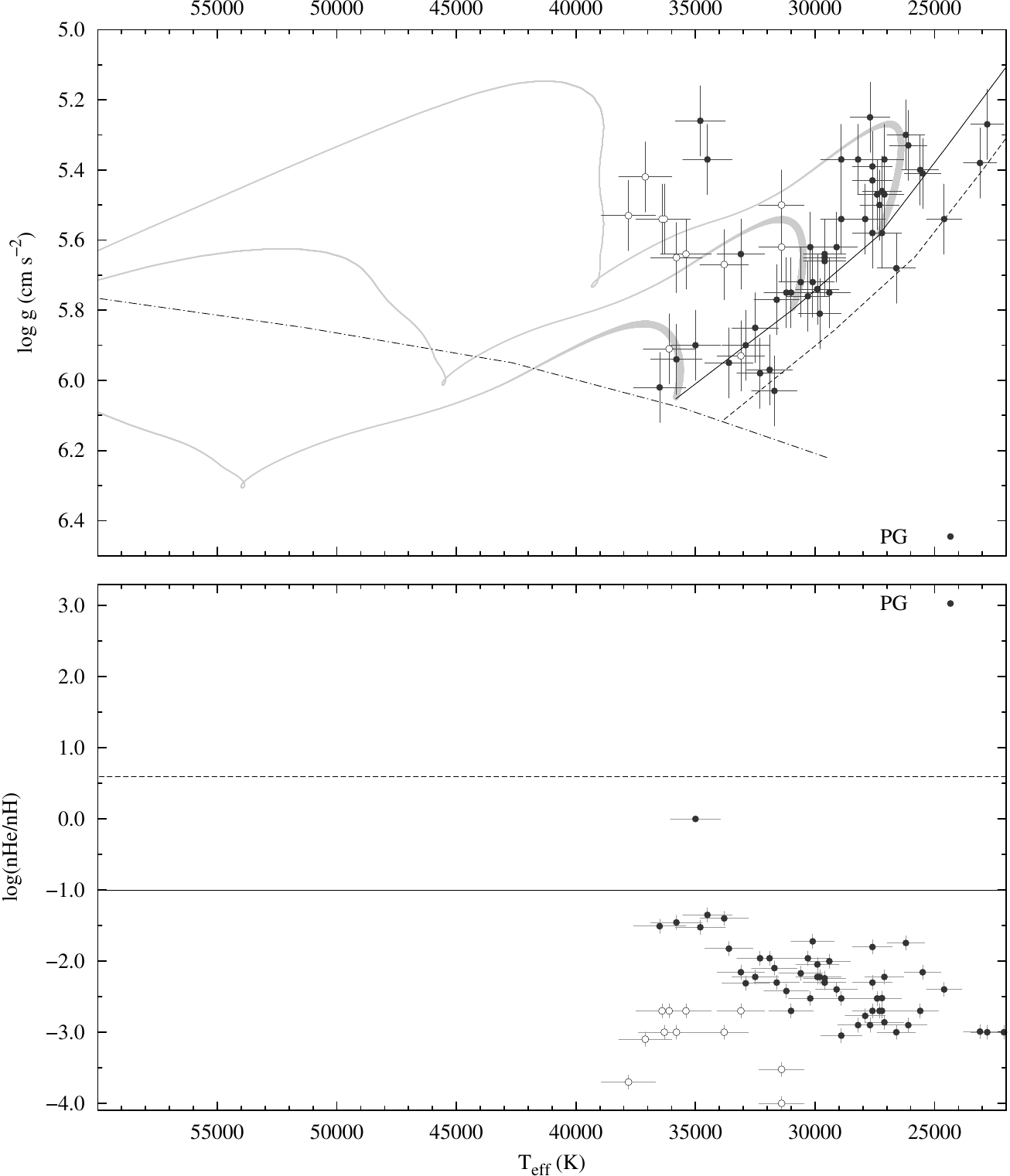}
\end{center}
\caption{
Same as Fig. \ref{fig:spy} but for the hot subdwarf sample drawn from the PG\&EC sample.
From N{\'e}meth (priv. comm.).
} 
\label{fig:pg}
\end{figure*}

The T$_{\rm eff}$ - $\log$ g - diagram  resulting from the Arizona-Montr\'eal Spectroscopic Program is shown in Fig. \ref{fig:arizona}, while Figs. \ref{fig:spy} to \ref{fig:pg} display it for the samples drawn from the ESO/SPY, {\it GALEX}, HQS and PG\&EC samples, respectively.

\begin{figure}
\begin{center}
\includegraphics[width=0.65\textwidth]{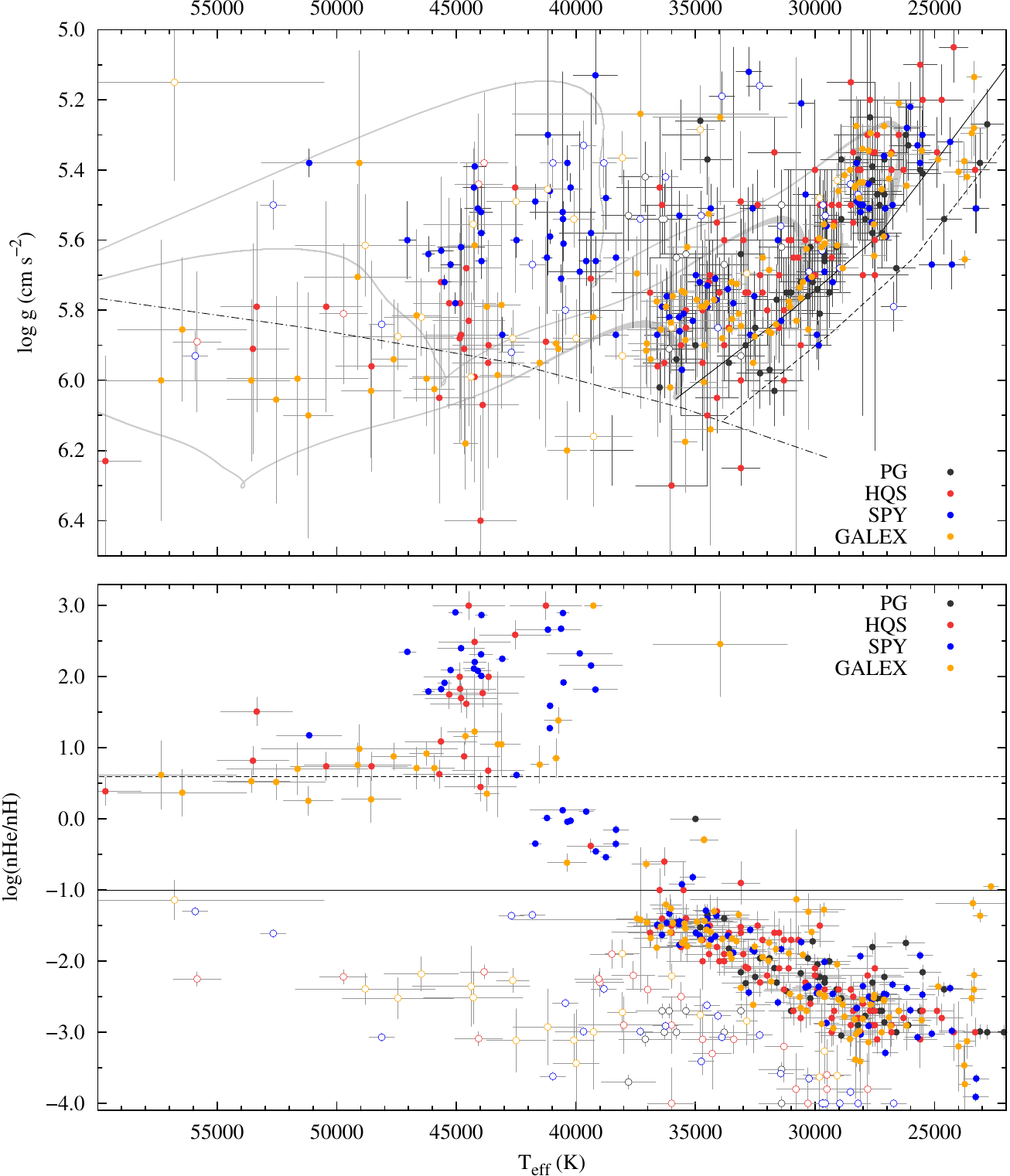}
\end{center}
\caption{
Same as Fig. \ref{fig:spy} but combining all samples shown in figures \ref{fig:spy} to \ref{fig:pg}.
From N{\'e}meth (priv. comm.).
} 
\label{fig:spy_hqs_galex_pg}
\end{figure}

\citetads{2012MNRAS.427.2180N} found that sdB stars concentrate in two groups in the $T_{\rm eff}-{\log g}$ and $T_{\rm eff}-{\rm He}$ diagrams 
suggesting two typical H envelopes with different masses and compositions.
However, this clustering is not seen in the full set of sdB stars (see Fig. \ref{fig:spy_hqs_galex_pg}). 
There is  a clear gap between He-sdO and He-weak sdO stars in the $T_{\rm eff}-{\rm He}$ diagram and a clustering of He-sdO stars in the temperature range from $\approx$40000K to $\approx$47000K (see Figs. \ref{fig:arizona} to \ref{fig:spy_hqs_galex_pg}). 

The most intriguing difference between the various T$_{\rm eff}$, $\log$ g - diagrams is the distribution of the He-sdO stars. The Arizona-Montr\'eal Spectroscopic Program finds a large population of high-gravity He-sdO stars
with $\log$ g from 6.0 to 6.4;  that is, they seem to lie below the helium main sequence. Their number even exceeds the number of He-sdOs of lower gravity ($\log$ g $<$6), which are on or above the He main sequence. A few such high-gravity He-sdOs are also found in the {\it GALEX} sample. However, their uncertainties are such that most of them are still consistent with an He-main sequence nature. The  T$_{\rm eff}$, $\log$ g - diagram of the ESO/SPY sample
does not show He-sdOs located at gravities larger than 6.0. However, it is worth noting that   
\citetads{2007A&A...462..269S} 
in their early analysis of the ESO/SPY sdO sample found several sdO stars having gravities below the helium main sequence \citepads[see Fig. 6 of ][]{2007A&A...462..269S}. 
The reanalysis using NLTE models that include carbon or nitrogen line blanketing by \citet{phd_hirsch2009} resulted in atmospheric parameters that  differ significantly from those of \citetads{2007A&A...462..269S}, who used NLTE models that were composed of hydrogen and helium only. In particular, the gravities turned out, on average, to be lower than the ones previously derived. A very steep correlation between surface gravity and effective temperature has to be acknowledged leading to large systematic uncertainties for the He-sdOs\footnote{A significant source of systematic uncertainty comes from helium line broadening theory. Unfortunately sophisticated line broadening tables are not available, in particular for many He {\sc i} transitions. The tables of \citetads{1997ApJS..108..559B} 
for white dwarfs need to be adapted to the physical conditions in the atmospheres of hot subdwarf stars.}. The reason for these difference between the samples is still obscure. 

Another riddle is, that the rather large spread in gravities amongst the He-sdO stars can hardly be explained by evolutionary models (neither in the hot flasher nor the merger scenario, see Sect. \ref{sect:evolution}).

\subsection{Helium abundances}\label{sect:abu_helium}

Besides the effective temperature and surface gravity, the helium abundance is an atmospheric parameter that has an important impact on the temperature-density stratification of the atmosphere, in particular for helium rich stars. 

\begin{figure}
\centerline{\includegraphics*[width=0.9\textwidth]{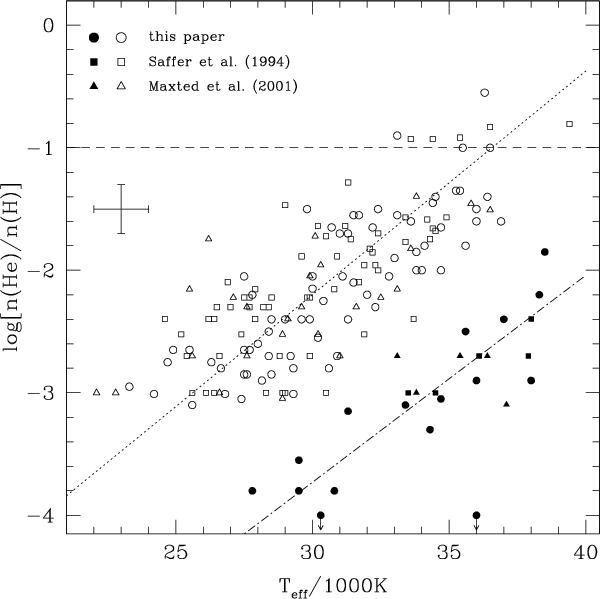}}
\caption{Plot of the helium abundance versus effective temperature from \citetads{2003A&A...400..939E}, \citetads{1994ApJ...432..351S}, and \citetads{2001MNRAS.326.1391M}  
 The dotted line indicates the linear regression for the bulk of the sdB stars (open symbols) and the dashed-dotted line shows the linear regression for the minority sdB stars (filled symbols). The dashed horizontal line denotes the solar helium abundance. From \citetads{2003A&A...400..939E}, copyright A\&A; reproduced with permission.}
\label{fig:he_teff}
\end{figure}

SdB star helium abundances tend to increase with increasing effective temperature, as first reported by \citetads{2003A&A...400..939E}
for the first time. Moreover two clearly defined sequences stand out  - a minority of stars follows a trend at a significantly lower helium abundance than the majority of stars (see Fig. \ref{fig:he_teff}). These trends appear in all samples as can be seen from Fig. 2 of \citetads{2014ASPC..481...83F} 
and Figs. \ref{fig:spy} to \ref{fig:pg}. Some He-sdOs from the ESO/SPY sample have very high helium abundances (up to n$_{\rm He}$/n$_{\rm H}$ =1000) much larger than any star in the {\it GALEX} sample, for instance. However, this may be due to the high spectral resolution  of the ESO/SPY (UVES) data, which allows small distortions of the He~{\sc ii} Pickering lines by very weak hydrogen Balmer lines to be detected, which cannot be detected at the lower spectral resolution of the {\it GALEX} data. 
Hence some of the {\it GALEX} helium abundances may have to be considered as lower limits only.

The existence of these trends in helium abundance has been confirmed by recent studies but still await an explanation \citepads[see][for detailed discussions of this issue]
{2008ASPC..392...67O,2014ASPC..481...83F}. 

\subsubsection{The $^3$He isotopic anomaly}

Another spectroscopic anomaly in sdB stars, which was first reported by \citetads{1987MitAG..70...79H}, 
is related to the $^3$He isotope. The $^3$He isotope can easily be identified in high resolution spectra, because the isotopic shifts with respect to the $^4$He isotope vary from line to line. While some lines like $^3$He~{\sc i}, 5876\AA\ are shifted only slightly (0.04 \AA) towards redder wavelengths, the shifts of $^3$He~{\sc i}, 4922\AA\ and $^3$He~{\sc i}, 6678\AA\ are significant (0.33 and 0.50 \AA, respectively). The observed helium lines are blends of lines arising from $^3$He and $^4$He isotopes. Hence the observed line shifts increase with increasing $^3$He/$^4$He ratio.  
In a few sdB stars, $^3$He was found to be strongly enriched and has almost completely replaced the $^4$He isotope, e.g. for SB~290 \citepads{1987MitAG..70...79H,2013A&A...557A.122G}. It is generally believed that diffusion is causing this anomaly. By now eight sdB stars are known to show this anomaly, which cluster on the EHB in a narrow temperature regime from
T$_{\rm eff}$ = 27,000~K to 31,000~K  \citepads{2013A&A...557A.122G}. 
The $^3$He isotopic anomaly has also been observed amongst blue horizontal branch stars \citepads{1979ApJ...231..161H} 
as well as chemically peculiar B-type main sequence stars \citepads{1961ApJ...134..777S,2014A&A...572A.112M}, although at lower temperatures. 

\clearpage
\subsection{The chemical composition of sdB stars} \label{sect:abu_metal} 


Concerning elements heavier than helium, information from optical spectra of sdB stars is quite comprehensive for carbon, nitrogen, magnesium, silicon, sulfur, and to a lesser extent for iron, because those elements show prominent lines in the optical. 
Ultraviolet spectra gave access to the iron group and trans-iron elements, which
have been detected in optical spectra in exceptional cases only. 
Since the spectra of hydrogen-rich subdwarfs differ considerably from those of hydrogen-poor ones, we shall discuss them separately starting with the hydrogen-rich ones.
 

The vast majority of subluminous B stars are hydrogen-rich and populate the 
EHB from $\approx$ 20000K to the helium main-sequence. Their anomalous chemical composition is due to atmospheric diffusion.
A very small group (termed He-sdB) is helium-rich, with helium abundances ranging from solar to almost pure helium. {{\citetads{2013MNRAS.434.1920N} subdivided the helium rich sdB stars into extreme and intermediate He-sdBs drawing a line at (n(He)/n(H)$=$4).}}
The subclass of intermediate He-sdB stars (n(He)/n(H)$<$4) are of particular interest, because they are regarded as transition objects which might link the evolution of sdB stars to that of helium-poor sdO stars \citepads[for a review see ][]{2012ASPC..452...41J} and shall be discussed separately (Sect. \ref{sect:abu_intermediate}). 

\subsubsubsection{Optical Spectroscopy}
High-resolution optical spectra, {{mainly}} from the ESO/SPY project, have been used to determine the abundance pattern.
 Early analyses \citepads[e.g.][]{2000A&A...363..198H,2004Ap&SS.291..341H,2001A&A...378L..17N,1999ASPC..169..546E,2001AN....322..401E,2006BaltA..15..103E} 
 have now been extended to more than 100 stars  
and to elemental abundances of up to 24 different ions per star \citepads{2013A&A...549A.110G}.
This large progress became possible thanks to the excellent spectra from the ESO/SPY project that contributed two thirds of the sample \citepads{2008MmSAI..79..723G}. To do this in an efficient way \citetads{2013A&A...549A.110G} designed a semi-automatic analysis pipeline to fit synthetic spectra computed from LTE models to a standard set of spectral lines.

\begin{figure}
 \begin{center}
\includegraphics[width=0.8\textwidth]{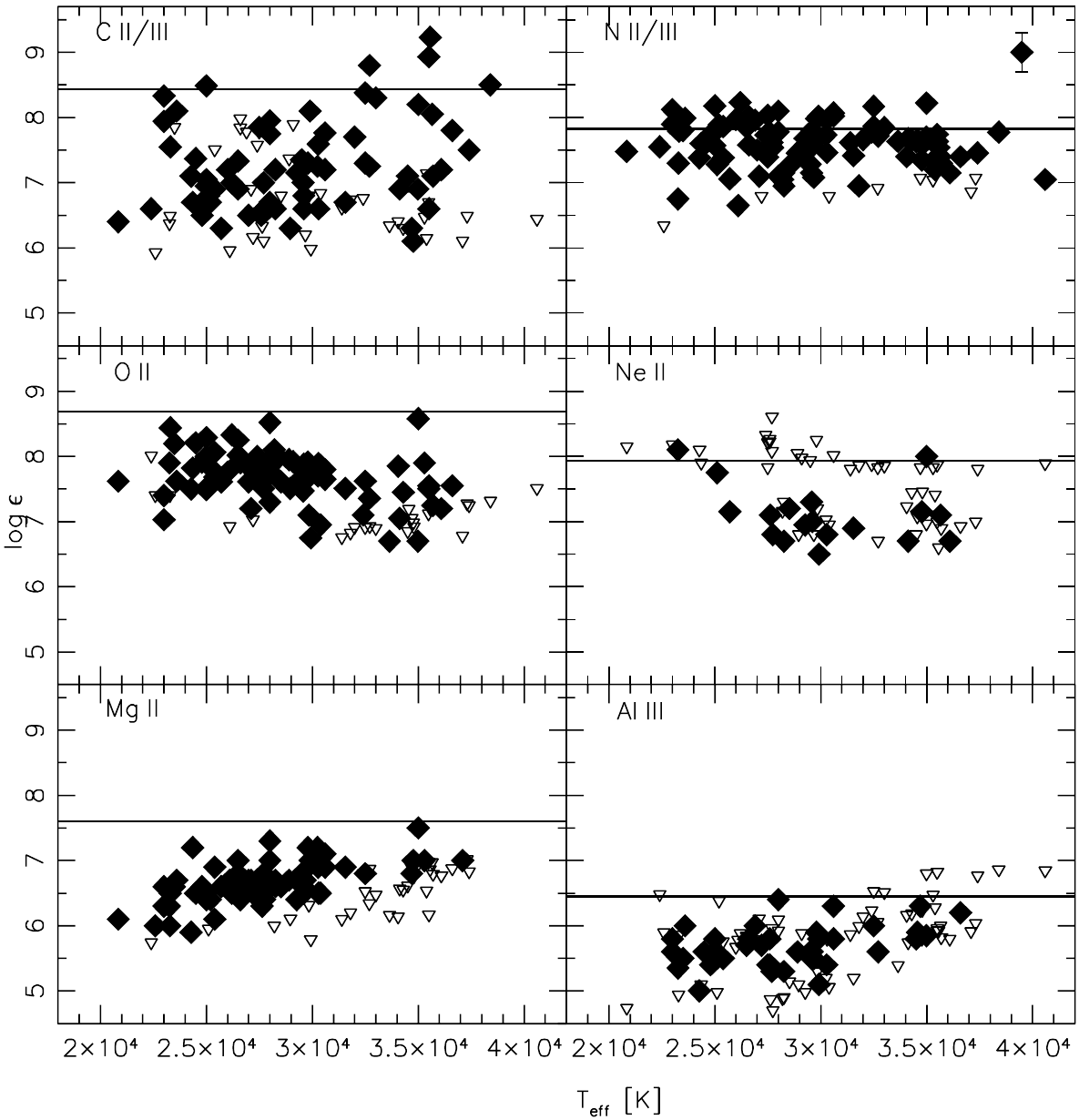}
\end{center}
\caption{a) Elemental abundances from carbon to aluminum plotted against effective temperature \citepads{2013A&A...549A.110G}. The filled diamonds mark measured abundances while the open triangles mark upper limits. Typical error bars are given in the upper right corner. The solid horizontal lines mark solar abundances \citepads{2009ARA&A..47..481A}. 
From \citetads{2013A&A...549A.110G}, copyright A\&A; reproduced with permission.
}\label{fig:geier_metal}
\end{figure}

\begin{figure}
 \begin{center}
\includegraphics[width=0.8\textwidth]{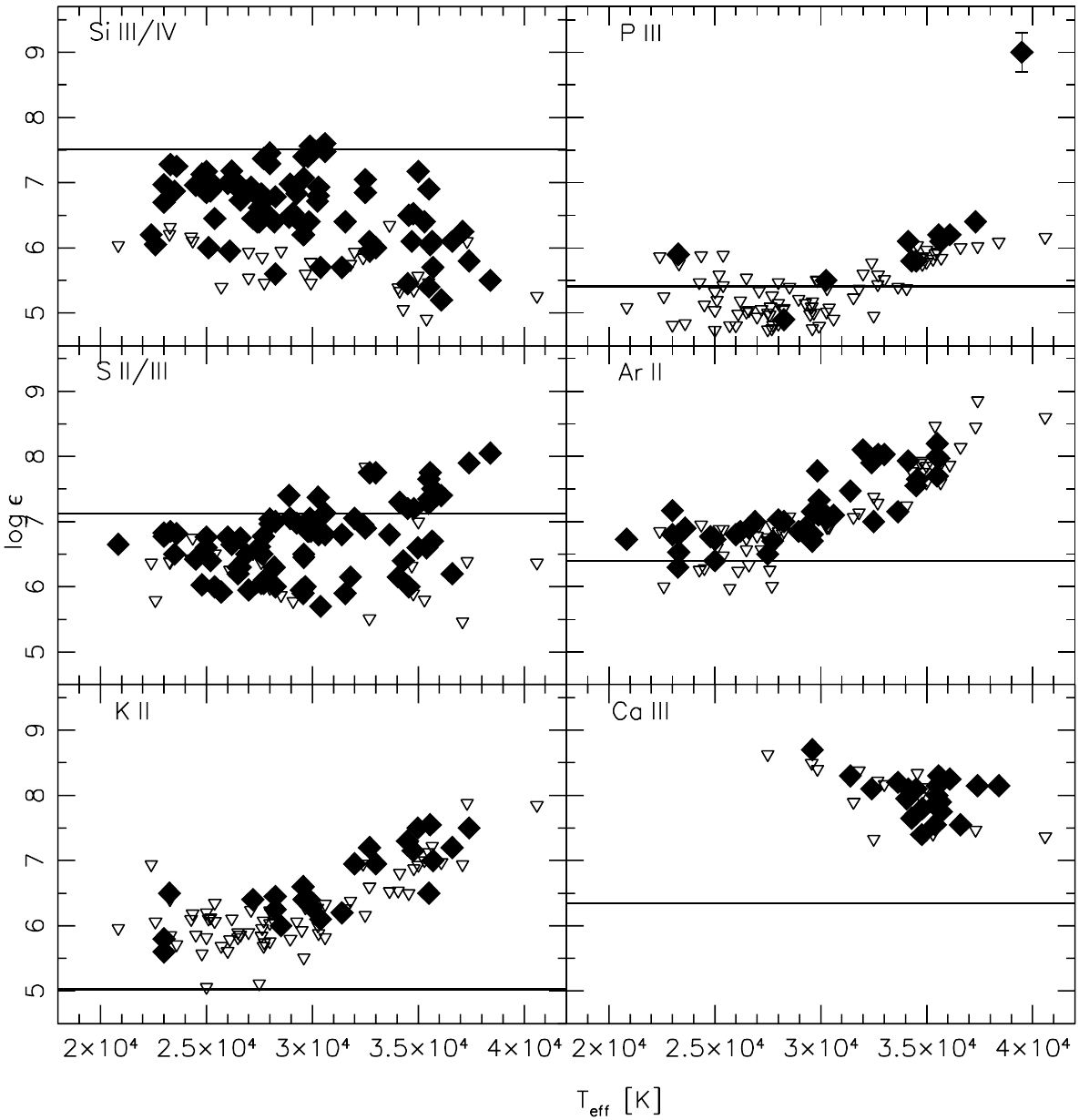}
\end{center}
\caption{b) Elemental abundances from silicon to calcium plotted against effective temperature \citepads{2013A&A...549A.110G}. The filled diamonds mark measured abundances while the open triangles mark upper limits. Typical error bars are given in the upper right corner. The solid horizontal lines mark solar abundances \citepads{2009ARA&A..47..481A}. 
From \citetads{2013A&A...549A.110G}, copyright A\&A; reproduced with permission.  }\label{fig:geier_metal1}
\end{figure}

\begin{figure}
 \begin{center}
\includegraphics[width=0.8\textwidth]{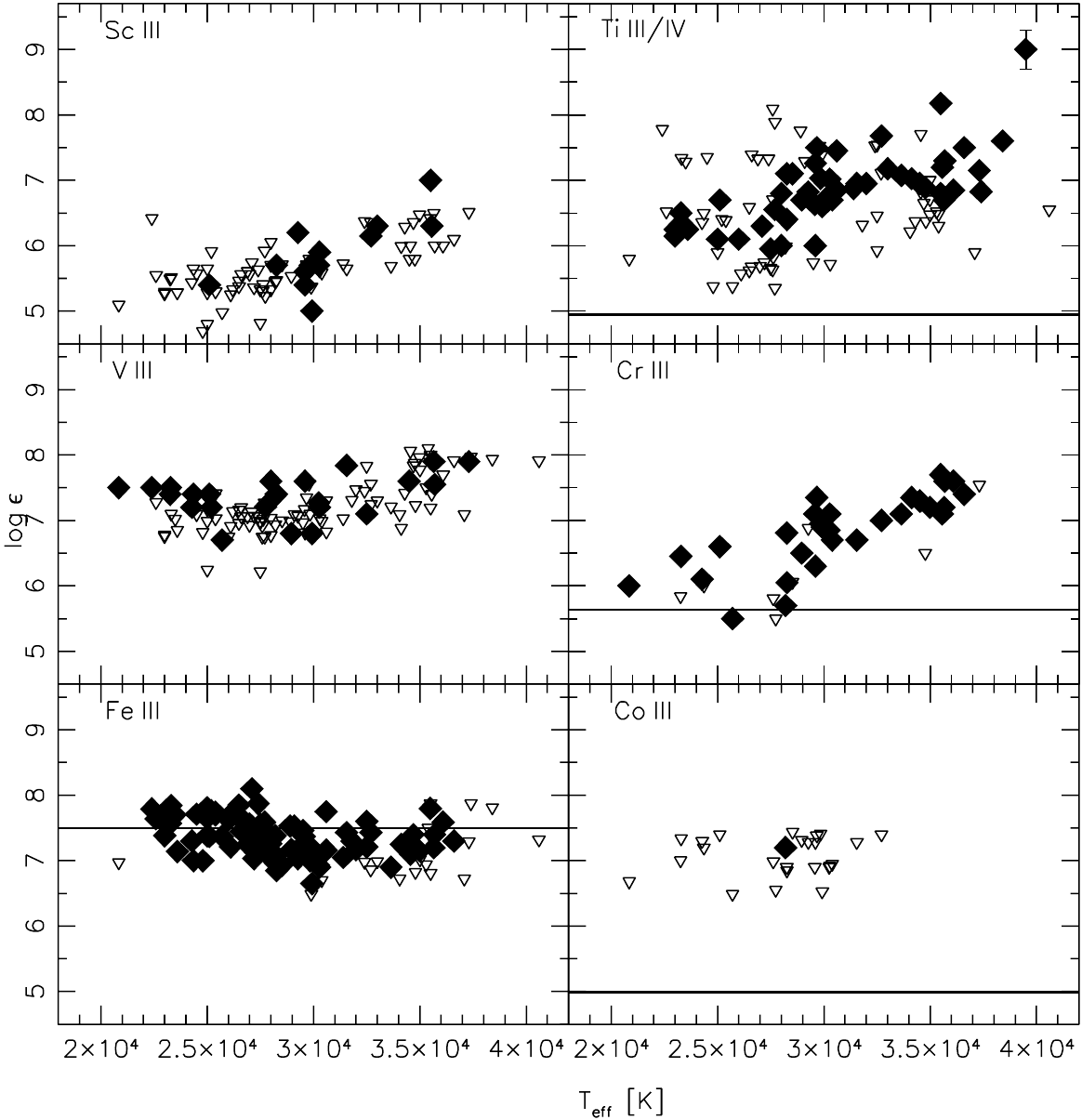}
\end{center}
\caption{c) Elemental abundances from scandium to cobalt plotted against effective temperature \citepads{2013A&A...549A.110G}.The filled diamonds mark measured abundances while the open triangles mark upper limits. Typical error bars are given in the upper right corner. The solid horizontal lines mark solar abundances \citepads{2009ARA&A..47..481A}. 
From \citetads{2013A&A...549A.110G}, copyright A\&A; reproduced with permission. } \label{fig:geier_metal2}
\end{figure}

Despite the large star-to-star variations, some similarities became obvious (see Figs. \ref{fig:geier_metal} to \ref{fig:geier_metal2}). Unlike for helium, no trends with effective temperatures can be found.\footnote{Note that in a few cases such as Ar {\sc ii} the trends appeared likely to be caused by NLTE effects unaccounted for.}
It has to be noted that for several elements in many stars only upper limits could be derived, which means that the star-to-star scatter for some elements might be larger than it appears in Figs. \ref{fig:geier_metal} to \ref{fig:geier_metal2}. 

{\it Carbon and nitrogen} show very different distributions
(see Fig. \ref{fig:geier_metal}) although radiative levitation is predicted to be very similar for C and N \citepads[as well as O, ][]{2001A&A...374..570U} irrespective of the assumed mass loss rates \citepads[see also ][]{2009AIPC.1135..148C}. While the nitrogen abundance is slightly subsolar throughout the entire temperature range, the abundance of carbon varies by orders of magnitudes from star to star ranging from strongly subsolar to slightly super-solar. {\it Oxygen} is depleted on average by 1 $\ldots$ 2~dex, i.e.  the scatter is less extreme than for C but not as small as for N.

\citetads{2012MNRAS.427.2180N} determined C, N, and O abundances for sdB and sdO stars and find a positive correlation of C and N with helium abundance, when combining both sdB and sdO stars \citepads[see Fig. 9 of][see also Sect. \ref{sect:cluster}]{2012MNRAS.427.2180N}. 
 
{\it Magnesium and iron} show little scatter (like nitrogen), Mg being subsolar by  one order of magnitude on average while Fe has almost solar abundance. It is conspicuous that the intermediate mass elements like Si, Al, and S are depleted, Si showing a particularly large star-to-star scatter. 
All the heavier elements (K to Co) are enriched with respect to solar.
{\it Titanium and Vanadium} are found in at least half (Ti) or one third (V) of the sample, both strongly enhanced by +2 dex and +3 dex, respectively. 
{\it Scandium and Chromium} are detected in a few stars, only, and strongly   enriched.    

\begin{figure}
\begin{center}
\includegraphics[width=0.6\textwidth]{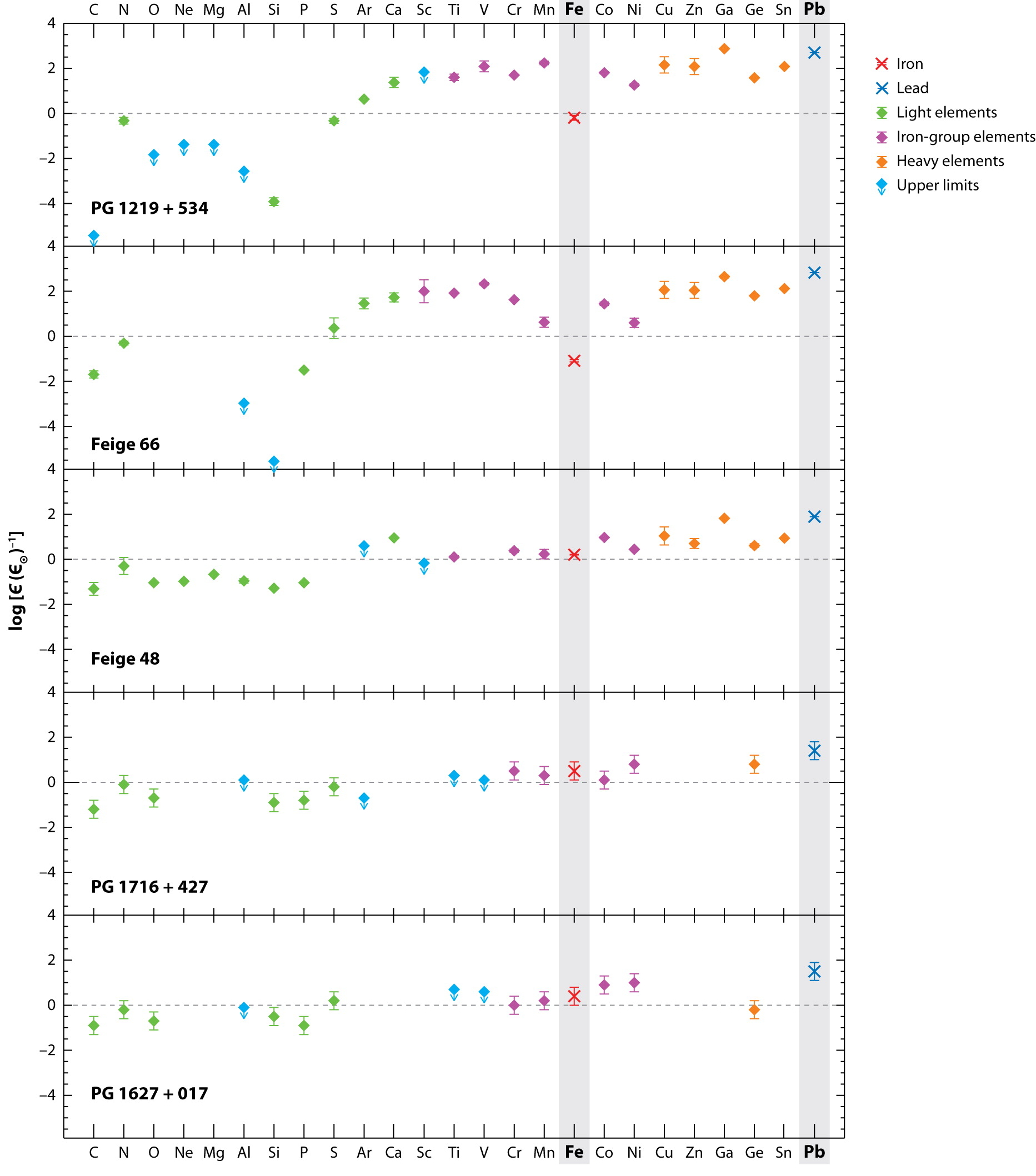} 
\caption{Abundances derived for three sdB stars  
from UV spectra measured with {\it HST}-STIS 
\citepads[top three panels, from][]{2006A&A...452..579O}
and derived from far-UV spectra
measured with {\it FUSE} \citepads{2008ApJ...678.1329B}. 
The stars are arranged with decreasing 
T$_{\rm eff}$ from top to bottom. 
PG~1219$+$534 (T$_{\rm eff}$=33,500~K) and Feige~48 (T$_{\rm eff}$=29,500~K) 
are rapid $p$-mode pulsators
of the V361~Hya type,
while PG1716$+$427 (T$_{\rm eff}$=27,400~K) and PG1627$+$017 (T$_{\rm eff}$=22,800~K) 
are slow $g$-mode pulsators of 
V1093~Her type, and Feige 66 is a constant star of T$_{\rm eff}$=34,000~K, very
similar to PG~1219$+$534. 
Light elements are marked by green symbols, the iron group elements in magenta, 
iron by red crosses, and lead by blue crosses. 
Upper limits are given by cyan symbols.
N is almost solar irrespective of the T$_{\rm eff}$ of the star.
Fe also changes little (around the solar value) whereas 
most other elements show large differences from star to star, while lead is
found to be strongly overabundant irrespective of T$_{\rm eff}$.
The overall enrichment of heavy elements in the hot stars ($\approx$2 dex) is
much larger than the mild one in the cooler stars. From \citetads{2009ARA&A..47..211H}; copyright ARAA; reproduced with permission.
}
\label{fig:uv_metals}
\end{center}
\end{figure}

\subsubsubsection{Ultraviolet spectroscopy}
Spectral analyses of FUV and UV spectra offer a plethora of spectral lines from the same ions as observed in the optical, but are much more sensitive to the elemental abundance, that is detection limits are at a much lower abundance level. In addition spectral lines from elements can be detected that are not represented in the optical.
However, spectral analyses of UV spectra are available for a few stars only. They  corroborate the findings (see Fig. \ref{fig:uv_metals}) from optical studies, extend them for elements that have upper limits from optical analyses (e.g. Cr), and add a lot of information on heavier elements (e.g. Ni), and trans-iron group elements such as Cu, Zn, Ga, Ge, Sn, Zn, and Pb
\citepads[see Fig. \ref{fig:uv_metals},][]{2006A&A...452..579O,2008ApJ...678.1329B,2006BaltA..15..131C}

In general, all elements beyond Ar are enriched with the notable exception of iron. The enhancements vary from star to star but increase with increasing effective temperatures to as much as a factor of ~1000 (see Fig. \ref{fig:uv_metals}). Whether a star is pulsating or not does not appear to matter; that is, no correlation of the stellar abundance pattern with pulsation type has been found \citepads{2006A&A...452..579O,2008ApJ...678.1329B}.

\citetads{2006BaltA..15..131C} analyzed far-UV spectra of 18 sdB stars taken with the {\it FUSE} satellite in the temperature range from 24,000~K to 34,000~K and derived abundances of germanium, zirconium, and lead, almost all of them being enriched by 0.5 to 3 dex relative to solar. It is worth noting that lead is present in large quantities in many sdB stars.
 
\subsubsubsection{Lead and lead isotopes.} The presence of lead in the UV spectra of many sdB stars, whether helium poor or helium rich, 
opens an interesting opportunity to study neutron-capture nucleosynthesis.  Lead is effectively the terminal product of the s-processing and therefore one of the most important elements in stellar nucleosynthesis modelling. Due to the lack of Pb spectral features in the cool metal-poor stars, it is hard to measure reliable abundances 
\citepads{2008ARA&A..46..241S}. 
The significant overabundances of lead in sdB atmospheres are due to diffusion, which acts like an amplifier and enriches otherwise non-detectable trace elements in the atmosphere, where their spectral features can then be analyzed.
A much better indicator for nucleosynthesis processes than the abundance of lead is the mixture of the lead isotopes. Stellar nucleosynthesis models predict different isotope mixtures depending on which of the neutron capture processes is dominant, as well as the metallicity of the source stars.
\citetads{2007ASPC..372..209O} 
have already been able to detect the individual lead isotope lines in very high resolution {\it HST}-STIS spectra thanks to the very slow rotation of the stars. In the two stars studied, the lead isotope ratio is consistent with the solar system value.

\subsubsection{Intermediate He-sdB stars: the transition population}\label{sect:abu_intermediate}

\begin{figure}
\begin{center}
\includegraphics[width=0.8\textwidth,angle=270]{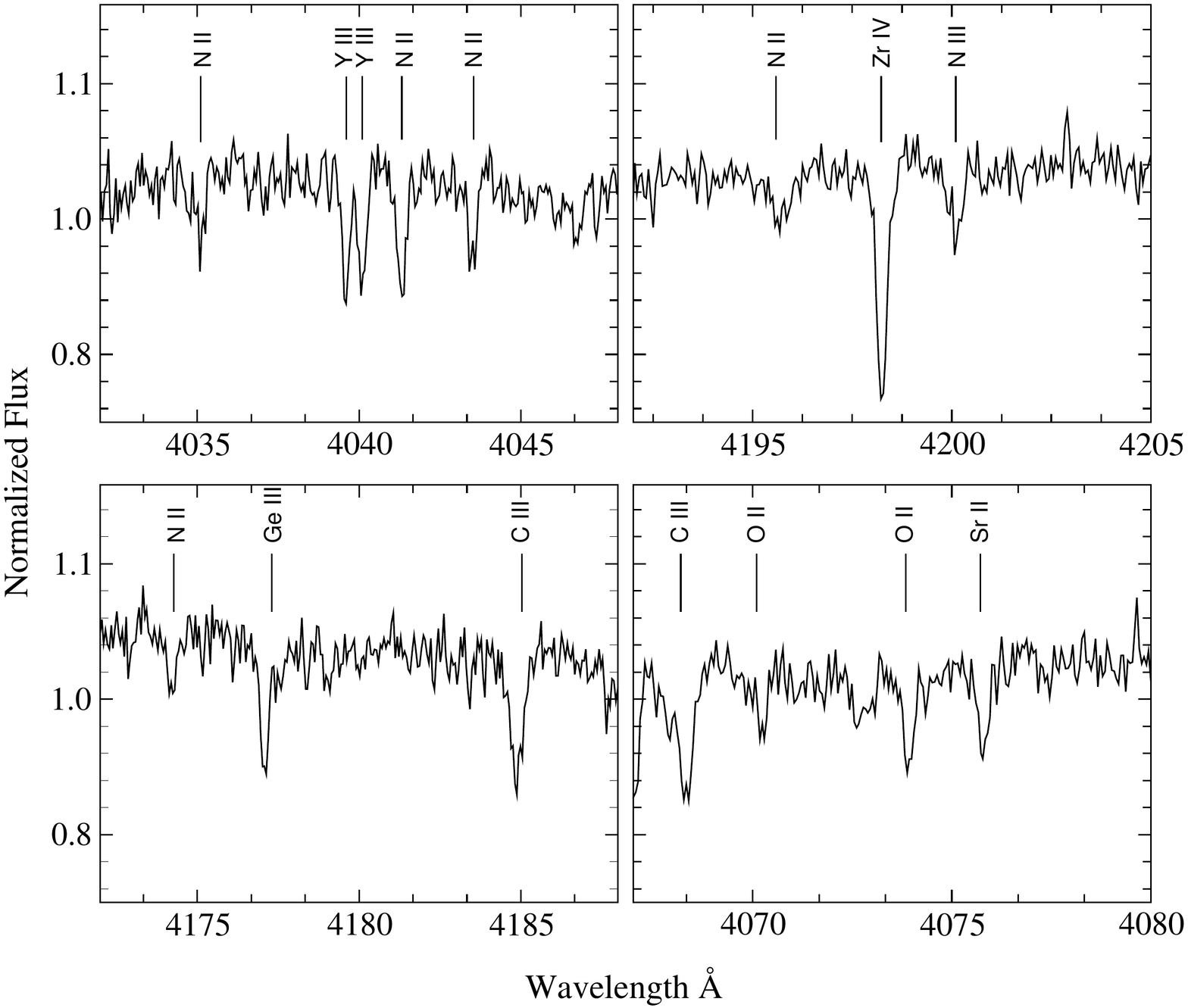} 
\caption{
Y{\sc iii},  Zr{\sc iv}, Sr{\sc ii} and  Ge{\sc iii} lines in LS\,IV$-14^{\circ}116$. From \citetads{2011MNRAS.412..363N}; copyright MNRAS; reproduced with permission.}
\label{fig:naslim_spec}
\end{center}
\end{figure}

\begin{figure*}
\begin{center}
\includegraphics[width=0.7\textwidth,angle=90]{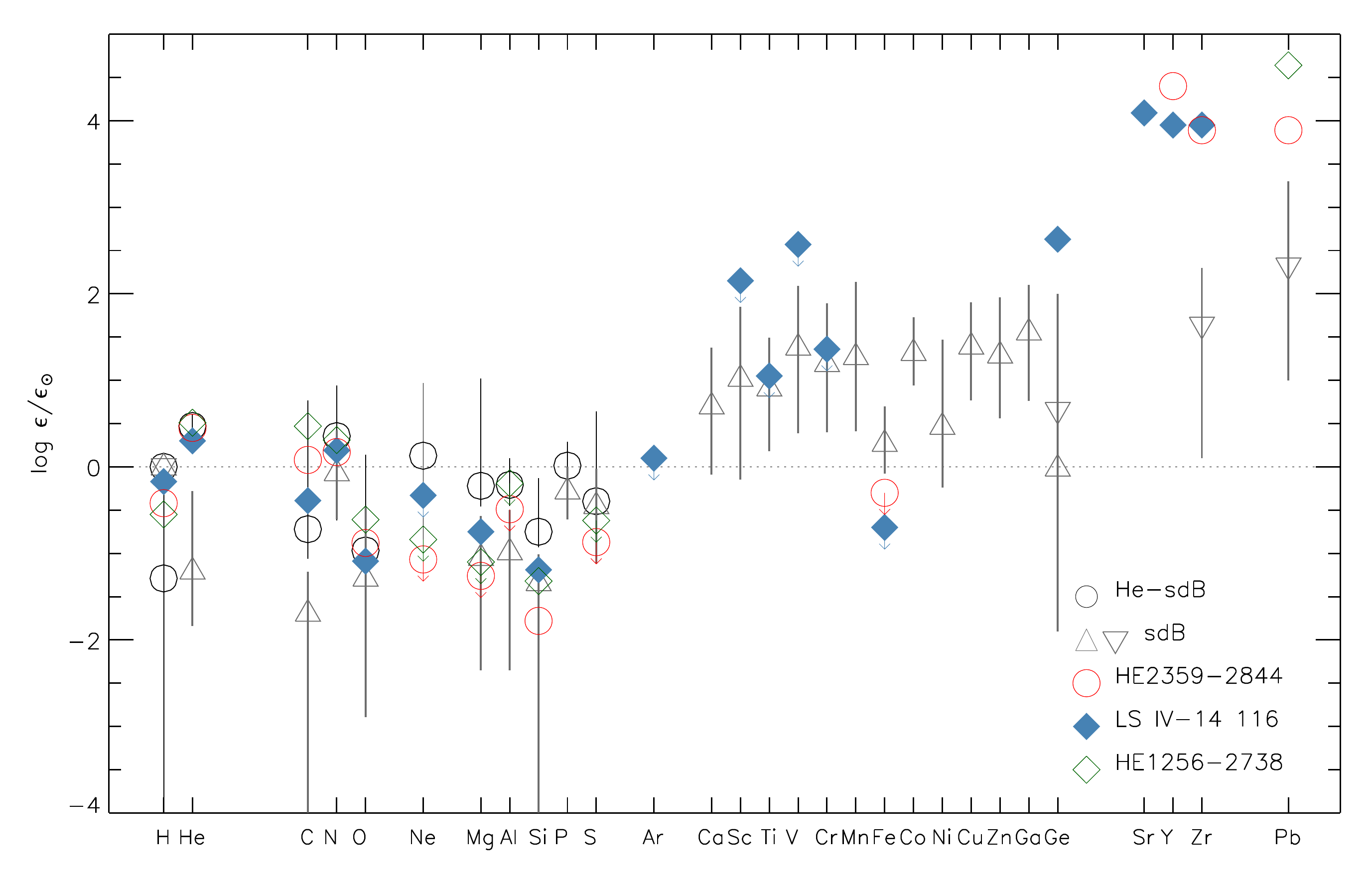} 
\caption{Surface abundances of the zirconium star LS IV$-$14 116 and the lead stars HE 2359$-$2844 and HE 1256$-$2738 relative to solar values. Mean abundances and ranges for other helium-rich subdwarfs and normal subdwarfs are also shown. 
This is an enhanced version of Figure 5 in \citetads{2015A&G....56b2.32J}, provided by C. S. Jeffery (2015, private communication).}
\label{fig:naslim_abu}
\end{center}
\end{figure*}

\citetads{2010MNRAS.409..582N,2011MNRAS.412..363N,2012MNRAS.423.3031N,2013MNRAS.434.1920N} derived abundances for He-sdB stars from high-resolution optical spectra (see Fig. \ref{fig:naslim_spec}). 
The resulting abundance pattern is shown in Fig. \ref{fig:naslim_abu}. The enrichment of some trans-iron elements is compelling. They are much larger than 
that of the normal sdB stars of similar temperature. 

Three stars are particularly interesting.
The intermediate He-sdB \object{LS~IV$-$14$^\circ$116}\footnote{An illustrative account of the properties of this enigmatic star and the history of discovery can be found in \citetads{2015A&G....56b2.32J}.} may be the most remarkable and actually a unique object up to now. It is distinct from any other hot subdwarf, not only by the presence of Ge {\sc iii}, Sr {\sc ii}, Y {\sc iii} and Zr {\sc iv} in its optical spectrum (see Fig. \ref{fig:naslim_spec}), but it shows a kind of oscillations that has not been seen in any other pulsating hot subdwarf \citepads{2005A&A...437L..51A}. 
Last but not least, \object{LS~IV$-$14$^\circ$116} was found recently 
to have kinematics typical of the Galactic halo population \citepads{2015A&A...576A..65R}. 

While Ge and Zr have been found in the FUV and UV spectra of many sdB stars
\citepads{2004A&A...423L..25O,2006BaltA..15..131C}, they have not been seen in optical spectra. This is because the overabundances of these elements in \object{LS~IV$-$14$^\circ$116} is up to 4 orders of magnitude in the line-forming region of the photosphere, much higher than in the hydrogen-rich sdB stars.  The He-sdB HE 2359$-$2844 also shows zirconium and yttrium abundances similar to those of
\object{LS~IV$-$14$^\circ$116} \citepads{2013MNRAS.434.1920N}. One hypothesis invoked was that a strong magnetic field may be responsible. FORS2 spectro-polarimetry of \object{LS~IV$-$14$^\circ$116}, however, ruled out a mean longitudinal magnetic field of more than 300 G \citepads{2015A&A...576A..65R}. 

Lead has been detected in the FUV and UV spectra of many hydrogen-rich sdBs
but never have any lines been found in optical spectra until \citetads{2013MNRAS.434.1920N} found 
two He-sdBs, HE 2359$-$2844 and HE 1256$-$2738, to show absorption lines due to triply ionized lead (Pb {\sc iv}) in the optical spectrum. From these lines, the atmospheric abundance of lead was found to be nearly 10 000 times solar, that is 10$-$100 times higher than that previously measured in normal hot-subdwarf atmospheres.

The strong chemical peculiarities, in particular for the heavy metals, as seen in the atmospheres of some sdB stars, e.g.  LS IV$-$14$^\circ$116, suggest that some diffusion process may have led to clouds of high concentration in the line-forming region.  

\subsection{The chemical composition of subluminous O stars}\label{sect:sdo}

Most of the hydrogen-rich sdO stars have been identified as the progeny of the normal
sdB stars (hydrogen-rich EHB stars). Hence core helium burning has ceased, {{but helium continues to burn in a shell shell surrounding the C/O core. These}} stars have evolved off the EHB to high effective temperature and are destined to end up on the white dwarf graveyard. 

Amongst the apparently brightest 
sdO stars, several belong to the luminous subclass \citepads[e.g.][]{1987Msngr..47...36H}, share properties with some central stars of planetary nebulae and DAO white dwarfs
\citepads{1999A&A...350..101N,2010ApJ...720..581G}, 
and are, therefore, likely post-AGB stars. 
This subclass shall 
not be discussed here \citepads[for a review see][]{2003IAUS..209..169W}
with the exception of BD+28$^\circ$ 4211, which 
is one of the brightest sdO stars, actually the first classified as such \citepads{1951ApJ...113..432M,1952PASP...64..256G} and, therefore, warrants a closer inspection here.

\subsubsection{BD+28$^\circ$ 4211 and the ``Balmer line problem''}

\object{BD+28$^\circ$ 4211} is frequently observed as a flux standard star, because its spectrum is simple at first glance. However, at high spectral resolution    
emission lines of highly ionized carbon, nitrogen, and oxygen in its optical spectrum \citepads{1999PASP..111.1144H} point to a very high temperature and strong NLTE effects. Its atmospheric parameters come close to that of DAO stars except for its somewhat lower gravity ($\log$ g=6.2). LS V+46$^\circ$21, the central star of the planetary nebula Sh 2-216 \citepads[T$_{\rm eff}$=95000 K, $\log$ g=6.9,]{2007A&A...470..317R} comes close to BD+28$^\circ$ 4211 in the (T$_{\rm eff}$-$\log$ g) plane \citepads[see Fig. 2 of][]{1999A&A...350..101N}. Hence, it is tempting to identify BD+28$^\circ$ 4211 as a post-AGB star although no nebula has been detected.

Indeed, the star challenged quantitative spectral analyses.
In early attempts to model the optical spectrum, using NLTE model atmospheres, \citetads{1992LNP...401..310N,1993AcA....43..343N} came across a problem with the Balmer lines. 
The observed Balmer lines could not be simultaneously reproduced with a unique set of fundamental parameters (log $g$ -
T$_{\rm eff}$). Individual lines needed different
temperatures in order to be matched properly. The higher lines of
the series require higher temperatures. The 
H$\alpha$ line requires T$_{\rm eff}$ $\simeq$~50,000 K while a much higher temperature (85,000 K) was necessary to match 
H$\epsilon$. 
The helium ionization equilibrium, a very sensitive temperature indicator, required a high temperature (82,000~K), in agreement with the high Balmer lines.

Subsequent spectroscopic analyses of very hot central stars of planetary nebulae and DAO white dwarfs showed that this ``Balmer
 line problem'' is often encountered for such stars hotter than  T$_{\rm eff}$~$\gtrsim$ 50,000 K \citepads{2010ApJ...720..581G}.
 
Although the modelling of spectra became more sophisticated by including line blanketing effects  \citepads{1993A&A...278..199D} 
and ion-dynamical effects on the Stark broadening of hydrogen and singly ionized helium lines
\citepads{1994A&A...285..603N} 
the problem persisted.  
\citetads{1996ApJ...457L..39W} 
suggested that the Balmer line problem in the case of BD+28$^\circ$4211 can be solved when surface cooling by photon escape from the Stark wings of CNO lines is accounted for. However, even with the most elaborate NLTE model atmospheres the Balmer lines of BD+28$^\circ$4211 cannot be matched simultaneously with the same parameter set \citepads{2015A&A...579A..39L}.
In their analysis of 29 DAO stars \citetads{2010ApJ...720..581G} 
found a correlation between higher metallic abundances and instances of the Balmer-line problem. This hints at a ``missing opacity problem''. Indeed, the knowledge of atomic data for highly ionized metals of interest is quite incomplete. To account for missing opacity \citetads{2015A&A...579A..39L} arbitrarily increased the metal abundances by a factor of ten\footnote{This work-around has also been used by \citetads{2006A&A...452..579O} 
to self-consistently match Balmer lines and helium ionization equilibrium in sdOB star spectra.} 
and successfully managed to overcome the Balmer line problem for BD+28$^\circ$4211. The missing opacity, though, has still to be identified (see also Sect. \ref{sect:sdo}).

\subsubsection{The helium-poor sdO stars Feige 110, AA Dor, and EC~11481$-$2303} 
   
\begin{figure*}
\begin{center}
\includegraphics[width=\textwidth]{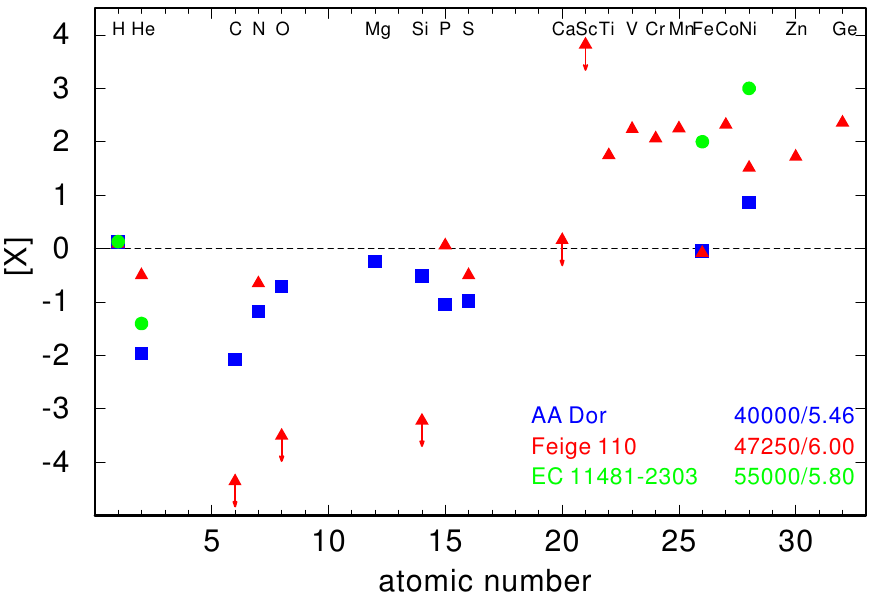}
\caption{
Comparison of the photospheric abundances relative to the solar values (arrows indicate upper limits) determined for the three OB-type subdwarfs AA~Dor \citepads{2011A&A...531L...7K}, EC~11481$-$2303 \citepads{2010Ap&SS.329..133R}, 
and Feige~110 \citepads{2014A&A...566A...3R}. Their T$_{\rm eff}$ and $\log$~g are shown in the legend. 
From \citetads{2014A&A...566A...3R} copyright A\&A; reproduced with permission.}
\label{fig:rauch_metal}
\end{center}
\end{figure*}

Amongst the helium-poor sdO stars three stars have been studied in great detail from far-UV spectra taken with the {\it FUSE} satellite, Feige~110, AA Dor, and EC~11481$-$2303  \citepads{2011A&A...531L...7K,2010Ap&SS.329..133R,2014A&A...566A...3R}. Metal abundances for BD+28$^\circ$4211 have also been derived from {\it FUSE} and {\it HST}-STIS spectra \citepads{2013ApJ...773...84L}. 
 The results are summarized in Fig.~\ref{fig:rauch_metal}. 
 The star-to-star scatter is large as is the case for the sdB stars. 
 The lighter elements (C to S) are mostly subsolar (-2 dex) to solar, except for C, O, and Si, which are strongly depleted in Feige~110. Iron is solar in Feige 110 and AA Dor but strongly enriched in EC~11481$-$2303.
 Comparing the resulting patterns to that of the cooler siblings shown in Fig.~\ref{fig:uv_metals} reveals striking similarities. 
Most remarkable is the strong enrichment of iron-group and trans-iron elements in both sdB and sdO stars. The enhancement of heavy metals increases with increasing temperature but the overall abundance indicate that the diffusion processes act in a similar way over a wide range of temperature, from ~23,000K (PG~1627$+$047, Fig. \ref{fig:uv_metals}) to 55,000K (EC~11481$-$2303, Fig. \ref{fig:rauch_metal}), although the degree of ionization changes and the luminosities increase, which should have a strong influence on radiative levitation and the strength of stellar winds. 

BD+28$^\circ$4211 does not fit into this scheme, because its heavy metal abundances are on average subsolar \citepads[see Fig. 8 of][]{2013ApJ...773...84L}. 
 However, the star is far hotter (82000K), of higher gravity (log g=6.2), solar helium content, and possibly a post-AGB star, and may therefore not be comparable to the other helium-poor sdO stars.

\subsubsection{He-sdO stars}

For He-sdO stars most of the abundance information on carbon and nitrogen came from the sample drawn from the ESO/SPY project.

\begin{figure}[tb]
\centerline{\includegraphics[width=0.75\textwidth]{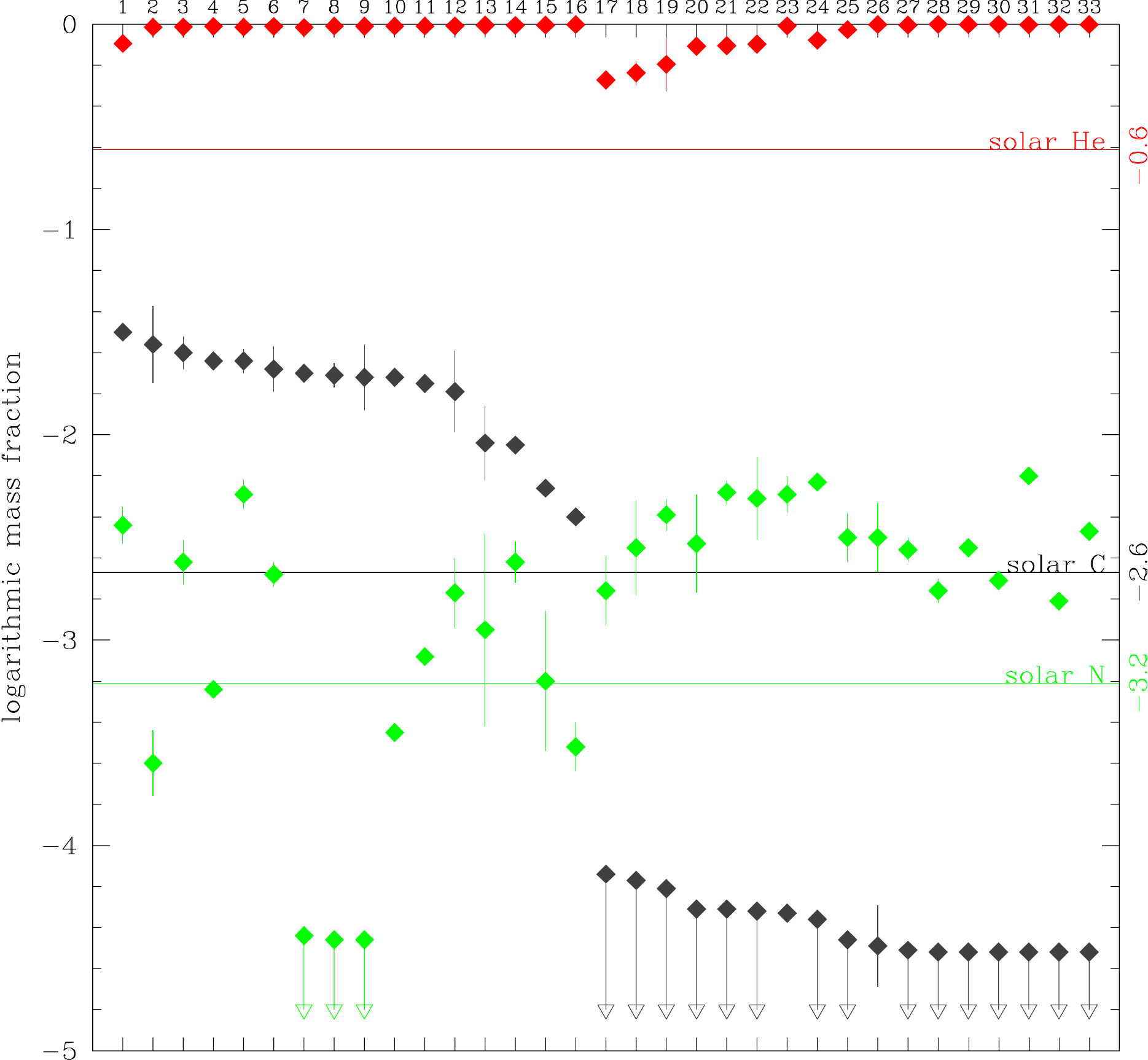}}
\caption{Abundances of He-sdO stars from the ESO/SPY project in logarithmic mass fractions of the programme stars, sorted by descending carbon abundance.
Red is the helium, grey is the carbon, and green is the nitrogen mass fraction.
The solar values are marked by the horizontal lines.
Objects, for which only an upper limit is available are marked by an arrow pointing down. {{For the sake of brevity the stars listed in table \ref{tbl:spyresult} of the appendix  
are idenfied with their running number.}} From \citet{phd_hirsch2009}.}
\label{fig:spyabun}
\end{figure}

In Fig.~\ref{fig:spyabun} the He, C, and N abundances of the 33 He-sdO stars from SPY are plotted, sorted by descending carbon mass fraction \citep{phd_hirsch2009}. While the variation in the helium abundance is very small, a bimodal distribution of carbon abundances is clearly seen: either the star has enhanced carbon abundances above the solar value, or the carbon is strongly depleted.
Nitrogen is super-solar for two thirds of the stars (enriched by a factor 3 to 10 with respect to the sun). While four stars have solar nitrogen
three others are mildly depleted in nitrogen, and another three strongly depleted by more than a factor of 15.

\begin{figure}
\centerline{\includegraphics[width=0.7\textwidth]{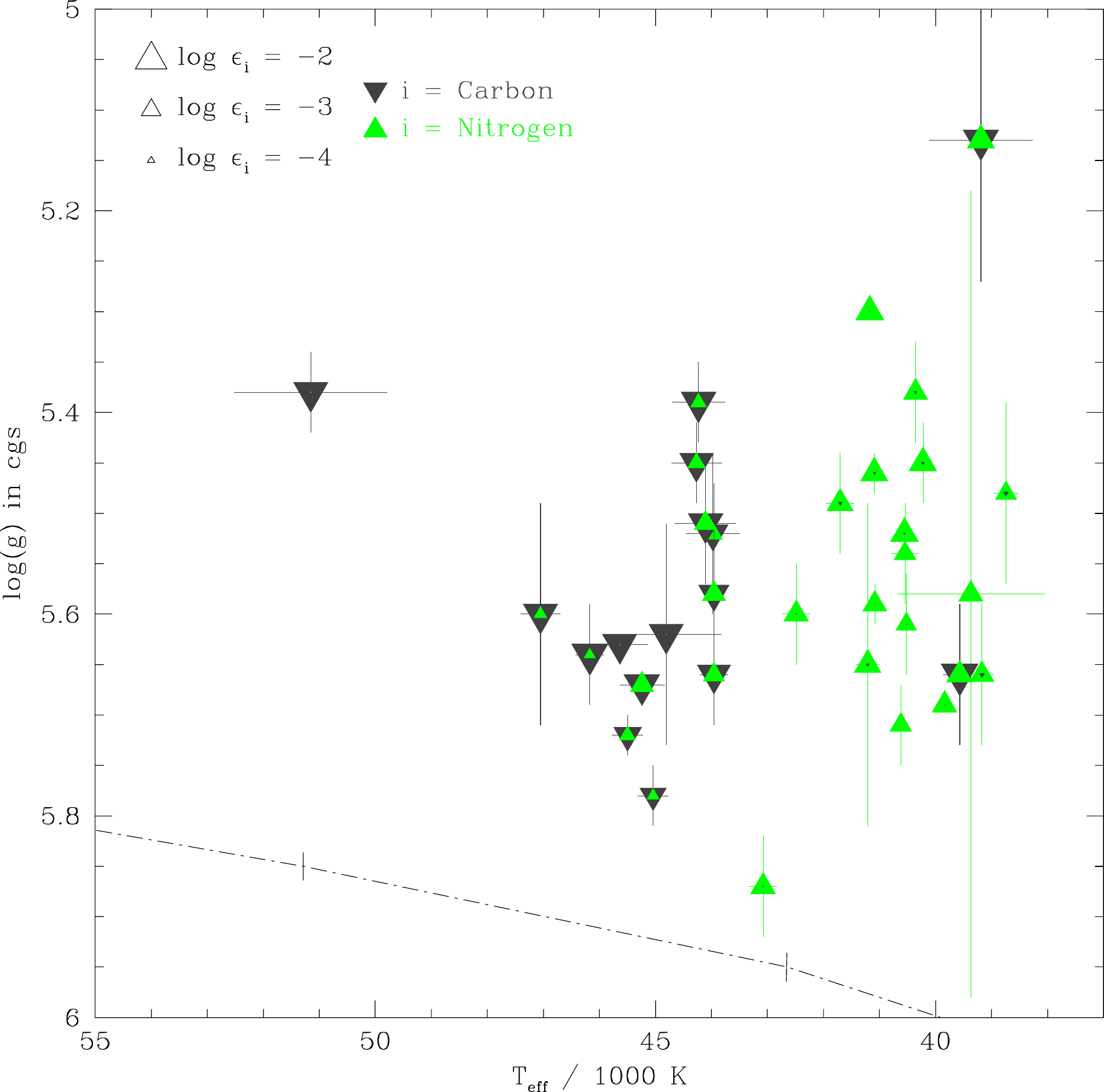}}
\caption{T$_{\rm eff}$-$\log$ g-diagram for the He-sdO stars from the ESO/SPY sample. 
The logarithmic mass fractions of carbon (dark grey) and nitrogen (green) are coded in the symbols' sizes. The dashed-dotted line marks the zero-age helium main-sequence.
From \citetads{2010AIPC.1314...79H}; copyright AIPC; reproduced with permission.
}
\label{fig:spytefflogg}
\end{figure}

The distribution of the He-sdOs from ESO/SPY in the T$_{\rm eff}$-$\log$ g-plane is shown in Fig.~\ref{fig:spytefflogg}.
The bimodality of the distribution becomes obvious immediately:
with only two exceptions, all carbon \emph{dominated} objects ($\log \beta_\mathrm{C} > \log \beta_\mathrm{N}$) are found at the hotter end of the distribution, i.e. T$_{\rm eff} > 43,900$ K.
Nitrogen is present in nearly all objects, and no clear statement concerning a correlation between nitrogen abundance and effective temperature becomes obvious.
Also no correlation between surface gravity and abundances can be drawn from Fig.~\ref{fig:spytefflogg}. These results pose important constraints on evolutionary scenarios as shall be discussed in Sects. \ref{sect:flasher} and 
\ref{sect:bin_evol}.

\subsection{Diffusion theory}\label{sect:diffusion}
    
The abundance anomalies of sdB stars and some sdO stars 
are caused by atomic diffusion processes in the stellar envelope.       
In a simplistic model the elemental 
abundances are set by a balance between gravitational settling and radiative levitation, because the radiative
acceleration depends on the elemental abundances through the incidence and strength of their spectral lines. {{Saturation of the spectral lines limits the radiation pressure which eventually allows an equilibrium elemental abundance to be reached. This is usually established on time scales that are short compared to the evolutionary time.}}

For helium atoms and ions the radiative support is rather small due to a scarceness of lines in the appropriate spectral range (UV), where the radiation flux is highest. 
Hence, helium should be depleted to very low abundances, on time scales much shorter than the evolutionary one. Actually, the equilibrium helium abundance is predicted to be \emph{lower} by two orders of magnitude than the average observed helium abundance
At such low abundances \emph{no} helium lines should be 
observable at all in optical spectra of sdB stars. 
On the other hand, radiative acceleration may be so large for those ions that have a plethora of UV lines, that the equilibrium abundance is super-solar.

\subsubsection{Slowing down diffusion: Mass loss}

Because the predictions for helium from simplistic models failed to match the observed abundances, additional processes were considered in order to slow down the diffusion process and support the chemical elements against gravitational settling. First, a radiatively driven stellar wind has been
suggested \citepads{1997fbs..conf..169F,2001A&A...374..570U} 
 which may explain the observed helium
abundances if the mass loss rate is of the order 10$^{-13}$ to 
10$^{-14}$\,M$_\odot$/yr. 
The radiation driven wind theory was used by \citetads{2002A&A...392..553V} 
to derive analytical formulae to estimate the mass-loss rates of hot horizontal-branch and sdB stars, assuming that the wind plasma behaves as a single fluid. 
 However, 
\citetads{2008A&A...486..923U} 
demonstrated that the densities are so low that the metals do not share their momentum with the hydrogen and helium ions through collisions, i.e. the wind would fractionate and becomes metallic. 

\subsubsection{Slowing down diffusion: Turbulence}

\begin{figure}
\begin{center}
\includegraphics[width=0.6\textwidth]{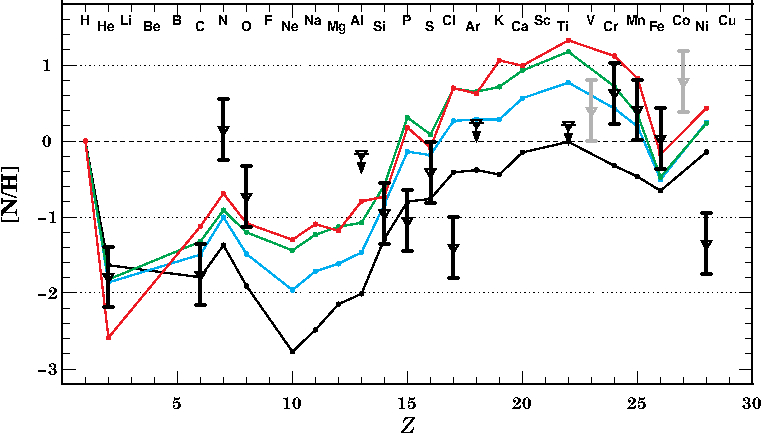}
\includegraphics[width=0.6\textwidth]{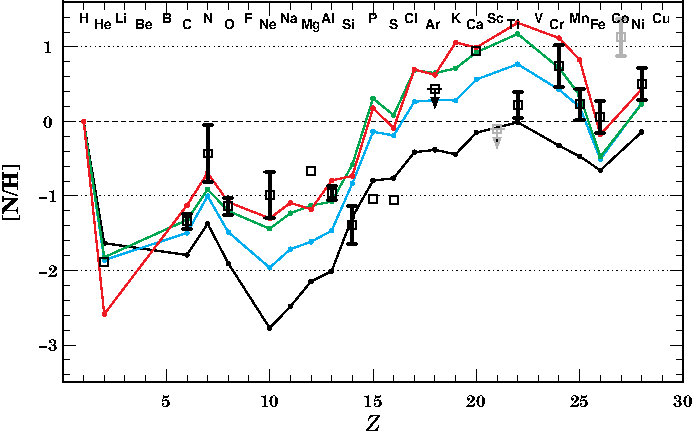}
\end{center}
\caption{Comparison of the observed abundance pattern of the sdB stars PG~0101+039 \citepads[][top]{2008ApJ...678.1329B} and 
Feige 48 \citepads[][bottom]{2006A&A...452..579O}
with prediction by diffusion models with turbulence of \citetads{2011A&A...529A..60M} after 25 Myr for models with 
original metallicities of Z$_0$ = 0.0001 (black), 0.001 (cyan), 0.004 (green), and 0.02 (red), respectively. Cobalt (grey symbol) was not included in the diffusion model due to the lack of atomic data.
This is a modified version of Figs. 5 and 13 of \citetads{2011A&A...529A..60M}}
\label{fig:michaud}
\end{figure}

Diffusion is taking place not only in the atmosphere of sdB stars, but also in subphotospheric layers, where it plays a vital role in order to drive pulsations via an opacity bump created by iron-group elements  
(see Sect. \ref{sect:asteroseismology}). A prerequisite is sufficient radiative levitation of those elements in the appropriate layers. Mass loss would weaken the opacity bump and pulsations would stop once the mass loss exceeds a certain critical limit.
\citetads{2011MNRAS.418..195H} 
found that the mass-loss rates required to
match the observed He abundances are not consistent with observed pulsations.
Weak turbulent mixing of the outer 10$^{-6}$ M$_\odot$
could also explain the He abundances while still allowing pulsation
modes to be driven.
A major caveat of the turbulence scenario is that it is not supported by a physical model. \citetads{2011MNRAS.418..195H} speculated that thermohaline mixing could play a role if a mean molecular weight inversion occurs.
\citetads{2008ApJ...675.1223M,2011A&A...529A..60M} adopted the turbulence model and carried out detailed diffusion calculations for metals in blue horizontal-branch stars and hot subluminous stars. 
They reproduce moderately well the surface abundances of normal sdB stars (see Fig. \ref{fig:michaud}).  Diffusion calculations for the trans-iron elements with excessive overabundances, have, however, not been possible yet due to the lack of atomic data. 

\subsection{Surface rotation}\label{sect:rotation}

Projected rotation velocities can be derived from high-resolution spectra along with the elemental abundances. Thermal broadening is the main limiting factor for stars as hot as sdB stars. Because the thermal velocity scales inversely with the square root of the atomic mass, it is advisable to make use of metal lines rather than of the helium and hydrogen lines. Unfortunately, metal lines are scarce in the optical spectra of many sdB stars, in particular in the hotter ones. The detection limit in most published studies is v$_{\rm rot}\sin$ i $\approx$ 5 km\,s$^{-1}$. 
However, processes other than rotation may lead to additional line broadening,
in particular, pulsations and magnetic fields. No indication for magnetic line broadening has been found yet in any sdB star. Pulsations cause the shape of the line profiles in the disk-integrated spectra to change with time \citepads[e.g.][]{2008A&A...492..815T}. The pulsation periods of the V361~Hya stars  are mostly shorter than the integration times of the spectral observations leading to a smearing of the lines that can easily be mistaken for rotation \citepads[see ][for an example]{1999A&A...348L..25H,2005A&A...442.1015K}. Indeed, substantial line broadening due to pulsational smearing has  
been found in several rapidly pulsating sdB stars.

Asteroseismology provides another option to determine rotational velocities, if the rotation splitting of pulsation frequencies can be measured. This technique is able to measure rotation rates too slow to be detectable from spectral line fitting. \citetads{2014MNRAS.440.3809R}, for instance, detected a rotation period as low as 88$\pm$8\,d for a 
g-mode pulsator in the {\it Kepler} field (see Sect. \ref{sect:puls_rotation}).

\subsubsection{Close binaries and synchronous rotation}\label{sect:bin_synchro}

Tidal interaction plays a crucial role in binary stars and exoplanets. In particular the tidal synchronization times are important, but present estimates are quite uncertain, especially for stars with radiative envelopes such as sdB stars. 

In the case of single-lined close binary sdB stars 
the analysis of radial velocity curves allows us to determine minimum masses from the mass function (see Sect. \ref{sect:binaries}).
While the mass of an sdB star can be reasonably well estimated, the mass of the unseen companion cannot be determined unless the system is eclipsing. However, there is a way to determine the inclination angle if the rotation velocity and the stellar radius are known. The latter can be calculated from surface gravity and mass. Spectroscopically the rotation velocity can be determined only in projection to the line of sight. In a tight binary, however, rotation may be tidally locked to the orbit. In such a case, the inclination results from the orbital period, the stellar radius, and the projected rotational velocity. Therefore, \citetads{2010A&A...519A..25G} 
embarked on a study to determine the distribution of
projected rotational velocities of a sample of 31 close sdB binaries by measuring the broadening of unblended metal lines and derived companion masses. Surprisingly in as many as eight cases (including the well known system KPD~1930+2752) the companion mass may exceed 0.9~M$_\odot$, four of which even exceed the Chandrasekhar limit; they may be neutron stars or even black holes. The distribution of the inclinations of the systems with low mass companions appears to be consistent with expectations. A lack of high inclinations for the massive systems signals a warning that the assumption of tidally locked rotation may be incorrect. This could be a matter of stellar age. If the EHB star is too young, synchronization might not yet have been reached. Taking this into account two objects, PG 1743+477 and, in particular, HE 0532-4503 remained whose companions may have masses close to or above the Chandrasekhar limit. However, no X-rays have been detected from both of them using Swift/XRT \citepads{2011A&A...536A..69M} 
nor has radio emission from HE 0532$-$4503 been discovered \citepads{2011A&A...531A.125C}, 
see also Sects. \ref{sect:bin_xray} and \ref{sect:bin_radio}. 

\begin{figure}
\begin{center}
	\includegraphics[width=0.6\textwidth]{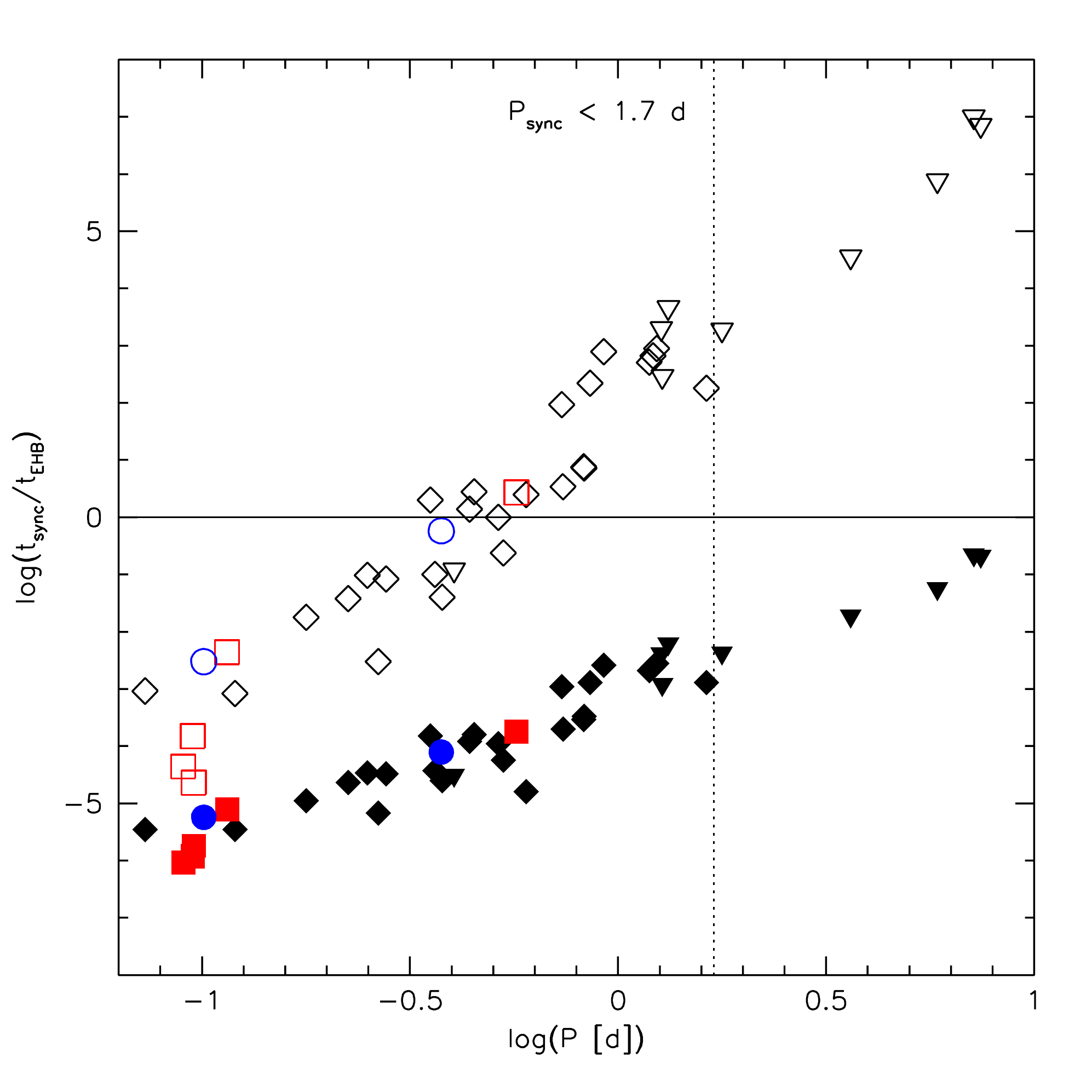}
	\end{center}
	\caption{The observed orbital period is plotted against the 
	synchronization times calculated with the theory of \citetads[][open symbols]{1977A&A....57..383Z}, and by that of \citetads[][filled symbols]{1992ApJ...395..259T}, both in units of 
	the average lifetime on the EHB \citepads[$10^{8}\,{\rm yrs,}$][]{1993ApJ...419..596D}. The solid horizontal line marks the border between
	synchronization within the EHB lifetime and synchronization times 
	longer than the EHB lifetime. The squares mark sdB binaries, where the 
	primaries have been proven to be synchronized by light curve analysis 
	of eclipsing or ellipsoidal variable systems. The circles mark 
	binaries where synchronization could be shown by asteroseismology. The 
	systems marked with diamonds could be solved consistently under the 
	assumption of synchronization, while the systems marked with triangles 
	rotate faster than synchronized. From \citetads{2010A&A...519A..25G}, copyright A\&A; reproduced with permission.}
	\label{fig:tsync}

\end{figure}

However, the assumption of tidal locking may not be correct. Because the envelopes of hot subdwarf stars are radiative, the spin-up by tidal forces is inefficient and very difficult to model. The results from rivalling theoretical concepts of tidal interaction \citepads{1977A&A....57..383Z,1992ApJ...395..259T}
 differ enormously (see Fig. \ref{fig:tsync}). A detailed discussion of the synchronization of sdB+dM binaries through coupling via dynamical tides can be found in \citetads{2012MNRAS.422.1343P}. In the least efficient case \citepads{1977A&A....57..383Z} tidal synchronization should be established within the evolutionary lifetime of an sdB star (10$^8$ yrs) for all systems with periods of less than half a day. 

Rotation velocities can be derived via asteroseismology by making use of the frequency splitting of pulsation modes (see Sect. \ref{sect:asteroseismology}). Hence pulsating sdBs in binaries provide an important testbed. Observational tests were scarce and lead to contradictory conclusions.
In the case of the pulsating sdB binary Feige\,48, for instance, \citetads{2008A&A...483..875V}
derived a spin period of 9.028 $\pm$ 0.480 h from the light curve, remarkably similar to the system's orbital period of 9.024 $\pm$ 0.072 h, from radial-velocity variations. Hence, Feige 48 rotates as a solid body in a tidally locked system. On the other hand 
\citetads{2012MNRAS.422.1343P} 
studied the {\it Kepler} light curves of two pulsating subdwarf B stars in close binaries with dM companions of orbital periods (~9h) similar to that of Feige~48 and measured the rotational splitting of the pulsation frequencies.
Both systems have several triplet spacings, which imply rotation periods of 10.3 and 7.4 d, respectively, indicating the sdB components rotate too slowly for synchronous rotation, because the orbital periods are much shorter.

We shall revisit stellar rotation in the context of asteroseismology in Sect. \ref{sect:asteroseismology} in the light of new {\it Kepler} measurements.

\subsubsection{Single sdB stars and blue horizontal branch stars}

Projected rotation velocities for single sdB stars\footnote{Subluminous B stars in binaries were also considered single if the separations of the components are so wide that tidal interaction can be neglected, that is their orbital periods exceed 1.2~days \citepads{2010A&A...519A..25G}.} were determined from high resolution spectra by
\citetads{2012A&A...543A.149G}. All stars in their sample have low projected rotational velocities (v$_{\rm rot}\sin i$ $<$ 10\,kms$^{-1}$). The distribution of projected rotational velocities is consistent with an average rotation of 8\,kms$^{-1}$ for the sample. 

\begin{figure*}
\plottwo{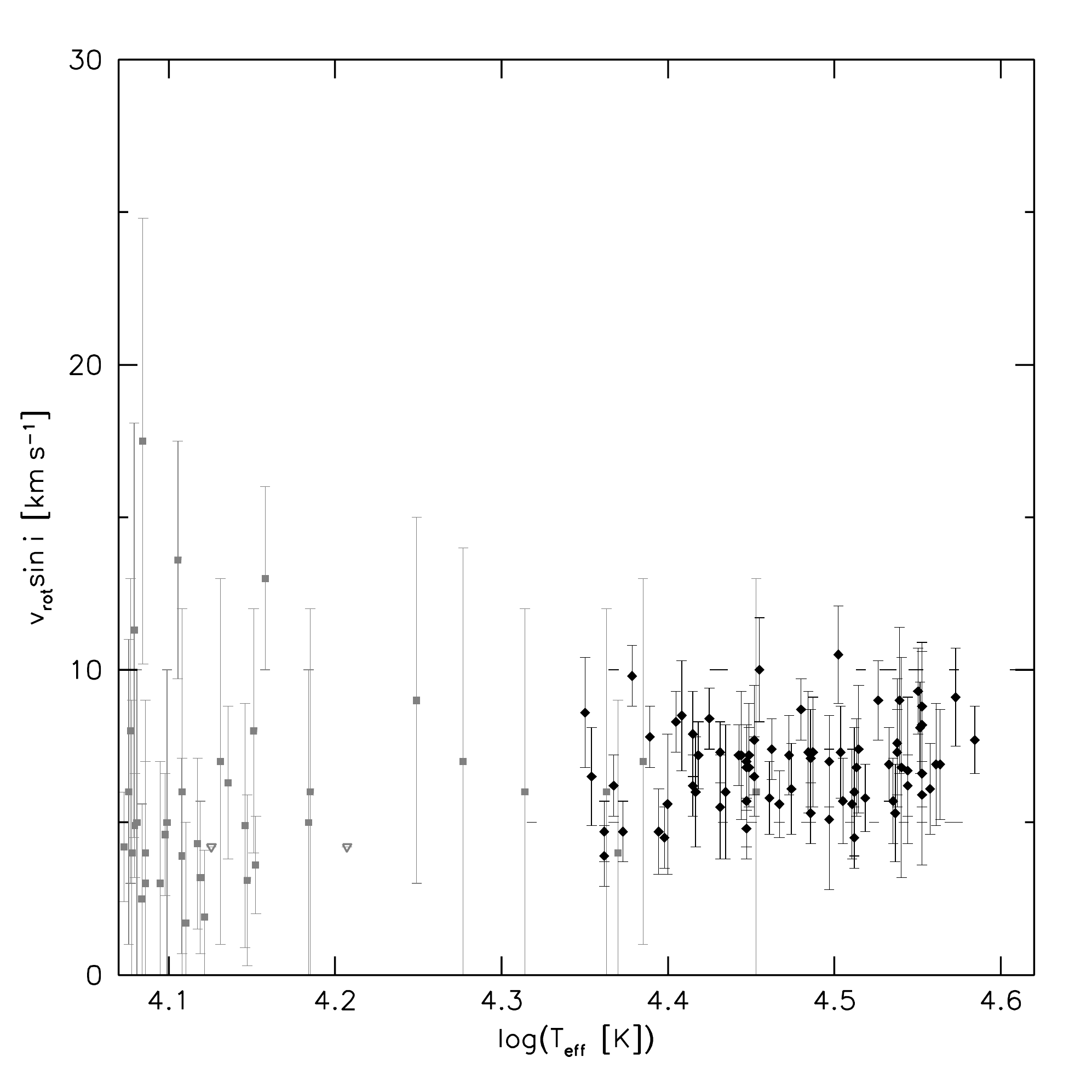}{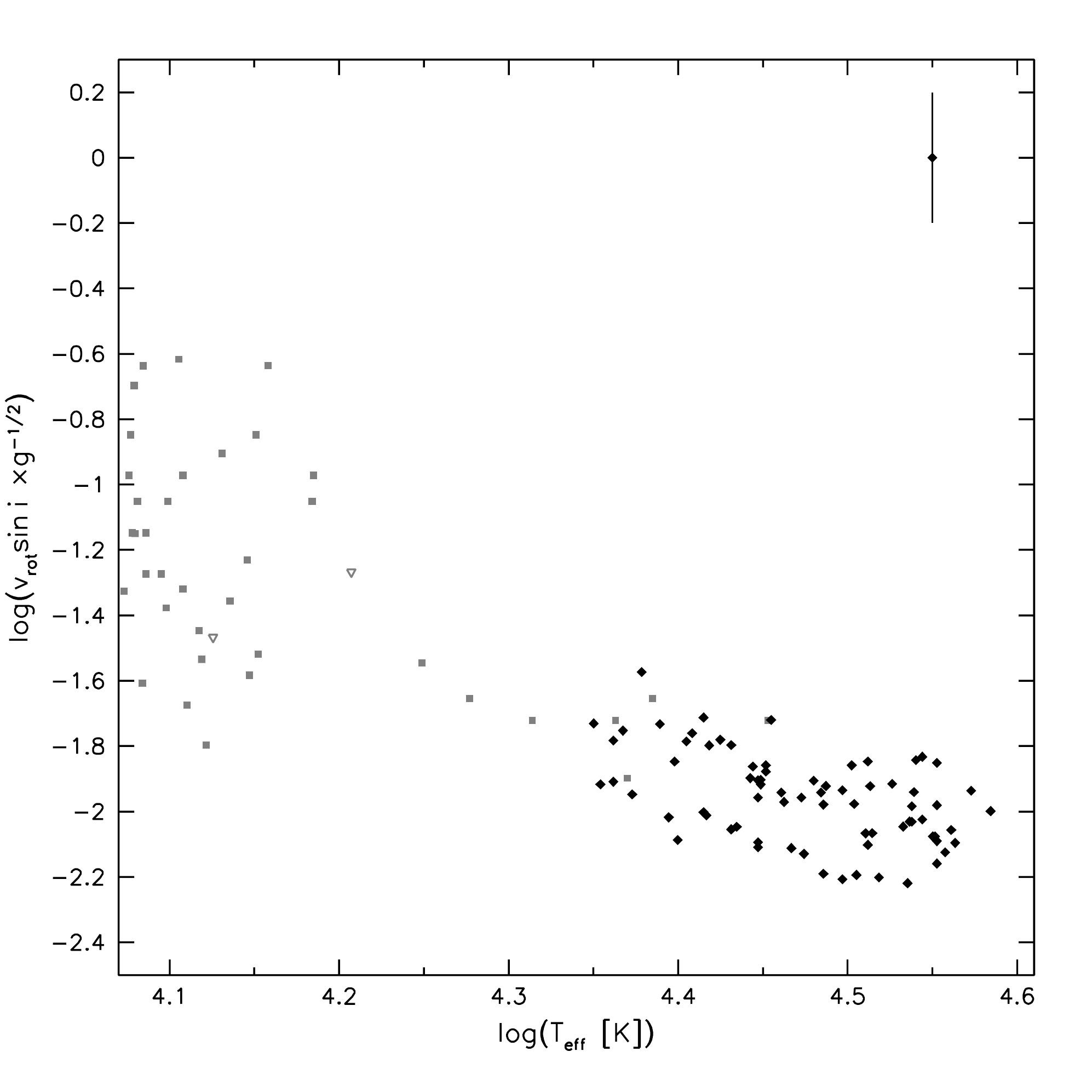}
\caption{left hand panel: Projected rotational velocity plotted against effective temperature.
Right hand panel: vrotsin $\times$ g$^{-1/2}$, a proxy for the angular momentum, plotted against effective temperature. The grey squares mark BHB and some sdB stars taken from 
\citetads{1995ApJ...453..214P}, 
\citetads{2003ApJS..149...67B,2003ApJS..149..101B}, \citetads{2000A&A...364..102K}, and \citetads{2002ApJ...572L..71R}.
Upper limits are marked with grey triangles. The black diamonds mark the sdBs from the sample of \citetads{2012A&A...543A.149G}. The vertical line marks the jump temperature of 11,500K. Typical uncertainties for the sdBs are given in the upper right corner. These are modified versions of Figure 6 and 7 in \citetads{2012A&A...543A.149G}, provided by S. Geier (2015, priv. comm.).\label{fig:geier_rotation}}
\end{figure*}

It is tempting to compare the rotational properties of sdB stars to those of the Blue Horizontal Branch  (BHB) stars, because the hot subdwarf stars 
form the extension of the horizontal branch. In Fig. \ref{fig:geier_rotation} rotation velocities of BHB stars from literature are compared to those of sdB stars from \citetads{2012A&A...543A.149G}. BHBs with diffusion-dominated atmospheres are slow rotators as well \citepads{2003ApJS..149...67B}. 
As can be seen from Fig. \ref{fig:geier_rotation} the sdB sequence extends the BHB trend to higher temperatures. The ${v_{\rm rot}\sin\,i}$ values remain at the same level as observed in hot BHB stars.  
Because BHB stars have larger radii than sdB stars it is important to compare their angular momenta.
The quantity v$_{\rm rot}\sin i$ $\times$ g$^{-1/2}$ is a proxy for the angular momentum and a comparison of the BHB and sdB stars is shown in the right hand panel of Fig. \ref{fig:geier_rotation}. The transition between BHB and EHB stars is smooth. Since the progenitors of the EHB stars lost more envelope material on the RGB, the EHB stars are expected to have lower angular momenta than the BHB stars. This is consistent with the trend seen in Fig.~\ref{fig:geier_rotation}. Hence single sdB stars as well as sdB stars in wide binaries are related to 
blue horizontal branch stars in terms of surface rotation and angular momentum evolution. This is remarkable because in the case of the binaries, a transfer of mass and angular momentum is likely to have occurred. 

\subsubsubsection{SB~290 and EC~22081-1916 -- the exceptions to the rule}

Despite the evidence presented above, that sdB stars are slow rotators unless they are spun-up by tidal forces from a close companion, two apparently single sdB stars, SB~290 \citepads{2013A&A...551L...4G} and \object{EC~22081-1916} \citepads{2011ApJ...733L..13G}, have been found to be rapidly rotating, indicating that both stars may be the result of He-WD mergers. 

While SB~290 is located on the EHB band and considered consistent with a He-WD merger by \citetads{2013A&A...551L...4G}, \object{EC~22081-1916} is of lower gravity and, hence, higher luminosity, suggesting that the hydrogen envelope may be unusually thick. This, however, would be at variance with the He-WD merger scenario, but consistent with a common envelope merger of a low-mass, possibly substellar object with a red-giant core. 

\subsection{Magnetic fields of hot subdwarfs: sdBs vs. He-sdOs}\label{sect:mag}

Strong magnetic fields can be easily detected in high resolution spectra due to Zeeman splitting or corresponding broadening of the lines. Magnetic line broadening can be distinguished from rotational broadening because the former scales with the square of the wavelength of the lines.
{{Although many}} sdB stars have been studied at high spectral resolution, none 
{{have}} been found to show magnetic broadening or splitting. 
Spectropolarimetry allows weaker fields to be measured. Indeed, the presence of magnetic fields at the 1kG level had been concluded for a handful of sdB stars \citepads{2005A&A...437..227O} from their spectrophotometric observations with FORS$@$ESO-VLT, although close to the instrumental detection limit. \citetads{2012A&A...541A.100L} reanalyzed the same data set using an improved wavelength calibration and showed that the detection was an instrumental artefact. The reanalysis resulted in upper limits to the magnetic field strength of a few hundred Gauss. Recently, \citetads{2015A&A...576A..65R}
 derived an upper limit of 300~G for the intermediate He-sdB \object{LS~IV-14$^\circ$ 116}. 
Claims for the existence of magnetic fields 
in Feige~66, HD~76431 and the pulsating sdB Balloon 090100001 have also been shown to be spurious \citepads{2012ASPC..452...87P,2013ARep...57..751S}.

More recently, though,  
Zeeman triplets in an He-sdO star were reported by \citetads{2013EPJWC..4304002H}, which was serendipitously discovered in the search for targets for the MUCHFUSS project in the SDSS spectral data base by B. G\"ansicke. In the same project two additional He-sdO  stars were found to show similar splittings indicating the presence of magnetic fields of about 500kG (N{\'e}meth, priv. comm.). Because the population of He-sdO stars has been much less extensively studied, these discoveries indicate that the magnetic properties of the He-sdO population is substantially different from that of the sdB stars. Magnetic He-sdO stars might be progenitors of strongly magnetic white dwarfs 
($>$1\,MG) if the magnetic flux is conserved during the contraction of the He-sdO to the white dwarf state 
\citepads[see][for a recent review on magnetic white dwarfs]{2015SSRv..191..111F}. 

\clearpage

\newpage
\section{Formation and evolution of hot subdwarf stars}\label{sect:evolution}

The main difficulty to explain the formation of EHB stars is the large amount of mass lost prior to or at the start of core helium burning. 
In order to resolve the puzzle of the origin and evolution of sdB and sdO 
stars and the potential
linkage between both classes of stars, several scenarios have been developed. 

Binary evolution through mass transfer and 
common-envelope ejection must be important for sdB stars because of the
high percentage of close binaries with periods of less than ten days.
The merger of two helium white dwarfs is another 
vital option to explain the
origin of single hot subdwarfs. Alternatively, the origin of hot subdwarf stars could be intrinsic to the star; that is, internal processes 
{{may decrease the hydrogen content of the envelope,}} 
e.g. through delaying the core helium flash (the so called    
hot-flasher scenario, see Sect.~\ref{sect:flasher}), during which surface hydrogen is burnt after mixing into deeper layers.
Finally it cannot be taken for granted that sdB stars are core helium-burning
objects. There is observational evidence that some sdB stars have  
masses too low to ignite core helium burning \citepads{2003A&A...411L.477H}. 
We shall come back to this class of stars in Sect. \ref{sect:lmwd_elm}.
  
\subsection{Canonical HB and post-HB evolution}\label{hb_evolution}

Horizontal-branch stars are in the core helium-burning phase of evolution following
the red-giant branch (RGB). The ignition of helium burning takes place under electron-degenerate
conditions for stars of $\leq 2.3M_\odot$ leading to the helium core flash.
The distribution of stars along the HB
can be explained by a constant core mass of slightly less than half a solar mass and a 
spread in envelope and
thus total mass. 
The smaller the envelope the bluer the
star ending at the theoretical helium main-sequence when the envelope mass is 
zero. The core
mass is fixed at the onset of the core helium flash at the tip of the RGB
 and depends only slightly on metallicity and helium abundance \citepads{1987ApJS...65...95S}. 
 Accordingly the  core mass is restricted to a narrow range from 0.46 to 0.5~$M_\odot$.
The post-EHB
evolution proceeds towards higher temperatures until the white-dwarf cooling 
track is reached and gravity increases. Hence the star will avoid the AGB stage. 

Only very few calculations have been able to follow stellar evolution 
through the 
violent helium core flash. Most HB models were calculated by starting new
sequences on the HB, where the initial structure is taken from that of
a red-giant progenitor, but the core composition
on the ZAHB has to be adjusted to account for the
modest carbon production during the flash \citepads[see][]{1997ApJ...474L..23S,1997fbs..conf....3S}. 
Widely used models \citepads[e.g.][]{1993ApJ...419..596D} are 
calculated in such an approximate way, but have been shown to be reliable,
because recent full evolutionary models following the star from the ZAMS 
through the helium core-flash to the 
ZAHB \citepads[e.g.][]{2005A&A...442.1041S} differ only slightly from the approximate 
ones. More recent two and three dimensional models of the evolution of a star through a helium core flash also confirm that the structure of the star is not significantly altered, because convection plays an important role {{in establishing}} hydrostatic equilibrium, but show that overshooting, indeed, occurs \citepads{2008A&A...490..265M,2009A&A...501..659M}. 
{{The MESA\footnote{\emph{Modules for Experiments in Stellar Astrophysics}} code allows to
consistently evolve stellar models through the He core flash \citep{2011ApJS..192....3P}.}}    


Post-EHB evolution has been computed by many groups, e.g. \citetads{1993ApJ...419..596D}, 
\citetads{2003MNRAS.341..669H}, and \citetads{2008A&A...490..243H} 
and evolutionary time scales
 are found to be shorter by an order of magnitude than the EHB lifetime.
The post-EHB tracks nicely connect the
(helium-deficient) sdB stars on
or near the EHB band to the helium-deficient sdO stars.  

However, the formation of EHB stars remains unexplained by canonical models, because there is no straight-forward way for a single red giant to remove all of its envelope.

\subsection{Helium mixing on the red giant branch}

\citetads{1997ApJ...474L..23S,1997fbs..conf....3S} 
presented a series of non-canonical models to explain the formation of BHB and EHB stars. Besides the hot flasher scenario, which will be discussed in the next section, he investigated helium mixing on the red giant branch by assuming that the outer convective envelope can penetrate into the hydrogen burning shell
and that some helium is mixed into the stellar envelope. Evolutionary sequences were calculated assuming different penetration depth. The most important consequence of helium mixing is that the luminosity of the tip of the RGB increases, which causes stronger mass loss. A caveat of the helium-mixing scenario is that the physical mechanism that causes the mixing remains obscure\footnote{The core-envelope coupling of red giants must be much stronger than predicted by theory to explain the results of {\it Kepler} light curve analyses  \citepads[][see Sect. \ref{sect:differential_rotation}]{2015AN....336..477A},
which could also lead to helium mixing.}. 
\citetads{1997fbs..conf....3S} 
conjectured that rapid rotation may be responsible and a spread in rotation rates may explain the distribution of stars along the horizontal branch. This scenario has recently been revived by \citetads{2015Natur.523..318T} to explain EHB and blue hook stars in globular cluster, who suggest that rapid rotation of second generation stars may result from the star-forming history and the early dynamics of very massive globular clusters, such as $\omega$ Cen
(see Sect. \ref{sect:cluster}). 

\subsection{The hot-flasher scenario}\label{sect:flasher}

Low-mass stars undergo the helium core flash at the tip of the red giant branch. 
However, \citetads{1993ApJ...407..649C} 
showed that, if sufficient
mass is lost on the RGB, the star can depart from the RGB and
experience the helium core flash while descending the hot white-dwarf cooling
track. The
remnants of these ``hot flashers''  \citepads[e.g.][]{2001ApJ...562..368B,1996ApJ...466..359D} 
are found to lie 
close to the helium main-sequence, i.e. at the very hot end
on the extreme horizontal branch.

\begin{figure*}
\begin{center}
\includegraphics[width=0.7\textwidth]{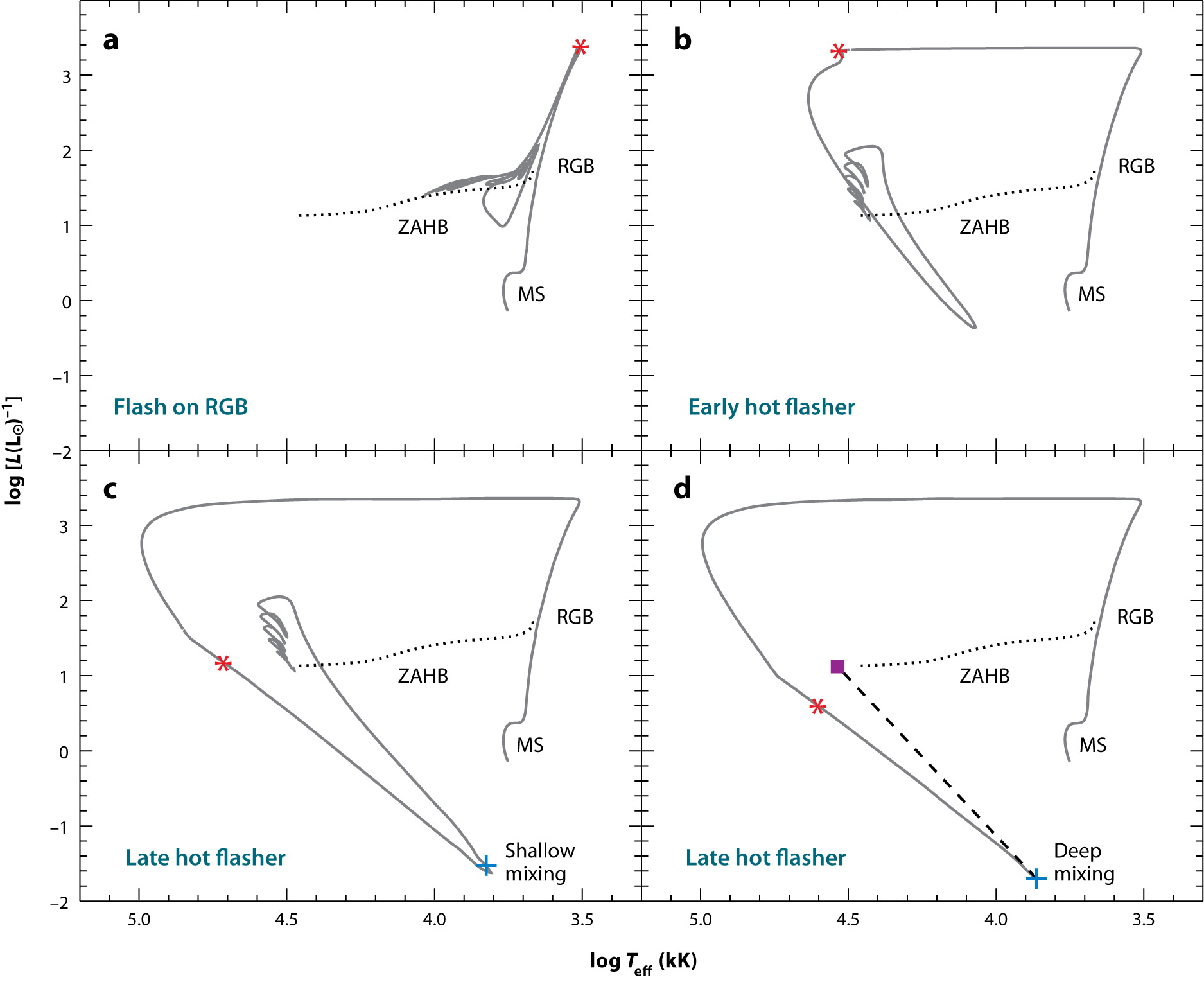} 
\end{center}
\caption{Evolution of a solar metallicity star from the main sequence
through the helium flash to the zero-age horizontal-branch 
(ZAHB, dotted curve) for
different amounts of mass loss on the red-giant branch \citepads{2004ApJ...602..342L}. 
Flash mixing did not occur for the canonical sequences in
panels {\it a\/} and {\it b\/}. 
 For
sufficiently
large mass loss, a star evolves off the RGB to high effective
temperatures
before igniting helium as either an early or late hot flasher. The
peak of the helium flash is indicated by an asterisk. The flash
convection zone reached the hydrogen envelope at the plus sign
along the tracks in panels {\it c\/} and {\it d\/}. These panels
illustrate the two types of flash mixing: shallow mixing in which
the hydrogen envelope is mixed only with the convective shell in the
outer part of the core and deep mixing in which the hydrogen envelope
is mixed all the way into the site of the flash. The
model calculations in panel {\it d\/} were stopped at the onset
of deep mixing, and a ZAHB model (solid square) was then computed
assuming a helium- and carbon-rich envelope composition.  The
evolution during this phase is shown schematically by the dashed
line. From \citetads{2009ARA&A..47..211H}, 
copyright Annual review Astronomy \& Astrophysics, reproduced with permission.  
}\label{fig:flasher}
\end{figure*}

The ``hot flasher'' phase bears a striking similarity to the late
helium shell flashes that produce ``born-again AGB stars'' \citepads{1984ApJ...277..333I}. 
As in the born-again scenario, the high luminosity 
($L_{\rm He}\sim 10^{10}L_\odot$) during the flash generates a convection zone,
which might engulf the H-rich envelope \citepads{1997fbs..conf....3S,1997ApJ...474L..23S}. Hence, hydrogen might
be mixed into hotter layers and be burnt there leading to a He-enriched surface.  \citetads{2003ApJ...582L..43C}, \citetads{2004ApJ...602..342L}, and \citetads{2008A&A...491..253M} 
showed that the outcome of a hot flasher depends on the 
evolutionary phase during which it occurs \citepads[][see also Fig. \ref{fig:flasher}]{2004ApJ...602..342L}:
\begin{itemize}
\item[i)] {\em Early hot flasher:} If the helium core flash occurs early after departure {{from}} the RGB, i.e. 
during the evolution at
constant luminosity, hot subdwarf stars with standard H/He envelopes result.

\item[ii)] {\em Late hot flasher (shallow mixing):} Mixing is found to occur 
only when the star has 
already embarked on the white-dwarf cooling track. If the flash occurs early 
on that track, mixing is
shallow and the atmosphere of the resulting hot subdwarf is somewhat enriched 
in helium and nitrogen due to convective dilution of the envelope. 

\item[iii)] {\em Late hot flasher (deep mixing):} Deep mixing, however, occurs  
in late hot flashers in which the H-rich envelope is engulfed and burnt in 
the convective zone generated by the primary helium core flash leading to strong
enrichment of He, C, and N.  

\end{itemize}

\subsubsection{The rapid evolution of the exciting star of the Stingray nebula}

Late thermal pulses proceed on time scale so short that stellar evolution may be observed in ''real time''. This has been witnessed for three post-AGB stars (FG Sge, V 605 Aql, and Sakurai's object V4334 Sgr) during the last century \citepads[see][for a review]{2003ARA&A..41..391V}. 

\begin{figure}
\includegraphics[width=0.9\textwidth]{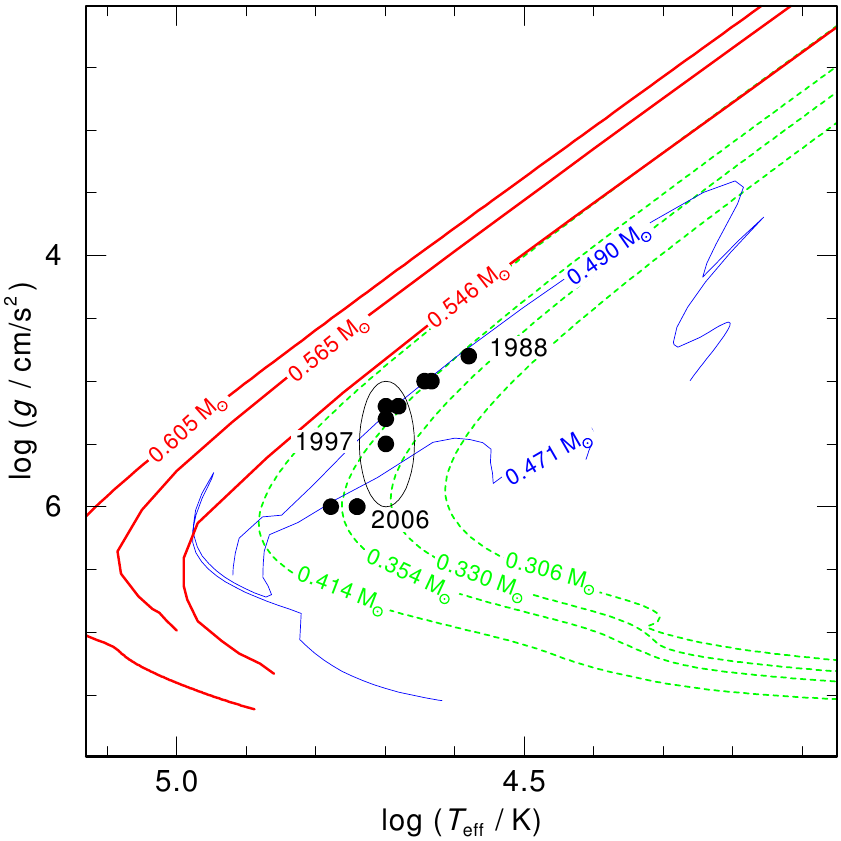}
\caption{Evolution of the central star of the Stingray nebula (black dots) in the $\log$ T$_{\rm eff}$, $\log$ g plane compared to post-AGB (red, thick) by \citetads{1995A&A...299..755B},
post-EHB (blue, thin) by \citetads{1993ApJ...419..596D},
and post-RGB (green, dashed) evolutionary tracks by \citetads{2013MNRAS.435.2048H}. 
The tracks are labeled with stellar masses. The ellipse indicates the errors of T$_{\rm eff}$ and $\log$ g in 1997. From \citetads{2014A&A...565A..40R}; copyright A\&A reproduced with permission.}
 \label{fig:stingray_evol}
\end{figure}

The central star of the Stingray nebula might be an exciting new such case.
\citetads{1995A&A...300L..25P} 
first discovered a rapid increase in effective temperature and a drop of luminosity, far too fast for canonical post-AGB evolution. 
From a detailed quantitative spectral analysis \citetads{2014A&A...565A..40R} 
found that from 1988 to 2002 the central star has steadily increased in effective temperature from 38 kK to 60kK while contracting; that is, the surface gravity increased during this period of time from $\log$ g = 4.8 to 6.0 (see Fig. \ref{fig:stingray_evol}). 
By the year 2002 its evolution reversed into cooling down to 55 kK in 2006. 
Unlike e.g. in Sakurai's object, which showed a dramatic change in chemical composition in only a few months time (H decreasing while Li and s-process elements were increasing) no such change of the (nearly) solar composition of the surface of the Stingray's central star has yet been observed. Furthermore, \citetads{2014A&A...565A..40R} showed that the Stingray's central star is a low mass star, consistent with a post-EHB or post-RGB nature (see Fig. \ref{fig:stingray_evol}). However its speed of evolution is much too fast for canonical post-EHB or post-RGB evolution suggesting that the star has just suffered from a late shell flash. Although no evidence for binarity has yet been found, \citetads{2014A&A...565A..40R} proposed that binary common envelope
evolution might lead to a remnant that has not yet re-established thermal equilibrium after the common-envelope ejection. 
 Continued monitoring will show whether the star is evolving back to the RGB or heading towards the white dwarf graveyard and whether a companion exists.

\subsection{Close-binary evolution}\label{sect:bin_evol}

The high fraction of sdB stars in close binaries implies that they are formed by binary-interaction processes.
The three main formation channels that have been proposed to produce
sdB stars are common-envelope (CE) evolution \citepads{1976IAUS...73...75P}, 
Roche Lobe Overflow (RLOF) evolution and a white-dwarf merger \citepads{1984ApJ...277..355W}. 
A detailed description of the binary evolution forming hot subdwarf stars can be found in 
\citetads{2002MNRAS.336..449H,2003MNRAS.341..669H}. 

In the CE formation model, the sdB progenitor
fills its Roche lobe near the tip of the RGB. If the mass transfer rate is sufficiently high the companion star will not be able to accrete all the matter. In consequence a common envelope is formed. Due to friction with the gas the two
components will spiral in until enough orbital energy is transferred to the common envelope to eject it. The remaining core of the red giant will become the sdB star. Because the CE phase is short the companion will remain almost unchanged. If the companion is a main-sequence star (MS) the resulting close binary is an sdB + MS with a period between 0.1 and 10 days. Eventually the main -sequence star
will evolve into a red giant. When it fills its Roche lobe, a
second CE ejection phase occurs resulting in a close sdB + WD binary, because the red giant fills its Roche lobe before the tip of the RGB is reached.

In the RLOF channel no common envelope will form because the mass transfer is dynamically stable and the companion slowly accretes the matter. The red giant loses its entire
envelope during RLOF to become an sdB star in a long-period binary
with a main-sequence component.
\citetads{2003MNRAS.341..669H} 
predicted that the orbital periods of such systems should be in the range 10 to 500 days. This is at odds with the period distribution of sdB+MS systems, recently found to have periods ranging from ~700 to ~1300 days  (see Sect. \ref{sect:binaries}). This discrepancy triggered new binary population synthesis calculations based on Han's models  
\citepads{2013MNRAS.434..186C} 
which included a more sophisticated treatment of angular-momentum loss and also considered atmospheric RLOF. Periods up to 1100
days result from solar-composition models. 
In addition, sdB models predict periods up to 1200 days, if atmospheric RLOF is included, close to what has been observed.
 
\begin{figure*}
\begin{center}
\includegraphics[width=0.75\textwidth]{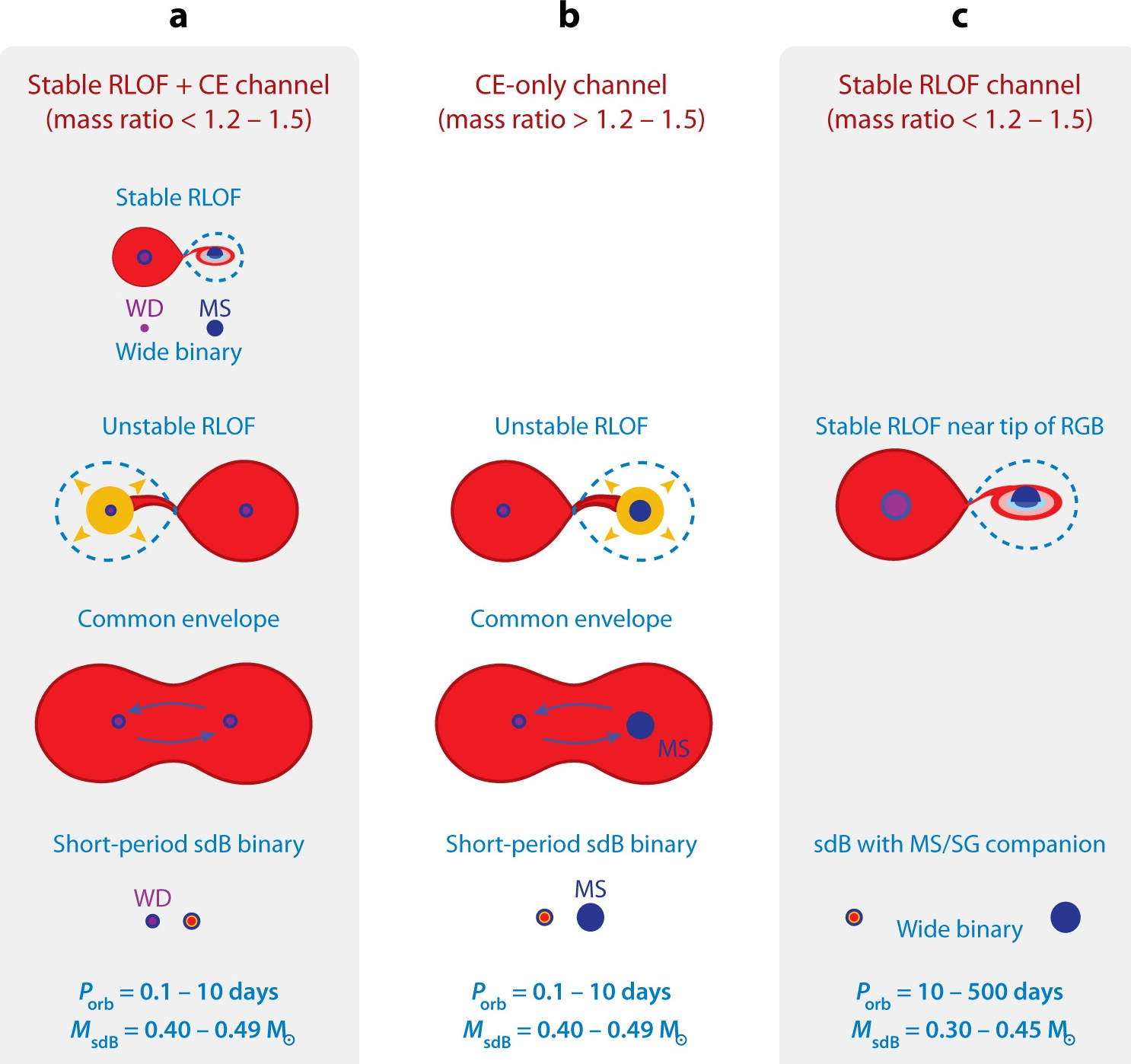}
\end{center}
\caption{ Some formation channels of sdB stars in close binaries 
\citepads{2008ASPC..392...15P}. 
The evolution of the system proceeds from top to bottom. (a) For small initial mass ratios (q $<$ 1.2 -- 1.5), two phases of mass transfer occur. The first Roche-lobe overflow (RLOF) is stable, whereas the second one is unstable, leading to the ejection of the common envelope (CE). The resulting binary consists of an sdB star and a white dwarf (WD) in a short-period orbit. (b) For initial mass ratios larger than 1.2--1.5 the first mass-transfer phase is unstable and the common envelope is ejected, producing an sdB star with a non-degenerate (mostly a main sequence star, MS) companion. (c) For low initial mass ratios the sdB star may also form in the first stable RLOF, resulting in a wide, long-period sdB binary with a non-degenerate companion, a main sequence or subgiant star. From \citetads{2009ARA&A..47..211H} modified by M. Schindewolf; copyright Annual Review of Astronomy \& Astrophysics; reproduced with permission. }
\label{fig:podsi}
\end{figure*}

Binary population models do not predict a second phase of stable RLOF
as contributing to the sdB population. After a RLOF phase,
a common envelope phase can occur when the companion evolves to giant structures and fills its Roche lobe. When the envelope of the red giant has been transferred, 
a close sdB + WD binary emerges with an orbital period between 0.1 and 10 days
(see Fig. \ref{fig:podsi}).

The physics of the common envelopes is still poorly known despite hydrodynamical simulations \citepads[e.g.][]{2012ApJ...746...74R,2012ApJ...744...52P}. 
 Binary population syntheses of post common envelope 
binaries still struggle to explain observation. New investigations by e.g. \citetads{2010MNRAS.403..179D} 
and \citetads{2013A&A...557A..87T} 
are aimed at explaining the white dwarf binaries, while those of  
\citetads{2012ApJ...746..186C} focused on sdB binaries. 

\paragraph{Double-core common-envelope evolution}
is a special case of common envelope evolution where both stars have expanded to giant-type structures by the onset of common-envelope formation. When the envelope is ejected both cores emerge; that is, a binary composed of two hot subdwarf stars is formed.  The binary He-sdB \object{PG~1544+488}  \citepads{2014MNRAS.440.2676S} might be an example of double-core common-envelope evolution  
\citepads{2011MNRAS.410..984J}. 

\paragraph{Hot subdwarf stars and common-envelope planetary nebulae.}

About a fifth of all  planetary nebula are understood to form by common-envelope ejection \citepads{2013A&ARv..21...59I}. 
Most of the central stars of those nebulae are likely to be in their post-AGB phase of evolution. A few systems 
have been suggested as possible post-RGB objects.  
However, \citetads{2013MNRAS.435.2048H} studied the formation of planetary nebulae arising from RGB common-envelope ejection, and concluded that none of the suggested planetary nebulae can be unambiguously identified as a post-RGB system. The best candidate is the central star of EGB~5, because it has been identified as a close sdB binary \citepads{2011A&A...528L..16G}.
However, additional observations are required to clarify whether the nebula is actually associated with the sdB and not an unrelated H~II region in the interstellar medium as has been found for the apparently single sdB  star PHL~932 \citepads{2010PASA...27..203F}. 

Common-envelope nebulae will be visible for only a few 10$^4$ years, before they disperse. Because of their long evolutionary lifetimes only a very small fraction of the sdB stars will be young enough for their common-envelope nebulae to be detectable. Hence, large samples of sdB stars need to be inspected to eventually discover a post-RGB nebulae associated with an sdB star.

\subsubsection{Helium white dwarf mergers}\label{sect:he_wd_merger}

White dwarf mergers are promising processes to explain the formation of several classes of peculiar stars, e.g. R\,CrB stars, extreme helium stars and last but not least the single hot subdwarf stars. The focus of hydrodynamical modelling, though, lies with mergers leading to type Ia supernovae 
\citepads[e.g.][]{2014MNRAS.438...14D}. 

The merger of helium white dwarfs has been suggested to explain the origin of the hot subdwarf stars \citepads{1984ApJ...277..355W}. 
Indeed, some He white dwarf binaries are promising candidates to form single sdB stars when they will merge. Arguably the best known candidate is CSS~41177, an eclipsing system (orbital period P = 2.78 hr) consisting of two He white dwarfs of M$_1$ = 0.38 $\pm$ 0.02M$_\odot$ and M2 = 0.32 $\pm$ 0.01 M$_\odot$  that will merge in 1.14 $\pm$ 0.05 Gyr due to gravitational wave radiation \citepads{2014MNRAS.438.3399B}.

\begin{figure}
\begin{center}
\includegraphics[width=0.9\textwidth]{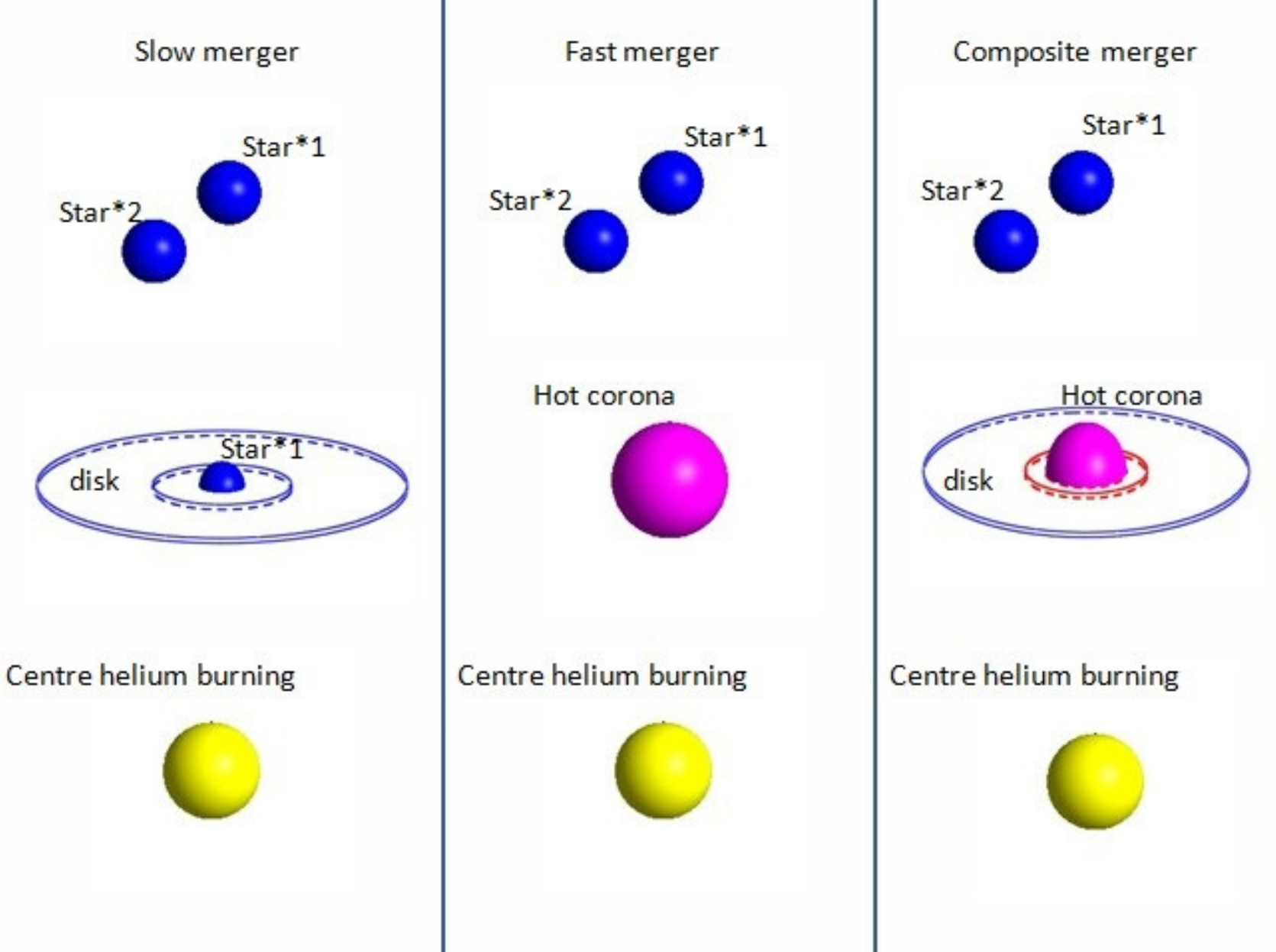} 
\end{center}
\caption{Schematic of three possible ways in which two helium white dwarfs might merge. 
From \citetads{2012ASPC..452...13Z}; copyright PASP reproduced with permission.
} \label{fig:merger_scheme}
\end{figure}

\citetads{2012MNRAS.419..452Z} 
computed several evolution models following the merger of two equal mass helium white dwarfs for different masses from 0.25 to 0.4 M$_\odot$ and three modes of merging, including slow and fast mergers, as well as a combination of both. A sketch of the three processes is shown in Fig. \ref{fig:merger_scheme}.

In the slow merger model the less massive and, hence, larger white dwarf fills its Roche lobe and its entire mass is transferred  to form a disk around the more massive one in a few minutes. The material remains cold and can be accreted slowly onto the primary surface at a rate comparable to the Eddington accretion rate, which could last for several million years. Angular momentum is dissipated towards the disk circumference \citepads{2012MNRAS.419..452Z}. 


In the fast merger model, on the other hand, no disk forms, but the entire mass of the less massive white dwarf directly falls onto the primary's surface. Because the material is strongly heated to 10$^8$K it expands to form a hot corona within a few minutes. 

The third model considered by \citetads{2012MNRAS.419..452Z}, based partly on the results of 3D hydro simulations by \citetads{2009A&A...500.1193L}, 
and others, combines both slow and fast merging to a composite merger model. One part of the disrupted companion forms a hot corona (30\% to  50\% of its mass) while the rest forms a cold disk from which the surviving white dwarf accretes at half the Eddington rate, i.e. 10$^{-5}$M$_\odot$/yr.

 \citetads{2012MNRAS.419..452Z} simulated the post-merger evolution. In the slow (cold) case the accreted material in the envelope of the survivor is that of the former companion and, therefore, nitrogen rich and carbon poor, whereas in the fast (hot) case carbon is predicted to be produced via the triple-$\alpha$ process and could be dredged-up to the surface by an opacity-driven convection zone that occurs when the material is heated up. 
The surface composition predicted by the composite model depends on the final mass of the merger. For low-mass mergers (M$<$0.65 M$_\odot$) the surface remains nitrogen-rich because no carbon-rich material is mixed into freshly accreted nitrogen-rich material. For higher masses, however, strong convection zones are predicted, which mix N-rich  with C-rich material. Accordingly the N-rich He-sdOs should be less massive than the C-rich ones.

\begin{figure*}
\includegraphics[width=0.99\textwidth]{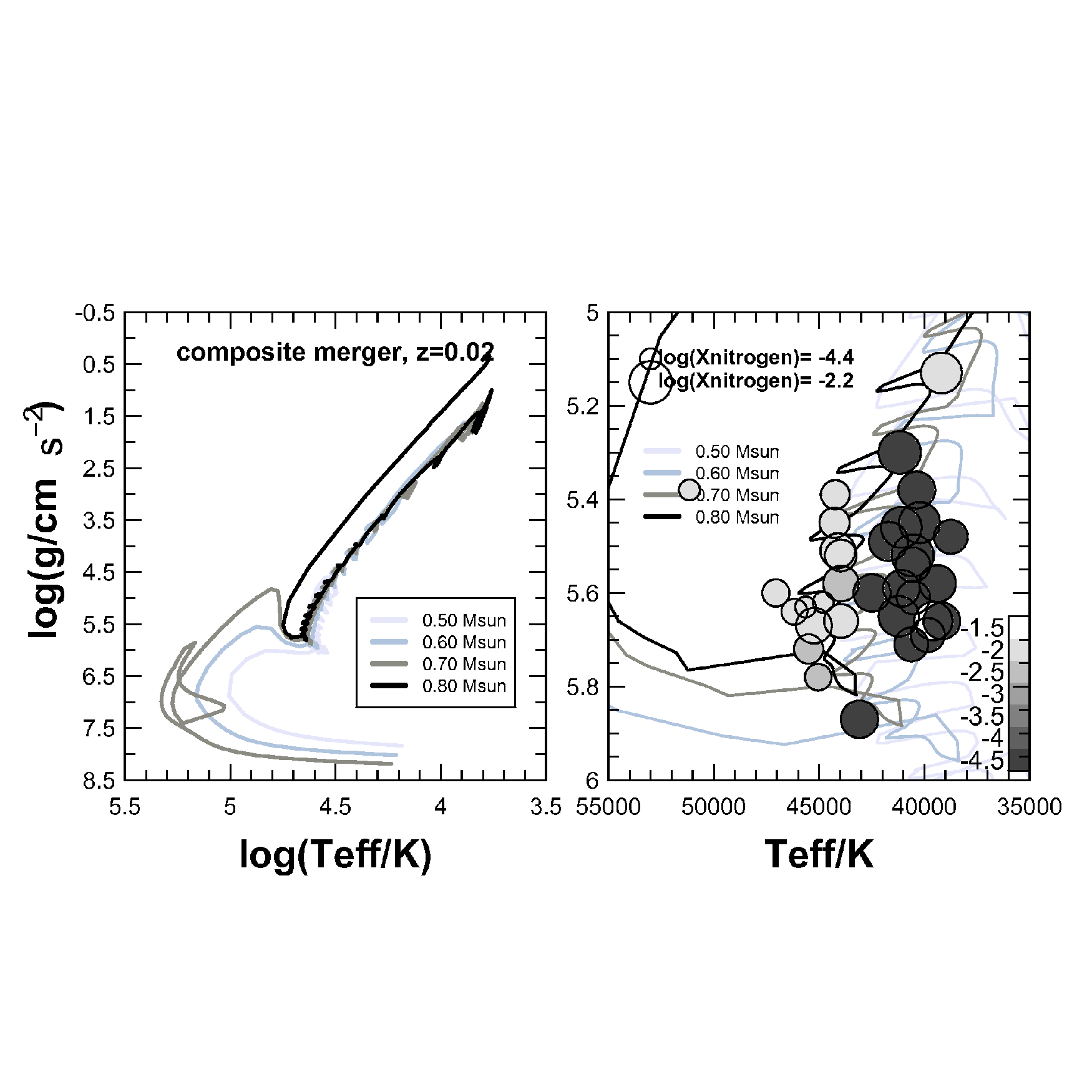}
\caption{Left Panel: Evolutionary tracks for
different masses on the gravity-temperature diagram for the composite merger.
Right panel:
Enlarged area of left panel.
The tracks plotted
from grey to dark have different masses ranging from 0.5 to 0.8
M$_\odot$. The circle symbols show the He-sdO stars from Hirsch (2009).
Grey to dark symbols depict the abundance of carbon (log $\beta_{\rm  C}$) decreasing 
from $-$1.5 to $-$4.5.
The abundance of nitrogen is shown by the size of the symbol.
From \citetads{2012ASPC..452...13Z}; copyright ASPC, reproduced with permission.} \label{fig:merger_hirsch}
\end{figure*}

The post-merger evolutionary tracks all end up evolving towards the helium main sequence through those regions of the T$_{\rm eff}$--$\log$ g diagram where the helium-rich hot subdwarf stars are found
(see Fig. \ref{fig:merger_hirsch}).

 \subsubsection{Single sdBs}

The existence of single sdB stars remains a conundrum. The merger of helium white dwarfs is an attractive option, although the predicted broad mass distribution does not seem to be consistent with the narrow mass distribution of sdB
stars determined from asteroseismic analysis. 
Other merger processes are also conceivable. A low mass star or brown dwarf may merge with a red-giant core \citepads{1998AJ....116.1308S,2008ApJ...687L..99P} to form an sdB star.  
The population synthesis calculations of \citetads{2008ApJ...687L..99P} predicted that this would lead to rapidly rotating horizontal branch stars with a core mass distribution strongly peaked  between 0.47 and 0.54 M$_\odot$, some of which may be single sdB stars if centrifugally enhanced mass loss driven by the fast rotation removed a sufficient amount of the envelope. 
 
\citetads{2011ApJ...733L..42C}  
suggest that the coalescence of a helium white dwarf with a low-mass, hydrogen-burning star would create a star with a helium core and a thick hydrogen envelope that evolves into an sdB star in a few Gyrs, 
which would also naturally explain the sdBs' slow rotation rates. Candidate systems might exist among low mass white dwarfs with dM companions in short-period systems \citepads{2013MNRAS.429..256P}. 

However, we cannot exclude that apparently single sdBs do have companions that have not been detected yet. {{Firstly, compact massive companions such as white dwarfs could have escaped detection because their orbital planes are orthogonal to the line of sight. }} 
Secondly, low mass main sequence stars and brown dwarfs on wide orbits would have very small radial velocity amplitudes, hard to detect. Such a population of binaries, however, is not predicted to exist by binary population synthesis models. {{{\it Gaia} observations are eagerly awaited.}}

\clearpage

\newpage
\section{Hot subluminous stars in clusters}\label{sect:cluster}

Stars in a cluster are expected to have common properties and are thought to have formed at the same time from the same interstellar cloud. Moreover, they are at almost the same distance from us, 
and, therefore, provide an important benchmark for stellar evolution theory allowing us to determine fundamental parameters, such as stellar mass, metallicity, distance, and age.

Hot subdwarfs stars have been found as extreme horizontal branch stars in several globular clusters as well as in the old, metal-rich open clusters NGC 6791 and NGC 188.

\subsection{The horizontal branch morphology of globular clusters and the second parameter problem}

The widely different morphology of the horizontal branches of globular clusters still awaits an explanation. Metallicity has been recognized as the main parameter, because metal-rich {clusters} have red horizontal branches, whereas metal-poor {ones} show blue HBs. However, the presence of clusters with the same metallicity but different horizontal branch morphologies requires one or more additional parameters. The list of candidate second parameters is long and includes properties of the globular clusters: age \citepads[e.g.][]{1994ApJ...423..248L}, 
mass \citepads[e.g.][]{2006A&A...452..875R},
core density/concentration \citepads[e.g.][]{1993AJ....105.1145F}, 
as well as 
internal rotation and helium mixing of (post-)red giant branch stars \citepads{1997fbs..conf....3S}, 
helium self-enrichment \citepads[e.g.][]{2002A&A...395...69D}, 
the presence of planets \citepads[e.g.][]{1998AJ....116.1308S} 
among others. 
The age has been favored for a long time \citepads{1994ApJ...423..248L,2010A&A...517A..81G,2010ApJ...708..698D}, 
but the helium content may play a role if a significant fraction of helium-enriched stars exist \citepads[for reviews see][]{2009Ap&SS.320..261C,2010A&A...517A..81G}. 

It is now well established that globular clusters (e.g. the massive $\omega$ Cen and NGC~2808) host multiple stellar populations, because their color--magnitude diagrams display two or more continuous sequences of stars from the main sequence to the red giant branch
\citepads[for a review see ][]{2012A&ARv..20...50G} 
and the list of multi-population globular clusters is growing steadily, both for the Galaxy \citepads[e.g.][]{2012ApJ...760...39P,2015MNRAS.447..927M} as well as for others \citepads[Fornax dSph,][]{2014ApJ...797...15L}. 
The ongoing {\it HST} UV legacy survey \citepads{2015AJ....149...91P} will allow {{the populations to be disentangled}} and improve 
our understanding of globular cluster formation and evolution. 
The differences {{among}} the stellar population with respect to their content of helium and light elements point towards 
two or more episodes of star formation. This scenario assumes that the second generation of stars formed from material polluted by the first one \citepads[see e.g.][and references therein]{2008MNRAS.391..825D,
2015MNRAS.446.1672M}. 

  The helium content has been identified as a viable second parameter.
However, it is difficult to test because helium lines can be observed in early-type stars, only. Blue horizontal branch stars are the best choice for a direct determination of the helium
abundance from its spectral lines. However, for stars hotter than 11,500 K the atmospheric composition is affected by the same diffusion processes that govern the atmospheres of sdB stars and the information on the original surface abundance has been erased \citepads{2003ApJS..149...67B,2008ApJ...675.1223M}. 
 Hence, the technique can be applied to horizontal branch stars in a limited temperature range ($\approx$8500\,K to 11,500\,K), where atmospheric convection impedes diffusion, so that the helium abundance should be {{the same as that from which the star was formed}}.
Indeed, \citetads{2014MNRAS.437.1609M} 
found an enrichment of helium with respect to the original value in  BHB stars with Teff $<$ 11 500K in the massive globular cluster NGC~2808. It has, therefore, been suggested that the evolution, stellar and atmospheric properties of hot subdwarf stars may also depend on the helium content of the cluster population(s) \citepads{2002A&A...395...69D,2010MNRAS.405.2295D}. 

\begin{figure*}
\centering
\includegraphics[width=0.99\textwidth]{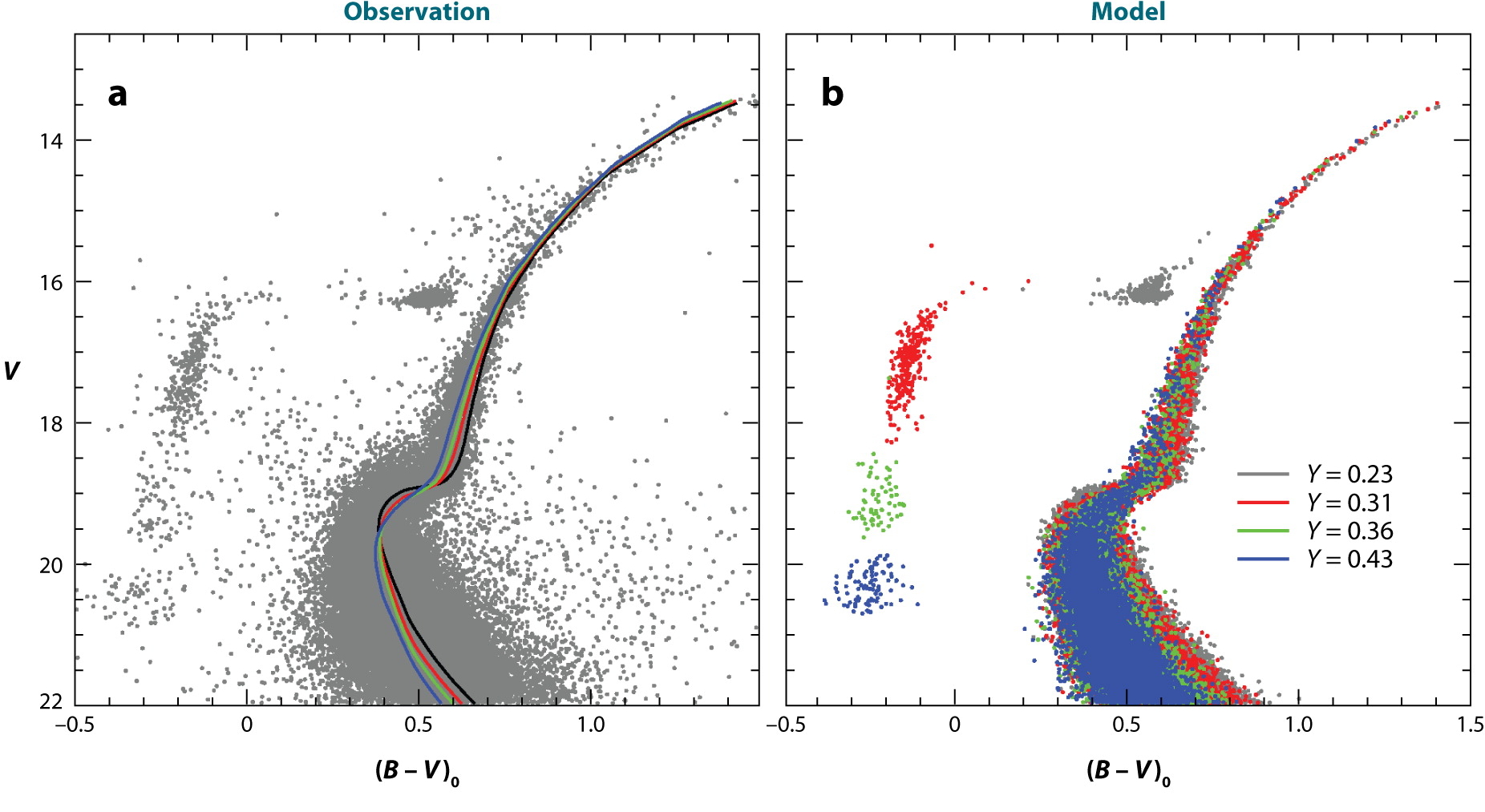}
\caption{The observed and modeled color--magnitude diagrams of
the globular cluster NGC\,2808 \citetads{2008ASPC..392....3Y}. Left-hand panel: Observed color-magnitude diagram, showing an exceptionally
wide distribution of horizontal branch stars. 
Right-hand panel: Theoretical color-magnitude derived from evolutionary calculations for four different helium compositions. 
The observations can be reproduced if a large range of helium
abundance is assumed. From \citetads{2009ARA&A..47..211H}; 
 Copyright ARA\&A, reproduced with permission. 
}
\label{fig:yi_2808}
\end{figure*}

\citetads{2008ASPC..392....3Y} computed stellar evolution sequences for different helium contents  to model the color-magnitude diagram
of NGC~2808. Using four sub-population {{having}} different helium enrichments, they succeeded to reproduce the observed horizontal branch morphology well. Accordingly, the bluest HB stars are the most helium-rich ones (see Fig.~\ref{fig:yi_2808}).

\subsection{The formation of EHB and blue hook star in globular clusters} 

The color-magnitude diagrams of several globular clusters show blue tails and blue hooks to their horizontal branches, which are supposed to consist of hot subdwarf stars \citepads[see][for reviews]{1999RvMA...12..281M,2001PASP..113.1162M,2010MmSAI..81..838M}. 
Blue hook stars are on the HB but with fainter luminosity than the normal EHB stars
\citepads{1998ApJ...495..284W,2000ApJ...530..352D} 
and populate the very blue end \citepads[T$_{\rm eff} >$ 32000 K, ][]{2012A&A...547A.109M}, for a detailed discussion of the occurrence of blue hook stars see \citetads{2010MmSAI..81..838M}. 

We have discussed the formation and evolution of sdB stars and compared them with observations of field stars extensively in 
Sect. \ref{sect:evolution}. Briefly,  the frequency of binary sdB stars is so large, 50\% of the single-lined systems have periods of less than 30 days, that  common envelope ejection is considered the only viable process to form such systems. Double-lined systems may form as wide binaries through stable RLOF and periods of about 1000 days. Single sdB stars may result form mergers of double He-core white dwarfs. For single stars the most popular scenario rivalling the close binary scenario is the late hot flasher scenario that explains the reduction of the hydrogen envelope of the progenitor star through internal mixing and burning.   

Do these {{scenarios}} also apply to EHB and blue hook stars in globular clusters?
The low frequency of close binary sdB stars in globular clusters \citepads{2009A&A...498..737M,2011A&A...528A.127M} 
argues against the presences of a large populations of close binary sdB stars.
{The recent discovery of a unique close binary consisting of a K-type main sequence star and an sdB star in a 1,6 day orbit \citepads{2015ApJ...812L..31M} comes much to our surprise, because no counterpart is known in the field.} 
Wide binaries with periods of 1000 days or more may be disrupted by dynamical interaction in the dense cluster environment. Therefore, the 
hot flasher scenario is often favored to explain the formation of EHB and blue hook stars in globular clusters.

An alternative scenario explains the EHB and blue hook stars by
the presence of a second generation of helium-enriched stars. 
Recently, three additional scenarios were added: (i) the He white dwarf merger scenario \citepads{2008A&A...484L..31H}, 
(ii) the enhanced mass loss by rapid rotation \citepads{2015Natur.523..318T}, 
and (iii) tidally enhanced mass loss in binaries \citepads{2013A&A...549A.145L,2013A&A...554A.130L,2015MNRAS.449.2741L}. 

\subsubsection{Late hot-flasher}

Many studies of extreme horizontal branch and blue hook stars in globular clusters explain their formation via the flasher scenario \citepads[e.g.][]{2008A&A...491..253M} 
although they cannot explain the range of colors observed in such stars, in particular for the most metal-rich clusters \citepads{2010ApJ...718.1332B}. 
The consequences for their location in the T$_{\rm eff}-\log$ g plane and the chemical compositions has already been discussed in Sect. \ref{sect:flasher}. 

\subsubsection{Helium-enriched second-generation stars}

The enhanced helium enrichment model assumes that the second generation stars are enriched in helium up to a mass fraction of Y$\approx$0.4. Otherwise, the evolution is considered canonical. The resulting horizontal branches are blue and more luminous than those for normal He content.

\subsubsection{Double helium white dwarf mergers}

The low fraction of close sdB binaries in globular clusters may result from a high fraction of mergers. Binary population synthesis calculations by 
\citetads{2008A&A...484L..31H} 
suggest that the merger rate for double white dwarfs increases with age
and after only about 8 Gyrs the merger channel dominates
over all others. Accordingly, the fraction of close binaries with short orbital 
periods for a stellar population {{age}} of 10 Gyr is only $\approx$2.5\% in the standard model and ranges 
between 0 and 20\% depending on the choice of {{common envelope ejection}} efficiency parameter $\alpha$. Hence, this scenario 
provides a straightforward explanation for the scarcity of short-period sdB
binaries in globular clusters. The predicted location of the stars in the
T$_{\rm eff}-\log$ g plane and the chemical composition should be similar to that discussed in Sect. \ref{sect:he_wd_merger}.

\subsubsection{Enhanced mass loss by rapid stellar rotation}

Because a late hot flasher is supposedly a rare event 
it is surprising to find such a high fraction of EHB stars in $\omega$ Cen and other massive clusters. 
Therefore, \citetads{2015Natur.523..318T} revisited the RGB helium-mixing scenario of \citetads{1997ApJ...474L..23S,1997fbs..conf....3S}, 
in which rapid rotation of a red giant might lead to a dredge-up of helium from the hydrogen-burning shell into the envelope. 
 \citetads{2015Natur.523..318T} suggest that rapid rotation of second generation stars may result from the star-forming history and the early dynamics of very massive clusters, such as $\omega$ Cen
(see Sect. \ref{sect:cluster}). 
Hydrodynamical models of the gas dynamics combined with N-body simulation \citepads{2008MNRAS.391..825D} 
showed that second generation stars that formed from material ejected from super-AGB and AGB stars of the first generation, are located in a centrally more concentrated subcluster. Hence, the second generation main sequence stars are born in very dense stellar environment and rotate faster because their disks are destroyed early-on. Core rotation is assumed to persist during RGB evolution, because angular momentum transport is considered to be slow. However, asteroseismology of red giants showed that angular momentum transfer is a lot faster than previously assumed \citepads{2015AN....336..477A}, which may challenge this scenario. 

\subsubsection{Tidally enhanced mass loss in binaries}

\citetads{2013A&A...549A.145L,2013A&A...554A.130L,2015MNRAS.449.2741L} 
investigated the role of binary evolution for the formation of sdB stars in globular clusters by considering a tidally enhanced wind of a red giant star to cause the required huge mass loss. The model binary was chosen to consist of 
0.85 M$_\odot$ star and a 0.53 M$_\odot$ companion. For initial orbital period longer than 2200 d, no effect was found and the evolution of the primary would be canonical.
For initial periods between 2200 and 2000 days, the primary star would experience an early hot flash and become a canonical EHB
star. However, the primary undergoes a late hot flash on the white dwarf cooling curve and ends up as a blue hook star if the initial period is between 1600 and 2000 days. For even shorter periods helium is not ignited and the red giant evolves into a helium white {{dwarf}}. These results are similar to those presented by \citetads[]{2014A&A...565A..57S} 
to explain the long-period low-mass WD \& K0~III/IV binary IP Eri (see Sect. \ref{sect:elm}). However, such wide binaries may be disrupted well before the primary {evolves} into a red giant. Moreover, the results may be difficult to reconcile with the observed orbital properties of wide sdB + F/G/K binaries in the field \citepads[see][]{2015A&A...579A..49V}. 
 
\subsection{Observational tests: Spectroscopic analyses of EHB and blue hook stars in globular clusters}

In order to clarify the formation of EHB and blue hook stars in globular clusters quantitative spectral analyses are important to place the stars in the T$_{\rm eff}-\log$ g plane, derive their chemical composition, and compare them to those of the field stars. 

\subsubsection{Classical globular clusters hosting EHB stars}
 
Because of its very blue horizontal branch morphology and its proximity, NGC\,6752 has been targeted for quantitative spectral analyses of hot subdwarf stars for a long time \citepads{1986A&A...162..171H,1997A&A...319..109M,2007A&A...474..505M}.  
More recently, these studies have been extended to the similar clusters  M80 and NGC 5986  \citepads{2009A&A...498..737M}. \citetads{2013A&A...559A.101S} present the spectroscopic analyses of blue horizontal branch stars in M22, an in-depth comparison with NGC\,6752, M~80 and NGC~5986,  
and conclude that the gravities and masses of  HB  stars in M 22  with T$_{\rm eff}$=7000 to 25000 K match those for NGC 6752, M 80, and NGC 5986. For all four clusters both the location of the HB band in the T$_{\rm eff}-\log$ g plane, their helium abundances as well as their masses agree with theoretical expectations \citepads[see ][ for details]{2013A&A...559A.101S}. Information on the chemical composition beyond helium remains scarce for EHB stars in globular clusters, while there is more detailed information for BHB stars \citepads[e.g.][]{2003ApJS..149...67B,2005A&A...434..235F,2006A&A...452..493P,2009A&A...499..865H}. 

\citetads{2014ASPC..481...59C} 
 carried out an analysis of
far-UV spectra obtained with the {\it FUSE} satellite of  three hot subdwarf stars in NGC 6752 and derived NLTE
abundances for the C,N \& O elements as well as for Si, P, and S.  The abundance patterns for all three stars turned out to be very similar to those of the field stars \citepads{2013A&A...549A.110G}, nitrogen being slightly subsolar, while all other elements are depleted by more than one order of magnitude with respect to the Sun. Although the number of stars studied in metal-poor clusters is small, the similarity of their chemical abundance pattern may hint that the outcome of atmospheric diffusion does not depend on metallicity.


\subsection{The enigmatic globular clusters NGC~2808 and $\omega$~Cen and their blue hook stars}

The massive globular clusters NGC~2808 and  
$\omega$ Cen have caught a lot of attention in recent years because they are the most extreme cases of globular clusters with multiple stellar populations and  host large populations of blue horizontal branch stars. In addition their horizontal branches show a so-called blue hook at the blue end in UV color magnitude diagrams, that cannot be
explained by canonical stellar evolution theory.
 \citetads{2004A&A...415..313M} 
pointed out that the frequency of blue hook stars is related to the clusters' total mass rather than to the population of the EHB.  NGC 6752 for instance shows a well populated EHB, but not a single blue hook star has been found, whereas there are roughly as many blue hook stars as EHB stars in NGC 2808 \citepads{2001ApJ...562..368B}, 
which is a hundred times more luminous than NGC 6752.

No less than five populations of main sequence and red giant stars have been identified from {\it HST} multi-wavelength photometry of NGC 2808, four of which might be enriched in helium by $\delta$Y=0.03 to 0.13 \citepads{2015ApJ...808...51M}. 
Many EHB and blue hook stars have been identified from UV photometry. 
Likewise, $\omega$ Cen shows complex multiple stellar population patterns and hosts blue hook stars \citepads{1998ApJ...495..284W,2000ApJ...530..352D} 

Both clusters 
have recently been targeted in several spectroscopic studies of blue horizontal branch stars \citepads[NGC~2808,][]{2004A&A...415..313M} and 
\citepads[$\omega$ Cen,][]{2011A&A...526A.136M,2012A&A...547A.109M,2014ApJ...795..106L}.
Multi-object spectroscopy is the most efficient way to obtain spectra in crowded fields such as in globular clusters. Therefore, \citetads{2004A&A...415..313M} and 
\citetads{2014ApJ...795..106L} used the multi-object mode of FORS at the ESO-VLT
to secure optical spectra of EHB and blue hook stars in NGC 2808 and $\omega$ Cen, respectively. Spectra of stars in $\omega$ Cen were also obtained by \citetads{2011A&A...526A.136M} using the multi-fiber instrument FLAMES + GIRAFFE at the ESO-VLT, which provides higher spectral resolution (R=6400) than FORS.
\citetads{2004A&A...415..313M} derived atmospheric parameters for 19 members of NGC 2808 using NLTE models of H/He composition for the hotter stars and Kurucz LTE models for the cooler ones, while \citetads{2011A&A...526A.136M} and \citetads{2014ApJ...795..106L}  determined atmospheric parameters and abundances of EHB and blue hook stars in $\omega$ Cen using different sets of NLTE and LTE models. 

\begin{figure*}
\begin{center}
\plottwo{subdwarfs_SPY.pdf}{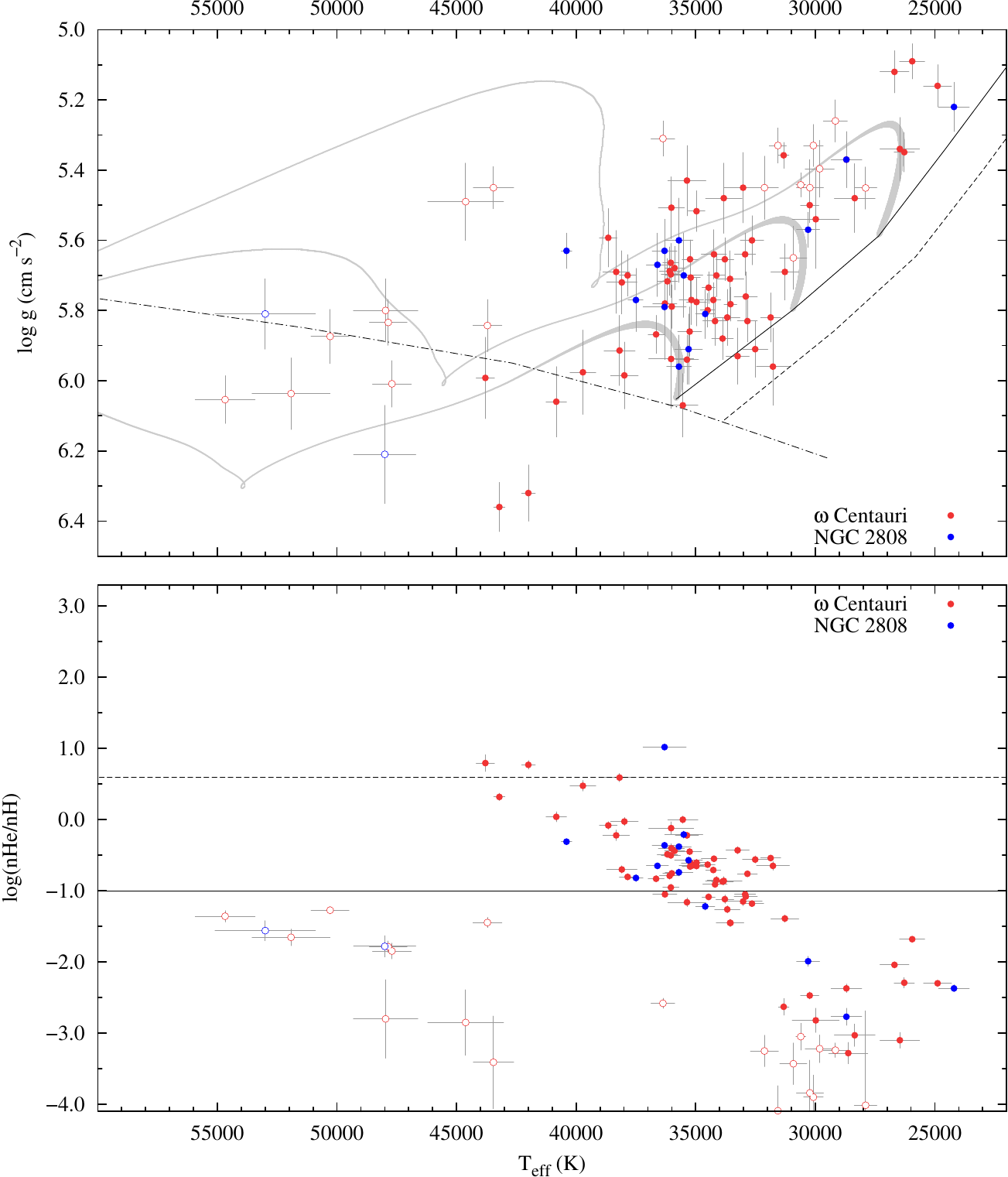}
\end{center}
\caption{{Upper, left hand  panel:} Distribution of hot subwarf stars in the T$_{\rm eff}$ / $\log g$- plane from the ESO-SPY project.
{Lower, left hand panel:} Same for the helium abundance vs. T$_{\rm eff}$ distribution.
{Upper, right hand  panel:} Distribution of hot subwarf stars in the T$_{\rm eff}$ / $\log g$- plane from different samples of stars in the globular cluster $\omega$ Cen (red symbols) and NGC 2808 (blue symbols). 
{Lower panels:} Same for the helium abundance vs. T$_{\rm eff}$ distribution.
From N{\'e}meth (priv. comm.).
} 
\label{fig:spy_ocen}
\end{figure*}  

The helium-poor hot subdwarfs form two groups, one on the EHB at typical sdB temperatures and another one at higher temperatures exceeding 40,000~K, typical for post-EHB stars (see Fig. \ref{fig:spy_ocen}). At intermediate effective temperatures  the majority of targets are He-rich (log N(He)/N(H)~$\gtrsim -1.0$) and cluster in the  temperature range from ~32000 K to ~43000 K along the bluest part of the EHB band (see Fig. \ref{fig:spy_ocen}). However, the enhancement of helium is moderate, i.e. the He/H ratio does not exceed unity except for a handful of stars with He/H $<$ 10. In comparison to the field stars, two striking differences become apparent from Fig. \ref{fig:spy_ocen}. Firstly, the helium-rich hot subdwarfs in the field are hotter than the cluster stars and lie beyond the EHB and secondly the helium enrichment is higher in many field stars than that of the most strongly enriched star in NGC 2808 and $\omega$ Cen. 

\citetads{2014ApJ...795..106L} concluded that ''these differences point toward fundamental differences between the helium-enriched EHB star population in the field and in $\omega$ Cen and are likely {{to be}} related to the fact that sdB and sdO stars in globular cluster have older (12--13 Gyr) and typically metal-poorer progenitors than their field counterparts.''

\begin{figure}
\begin{center}
\includegraphics[width=0.9\textwidth]{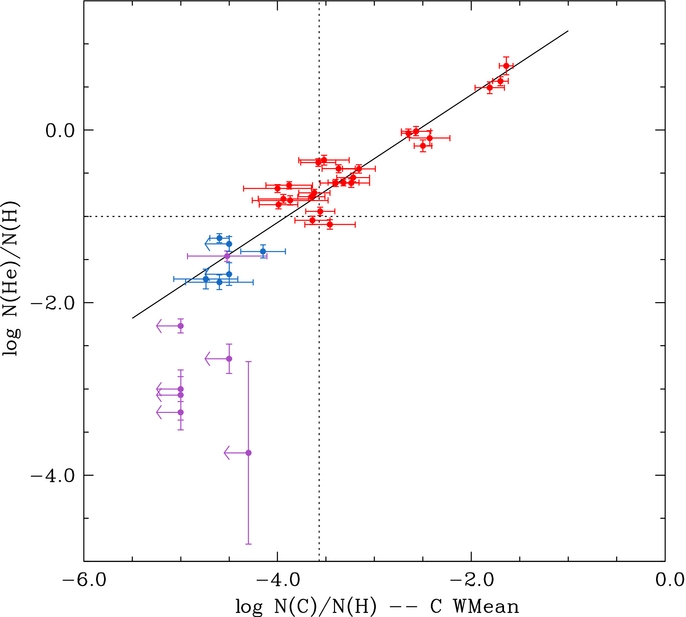}
\end{center}
\caption{Helium abundance vs. the mean carbon abundance for $\omega$ Cen stars
\citepads{2014ApJ...795..106L}. The ''cool'', helium-poor EHB stars are shown in magenta, the intermediate temperature, helium-rich ones in red, and the hottest, helium-poor ones in blue. The upper limits on the carbon abundance inferred for eight stars are indicated by arrows instead of error bars. This diagram shows an obvious relation between the abundances of the two elements, which is illustrated by the linear regression (black solid line). Dotted lines indicate the solar helium and carbon abundances.
From \citetads{2014ApJ...795..106L}; copyright ApJ; reproduced with permission. 
}
\label{fig:correlation_latour}
\end{figure}

Both studies find a strong positive correlation between the carbon and helium abundances (see Fig. \ref{fig:correlation_latour}). The carbon ranges from 1/10 solar for the cool, helium-poor stars to almost a hundred times solar for the most helium-rich stars of intermediate temperature.
Almost the same positive correlation of carbon (and nitrogen) with helium abundance, has been found for the hot subdwarfs in the GALEX sample \citepads[see Fig. 9 of][see also Sect. \ref{sect:atmos}]{2012MNRAS.427.2180N}. 
In this respect metal-poor and metal-rich populations do not differ.

\begin{figure}
\begin{center}
\includegraphics[width=0.9\textwidth]{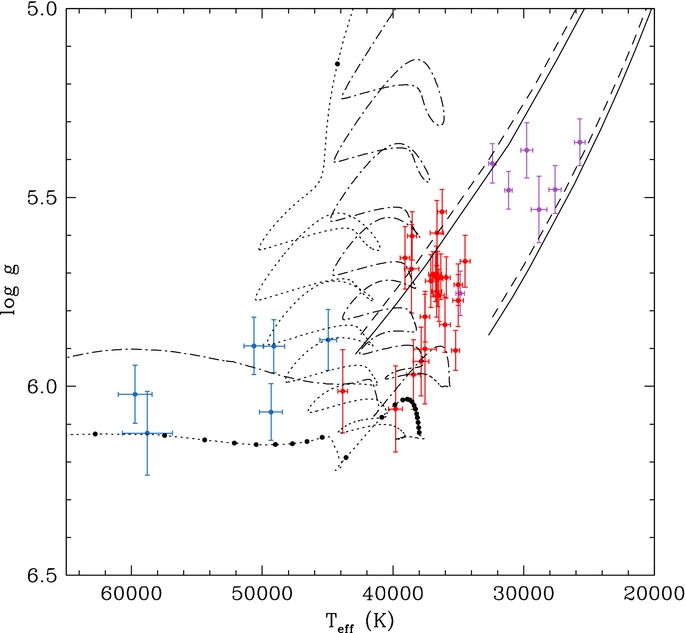}
\end{center}
\caption{Comparison of the sample of hot subdwarf stars of \citetads{2014ApJ...795..106L} in the globular cluster $\omega$ Cen with late-flasher evolutionary tracks by \citetads{2008A&A...491..253M}. 
The tracks are for a metallicity of Z=0.001 and refer to a deep mixing event (M=0.48150 M$_\odot$; dotted line) and a shallow mixing event (M=0.49145 M$_\odot$; dashed-dotted line), respectively. Points at 5 Myr intervals are shown on the first track, to give an idea of the evolutionary timescale in the different regions.
From \citetads{2014ApJ...795..106L}; copyright ApJ; reproduced with permission. }
\label{fig:kiel_latour}
\end{figure}

In order to understand the formation of the blue hook stars in $\omega$ Cen, 
\citetads{2011A&A...526A.136M} and \citetads{2014ApJ...795..106L} compared their results to the predictions of the He-enhanced scenario as well as to the hot flasher scenario (see Fig. \ref{fig:kiel_latour}).


The enhanced helium scenario can readily be dismissed, because the helium mass fraction should not exceed $\approx$0.4 and no enrichment of carbon is expected, nor for any other chemical element.

If the hot flasher scenario were correct, the burning and mixing in a typical hot flasher event should enrich the atmosphere not only with helium but also with carbon, and to a lesser extent, nitrogen \citepads{2003ApJ...582L..43C}. 
The observed correlation between helium and carbon enrichments, therefore, is in qualitative agreement with the hot flasher scenario. At the quantitative level, however, \citetads{2011A&A...526A.136M} and  \citetads{2014ApJ...795..106L} noted some inconsistencies, in particular they find that the predicted helium (96\% by mass) and carbon content (3\% to 4\% by mass) is too high, that is only the three most helium-rich stars in the sample come close to the prediction. The helium and carbon abundances of all other {{stars}} are a lot lower than predicted. Another inconsistency becomes apparent when comparing the observed locations of the stars in the T$_{\rm eff}$-$\log$ g - plane to the prediction for the hot flasher  evolutions of both a deep and a shallow mixing event (Fig. \ref{fig:kiel_latour}). The observed He-rich stars are cooler than predicted by both tracks, even for the case of shallow mixing only.
This discrepancy was already noted by \citetads{2002A&A...395...37M} 
who considered that up to 10\% of hydrogen may survive the hot flasher evolution and calculated the position of stars in the T$_{\rm eff}$ - $\log$ g - diagram by adding hydrogen layers of different thickness to post-flasher models. Indeed, the models predict the stars to be somewhat cooler and, hence, closer to the observed position of blue hook stars.  


\subsection{Post-EHB stars in globular clusters: The UV-bright stars}\label{sect:uvbright}

For more than a hundred years it is known that some globular clusters host 
extremely blue stars of high luminosities \citepads{1900ApJ....12..176B,1921VeBon..15....1K} that stand out against the large population of red stars. 
These so-called UV-bright stars are rare; \citetads{1987fbs..conf...95D}
lists 45 UV-bright stars in 36 globular clusters. 
Quantitative spectral analyses have been presented by e.g. by \citetads{1986A&A...169..244H,1998A&A...339..537M,1998A&A...335..510M,2002A&A...381.1007R,2007MNRAS.378.1619T,2015MNRAS.452.2292C}. 
The most striking object is the helium-rich sdO star ZNG-1 in the globular cluster M\,5 because it rotates so fast (170 km\,s$^{-1}$). \citetads{2004ApJ...600L..43D} favor a ``born-again'' origin but previous binary interaction may offer a straightforward explanation for the rapid rotation of the star.
The UV-bright stars represent a mixed bag of post-AGB stars, including four central stars of planetary nebulae \citepads{1928PASP...40..342P,1989ApJ...338..862G,
1993ApJ...411L.103H,1997AJ....114.2611J}, 
post-early AGB stars and post-EHB stars, %
amongst them also stars of relatively low temperatures ($<$10000K) \citepads[e.g.][]{2004A&A...417..293A,2004A&A...423..353J}. 
Optically selected UV-bright stars tend to be mostly post-AGB and post-EAGB stars, whereas UV-selected stars turn out to be less luminous (''supra-HB'' stars) and are mostly identified as post-EHB stars 
\citepads{1998A&A...335..510M}, which share the helium deficiency with their counterparts of the field populations \citepads[for reviews see ][]{2010MmSAI..81..838M,2003ARA&A..41..391V}. 

\subsection{Hot subdwarf stars in open clusters}

The concept that globular clusters host single populations of coeval stars of the same chemical composition has recently been challenged by the detection of 
multiple populations of stars with different properties in several massive clusters. Therefore, \citetads{2015ASSP...39...43S} 
argues that open clusters probably remain the only example of simple single population environments and provide the best laboratory available to test stellar evolution. However, EHB stars have been found in two open clusters only,
i.e. in the near-by 
old 
metal-rich
 clusters, 
 NGC 6791 \citepads{1992AcA....42...29K,1994AJ....107.1408L}, one of the richest open clusters known, 
 and NGC~188 \citepads{1997fbs..conf..271G,2004Ap&SS.291..267G}. 
While NGC~188 hosts only one sdB star (a binary with an 2.15 d orbital period), NGC 6791 harbors 5 (perhaps 6) sdB stars \citepads{2015ApJ...806..178S}. 
NGC~6791 has received great attention in recent years, because it is one of only two open clusters in the {\it Kepler} field.
The presence of EHB stars make this cluster an important laboratory to study their evolution. 
  
\citetads{2015ApJ...806..178S} used NGC~6791 and NGC 188 as template clusters to understand the formation of the EHB stars by 
investigating 15 open clusters  that come close to NGC~6791 and NGC 188 with respect to age and metallicity. Four of them have similar age but lower metallicities, three are of similar metallicity but slightly younger age, and eight clusters are of slightly lower metallicity and age.
Combining the color magnitude diagrams of all 15 clusters yields four times as many red giant clump stars as NGC 6791 but not a single EHB star.  
\citetads{2015ApJ...806..178S} 
conclude that older stellar populations (6--9 Gyrs, turn-off masses of $\approx$ 1.1 to 1.3 M$_\odot$) of very high metallicity produce a much larger fraction of EHB stars than younger ones. Accordingly red giant progenitors are preferentially of lower mass. 

However, it should be noted that NGC~6791 may be a rather unusual cluster and not suitable as the cluster of reference.
\citetads{2001ASPC..226..192G} 
point out that both NGC~6791 and NGC~188 have enormously high fractions of blue stragglers with respect to horizontal branch stars.
NGC~6791 also {{hosts}} a large number of interacting binaires \citepads[cataclysmic variables and contact binaries,][]{2007A&A...471..515D}.

Hundreds of white dwarfs have been identified on deep {\it HST} images of NGC 6791 \citepads{2005ApJ...624L..45B,2008ApJ...678.1279B}, reaching the very faint end of its white dwarf cooling sequence. 
Two bumps in the white dwarf luminosity function pointed to a cluster age of only 6 or 4 Gyrs, respectively, in strong conflict with the turn-off age of 8.3 Gyrs \citepads{2012A&A...543A.106B}.
\citetads{2007ApJ...671..748K} 
derived spectroscopic masses and find a substantial fraction of the white dwarfs to be undermassive and, therefore, to be of helium composition. The two peaks in the white dwarf luminosity function may, therefore, result from two populations of white dwarfs,
the brighter one being caused by helium-core white dwarfs \citepads{2005ApJ...635..522H,2008ApJ...678.1279B}. 
However, this conclusion was found to be at odds with evolution theory. 
\citetads{2008ApJ...679L..29B} revisited the problem and found that the bumps in the luminosity function can be naturally accounted for if $\approx$34\% of the white dwarfs are actually double degenerate systems. Such a large fraction is plausible in view of the many interacting binaries. \citetads{2008ApJ...679L..29B} 
conjectured that the high binary fraction may be related to the other peculiarities of this cluster, including the existence of the EHB. 
In fact, the sdB in NGC~188 and NGC6791/B4  are close binaries of 2.15 and 0.3985 days period, respectively
\citepads{2004Ap&SS.291..267G,2011ApJ...740L..47P}. 

Whether NGC~6791 and NGC~188 are just peculiar cases among the open clusters or represent a standard stellar population suitable as a reference for the field population and UV-upturn galaxies, remains an open-ended question. 

More information on the sdB stars in NGC 6791 has been gathered recently from the analysis of {\it Kepler} light curves that led to the discovery of multi-periodic oscillations in three of them, including the binary B4
\citepads{2011ApJ...740L..47P,2012MNRAS.427.1245R}.
Hence B4{, as well as B3 and B5, which are apparently non-binary sdBVs, are} of extra-ordinary interest, because cluster membership provides a stringent constraint on age, metallicity, and progenitor mass, which must be close to the turnoff mass of 1.1--1.2 M$_\odot$ \citepads{2011ApJ...740L..47P}. 
The analysis of the g-mode pulsations of B4 as well as its binary properties allows  
models of the interior structure and evolution of sdB stars to be constrained
as will be discussed in more detail in Sects. \ref{sect:binaries} and \ref{sect:asteroseismology}. 

\clearpage

\newpage
\section{Binaries}\label{sect:binaries}

In the course of the Palomar-Green survey \citepads{1986ApJS...61..305G} it became clear that a significant
fraction of sdB stars \citepads[at least 20\%,][]{1984ApJ...287..320F} 
have composite colors and spectra. In particular companions of F, G, or K type can easily be detected.
The 2MASS survey led to a better estimate of the fraction of 
 about 30\% in an approximately
volume-limited sample \citepads{2003AJ....126.1455S}. 

Since the year 2000 evidence is growing that many more  
hot subdwarfs reside in close binaries with invisible companions. About half of the single-lined sdB stars have invisible companions orbiting with periods of 30 days or less 
\citepads{2001MNRAS.326.1391M,2003MNRAS.338..752M,2004Ap&SS.291..321N,2011MNRAS.415.1381C}. The invisible companions are either low mass main sequence stars or white dwarfs.

\begin{figure*}
\centerline{\includegraphics[width=0.99\textwidth]{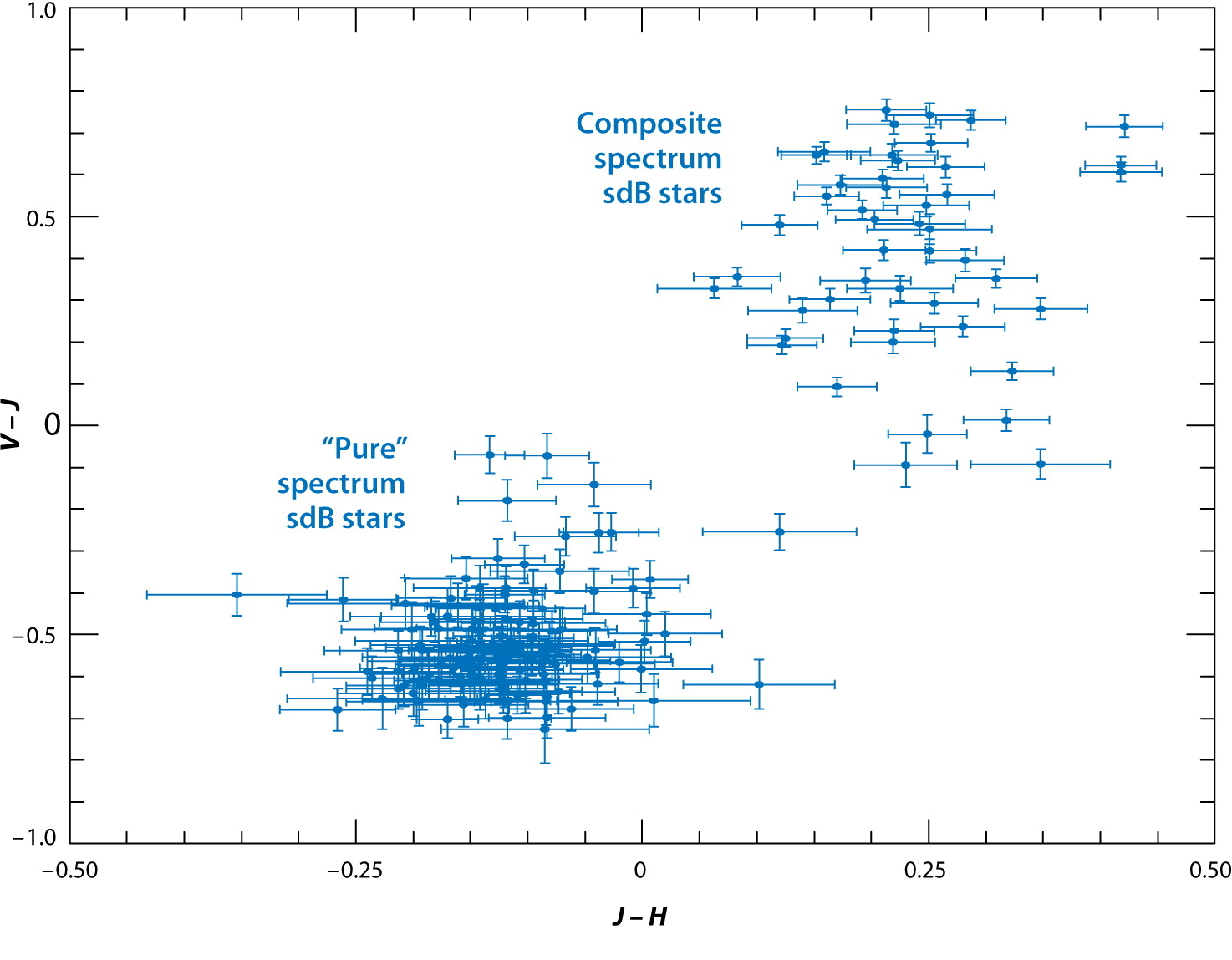}}
\caption{Two-color plot of $V-J$ vs.\ $J-H$ of sdB stars from the
sample of \citetads{2008ASPC..392...75G}. 
The composite-spectrum sdB-stars are in
the upper right and the stars with ``pure'' sdB spectra at the lower
left.  The latter group includes both the ``apparently single'' sdBs 
that have no detectable radial velocity (RV) variations above a level
of a few km~s$^{-1}$ over periods of many months, as well as sdB
binaries with invisible secondaries, either degenerate objects or dwarf M
companions too faint to affect the $V-J$ or $J-H$
colors. All known binaries that
fall at the lower left are post-common-envelope systems with periods
of a few hours to several days, while all of the composite-spectrum
binaries at the upper right appear to have much longer periods \citepads{2008ASPC..392...75G}. From \citetads{2009ARA&A..47..211H}; copyright ARAA; reproduced with permission.} 
\label{green_color}
\end{figure*}

The composite-color systems are well separated from the ''pure spectrum'' sdBs, whether radial-velocity (RV) variable or not, 
in two color diagrams (see Fig. \ref{green_color}), which points to a bimodal mass distribution for main-sequence companions.

While many radial velocity studies have provided orbit information for more than 140 sdB binaries, such information for composite-color sdB binaries have become available only recently and is still scarce. 
We shall begin with a discussion of the short-period single-lined systems in Sect. \ref{sect:single_bin}, highlighting the HW~Vir stars, which are eclipsing 
sdB systems hosting low mass stars and substellar objects (brown dwarfs), in 
Sect. \ref{sect:bin_hwvir}.  The {\it Kepler}  mission \citepads{2010ApJ...713L..79K} has provided light curves of unprecedented precision, which allowed the detection of low-amplitude effects, such as Doppler boosting and self-lensing (Sect. \ref{sect:bin_lowamp}). Evidence for massive compact companions comes from X-ray studies that will be summarized in Sects. \ref{sect:bin_xray} and \ref{sect:bin_radio}.
Hot subdwarf binaries with massive white dwarf companions are considered as viable progenitor systems for type Ia supernovae as shall be discussed in \ref{sect:sn_progenitors}. An outlook on the options to measure the orbital decay due to gravitational wave radiation follows in Sect. \ref{sect:grav_wave}
We turn to the composite-color binaries in Sect. \ref{sect:double_bin}, report evidence for the existence of triple or quadruple systems (Sect. \ref{sect:bin_mult}), and end this section with a glimpse at the zoo of giant and massive companions to hot subdwarf stars (Sect. \ref{sect:bin_massive}).

\subsection{Single-lined sdB binaries}\label{sect:single_bin}

The first sdB/O binaries were discovered from their light variability caused by eclipses, reflection effects or ellipsoidal deformations, such as the short-period sdB/O+dM binaries AA~Dor
\citepads{1978Obs....98..207K}, HW~Vir
\citepads{1986IAUS..118..305M} 
 and PG 1336$-$018 \citepads{1998MNRAS.296..329K} 
and in sdB+white dwarf binaries that show ellipsoidal
variability, e.g. KPD 0422+5421 \citepads{1998MNRAS.300..695K} 
and
KPD 1930+2752 \citepads{2000MNRAS.317L..41M}.

HD~49798 was the first sdO+WD system discovered from radial velocity measurements \citetads{1970MNRAS.150..215T}.
Indeed, surveys for radial velocity variations among apparently single hot subdwarf stars turned out to be very efficient in discovering short-period sdB*WD and sdB+dM systems.   
The early exploitation of the Palomar Green catalog \citepads{1986ApJS...61..305G} uncovered seven  
sdB+WD systems with periods between 0.25 and 2.5 day \citepads{1998ApJ...502..394S,1999MNRAS.304..535M}, but the number of systems grew quickly as the survey proceeded 
\citepads{2001MNRAS.326.1391M,2003MNRAS.338..752M,2004Ap&SS.291..321N}. 
Other surveys, e.g. the ESO/SPY and the SDSS based MUCHFUSS projects followed, which extended the sample of short-period sdB binaries to well above one hundred.

\subsubsection{The ESO/SPY project}

The ESO/SPY project was a large project to search for double degenerate systems and targeted mostly white { dwarfs} \citepads{2005ASPC..334..375N}, 
 but also a significant number of hot subdwarf stars (see Sect. \ref{sect:atmos}). While the frequency of radial velocity variable double white dwarfs was found to be low \citepads[5.7\%,][]{2009A&A...505..441K} 
 that fraction for sdB binaries was found to be 48\% \citepads{2004Ap&SS.291..321N}. 
Follow-up spectroscopy allowed the orbits of twelve sdB+WD binaries to be measured \citepads{2001A&A...378L..17N,2005ASPC..334..369K,2006BaltA..15..151K,2008ASPC..392..225G,2010A&A...515A..37G,2011A&A...528L..16G}. 
Spin-off studies from SPY included the quantitative spectral analyses of sdB  and sdO stars \citepads[][see Sect. \ref{sect:atmos}]{2005A&A...430..223L,2007A&A...462..269S} 

\subsubsection{The MUCHFUSS project}

A treasure chest for radial velocity studies of hot subdwarfs stars is the 
the Sloan Digital Sky Survey (SDSS) because its data base provides spectra for a large selection of faint hot subdwarfs. 
At least three individual spectra are available for each star, which provides a valuable starting point to search for radial velocity variations and to detect unseen companions via the Doppler reflex motion of the primary. 
Therefore, a collaboration, the  {\it Massive Unseen Companions to Hot Faint
Underluminous Stars from SDSS (MUCHFUSS)}, was established to explore the SDSS data base and in the first step $\approx$1400 hot subdwarf stars were selected by color and visual inspection of the spectra from Data Release 7 \citepads{2009ApJS..182..543A}. Finally a priority list was established for follow-up observations \citepads[see ][]{2011A&A...530A..28G,2015A&A...577A..26G}. 
 The main aim of the project is to search for the most extreme close hot subdwarf binaries; that is, those with (i) the shortest periods, (ii) the most massive as well as (iii) the least massive companions. Because such systems are rare, an observing strategy was developed to select targets with the highest radial velocity variations as well as the most rapid changes in radial velocity in order to suppress the typical sdB binaries with orbital periods of more than half a day and radial velocity half-amplitudes of less than 100 km\,s$^{-1}$.  
A detailed description of the selection procedure can be found in \citetads{2011A&A...530A..28G}. 
Atmospheric parameters have been determined for 190 hot subdwarf stars 
\citepads{2015A&A...577A..26G} 
and results for additional sdO stars were reported by \citetads{2008ASPC..392..131H}. 

\subsubsection{Single-lined sdB binaries}\label{sect:single}

\begin{figure}
\begin{center}
\includegraphics[width=\textwidth]{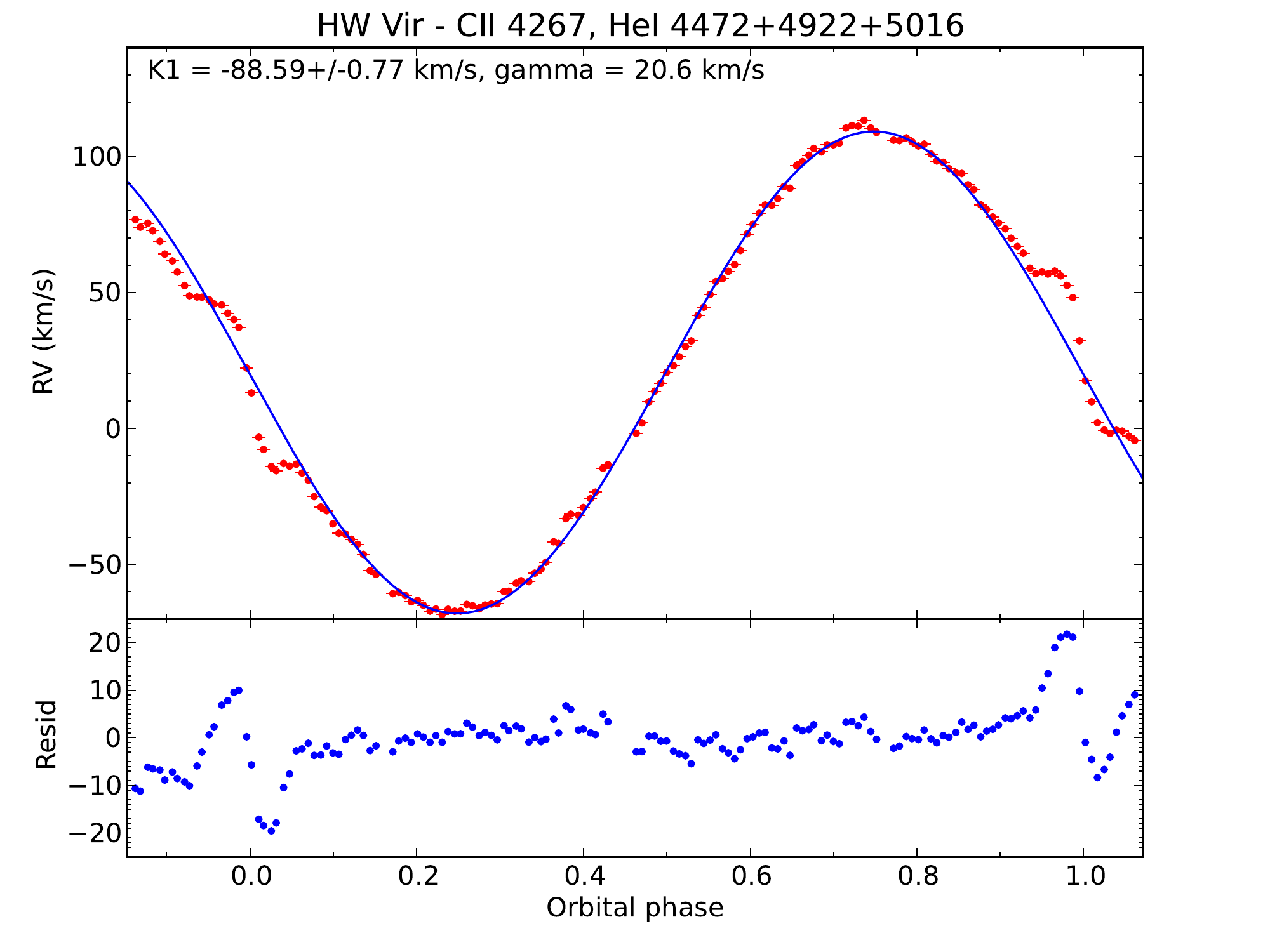}
               \end{center}
\caption{Radial velocity curve \textbf{(upper panel)} of HW~Vir. {{Red symbols represent the observation.}} Note the Rossiter-McLaughlin effect at phase 0 and 1 \textbf{which is} caused by the rotation of the star. {{The blue sinusoid is not a fit to the data. Therefore, the residuals (lower panel) near phase 0.0 and 1.0 seem to be different, which is actually not the case as can be seen from the radial velocity curve (upper panel).}} From \citetads{2014ASPC..481..259V}; copyright ASP; reproduced with permission.
}
\label{fig:hw_vir_rv}
\end{figure}

The sample of short-period sdB binaries that have known orbits has grown to more than 140 \citepads{2015A&A...576A..44K,2015MNRAS.450.3514K}. 
All systems are single-lined; that is, have unseen companions. 
The radial-velocity curve of HW~Vir is shown in Fig. \ref{fig:hw_vir_rv} as an illustrative example. Companion masses have been determined for eclipsing systems only, which comprise a little more than 10\% of the sample. 
For all others minimum companion masses were derived from the mass functions by adopting the canonical mass for the sdB component \citepads{2015A&A...576A..44K,2015MNRAS.450.3514K}.

\begin{figure*}
\begin{center}
\includegraphics[width=\textwidth]{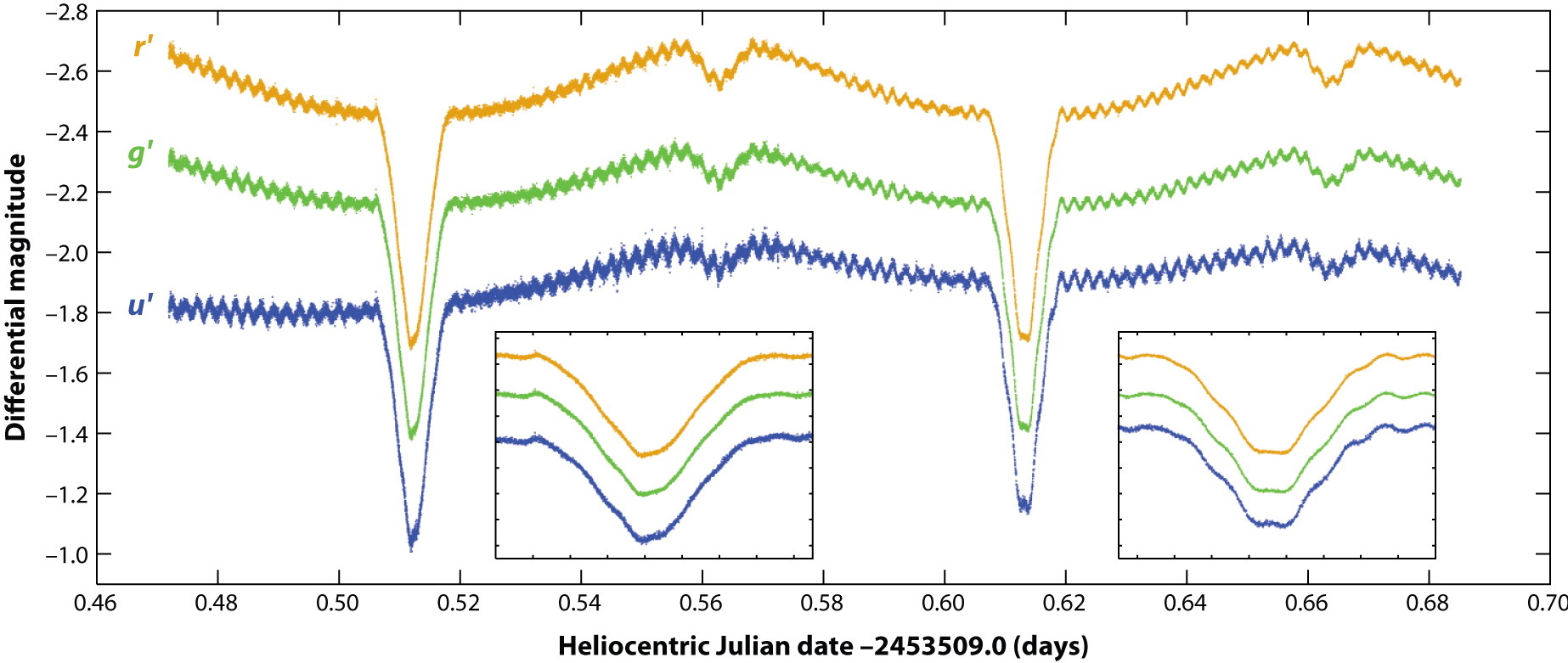} 
\caption{ULTRACAM/VLT r (upper), g (middle), and u (bottom) light curves of the eclipsing sdB star NY Vir \citepads{2007A&A...471..605V}.   
Light variations are due to reflection effect and eclipses as well as to multi-periodic pulsations. The insets show
enlarged sections of the two primary eclipses, where pulsations are clearly visible. The differences between the two consecutive primary
eclipses are due to the beating of the modes and different phases covered during the eclipse. This object provides an excellent opportunity
to derive the stellar mass in two independent ways: from light and radial velocity curve as well as from asteroseismology. Both results
are in excellent agreement. From \citetads{2009ARA&A..47..211H}; copyright ARA\&A; reproduced with permission. 
}
\label{fig:ny_vir}
\end{center}
\end{figure*}

As to the nature of the unseen companion, important information comes from photometry. A low-mass star or brown dwarf companion would reflect a significant amount of light, that varies with orbital phase (Fig. \ref{fig:ny_vir}). 

A white dwarf would be too small and the reflection effect is expected to be very small. 
The absence of a reflection effect may be a hint that the companion is not a main sequence star but a white dwarf \citepads[see e.g.][]{2002MNRAS.333..231M,2004Ap&SS.291..307M,2008ARep...52..729S}. 
However, tidal forces would distort the hot subdwarf star and produce ellipsoidal light variation with a period of half the orbital one (see Fig. \ref{fig:cd30_lc} for an example). 
An infrared excess could hint at the presence of a cool, low mass companion as well.

Spectroscopically the a main-sequence nature can be excluded, if the companion is more massive than $\approx$0.45 M$_\odot$ 
because otherwise spectral feature would be detectable in optical spectra \citepads{2005A&A...430..223L}. 

\begin{figure}
\begin{center}
\includegraphics[width=0.8\textwidth]{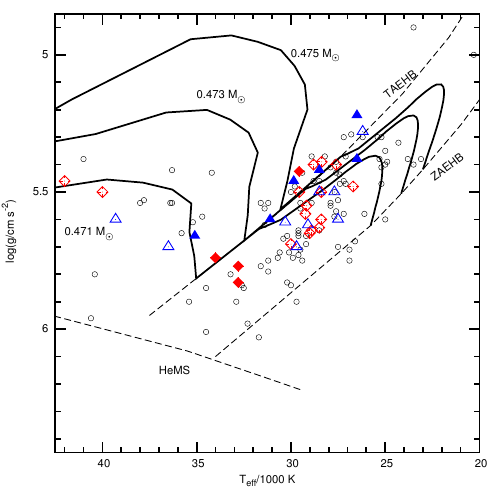}
                \end{center}
		\caption{Distribution of sdB stars in close binaries \citepads{2015A&A...576A..44K} 
		in the T$_{\rm eff}$--$\log$ g-plane. Reflection effect systems are highlighted in color, eclipsing (HW~Vir types) as red squares, non-eclipsing ones as blue triangles. Filled symbols mark pulsators.
This is a modified version of Fig. 5 from \citetads{2014A&A...570A..70S}; V. Schaffenroth (priv. comm.). 
}
\label{fig:hw_vir_teff_logg}
\end{figure}

The distribution of the sdB binaries of \citetads{2015A&A...576A..44K} 
in the T$_{\rm eff}$--$\log$ g-plane is shown in Fig. \ref{fig:hw_vir_teff_logg}. Most of the stars populate the extreme horizontal branch (EHB) band all the way down to the helium main sequence while about 10\% of the sdB sample has already evolved off the EHB. Such a homogeneous coverage is indeed expected from evolutionary models, which show a linear time-luminosity-relation while the star is in the EHB strip until close to central helium exhaustion. Thereafter, the evolution speeds up by a factor of ten consistent with the $\approx$10\% of post-EHB stars observed. 
Hence the observations are consistent with canonical evolution.

 \subsubsection{Distribution of orbital periods and minimum companion masses}\label{sect:p_m_dist}

The nature of the companion has been established for more than half of the sample only. In 30 systems low-mass stellar or substellar companions have been identified, while in another 52 cases the companion is a white dwarf \citepads{2015A&A...576A..44K}. In a few systems the companion mass may exceed the Chandrasekhar limit \citepads[][]{2010A&A...519A..25G}; that is, the companion could be a neutron star or black hole. 

\begin{figure*}
 \centering
 \plottwo{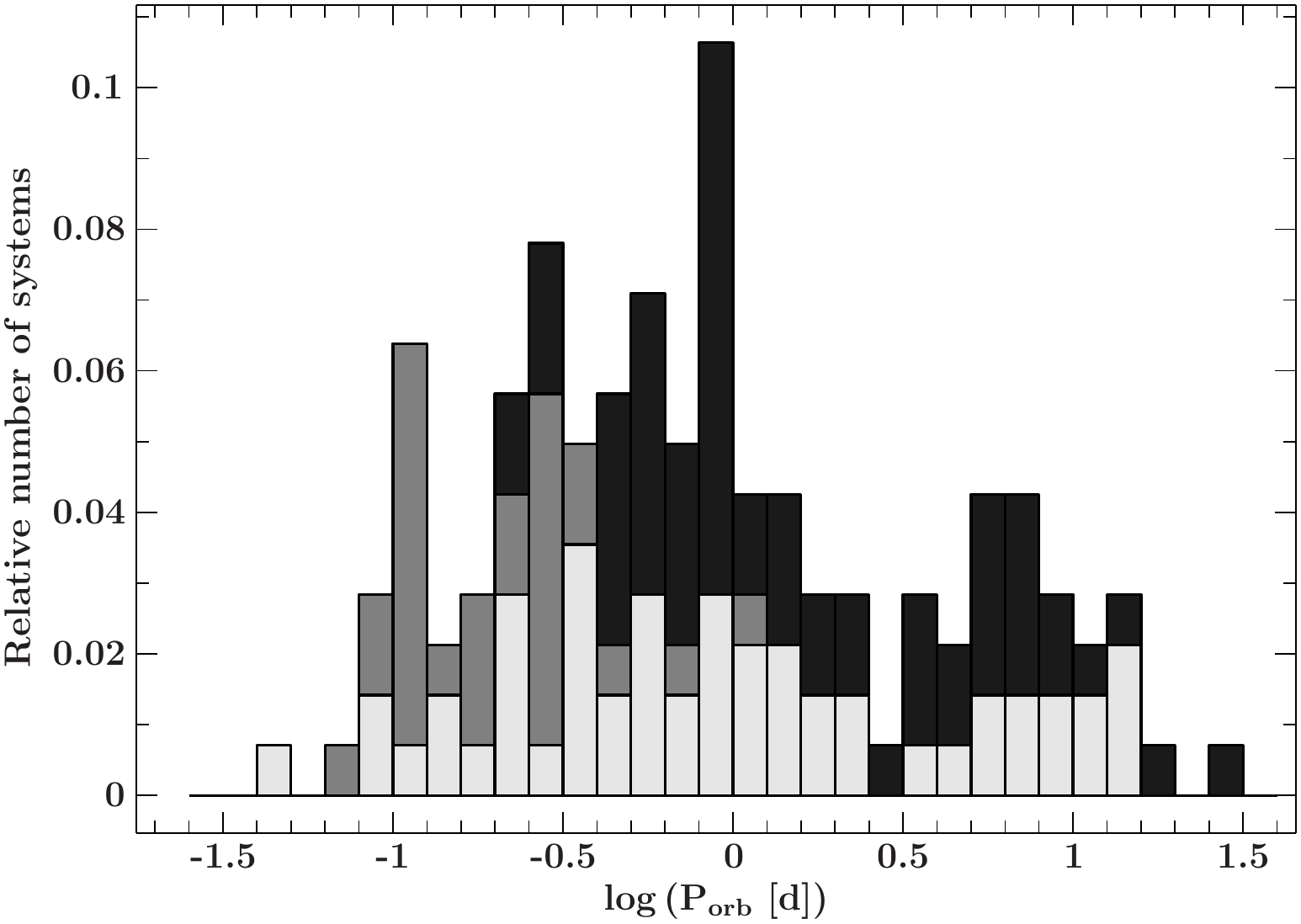}{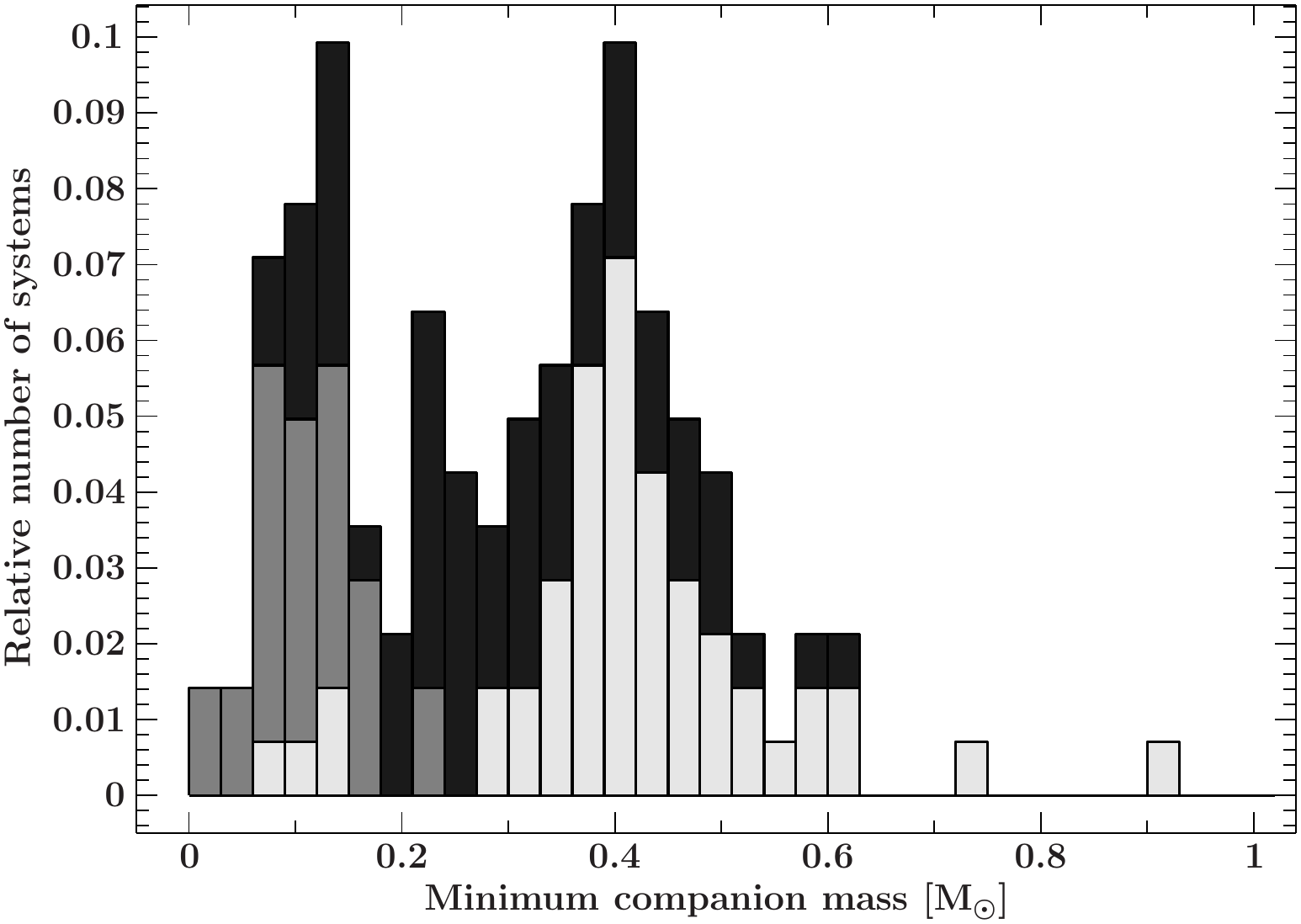}
         \caption{Distribution of periods and minimum companion masses. Light grey: WD companions, grey: dM companion, dark grey: unknown companion type.
	 { Left hand panel:} 
	 Period histogram of the full sample. 
	 { Right hand panel:}
	Histogram of minimum companion masses. At least two separated populations are present. The first one peaks at around $0.1$\,M$_\odot$ and consists mainly of systems with low-mass main-sequence companions. The second population peaks at around $0.4$\,M$_\odot$ and consists of WD companions if the nature of the companion has been revealed at all. Only two WD companions have masses exceeding 0.65 M$_\odot$. From \citetads{2015A&A...576A..44K}; copyright A\&A; reproduced with permission.}
\label{fig:period_comp_mass}

\end{figure*}

The distribution of orbital periods  of the single-lined sdB binaries is bimodal (Fig.~\ref{fig:period_comp_mass}).  
A wide peak around $P_{\rm orb}=0.3$\,days is found. The majority of systems in this group host dM companions detected from reflection effects in their light curves. Beyond half a day the contribution from the confirmed dM companions decreases significantly, because the probability for eclipses decreases as the separation of the stars increases and the reflection effect weakens and becomes  harder to detect. Another peak occurs at P = $0.8 - 0.9$\,days. The companions turn out to be  white dwarfs although the nature of many companions in this period range is still unknown. At longer periods the number of systems drops, but selection bias increases. 
White dwarf companions are found over the full period range, although a gap near $3$\,days might occur in Fig. \ref{fig:period_comp_mass}, whether real or not.

The distribution of the minimum masses of the companions is also bimodal (Fig.~\ref{fig:period_comp_mass}).
At the low mass end ($0.1$\,M$_\odot$; that is, close to the hydrogen burning limit) most companions were identified as dwarf M stars.
Another four stars have masses lower than the hydrogen-burning limit and therefore are brown dwarfs,
(see Sect. \ref{sect:bin_hwvir}).  Only four white dwarf companions have been found in this group, which could belong to the class of extremely low mass (ELM) white dwarfs (M$<$0.3 M$_\odot$, see Sect. \ref{sect:elm}) if the inclination of the orbit is sufficiently high. 

At higher minimum masses ($>$0.2 M$_\odot$) most of the companions are white dwarfs with minimum masses near $0.4$\,M$_\odot$.
The minimum masses of all confirmed white dwarf companions (except for the massive KPD\,1930+2752 and CD$-$30$^\circ$\,11223) are lower than the average mass of single C/O white dwarfs and might indicate that the white dwarfs need to lose a significant amount of mass during the evolution either during the first phase of mass transfer when the white dwarf is formed or during the common envelope phase when the sdB is formed. 

\begin{figure*}[t!]
\begin{center}
\includegraphics[width=0.8\textwidth]{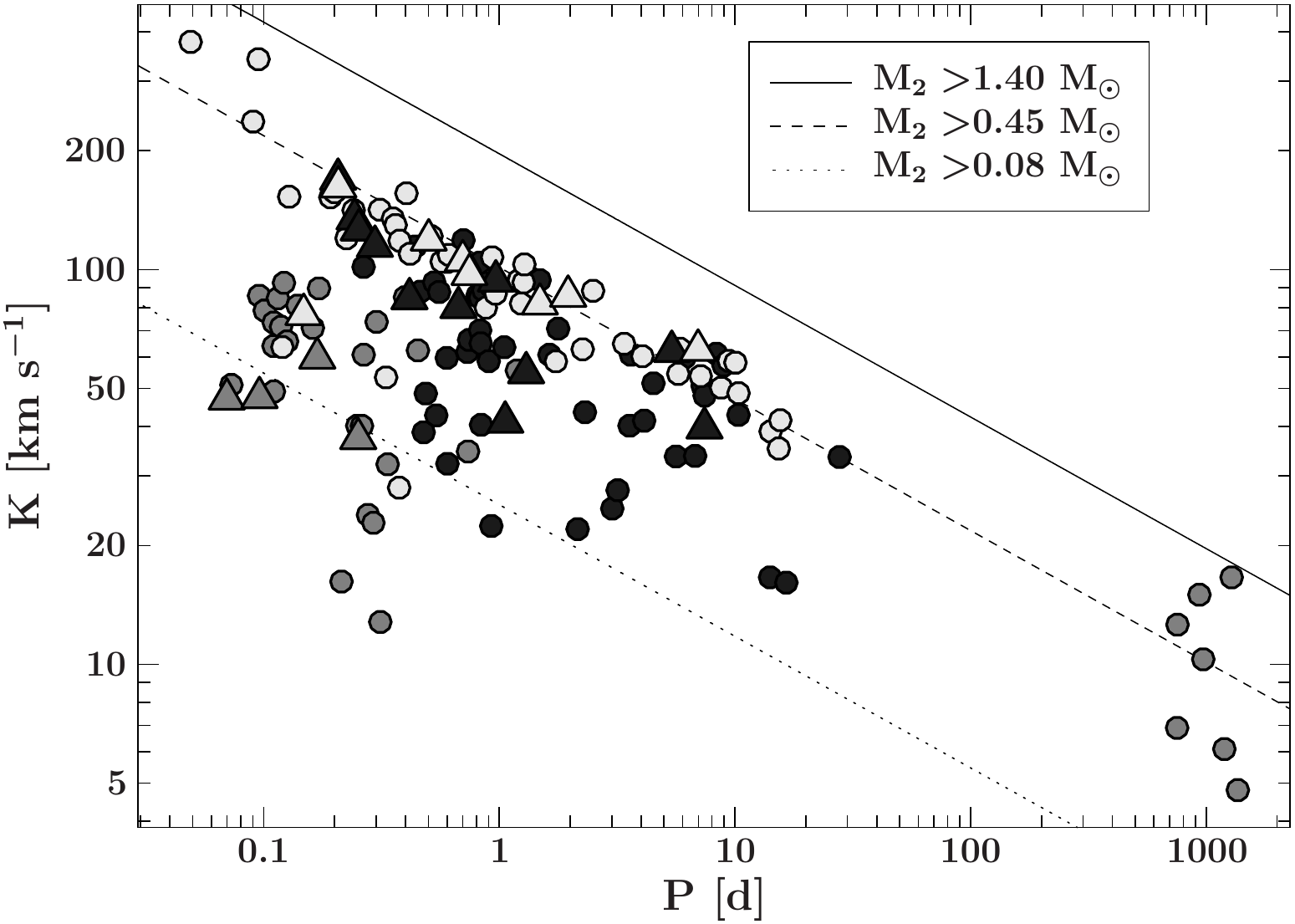}
                \end{center}
\caption{The RV semi-amplitudes of all known sdB binaries (single-lined systems from \citetads{2015A&A...576A..44K}, double-lined from Fig. \ref{fig:long_vos_ecc}) with spectroscopic solutions plotted against their 
orbital periods (light grey: WD companions, grey: M-dwarf companion, dark grey: unknown companion type). 
The lines mark the regions to the right where the minimum companion masses derived from the binary mass function (assuming $0.47$\,M$_\odot$ for the sdBs) exceed the hydrogen burning limit (dotted), the canonical core helium flash  mass (dashed) and the Chandrasekhar mass (full drawn). 
 From T. Kupfer (Priv. comm.)}
\label{fig:k_mass}
\end{figure*}

A proper discussion of the mass distribution for the companions, however, would require knowledge of the system inclinations, which is available for few stars only. Therefore, \citetads{2015A&A...576A..44K} 
preferred to discuss the radial velocity half amplitude K as a function of orbital periods (Fig.~\ref{fig:k_mass}). The distribution of the stars in this diagram is truncated near the 0.45 M$_\odot$ limit, which would indicate  
the masses of the white dwarf companions are below the average mass of a C/O white dwarf.

\begin{figure}[t!]
\begin{center}
\includegraphics[width=0.7\textwidth]{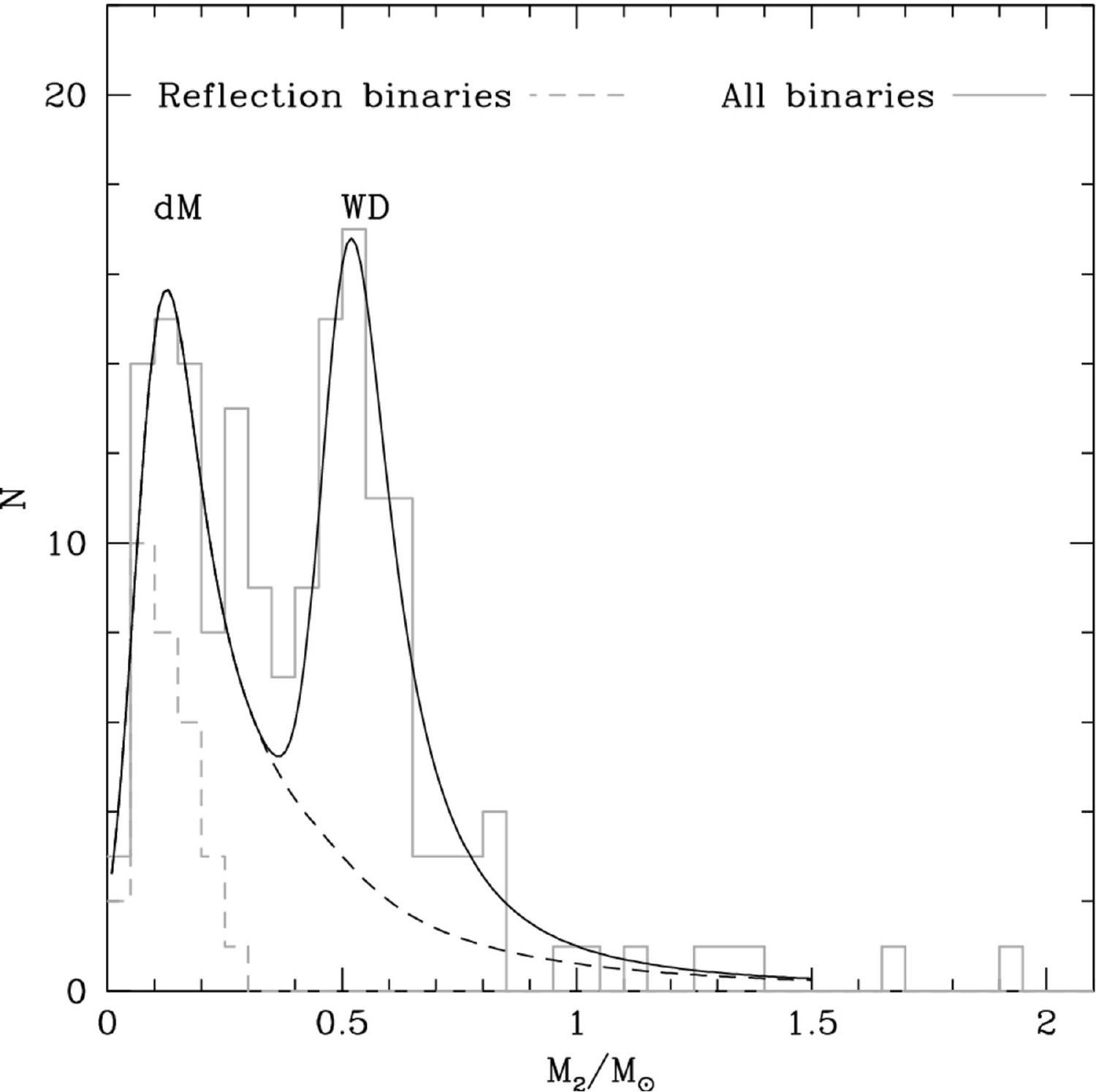}
                \end{center}
\caption{Mass distribution of all known binaries with a hot subdwarf primary star as a function of the secondary mass, assuming an average inclination of 57$^\circ$ (full histogram). The peak distribution of low-mass stars is marked dM and that of white dwarfs, WD. Binaries showing the reflection effect in their light curves are shown with a dashed histogram. The full lines show synthetic distributions smoothed to two-bins width for a combination of late-type stars and white dwarfs (double-peaked full line), and that excluding white dwarfs (dashed line). 
From \citetads{2015MNRAS.450.3514K}; copyright MNRAS; reproduced with permission.
}
\label{fig:m2_histo57}
\end{figure}

\citetads{2015MNRAS.450.3514K} took a different approach by assuming that no inclination bias exists and therefore a statistical average may applicable for the sample. The resulting mass distribution is shown in Fig. \ref{fig:m2_histo57}. Besides the peak at $\sim$0.1 M$_\odot$ a second peak occurs at $\sim$0.6 M$_\odot$, which would be consistent 
with the companions being C/O white dwarfs \citepads{2005ApJS..156...47L}. 
\citetads{2015MNRAS.450.3514K} point that a third peak occurs at $\sim$0.25 M$_\odot$ in Figs. \ref{fig:period_comp_mass} and \ref{fig:m2_histo57}, which might be helium white dwarfs.
 

\citetads{2015A&A...576A..44K}, however, argue that the average inclination is likely to be larger than the statistical one because of selection biases. The sample under study is an inhomogeneous collection of results from the literature. This means that targets  were pre-selected either from spectroscopy (radial velocity variations) or from light variations (reflection effect, ellipsoidal variation and/or eclipses) and, therefore, the sample is biased towards high inclinations, both for RV variables (large amplitudes preferred) and light variables (reflection effect and/or eclipses detected). Hence the minimum companion masses will be preferentially below the statistical mean, close to their true values in many cases. 

Whether the companion white dwarf is of helium or C/O composition is of utmost importance for understanding the evolution of hot subdwarf stars, because the 
primary (now the white dwarf) had to fill its Roche lobe on the RGB in the first case or the AGB in the second (cf. panel a of Fig. \ref{fig:podsi}).

White dwarfs with masses of $\sim\,0.4$\,M$_\odot$ are usually considered to be of helium composition and a significant fraction of the white dwarf companions in the WD peak in Fig. \ref{fig:m2_histo57} might be of helium- rather than of C/O composition. 

\clearpage

  \subsection{Reflection effects in sdB/O+dM binaries: the HW~Vir stars}\label{sect:bin_hwvir}
  
Close low-mass companions to hot subdwarf stars may be identified via the reflection effect. In a typical system a cool low-mass main-sequence star orbits a hot subdwarf at a distance of about one solar radius and is irradiated by an intense radiation field, which heats the facing hemisphere
and leads to reemission of light. This effect is misleadingly called ''reflection'' effect and in most light curve modelling tools (e.g. {\sc MORO} \citepads{1995A&A...294..723D} 
 or {\sc PHOEBE} \citepads{2005ApJ...628..426P} based on the Wilson-Divinney approach \citepads{1971ApJ...166..605W} 
 treated simply by albedos or allow for some kind of energy redistribution. The physics of an irradiated atmosphere, however, is far more complicated but difficult to model \citepads{2004ApJ...614..338B,2012ARep...56..867S,2015A&A...578A.125H,2016A&A...586A.146V}. 
Although this approach reproduces observed light curves in the pre-{\it Kepler}  era reasonably well,
it is challenged by the high-precision {\it Kepler} 
light curves of reflectionÃ-effect binaries.
Since the temperature stratification of an irradiated atmosphere is inverted, emission lines are expect to occur along with the continuum emission and are actually observed for HW~Vir and AA~Dor.

Eclipsing binary stars are important for deriving stellar masses and radii. Very few eclipsing sdB+WD binaries are known (see Sect. \ref{sect:elm}). However, there is a growing number of sdB/O+dM
systems, that do show eclipses accompanied by a strong reflection effect, which is characteristic for such so-called HW~Vir systems (see  Fig. \ref{fig:ny_vir} for an illustrative example).
An example of a radial velocity curve is shown in Fig. \ref{fig:hw_vir_rv}. 
Because the sdB stars in close binaries are spun up by tidal force, the Rossiter-McLaughlin effect is seen 
and can be used to determine the rotation velocity and the orbital obliquity.
   
Seventeen systems
\citepads[see ][for an overview]{2014MNRAS.442L..61J,2015A&A...576A.123S} and about an equal number of non-eclipsing sdB+dM binaries are known, which according to \citetads{2014MNRAS.442L..61J} should be named XY Sex stars after the prototype. 
A spectacular addition to the sample is \object{2M1938+4603} which was  discovered by 
\citetads{2010MNRAS.408L..51O} from an early {\it Kepler}  light curve to be a sdB binary with a dM companion in the {\it Kepler}  field showing strong reflection and shallow eclipses. The orbital period turned out to be relatively long (181.1 min) and the radial velocity amplitude of the sdB star in \object{2M1938+4603} to be K$_1$ = 65.7 $\pm$ 0.6 km\,s$^{-1}$. Extended {\it Kepler}  light curves
showed pulsations and hint at the { presence} of a circumbinary substellar object (see Sect. \ref{sect:eclipse_puls}). 

Actually, in three HW~Vir systems, AA~Dor \citepads{1982A&A...106..254K}, \object{V1828 Aql} \citepads{2012MNRAS.423..478A}, and Konkoly J064029.1+385652.2 \citepads{2015ApJ...808..179D},  
 the primary is an sdO star and, hence, has evolved beyond the EHB already.
 Nevertheless, we count these systems in the HW~Vir class\footnote{HW~Vir stars form a subgroup of pre-cataclysmic variables, which include very hot sdO and DAO white dwarf binaries \citepads[e.g.][]{2004AJ....127.2936S,2007A&A...469..297A}, some of which are associated with planetary nebulae \citepads[e.g. BE UMa,][]{1999ApJ...518..866F,
2008ARep...52..558S}. Many of them show very strong reflection effects with amplitudes exceeding one magnitude and emission line dominated spectra
\citepads[see][for an overview]{2006A&A...456.1069S}.
A compilation of pre-cataclysmic variables can be found in  \citetads{2003A&A...404..301R}. The primaries are probably very hot ($\sim$ 100kK) post-AGB stars \citepads{2008AstL...34..423S} rather than post-
 EHB objects and are, hence, not considered here in more detail.}. 
The distribution of the HW Vir systems as well as that of other non-eclipsing reflection effect systems in the T$_{\rm eff}$--$\log$ g-plane is shown in Fig. \ref{fig:hw_vir_teff_logg}.

\begin{figure*}[!tbp]
  \centering
  \begin{minipage}[b]{0.49\textwidth}
    \includegraphics[angle=270,width=\textwidth]{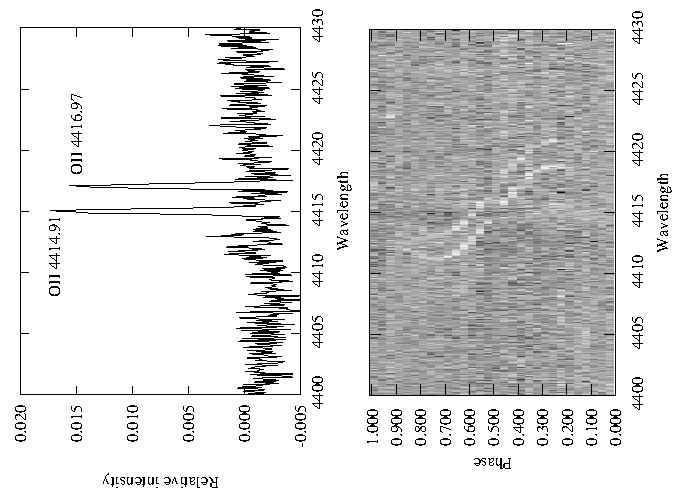}
  \end{minipage}
  \hfill
  \begin{minipage}[b]{0.49\textwidth}
    \includegraphics[angle=270,width=\textwidth]{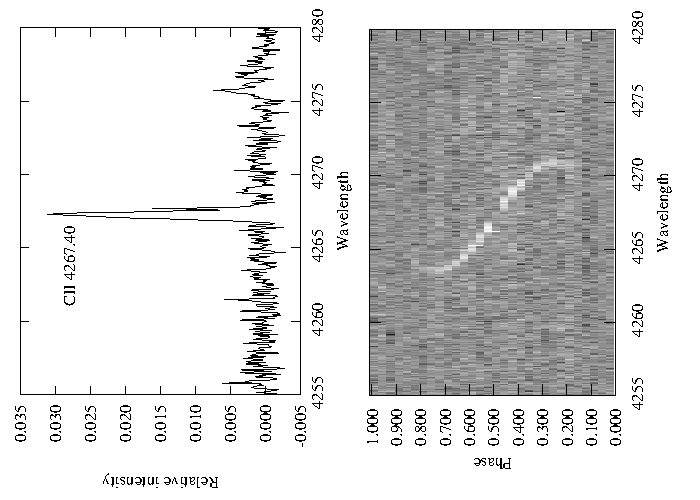}
  \end{minipage}
 \caption{Emission lines originating from the heated atmosphere of the companion of AA~Dor. 
 Top: The residual spectrum in the rest frame of the secondary after the contribution of the primary has been removed. The O {\sc ii} (left) and C {\sc ii} (right) lines are clearly visible.
 Bottom: trailed and phase-binned residual spectra in the rest frame of the primary. From \citetads{2008ASPC..392..199V}; copyright ASP; reproduced with permission.}
\label{fig:aa_dor_emission}
\end{figure*}

\begin{figure*}
\centering
\includegraphics[angle=0,width=0.9\textwidth]{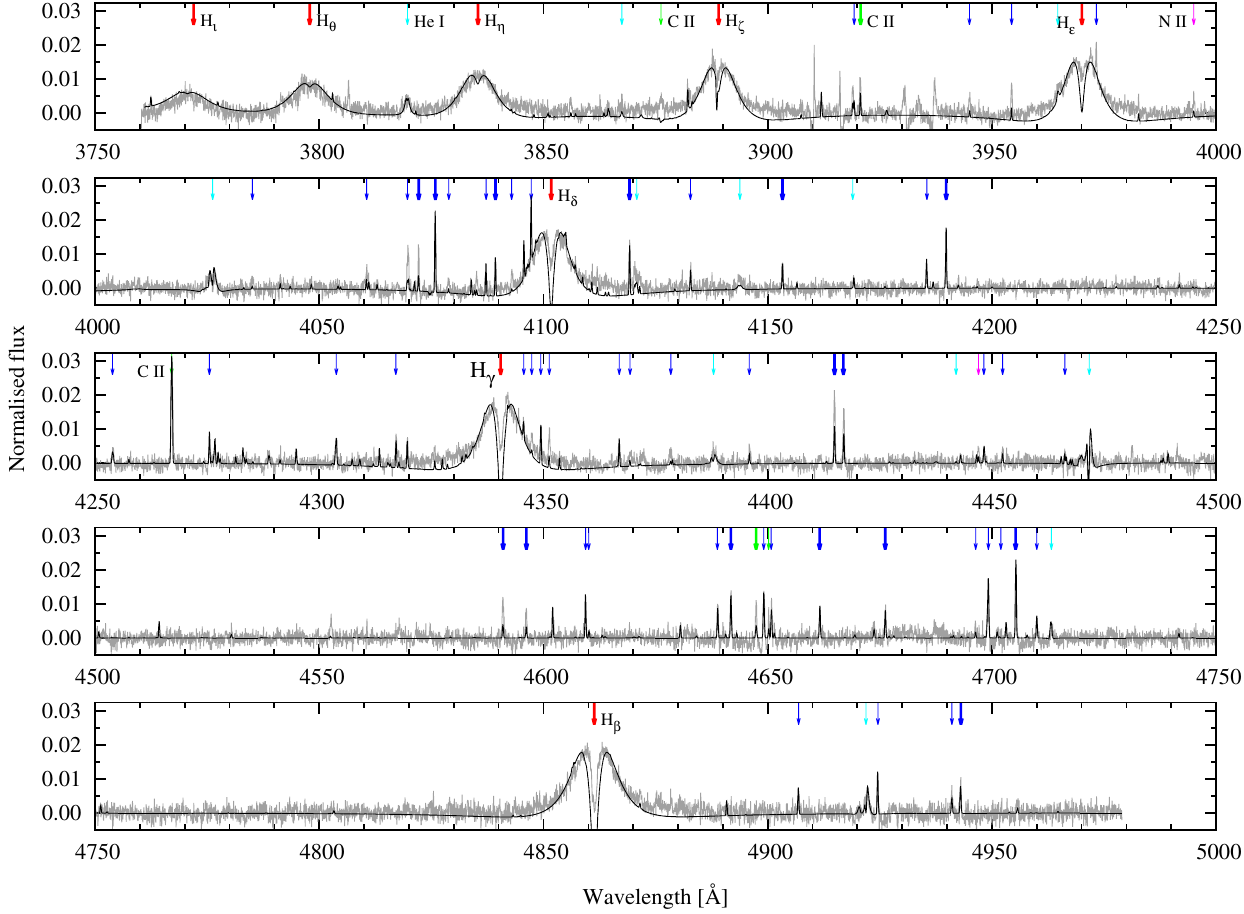}
\caption{Synthetic emission line spectrum of the dM secondary in AA~Dor compared to residual high-resolution spectrum after subtraction of the sdOB component \citepads{2016A&A...586A.146V}. Species are identified by the color coding of the arrows; Balmer lines: red,
         O {\sc ii}: blue, C {\sc ii} and C {\sc iii}: green,
         He {\sc i}: cyan, N {\sc ii}: purple). From \citetads{2016A&A...586A.146V}; copyright A\$A; reproduced with permission.
        }
\label{fig:aa_dor_reflected}
\end{figure*}

 The prototype HW~Vir and its hotter sibling, AA~Dor, have been studied from light- and radial-velocity curves. Their atmospheric parameters and chemical composition have been derived from high-resolution spectra (see Sect. \ref{sect:abu_metal}). Most importantly, in both cases the radial velocity curve of the companion has been measured from emission lines (see Fig. \ref{fig:aa_dor_emission}) arising from the heated hemisphere of the dM companion \citepads{2014ASPC..481..259V,2008ASPC..392..199V,2016A&A...586A.146V,2015A&A...578A.125H,2015ApJ...808..179D}.
 Recently, attempts to model the spectral and light variations  were successful to reproduce the spectrum of the irradiated hemisphere of AA~Dor \citepads[][see Fig. \ref{fig:aa_dor_reflected}]{2015A&A...578A.125H,2016A&A...586A.146V}  
 allowing tight constraints on their masses and radii to be derived.  
\citetads{2015A&A...578A.125H} used UVES and XSHOOTER spectra  
to determine the masses for AA~Dor and its companion to unprecedented precision
and obtain the mass of the primary to $M_\mathrm{pri}  = 0.470^{+0.0975}_{-0.0354}\,M_\odot$, in perfect agreement with the canonical mass from evolution theory.
The mass of the companion of $M_\mathrm{sec}  = 0.0811^{+0.0184}_{-0.0102}\,M_\odot$ places it right at the hydrogen-burning limit. An independent analysis of the same UVES spectra by \citetads{2016A&A...586A.146V} confirmed those numbers.

All-sky photometric surveys such as the ''All Sky Automated Survey'' \citepads[ASAS,][]{1997AcA....47..467P}, e.g. ASAS102322$-$3737.0 \citepads{2013A&A...553A..18S}, SuperWASP \citepads{2006PASP..118.1407P} 
 and NSVS \citepads{2004AJ....127.2436W}, e.g. NSVS 14256825 \citepads{2012MNRAS.423..478A}, provided new discoveries of  HW~Vir stars, almost as bright as the prototype, and suitable for high resolution spectroscopy.   
All transient surveys are bound to find such objects at fainter magnitudes.
For example the latest study \citepads{2015A&A...580A.117S} 
deals with an HW~Vir system discovered by the Palomar Transient Facility \citepads{2009PASP..121.1395L}. 
Recently, \citetads{2013AcA....63..115P} and \citetads{2015AcA....65...39S} 
announced the discovery of three dozen HW~Vir stars from the OGLE experiment.  
A detailed spectroscopic and photometric analysis by the recently formed EREBOS collaboration\footnote{''Eclipsing Reflection Effect Binaries from the OGLE Survey'', a large project at the ESO VLT led by V.Schaffenroth} of these systems will triple the sample  and will allow solid conclusions about the 
formation of HW Vir stars and sdB stars in general to be drawn.

\begin{figure}
\begin{center}
\includegraphics[width=0.7\textwidth,angle=270]{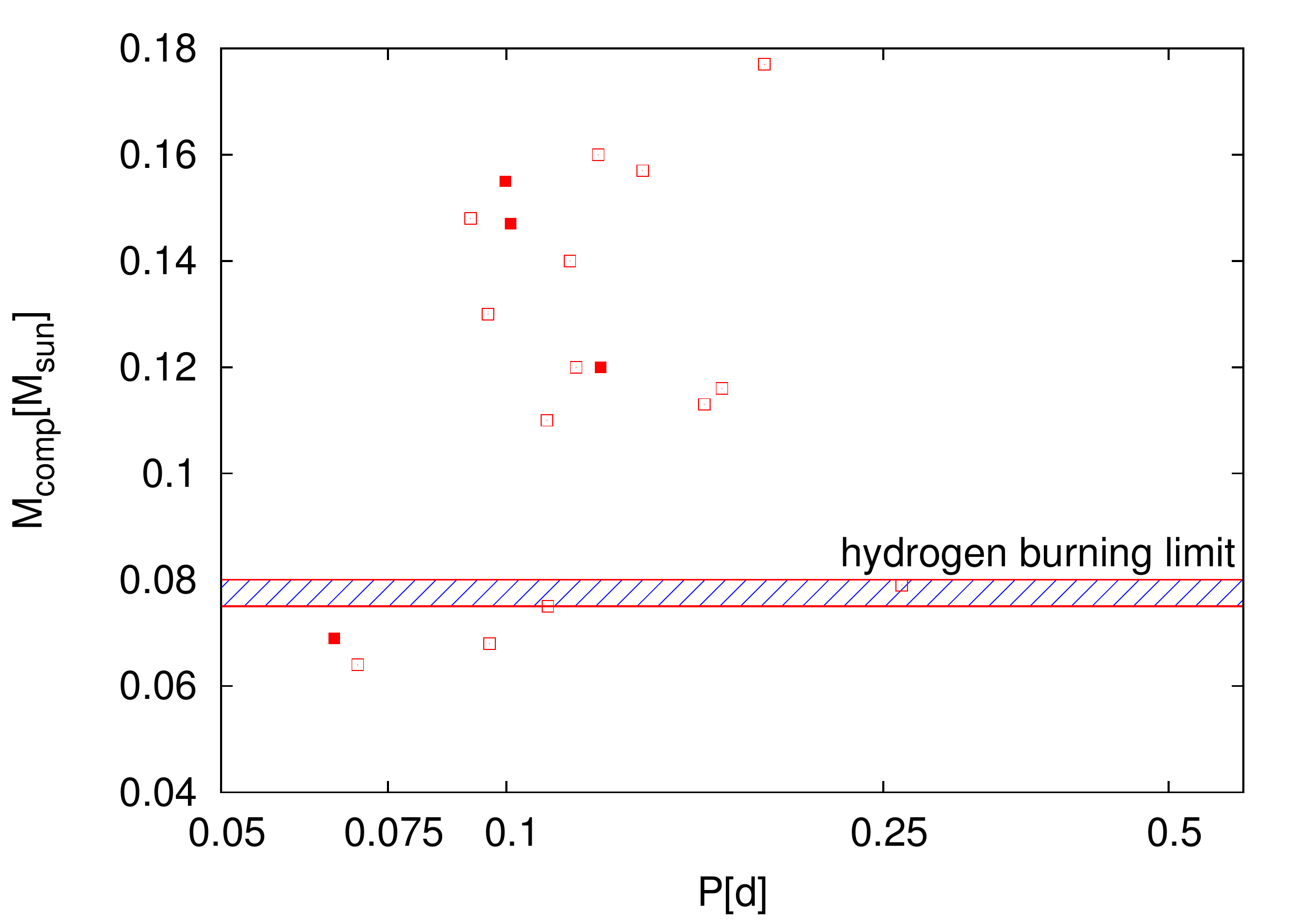}
\end{center}
\caption{Masses vs. orbital period of all known HW~Vir. Filled symbols mark pulsating stars. The hydrogen-burning limit is shown as a hatched band (Schaffenroth, priv. comm.)}
\label{fig:hwvir_logp_m}
\end{figure}

Figure \ref{fig:hwvir_logp_m} displays the companion masses in HW~Vir systems as a function of orbital period 
\citepads{2014A&A...564A..98S,2015A&A...576A.123S}. 
While most companions are late M-dwarfs with masses between 0.1\,M$_{\rm \odot}$ and 0.16\,M$_{\rm \odot}$, there is no sharp drop below the hydrogen-burning limit.  AA~Dor and HS~2231+2441 \citepads{2007ASPC..372..483O} 
have companion masses of 0.08\,M$_{\rm \odot}$, which is on the hydrogen-burning limit. 
Of particular interest are the binaries SDSS J082053.53+000843.4 \citepads{2011ApJ...731L..22G}, SDSS J162256.66+473051.1 \citepads{2014A&A...564A..98S}, and V2008$-$1753 \citepads{2015A&A...576A.123S},
because their companion masses are likely to be below that limit and, accordingly, they would be brown dwarfs. The primary is pulsating which makes V2008$-$1753 a { particularly interesting} system.

\begin{figure}
\begin{center}
\includegraphics[width=0.8\textwidth,angle=270]{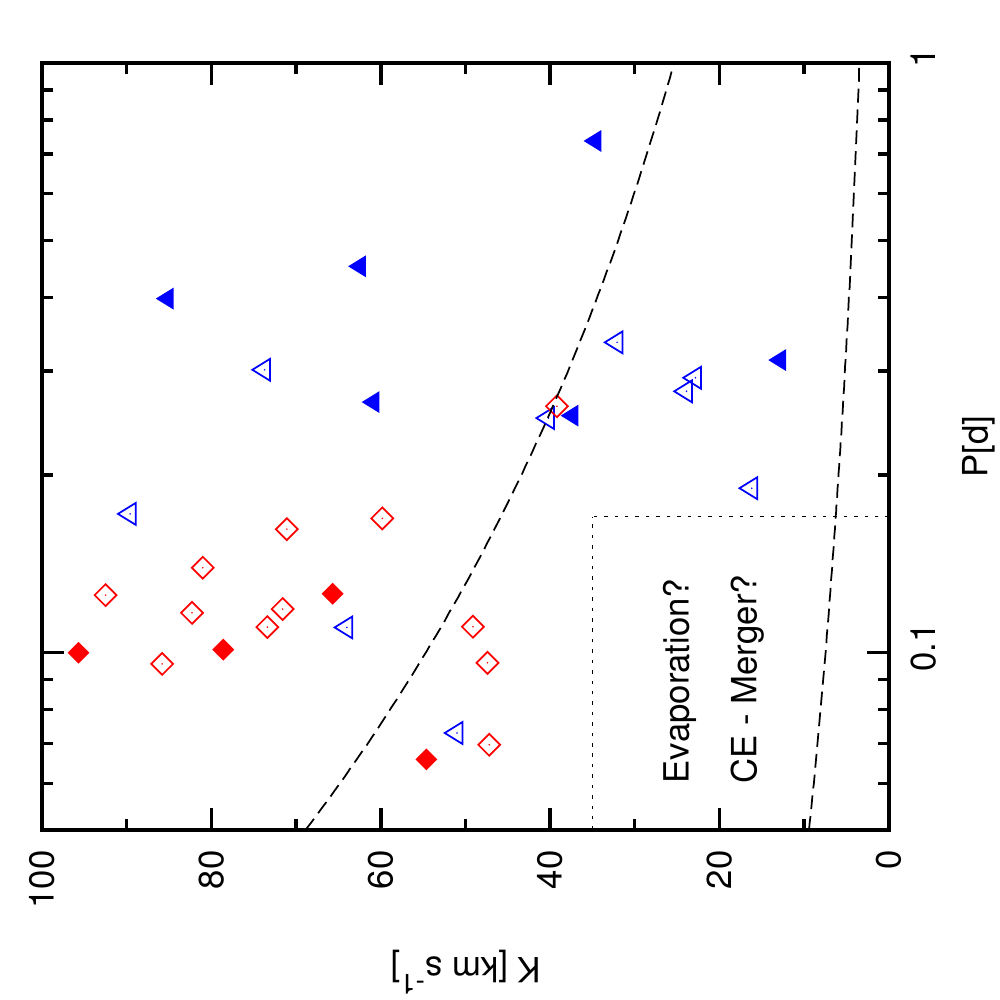}
                \end{center}
		\caption{The RV semiamplitudes of all known sdB binaries with reflection effect and RV curve solutions plotted against their orbital periods. 
The dashed lines mark the regions to the right
where the minimum companion masses derived from the binary mass
function exceed 0.01M$_\odot$ (lower curve) and 0.08 M$_\odot$ (upper curve).
Filled symbols are pulsators. Eclipsing binaries are shown as red diamonds (HW Vir). Blue triangles mark non-eclipsing reflection effect binaries.
This is a modified version of Fig. 4 from \citetads{2014A&A...564A..98S}; 
V. Schaffenroth (priv. comm.).
}
\label{fig:hw_vir_m2}
\end{figure}

Figure~\ref{fig:hw_vir_m2} compares the HW~Vir stars with their non-eclipsing siblings, which show a reflection effect only. As orbital inclinations of the latter are unknown, the radial velocity half-amplitudes are shown rather than the masses. 
There are four non-eclipsing systems { \footnote{One of them, PHL 457, is of particular interest, because it is a slow pulsator \citep{2016MNRAS.457..723K}.}} with periods between $\sim$0.2d and $\sim$0.4d and K$<30\,{\rm km\,s^{-1}}$, which corresponds to masses less than the hydrogen-burning limit if the inclination is sufficiently high. \citetads{2014A&A...564A..98S} calculate that the probability, that the companion in all four cases is of stellar type, is just 2\%. Hence it is likely that at least one of them is a brown dwarf. This suggests that a substantial fraction of the reflection effect binaries host a brown dwarf.
A lack of binaries with even shorter periods ($<0.2\,{\rm d}$) and $K<54\,{\rm km\,s^{-1}}$ is obvious from Fig. \ref{fig:hw_vir_m2}, which  corresponds to companion masses of less than $\sim0.06\,M_{\rm \odot}$. \citetads{2014A&A...564A..98S} suggest that such companions cannot exist because they are of too low mass to survive the common envelope phase but evaporate or merge with the red giant core; that is, those systems have evolved into single sdB stars.

\subsection{Low-amplitude signals in binary light curves: 
Doppler boosting, gravitational lensing and R{\o}mer delay} \label{sect:bin_lowamp}

Although only a few dozen of hot subdwarfs were discovered in the field monitored by the {\it Kepler}  satellite, the high precision light curves turned out to be a treasure chest for sdB asteroseismology (see Sect. \ref{sect:asteroseismology}) and binary research. Subtle effects such as Doppler boosting, R\o mer delay,
and gravitational microlensing were detected and used for analysis for the first time. 
These low-amplitude signals allow physical information to be derived without the need for spectroscopy. A review is presented by \citetads{2013EAS....64..269B}. 

\subsubsection{Doppler beaming (Doppler boosting)}

Doppler beaming, which means light gets beamed towards the direction of motion making the star appear brighter when it moves towards the observer, is a relativistic effect. However, non-relativistic effects add to the low-amplitude flux variations and can provide the dominant contributions to the flux variations observed. First the photon arrival rate is enhanced/decreased and the star appears brighter/fainter when it moves on its orbit toward/away from the observer. Secondly, the Doppler shift of the stellar spectrum with respect to the photometric bandpass leads to a change of the observed flux unless the spectrum is flat \citepads{1987A&A...183L..21S}. The beaming factor, therefore, depends on the shape of the spectrum.
 Since all three effects act at the same time the term ``Doppler boosting'' is often used instead of Doppler beaming, as we shall do in this paper.  
Because the combined effect scales linearly with the ratio of the orbital velocity to the speed of light the changes to the light curves are minute. However, it allows the radial velocity curve to be measured photometrically without any spectroscopic measurements \citepads[see e.g. ][ and Sect. \ref{sect:elm}]{2010ApJ...715...51V}.
   
Doppler boosting was first discovered in the light curve of \object{KPD~1930+2752} \citepads{2000MNRAS.317L..41M}, an sdB+WD binary, showing strong ellipsoidal variations. More recently the effect was found in ground-based light curves of  CD$-$30$^\circ$ 11223 (see Sect. \ref{sect:sn_progenitors}), and the eclipsing white dwarf binaries \object{NLTT~11748} \citepads{2010ApJ...725L.200S} 
and \object{SDSS J0651+38} \citepads{2011ApJ...737L..23B}. 

The {\it Kepler}  satellite has provided light curves of sdB binaries of unprecedented
precision, which allowed the Doppler boosting effect at lower amplitudes to be studied. Three sdB+WD binaries were of particular interest\footnote{Two other interesting Doppler boosting cases are 
KOI-74 and KOI-81, which are  binaries consisting of early type main-sequence stars orbited by helium-core objects to be discussed in Sect. \ref{sect:elm}.}. 

 The first one is the pulsating sdB binary KIC~11558725, for which  \citetads{2012A&A...544A...1T} 
detected light variations at the 238 ppm level with a clear signal of Doppler boosting,  consistent with the observed spectroscopic orbital radial-velocity amplitude of the subdwarf, despite its rather long orbital { period} (P = 10.05 d). 
Because the star is pulsating a third consistent estimate of the orbital radial-velocity amplitude was derived from the orbital light-travel time delay of 53.6 s, which causes aliasing and lowers the amplitudes of the shortest { -period} pulsation { modes}. In the same way \citetads{2014A&A...570A.129T} analyzed the pulsating sdB+WD binary KIC 7668647 whose orbital period is even longer (14.2 d). Again its radial-velocity amplitude of 39 km s$^{-1}$ was consistently derived from the orbital Doppler beaming of 163 ppm in the {\it Kepler}  light curve, the spectroscopic radial velocity curve and the orbital light-travel time delay of 27 s from timing of pulsation modes. 

Another Doppler-beaming system, KIC 10553698B, was found by \citetads{2014A&A...569A..15O} to host a normal white dwarf with a mass close to 0.6 M$_\odot$ { if the pulsation axis is aligned with the orbital axis, because the inclination of the former can be derived from the shape of the rotationally split multiplets}. { \bf T}he star is pulsating, slowly rotating  
(P$_{\rm rot}$=41 d, see Sect. \ref{sect:eclipse_puls}){ , and, most interestingly shows trapped pulsation modes as will be discussed in Sect. \ref{sect:g_mode_ident}}.

Another effect of even lower amplitude may become detectable for rotating stars. \citetads{2012ApJ...745...55G} 
suggested that Doppler boosting due to rotation, the photometric analog of the spectroscopic Rossiter-McLaughlin effect (see Fig. \ref{fig:hw_vir_rv}), might become detectable in systems such as NLTT 11748 and SDSS J0651+38, which would provide information about the sky-projected spin-orbit angle from gravity darkening of the rotating star \citepads{2013ApJ...774...53B}. 

\subsubsection{Gravitational lensing}
 
Transiting compact objects can gravitationally magnify their companions, as predicted by \citetads{1973A&A....26..215M}.  \citetads{2001MNRAS.324..547M} 
investigated the effect for white dwarf binaries.
The high quality of the {\it Kepler}  photometry allowed the detection a ``self-lensing'' system \citepads[KOI~3278,][]{2014Sci...344..275K}, a sun-like star and a white dwarf in a 88.18 day orbit. A 5-hour pulse of 0.1\% amplitude occurs every orbital period. The pulse is due to magnification by microlensing, which allowed \citetads{2014Sci...344..275K} to derive the mass of the white dwarf as 0.63 M$_\odot$, typical for a C/O white dwarf \citepads{2005ApJS..156...47L}.

\begin{figure}
\epsscale{0.6}
\plotone{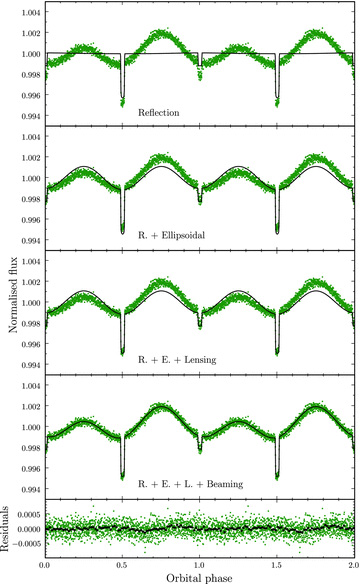}
\caption{Phase-folded light curve of KPD 1946+4340 (green, data points grouped by 30) from {\it Kepler}  data and the best-fitting model (black) of \citetads{2011MNRAS.410.1787B}. In the top panel, only the eclipses and reflection effects are modelled. In the second panel, ellipsoidal modulation is added. In the third panel, gravitational lensing is taken into account as well, which affects the depth of the eclipse at orbital phase 0.5. The bottom panels show the full model, taking into account Doppler beaming, and the residuals (grouped by 30 in green and grouped by 600 in black). From
\citetads{2011MNRAS.410.1787B}; copyright MNRAS; reproduced with permission.\label{fig:kpd1946}}
\end{figure}

A combination of ellipsoidal variations, Doppler boosting and self-lensing has been found in the {\it Kepler}  light curve of the sdB binary \object{KPD~1946+4340} \citepads{2011MNRAS.410.1787B} (see. Fig. \ref{fig:kpd1946}).  %
The former leads to unequal amplitudes of the light maxima caused by ellipsoidal deformation of the sdB, the latter weakens the eclipse depth.
The photometric radial velocity from Doppler boosting is consistent with the spectroscopic one, demonstrating the reliability of photometric radial velocities from Doppler boosting.

\subsubsection{R\o mer delay { from eclipse timing}}

The lightcurve of a binary on a perfectly circular orbit should show primary and secondary eclipses separated by half the period. Due to the finite speed of light, however, a travel time delay occurs, when the masses of the binary components are unequal. This so-called R\o mer delay depends on the size of the orbit and
allows the mass ratio of the binary to be constrained \citepads[see ][]{2010ApJ...717L.108K}. 

However, a slight ellipticity of the orbit would also shift the secondary minimum from phase 0.5. The two effects can be disentangled, because the R\o mer delay always shifts the eclipse to a larger phase, whereas the eccentricity can shift it either way. Hence,
the observed shift of the minimum can be a superposition of both effects.

While the R\o mer effect is familiar in radio pulsar timing, { such light travel time effects have been found in the light curves of pulsating sdB stars \citepads{2007Natur.449..189S,2012A&A...544A...1T} (see Sects. \ref{sect:planets} and \ref{sect:asteroseismology}). From eclipse timings the R\o mer delay has been discovered in the light curve} of an eclipsing binary white dwarf for the first time by \citetads{2014ApJ...780..167K} who also constrained the eccentricity of the orbit to be as small as $e\cdot\cos \omega = (- 4 \pm 5) \cdot 10^{-5}$.

Detection of the R\o mer delay { from eclipse timing} in an sdB binary was first claimed for \object{2M1938+4603} from the analysis of its {\it Kepler}  lightcurve \citepads{2012ApJ...753..101B} 
based on the delay of the secondary eclipse of 2.06$\pm$0.12 sec.
However, the analysis of a more extended {\it Kepler}  light curve called this conclusion into question. \citetads{2015A&A...577A.146B} 
 demonstrated that even the larger data set is insufficient to disentangle the R\o mer delay signal from ellipticity. Instead evidence for a circumbinary substellar companion was found (see also Sect. \ref{sect:hwvirplanets}).

\subsection{X-rays from hot subdwarf stars}

Main-sequence O stars emit X-rays originating from turbulence and shocks in their strong winds \citepads{1980ApJ...241..300L,1988ApJ...335..914O}. 
The mass-loss of hot subdwarf stars is much lower and, therefore, they are not expected to be X-ray sources. However, X-ray emission from the sdO star HD\,49798 was discovered with the ROSAT satellite 
\citepads{1997ApJ...474L..53I}. It turned out that the X-rays do not originate from the sdO star but from accretion onto a compact companion, either a neutron star or a massive white dwarf. Recently additional X-ray emitting sdO stars have been discovered from Chandra and XMM-Newton observations and limits on the mass loss rates of sdB { stars} have been derived \citepads[for a recent review see][]{2015arXiv151004173M}.      

\subsubsection{X-rays from wind accretion in hot subdwarf binaries} \label{sect:bin_xray}

\begin{figure}
\begin{center}
 \includegraphics[width=0.8\textwidth]{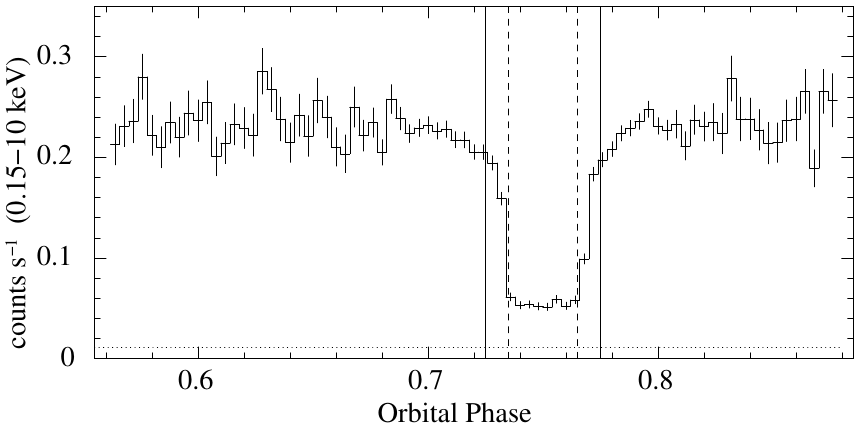}
 \caption{Light curve of HD~49798 in the 0.15 -- 10 keV energy range folded at the orbital period in bins of 535 s. The eclipse is clearly visible with totality lasting 4311s and gradual ingress and egress lasting $\approx$500s, each. The horizontal line indicates the background level. Note that there is residual X-ray flux at totality which might arise from light from the companion scattered in the wind of the sdO star or from HD~49798 itself, due to shocks and instabilities in the wind. The vertical lines delimit the phase intervals used for the spectral analysis. From \citetads{2013A&A...553A..46M}; 
 copyright A\&A; reproduced with permission.
 }
 \label{fig:hd49798}
\end{center}
\end{figure}

The compact companions of hot subluminous stars
are difficult to detect in optical/UV data. However, they might give rise to 
detectable X-ray emission if the subdwarf
can provide an adequate accretion rate. 
In this respect, luminous sdO stars offer the best prospects for X-ray detections because of their relatively high mass loss rates 
\citepads[$\dot M \sim10^{-7}-10^{-10} M_{\odot}$ yr$^{-1}$,][]{2010MNRAS.404.1698J}. 
This is well demonstrated by the case of the luminous sdO HD 49798, a single-lined
spectroscopic binary (P=1.5 days), which was discovered as a soft X-ray source
by the ROSAT satellite. Remarkably, the X-ray flux shows a strong periodic
modulation at 13.2 s \citepads{1997ApJ...474L..53I}, 
indicating the presence of a compact object. 
X-ray eclipses (which constrain the system's inclination, see Fig. \ref{fig:hd49798}) allowed a dynamical measurement of the masses of the two binary components.
This showed that the companion of HD 49798 is a very
massive (1.28$\pm$0.05 $M_{\odot}$) white dwarf \citepads{2009Sci...325.1222M,2011ApJ...737...51M,2013A&A...553A..46M}, 
making this binary a potential progenitor of a type Ia supernova (see Sect. \ref{sect:sn_progenitors}).
Prompted by the case of HD 49798, a search for X-ray emission from other hot subdwarf binaries was launched.
While searches with the Swift/XRT of a dozen sdB binaries hosting compact companions gave negative results \citepads{2011A&A...536A..69M}, 
pulsed X--ray emission in the luminous, hydrogen-deficient sdO BD\,+37$^\circ$442 was discovered \citepads{2012ApJ...750L..34L}.
The observed X-rays have a soft spectrum (a blackbody with temperature $\sim$45 eV
plus a weak power-law component), and
show a significant modulation at a period of 19.2 s. Hence  it is tempting to assume that BD+37$^\circ$ 442 is an X-ray binary very similar to HD~49798.
However, no radial velocity variations could be found despite of intense spectroscopic monitoring \citepads{2014ASPC..481..307H}. Therefore, the nature of the pulsing X-ray source remains a mystery.

Because sdB stars are less luminous than the sdO stars discussed and, therefore, have much lower mass loss rates, X-ray fluxes from wind accretion onto a compact companion are expected to be lower and may occur in the most favorable cases; that is, the shortest period sdB binaries with the most massive companions. CD $-$30$^\circ$ 11223 and PG~1232$-$136 are considered the best targets and were observed by \citetads{2014MNRAS.441.2684M} 
with XMM-Newton. However, no X-ray photons were detected, which  constrained  their mass loss rates to lower than $\sim 10^{-13}$ M$_\odot$\,yr$^{-1}$, 
which would be consistent with predictions from models of radiatively driven winds \citepads{2002A&A...392..553V} if the metallicities were low, but too low for solar metallicity. 

\subsubsection{X-rays from the winds of hot subdwarfs?}

A continued X-ray search with Chandra  \citepads{2014A&A...566A...4L} and XMM-Newton \citepads{2015A&A...580A..56L} resulted in the discovery of an additional X-ray emitting luminous sdO star, BD+37$^\circ$1977 (a spectroscopic twin to BD+37$^\circ$ 442), for which no radial velocity curve has been measured yet. It is, therefore, unknown whether the star hosts a compact companion or not. \citetads{2015A&A...580A..56L} were able to match the observed X-ray spectrum with the sum of two thermal plasma components, as in normal early-type stars, and conclude that the X-ray emission is intrinsic to the star and arises from shocks in the stellar wind. 

\citetads{2014A&A...566A...4L} detected X-rays from the very hot compact sdO stars Feige 34 and BD+28$^\circ$4211. There is no evidence for a compact companion in either case; no radial velocity variations, neither on short nor on long time scales have been found for both BD+28$^\circ$4211 \citepads{2015A&A...579A..39L} {{and BD+37$^\circ$ 442 \citepads{2014ASPC..481..307H}}.} 
{{If there are compact companions orbiting BD+37$^\circ$ 442 and BD+28$^\circ$4211 and accreting from their winds, they would need to be on orbits seen almost pole-on. {\it Gaia} observation will be tale telling.}}
Therefore, the X-ray emission is more likely to be intrinsic to the sdO star. 

\citetads{2015A&A...580A..56L}  and \citetads{2015arXiv151004173M} find that X-ray spectra and the X-ray to bolometric luminosities L$_{\rm X}$/L$_{\rm bol}$ of intrinsic emission of all the sdOs are {{ well-fitted by a}} multi-temperature plasma, once the proper elemental abundances are used. Despite their lower mass-loss rates,  the estimated L$_{\rm X}$/L$_{\rm bol}$ ratios of the sdO stars are consistent with those of the massive early-type stars and \citetads{2015arXiv151004173M} conclude that X-ray emission is caused by the same mechanism as for the massive stars; that is, by turbulence and shocks in their winds.

\subsubsection{Radio-emission from hot subdwarf stars}\label{sect:bin_radio}

If the most massive companions of sdB binaries were in fact neutron stars, they  could be pulsars. This idea has been tested by \citetads{2011A&A...531A.125C} who 
searched for radio pulsations from the potential neutron star binary companions to the subdwarf B stars HE~0532$-$4503, HE~0929$-$0424, TONS~183 and PG~1232$-$136 proposed by \citetads{2010A&A...519A..25G} 
 using the Green Bank Telescope. No pulsed emission was found to mean flux densities as low as 0.2 mJy.

\subsection{Progenitors of type Ia supernovae}\label{sect:sn_progenitors}

Hot subdwarf stars with white dwarf companions are viable progenitor systems
for type Ia supernovae \citepads{2000MNRAS.317L..41M}. 
Because recent observations show that the class of type Ia supernovae is more diverse than previously thought, a diversity of progenitor systems is expected too, which reinforced the search for progenitors of type Ia supernovae.
The various observational evidence on the nature of SN Ia as well as the rivalling progenitor and explosion models are reviewed by \citetads{2013FrPhy...8..116H}, \citetads{2014ARA&A..52..107M}, \citetads{2014LRR....17....3P}, and \citetads{2014NewAR..62...15R}. 

A type Ia supernova is caused by the thermonuclear explosion of a white dwarf in a close binary either with another white dwarf (double degenerate scenario, DD) or with a non-degenerate companion, either a normal star or a helium star (single degenerate scenario, SD). In a double degenerate system the merger of two C/O white dwarfs driven by gravitational wave radiation might lead to a SN Ia if the total mass is larger than a critical mass, which is mostly assumed to be the Chandrasekhar mass. If the orbital period of a DD system is shorter than about half a day, the system will merge within a Hubble time. 

 A sdB+WD system will evolve into a DD system within 
$\approx$10$^8$ years, which is shorter than the merger time for most known systems. If the system is sufficiently massive, it would qualify as a SN Ia progenitor system. However, a sdB+WD system may also become a SD progenitor system, if the sdB transfers parts of its helium envelope to the white dwarf companion before having reached the white dwarf cooling sequence.

The sdB+WD binary KPD 1930+2752 was the first DD system discovered to be 
a good candidate for the progenitor of a type Ia supernova \citepads{2000MNRAS.317L..41M}, 
because the white dwarf is sufficiently massive ($>$0.9~M$_\odot$) so that the total  mass will exceed the Chandrasekhar limit when the system will merge in $\approx$10$^8$ years.
\citetads{2007A&A...464..299G} revisited KPD 1930+2752 
by analyzing time-resolved high-resolution spectra and  
derived the masses of both components by assuming tidally locked rotation
(see Sect. \ref{sect:bin_synchro} and \ref{sect:synchro_puls}).
The resulting masses sum up to almost
exactly the Chandrasekhar mass. Hence, the analyses confirmed that KPD 1930+2752 is a viable SN Ia progenitor candidate. 

Several studies, notably from the SPY survey, have provided periods and 
masses for a couple of DD systems, mostly double white dwarfs but also 
several sdB+WD systems. Although no system has yet been found to exceed
the Chandrasekhar limit beyond any doubt, several have been found close to that
limit
\citepads[see Fig.~19 of][]{2014LRR....17....3P}. 
The resulting DD mass distribution 
implies that super-Chandrasekhar
systems must exist and that the DD scenario is viable.
    
More recently, the sdO+WD system HD~49798 has been proposed as a candidate SN Ia progenitor system because it hosts a white dwarf of very high mass \citepads{2010RAA....10..681W}. 
 
\paragraph{The double detonation scenario} predicts that an (underluminous) supernova may result from accretion of helium by a C/O white dwarf. A detonation of the accreted helium shell can occur once a sufficient amount of helium ($\approx$0.1 M$_\odot$) accumulated \citepads{1990ApJ...361..244L}.
The explosion of the He-shell triggers a
subsequent detonation in the C/O core which might produce SNe Ia explosions depending on its mass \citepads{2007A&A...476.1133F,2010A&A...514A..53F}. 
The most favorable accretor mass is $\approx$1.0 M$_\odot$ which is predicted to launch a normal SN Ia \citepads{2010ApJ...714L..52S}, but minimum mass models of 0.8 M$_\odot$ predict double detonation SN Ia, though underluminous \citepads{2010A&A...514A..53F}.

\begin{figure*}
\begin{center}
\includegraphics[width=0.8\textwidth]{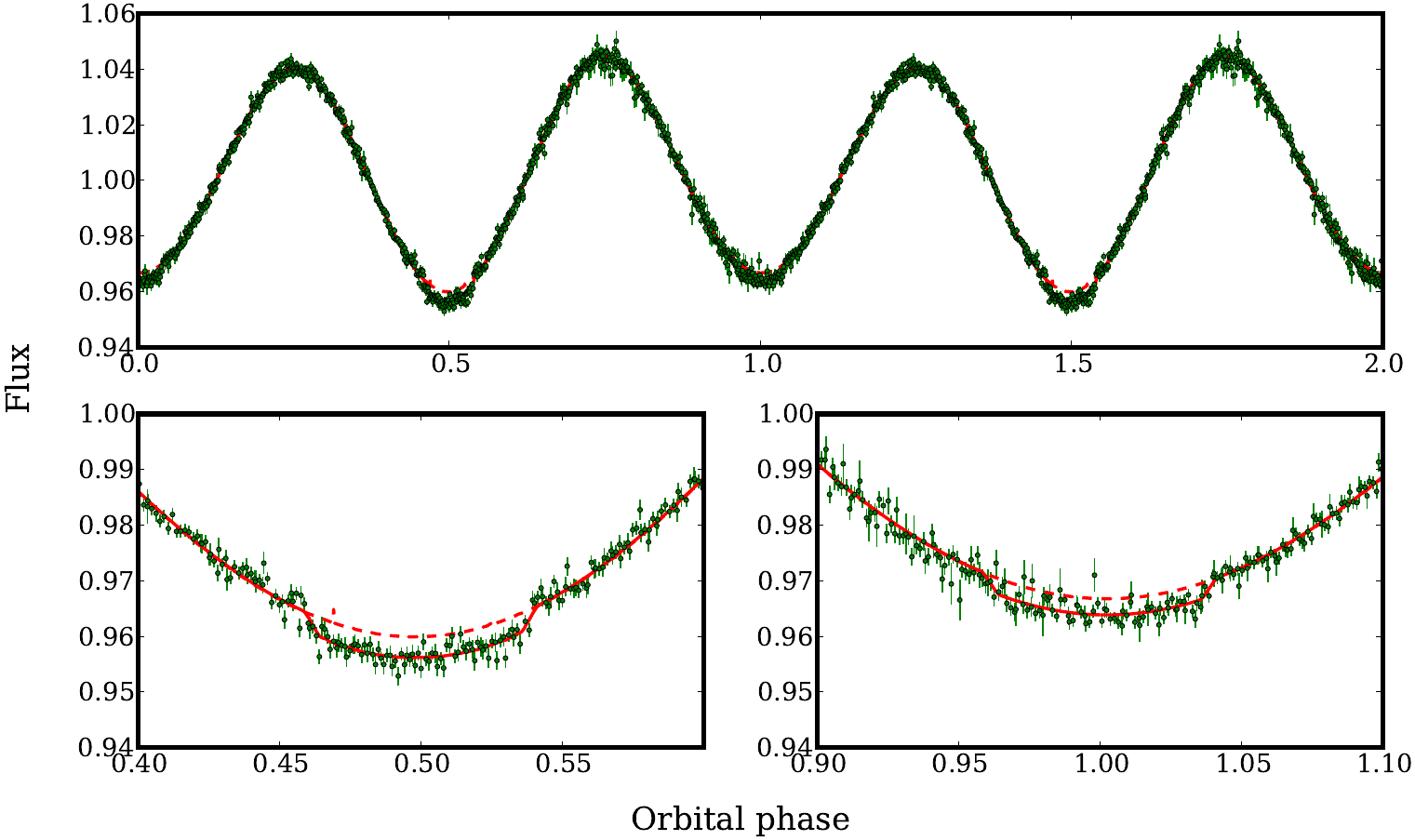}
\end{center}
\caption{
Upper panel: V-band light curve of the eclipsing sdB+WD system CD$-$30$^\circ$11223 (green) with superimposed model (red) plotted twice against the orbital phase for better visualization. The dashed red curve marks the same model without transits and eclipses. The sinusoidal variation is caused by the ellipsoidal deformation of the hot subdwarf as a result of the tidal influence of the compact white dwarf. The difference in the maxima between phase 0.25 and 0.75 originates from the Doppler boosting effect (see Sect. \ref{sect:bin_lowamp}), Lower panels: close-up on the transit of the WD in front of the sdB (left) and the eclipse of the WD by the sdB (right). From \citetads{2013A&A...554A..54G}; copyright A\&A; reproduced with permission.
}
\label{fig:cd30_lc}
\end{figure*}

\paragraph{CD-30$^\circ$11223} is the best candidate for for a binary evolving into a double detonation supernova. The system was discovered independently by 
\citetads{2012ApJ...759L..25V} and
\citetads{2013ASPC..469..373G} to be an eclipsing binary sdB in a 70 min orbit.
The light curve  displays ellipsoidal variations, eclipses of both companions, and the Doppler boosting effect. The companion must be a white dwarf because strong ellipsoidal variations are seen in the light curve (see Fig.  \ref{fig:cd30_lc}). Shallow secondary eclipses demonstrate that the white dwarf must be hot and relatively luminous. The Doppler boosting effect can be used to measure the radial velocity from photometry  and agrees well with the spectroscopic radial velocity curve.
\citetads{2013A&A...554A..54G} 
constrain the masses to  $0.47\,M_{\rm \odot}$ or $0.54\,M_{\rm \odot}$ for the sdB, and to $0.76\,M_{\rm \odot}$ or $0.79\,M_{\rm \odot}$ for the white dwarf by assuming tidally locked rotation or making use of the mass-radius relation for the white dwarf, respectively. The derived mass indicates that the white dwarf is of C/O composition.    
Its future evolution strongly depends on the stability of the mass transfer of helium to the white dwarf and the companion's mass. 
\citetads{2013A&A...554A..54G} investigated the future evolution of the system and find that stable mass transfer will occur once the sdB star fills its Roche lobe. The detonation in the accreted helium layer is sufficiently strong to trigger the explosion of the core. The helium star will then be ejected at such high velocity that it will escape the Galaxy. Such a so-called hyper-velocity star has, indeed, been discovered, the He-sdO US~708, which will be discussed in Sect. \ref{sect:kinematic}.

\subsection{Orbital decay by gravitational wave radiation}\label{sect:grav_wave}

Because the binaries emit gravitational wave radiation, their orbital periods have to decrease. The rate of period change strongly depends on the orbital period \citepads{1963PhRv..131..435P}, 
 and therefore will be easiest to detect for the shortest period systems. 
\citetads{2012ApJ...757L..21H} 
reported the detection of orbital decay in the 12.75-minute, detached binary white dwarf SDSS J065133.338+284423.37 at a rate of  -0.31 $\pm$ 0.09 ms yr$^{-1}$, consistent with the prediction from General Relativity for the orbital decay due to gravitational wave radiation (-0.26 $\pm$ 0.05 ms yr$^{-1}$).

\object{2M1938+4603} is the only sdB binary for which the period decrease has been derived  
\citepads{2015A&A...577A.146B} 
  to be
$\dot{P} = 4.13(2) \cdot 10^{-11} s/s$ using {\it Kepler}  data. The shortest period sdB+WD systems \object{KPD 1930+2752} and  \object{CPD$-31^\circ$11223} are the most promising candidates {{for measuring the}}  period decay due to gravitational wave radiation. 

\subsection{Composite color hot subdwarf binaries}\label{sect:double_bin}

Previously most sdB stars were discovered as very blue objects in surveys for  UV-excess objects. Because of color-cuts usually applied, such surveys are biased against sdB binaries with cool companions, because their colors are reddish and their composite spectra are flat. Hence, a large number of sdB stars may have been missing from current samples. An sdB star shines as bright as the Sun in the visual band. An F-type companion would dominate the optical spectrum while an  A-type dwarf (or earlier) would outshine a hot subdwarf completely (but see Sect. \ref{sect:bin_massive}).

Nevertheless, a large number of systems with  F- to K-type companions are known, but quantitative spectral analyses remained scarce \citepads[e.g.][]{2002A&A...385..131A} 
because of the difficulties of disentangling the composite spectrum.

Recent large-area ultraviolet ({\it GALEX}), optical (SDSS and several others) and infrared (2MASS, UKIDSS) photometric surveys offer 
a less biased option to search for new composite systems comprising subdwarfs plus MS star companions of mid-M-type and earlier. 
By combining the photometric measurements the spectral energy distribution (SED) can be constructed in a broad wavelength range. 
Cuts in color -- color space have to be employed to separate these objects from possible contaminants. 

\cite{2012MNRAS.425.1013G} constructed two complementary samples of composite sdB candidates by applying color cuts to the aforementioned photometric catalogs. A large but shallow one, and a smaller, but deeper sample, led to the discovery of a large number of composite subdwarf binaries. Their SEDs were matched with synthetic spectra using grids of appropriate model spectra for the sdB and the companion star, respectively. Effective temperatures of both stars as well as the distance were derived for all systems. The subdwarf effective temperatures primarily lie in the cooler part of the EHB (20\,000-30\,000 K), with 5-10 \% at even lower temperatures. 
The companions are mostly of spectral type F0 to K0, while subdwarfs with M-type companions appear much rarer. This is consistent with population synthesis models predicting a very efficient first stable RLOF channel.

\subsubsection{Double lined sdB binaries: Orbital periods found at last}

Spectral energy distributions are a very efficient tool to characterize binary systems. 
More information can be obtained from quantitative spectral analyses of the composite spectra, which however, require broad wavelength coverage and accurate flux calibration. From the spectra the atmospheric parameters of both components can be derived as well as their radial velocity curves. 

\paragraph{Atmospheric Parameters}\label{sect:double_bin_atmo}

\begin{figure*}
\begin{center}
\includegraphics[width=0.99\textwidth]{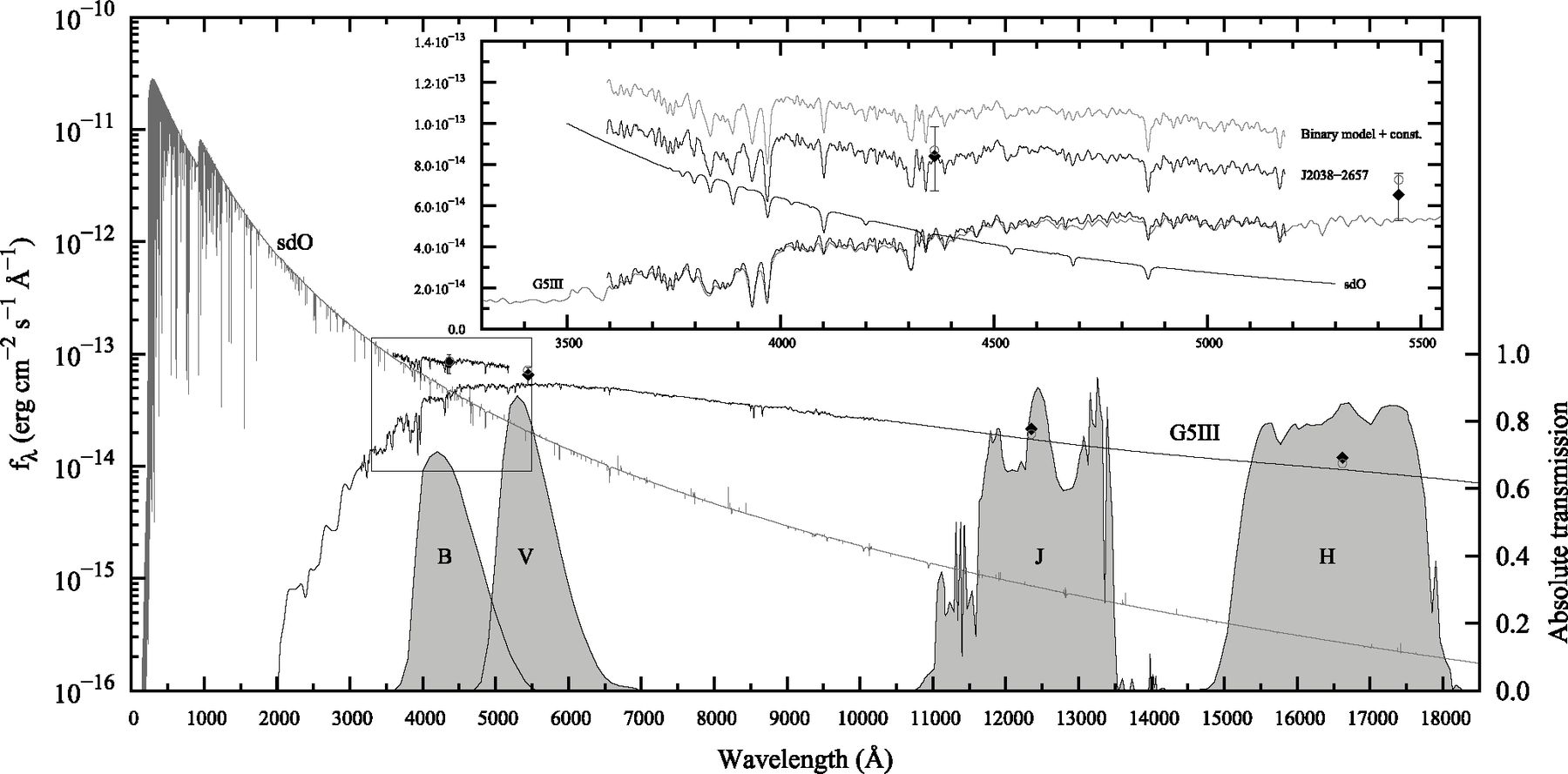}
\end{center}
\caption{Disentangling the composite spectrum of an sdO binary. The synthetic SED from an NLTE model for the sdO star (grey) and a G5III template are marked. Bessell B, V and 2MASS J and H filter transmission curves cut out the contribution of each star to the synthetic photometry. Inset: magnified part of the spectrum where decomposition was performed. The G5III templates and the sdO spectra add up nicely and this binary model fits the observation. The binary model is shifted up for clarity. Photometry measurements are plotted with black diamonds. Grey circles show the synthetic photometry.  From \citetads{2012MNRAS.427.2180N}; copyright MNRAS; reproduced with permission. }
\label{fig:double_Nemeth}
\end{figure*}

It is more difficult to derive atmospheric parameters of sdB binaries from composite spectra or spectral energy distributions. 
Substantial progress in spectral disentangling has been reported by \citetads{2012MNRAS.427.2180N} who analyzed composite spectrum sdBs from the {\it GALEX} sample based on NLTE model atmospheres and carefully selected template spectra for the cool companions (see Fig. \ref{fig:double_Nemeth}).

\begin{figure*}
\begin{center}
\includegraphics[width=0.9\textwidth]{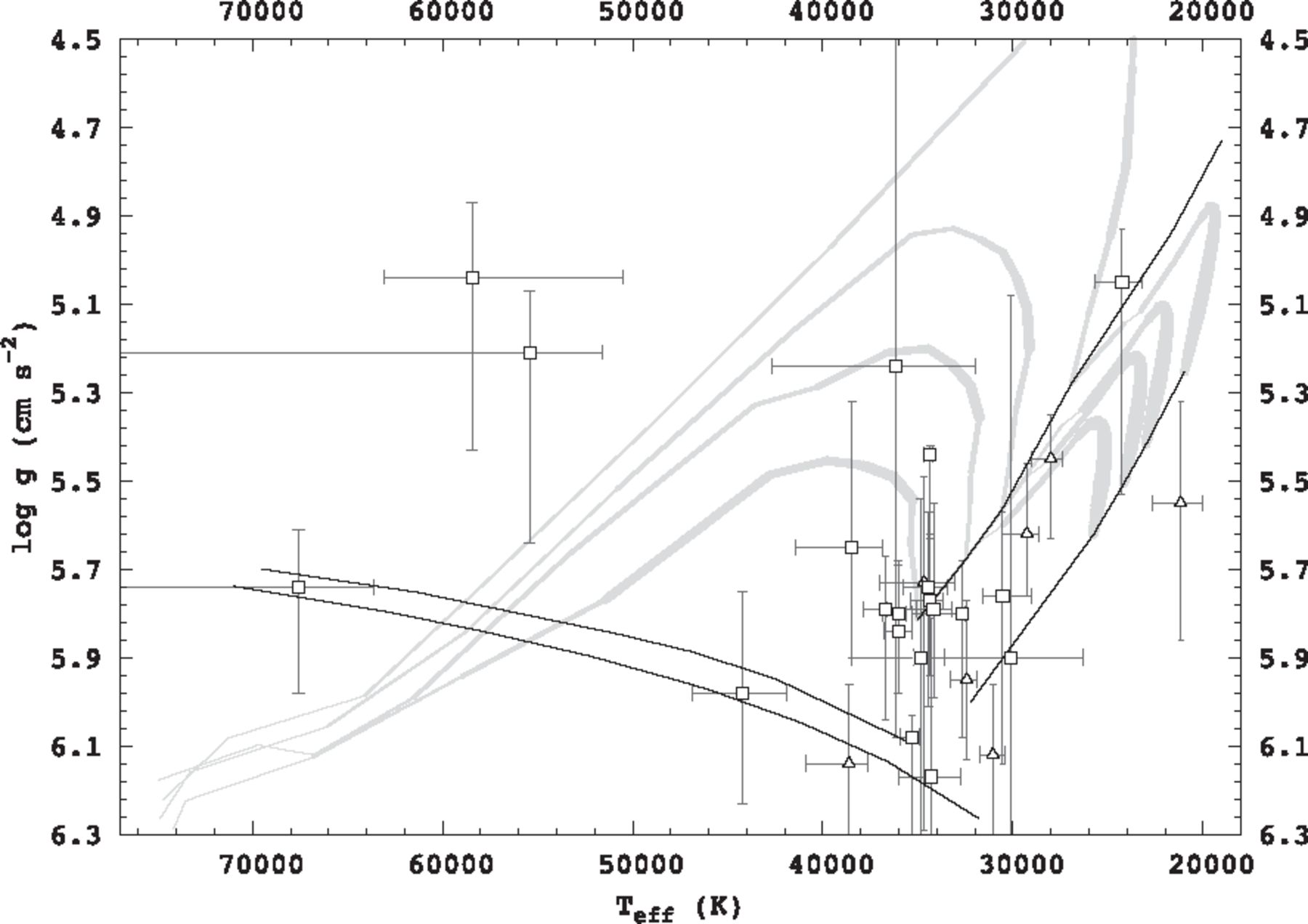}
\end{center}
\caption{T$_{\rm eff}$ - $\log$\,g diagram of sdB star composite spectrum binaries. A crowding of composites can be observed near the location of possible rapid pulsators at T$_{\rm eff}$ = 33\,500\,K, $\log$\,g = 5.8.  \citetads{2012MNRAS.427.2180N}; copyright MNRAS; reproduced with permission. }
\label{fig:Nemeth_tg_comp}
\end{figure*}

The companions turned out to be dwarfs with spectral types between G1V and F2V. One giant G3.5III companion to an sdO star was found as well as subgiant companions to two sdB stars. 
The distribution of the hot subdwarf primaries in the T$_{\rm eff}$ - $\log$ g  - plane  (see Fig. \ref{fig:Nemeth_tg_comp}) shows that they predominantly populate the hot end of the EHB unlike the sample studied by \citetads{2012MNRAS.425.1013G} from spectral energy distributions. 
They appear to cluster in the instability strip of rapid p-mode pulsators (28\,000-36\,000K, see Sect. \ref{sect:asteroseismology}). In fact, the first 
four p-mode pulsators discovered were sdB+F/G dwarfs \citepads{1997MNRAS.285..640K,1997MNRAS.285..645K,1997MNRAS.285..651S,
1997MNRAS.285..657O} 

\paragraph{Radial velocity curves}\label{sect:double_bin_rv}

Binary population synthesis models \citepads{2003MNRAS.341..669H} 
predicted that the orbital periods of such systems should be in the range 10 to 500 days. Early long-term radial-velocity monitoring at low spectral resolution did not find any variations indicating that the orbital periods must be larger than predicted  and accordingly the amplitudes of the RV-curves below the detection limits of low resolution spectroscopy. This called for high spectral resolution observations, which were initiated at the Hobby-Eberly telescope \citepads{2012ApJ...758...58B,2014ASPC..481..311W} 
as well as the Mercator telescope \citepads{2012A&A...548A...6V,2014ASPC..481..265V}. 
\citetads{2012MNRAS.421.2798D} 
were the first to derive the orbit of such a composite sdB system (\object{PG 1018-047}) and found the orbital period to be, indeed, longer (753 days) than predicted by the binary population models available at that time.
Additional discoveries followed quickly. 

\begin{figure*}
\begin{center}
\includegraphics[width=0.99\textwidth]{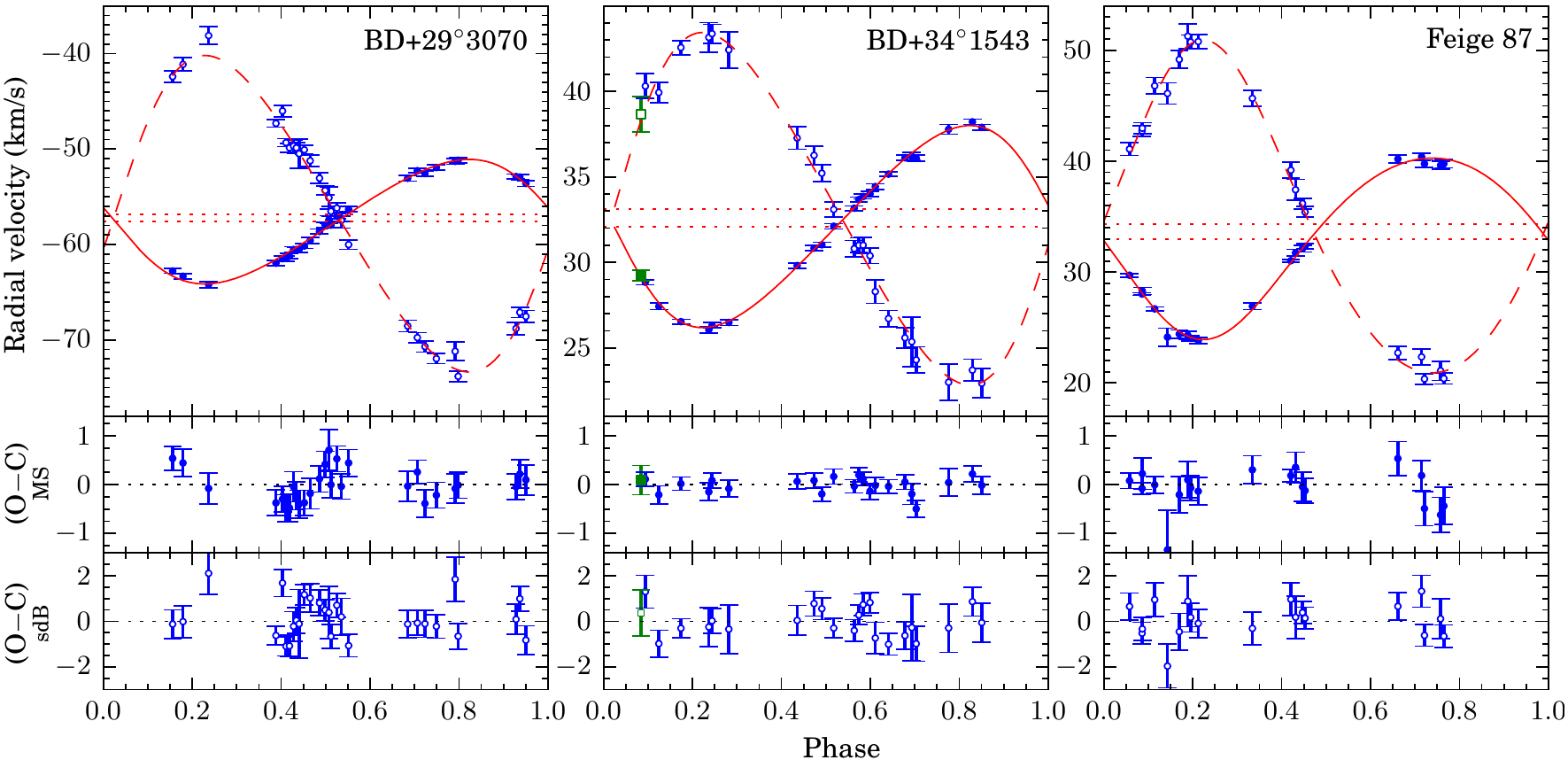}
\end{center}
\caption{Radial velocity curves for BD+29$^\circ$3070 (left), BD+34$^\circ$1543 (center), and Feige 87 (right). Top: spectroscopic orbital solution (solid line: MS, dashed line: sdB), and the observed radial velocities (filled symbols: MS component, open symbols: sdB component). The measured system velocities of both components are shown by a dotted line. The offset between the two lines results from gravitational redshift. Middle: residuals of the MS component. Bottom: residuals of the sdB component. From \citetads{2013A&A...559A..54V}; copyright A\&A; reproduced with permission. }
\label{fig:long_vos}
\end{figure*}

\citetads{2013A&A...559A..54V} derived orbits for  
\object{BD$+29{^\circ}3070$} (1283d), 
\object{BD$+34{^\circ}1543$} (972d), and \object{Feige 87} (936d). The radial velocity curves for those stars are displayed in Fig. \ref{fig:long_vos}. Note that the system velocities seem to be different for the primary and secondary in all three cases, which is caused by the difference in gravitational redshift. 
Although well known in white dwarf systems \citepads{1925PNAS...11..382A} 
 this is the first discovery of the gravitational redshift in a sdB star and allowed \citetads{2013A&A...559A..54V} to determine the surface gravities of the sdB stars.  
At about the same time long-term spectroscopic monitoring with the Hobby-Eberly telescope 
was successful and allowed orbits for \object{PG 1449+653} (P=909d) and \object{PG 1701+359} \citepads[734d, ][]{2013ApJ...771...23B} to be determined. 

By now, the orbits of nine composite spectrum sdB binaries have been solved, the periods of which range from 707 to $\approx$1370 days, exceeding the predicted upper limit of 500 days considerably. This triggered improved binary population studies, that, indeed, succeeded in producing systems of similarly long periods as observed (see also Sect.~\ref{sect:bin_evol}). 

\subsubsection{The eccentricity puzzle}

\begin{figure}
\begin{center}
\includegraphics[width=0.8\textwidth]{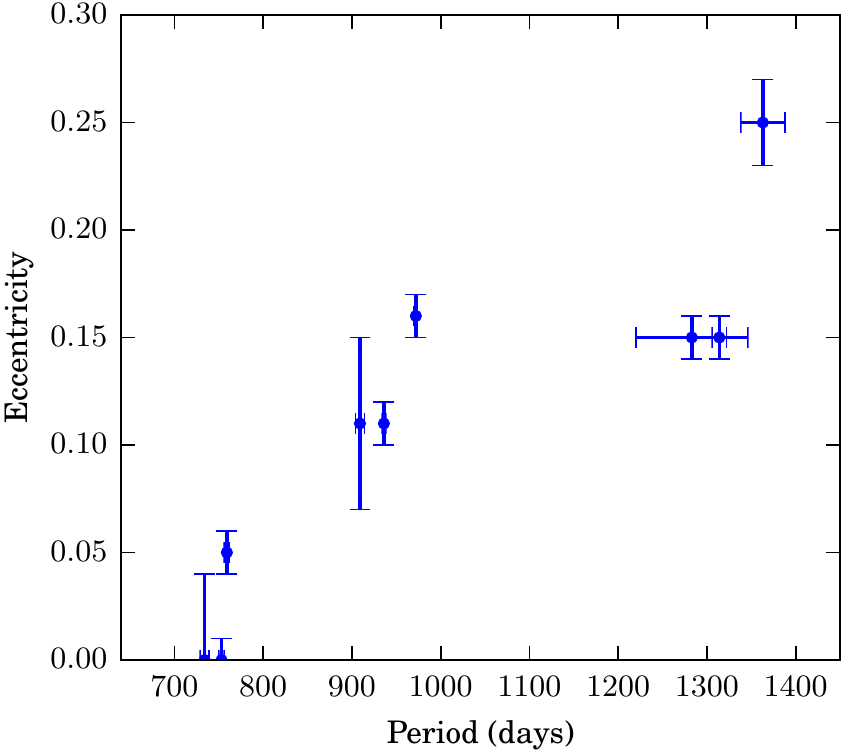}
\end{center}
\caption{The observed period-eccentricity diagram of all known long-period sdB binaries. There is a clear trend for higher eccentricities visible at higher orbital periods.
From \citetads{2015A&A...579A..49V}; copyright A\&A; reproduced with permission.}
\label{fig:long_vos_ecc}
\end{figure}

It came much as a surprise that the orbits turned out not to be circular as expected, because tidal interaction should lead to circularization of the orbit before RLOF occurs. The observed radial velocity curves, however, indicated eccentricities as large as 0.25 and a trend of the eccentricity to increase with increasing orbital period \citepads[see Fig. \ref{fig:long_vos_ecc}, ][]{2015A&A...579A..49V}.  

A similarly puzzling binary is IP Eri, a system consisting of a giant K0 and a
He WD with a period of 1071 d and an eccentricity of 0.23 \citepads[][see also Sect. \ref{sect:elm}]{2014A&A...567A..30M}. In order to explain the formation of such a system \citetads{2014A&A...565A..57S} suggested that the sdB progenitor avoided RLOF but ejected  its  envelope on the RGB due to
a tidally enhanced stellar wind. According to this scenario
the eccentricity  is  preserved  or  can  even  be
increased if the orbital period is long enough. \citetads{2015A&A...579A..49V} performed similar calculations for the long-period sdB binaries, but found that
they are insufficient to explain the high eccentricities, because the 
red giant would need to loose so much mass that it would not ignite helium and could not evolve into an sdB star.  Therefore, \citetads{2015A&A...579A..49V}
investigated two additional   
eccentricity pumping mechanisms; that is, phase-dependent RLOF on eccentric orbits and the interaction between a circumbinary disk and the binary. Assuming a certain minimal eccentricity, phase-dependent RLOF can reintroduce eccentricity and could produce binaries at the short period regime with eccentricities up to 0.15. When adding a circumbinary disk, higher eccentricities could be reached due to resonances between the binary and the disk\footnote{A similar scenario has been recently proposed by \citetads{2014ApJ...797L..24A} to explain eccentric orbits of milli-second pulsars with helium white dwarf companions.}. { Because such a disk would exist during the mass-loss era and then dissipate quickly thereafter it would hardly be observable.} However, the models predict that the eccentricities should decrease with increasing period, while the observed systems show the opposite trend. However, 
the models of \citetads{2015A&A...579A..49V} are of exploratory nature as many model parameters are poorly constrained yet.

\subsection{Multiple Systems}\label{sect:bin_mult}

More than 20\% of all binary stars are in fact members of triple systems \citepads{2013ApJ...768...33R} 
and stay bound on long time-scales if they have a hierarchical structure; that is, consist of a close binary accompanied by a relatively distant third star. 

Some sdB binaries may actually be hierarchical triples, such as PG~1253$+$284, a composite-spectrum sdB with a dwarf companion. The components were resolved by {\it HST} imaging  \citepads{2002A&A...383..938H}, 
which indicated that the dwarf companion is on a wide orbit. Nevertheless, RV variations of the sdB star were observed, which  must stem from another unresolved companion on a short-period orbit. Hence PG1253$+$284 is a triple system. 
Additional evidence of multiplicity (triples, quadruples) amongst sdB systems has recently been reported by \citetads{2014ASPC..481..301B}. 
New observations from the  MUCHFUSS project have identified another candidate triple system \citepads[J09510$+$03475,][]{2015A&A...576A..44K}. 

\subsection{Giant and massive companions to hot subdwarf stars}\label{sect:bin_massive}

Most cool companions to sdB stars are main-sequence stars. A giant companion has
been found to BD$-$7$^\circ$5477 on a wide orbit \citepads[orbital period P= 1195 $\pm$ 30 d,][]{2014ASPC..481..265V}
resulting from stable RLOF (see Sect. \ref{sect:double_bin_rv}). 
In addition subgiant or giant companions were found for the sdO star FF~Aqr \citepads[K0 III,][]{1977PASP...89..616E,1977MNRAS.181P..13D,2003MNRAS.342..564V,2011AstBu..66..332S} and the sdB HD~185510 \citepads[K~III--IV,][]{1992MNRAS.258...64J,1993AJ....106.2370F,1998A&A...333..205F} on short orbits
 (P=9.2d and 20.7d, respectively), but still detached\footnote{The hot subdwarf in HD 185510 is possibly a low-mass helium-core object \citepads{1997MNRAS.286..487J}  
(see Sect. \ref{sect:elm}).}.  
The systems are photometrically variable, caused by the reflection effect, chromospheric activity and spots.
\citetads{2012MNRAS.427.2180N} 
discovered a third such system, {\it GALEX} J2038$-$2657, which hosts a G8III companion. Its variability was discovered only recently by \citetads{2015MNRAS.450.3514K}. 
\citetads{1990PASP..102..912H} 
and \citetads{1993A&A...276..171R} 
showed that the sdO+G giant binary \object{HD~128220} (orbital period P= 871.78d) formed via case C mass transfer; that is, the sdO progenitor filled its Roche lobe while ascending the asymptotic giant branch. Accordingly the sdO component is a post-AGB star of 0.55 M$_\odot$.

Three sdO stars
are known as companions to Be stars: 59~Cyg \citepads{2008ApJ...686.1280P}, 
FY CMa 
\citepads{2013ApJ...765....2P}, 
and $\phi$ Per \citepads{1998ApJ...493..440G}\footnote{The Be star HR 7409 (7~Vul) may be a related object \citepads{2011MNRAS.413.2760V}, but the nature of the companion needs to be clarified.}. 
Such sdO stars were predicted by \citetads{1991A&A...241..419P} to form when the donor star begins mass
transfer during its shell-hydrogen
burning phase and evolves into an sdO star. The less-massive gainer is spun-up and evolves into a Be
star. 

\clearpage

\newpage
\section{Substellar and planetary companions to sdB stars}\label{sect:planets}

Most extra-solar planets known today have been discovered by either the radial-velocity method or the transit method.
In the case of hot subluminous stars, both methods have not yet been successful {\citepads{2011AIPC.1331..304J,2011ApJ...743...88N}}, although brown dwarf companions have been found.

However, a scientific breakthrough came 
with the discovery of sinusoidal variations of the pulsation frequencies of the sdB pulsator V391~Peg (=HS2201+2610), which are due to the light travel time effect. Accordingly, the sdB star is orbited by a
$M\,\sin i$ = 3.2~M$_{\rm Jupiter}$ planetary companion 
at about 1.7~AU in 3.2 years and implies that a planet
may survive the expansion of the red giant host star at distances of less 
than 2 AU \citepads{2007Natur.449..189S}.  

Light curves from the {\it Kepler} mission provided evidence for Earth-sized planets around two pulsating sdB stars. The analysis of the {\it Kepler} light curves from quarters Q5 to Q8 of KIC~05807616 revealed a rich pulsation spectrum 
\citepads{2011Natur.480..496C}. The frequency spectrum shows gravity modes as well as acoustic modes but no clear multiplet structures. Most importantly, 
\citetads{2011Natur.480..496C} 
detected two weak modulations at frequencies below the cut-off frequencies, which limits pulsational periods. Excluding other effects \citetads{2011Natur.480..496C} arrived at the conclusion that the signals are due to the reflection effect of two Earth-size planets orbiting with periods of 5.7 and 8.2 hours, respectively.
\citetads{2015A&A...581A...7K}  analyzed all available {\it Kepler} photometry from Q4 to Q17 and confirm the presence of both frequencies. Moreover, they found evidence that both amplitudes and frequencies are unstable, with amplitudes varying between non-detection (signal-to-noise S/N$<$4) to S/N=6--8. 
These variations can hardly be due to planetary reflection effects and  \citetads{2015A&A...581A...7K} {{conjectures}} that both frequencies are some sort of damped pulsation frequencies beyond the pulsational cut-off. 

Similar low-frequency modulations were found for KIC~10001893
in all available {\it Kepler} photometry (993.8 d of short and 147.9 d of long cadence) data and interpreted as caused by planetary reflection effects 
suggesting the presence of three planets
with orbital periods of P1 = 5.273, P2 = 7.807, and P3 = 19.48 h, respectively
\citepads{2014A&A...570A.130S}. 
Interestingly, the period ratios P2/P1 = 1.481 and P3/P2 = 2.495 are very close to the 3:2 and 5:2 resonances, respectively. One of the main pulsation modes of the star at 210.68 $\mu$Hz corresponds to the third harmonic of the orbital frequency of the inner planet, suggesting that g-mode pulsations of an sdB star are tidally excited by a planetary companion. This interpretation, though, should be taken with a grain of salt in view of the results for KIC~05807616. However,
phase and amplitude variations such as seen in the light curve of the latter have not been reported by \citetads{2014A&A...570A.130S}.

If confirmed such systems would be interesting {{for}} studying the survivability of planets during red giant evolution \citepads[see e.g.][]{2007ApJ...661.1192V,2009ApJ...705L..81V,2010MNRAS.408..631N,2013MNRAS.432..500N,2014ApJ...794....3V}\footnote{See also the proceedings of the conference ''Planetary systems beyond the main sequence'' \citepads{2011AIPC.1331.....S}.}.
Those planets must have lost a large amount of their initial few Jupiter masses during the CE phase. 
\citetads{2012ApJ...759L..30P} 
investigated whether such planets could have been stripped of significant amounts of mass during the CE phase and concluded that the Earth-mass planets of  KIC 05807616 could well be the remnants of one or two Jovian-mass planets that were stripped of their envelope when they orbited in the envelope of the red giant.  
\citetads{2012ApJ...749L..14B} suggested an alternative scenario,
which assumes a single massive planet of 5 Jupiter masses, that loses its entire gaseous envelope during spiral-in in the CE phase, its metallic core is then disrupted {{into}} two or more surviving Earth-size fragments that migrate to resonant orbits.

However, independent confirmations of those discoveries are still lacking \citepads[see ][for discussions]{2014ASPC..481....3S,2014ASPC..481...13S}. 
Hence, all substellar companions reported in this section should be regarded as 
candidates, only.  The more so, because recent  
deep SPHERE@ ESO-VLT observations 
 of the white dwarf plus brown dwarf candidate V471 Tau ruled out that the observed eclipse timing variation are caused by a proposed circumbinary brown dwarf at the expected position \citepads{2015ApJ...800L..24H}. 
 
\subsection{Circumbinary planets and brown dwarfs} \label{sect:hwvirplanets}

Circumbinary planets have been discovered around unevolved binaries such as the dG/dM system \object{{\it Kepler}~47} \citepads{2012Sci...337.1511O} 
or \object{{\it Kepler}~16} \citepads{2011Sci...333.1602D}. 
To find out whether such planetary systems can survive stellar evolution, it is important to search for planets around evolved binaries. Eclipsing sdB binaries provide a useful playground, because mid-eclipse timings would allow a search for period changes. HW~Vir stars are well suited for such an exercise, because their periods are short and some, in particular the prototype, have already been observed for decades.  \object{HW~Vir} was discovered by \citetads{1986IAUS..118..305M} 
 and the first orbital period change was found nine years later \citepads{1994MNRAS.267..535K}\footnote{The history (``saga'') of \object{HW Vir} observations is nicely summarized by \citetads{2014ASPC..481..259V}.}. 

 Moreover, the brightest HW~Vir systems have been covered by automatic all-sky surveys such as SuperWASP \citepads{2014A&A...566A.128L} that provide plenty of eclipse epochs. 

While \object{AA~Dor} did not show any evidence for period variations for decades
\citepads{2014MNRAS.445.4247K}, 
almost all HW~Vir systems, that have been monitored for a sufficiently long time show period changes. 
\citetads{2009AJ....137.3181L} 
announced the detection of two giant planets orbiting the prototype \object{HW Vir}
from systematic variations in the timing of eclipses. This finding, however, was questioned, because the planetary orbits were found to be mutually crossing. Strong gravitational interactions have to be expected so that the system would be 
unstable on the required long time scales.

\citetads{2012MNRAS.427.2812H} 
performed a detailed stability analysis of the proposed \object{HW~Vir} planetary system and confirmed that, indeed, the system is unstable with a mean lifetime of less than a thousand years. Hence, it was concluded that the exoplanets do not exist.

\begin{figure*}
\begin{center}
\includegraphics[width=0.9\textwidth]{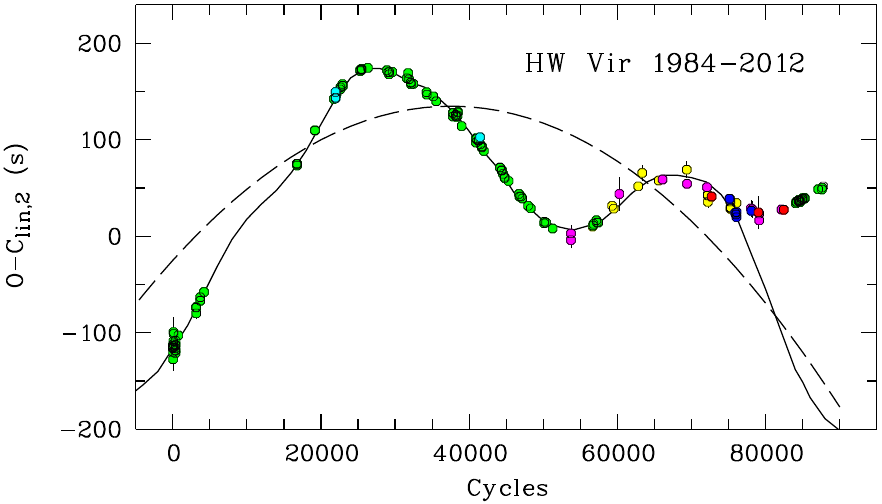}
                \end{center}
\caption{O-C diagram residuals of the mid-eclipse times from the linear ephemeris used by \citetads{2009AJ....137.3181L} along with their model curves for the two-companion model (solid) and the underlying quadratic ephemeris (dashed). Note that the new 2008--2012 epochs from \citetads{2012A&A...543A.138B} strongly deviate from the \citetads{2009AJ....137.3181L} model. 
From \citetads{2012A&A...543A.138B}; copyright A\&A; reproduced with permission.}
\label{fig:hwvir}
\end{figure*}

\begin{figure*}
\begin{center}
\includegraphics[width=0.9\textwidth]{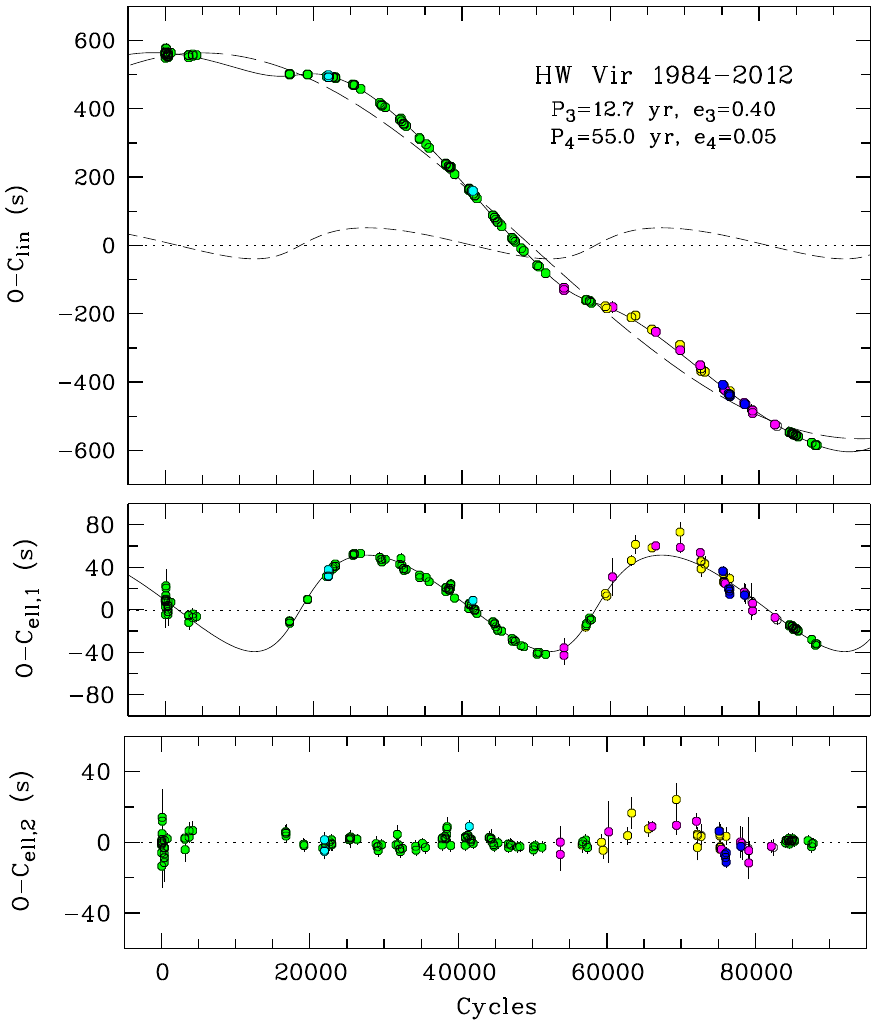}
                \end{center}
\caption{Fit of two Keplerian orbits to the eclipse-time variations of HW~Vir. Top: data of Fig. \ref{fig:hwvir} 
relative to the linear ephemeris of \citetads{2012A&A...543A.138B}. The curves denote the model light travel time effect (solid) and the contributions by the outer companion (long dashes) and the inner planet (short dashes). Center: data with the contribution by the outer companion subtracted and model for the inner planet (solid curve). Bottom: residuals after the subtraction of the contributions by both companions.
From \citetads{2012A&A...543A.138B}; copyright A\&A; reproduced with permission.}
\label{fig:hwvir_new}
\end{figure*}

The picture changed again, when new eclipse timings became available
\citepads{2012A&A...543A.138B}  
as demonstrated in Fig. \ref{fig:hwvir}.
The new timings deviate strongly from the model of \citetads{2009AJ....137.3181L}. Including observations up to 2012,
\citetads{2012A&A...543A.138B} presented a new secularly stable solution with two companions (see Fig. \ref{fig:hwvir_new}). The inner one is a giant planet of M$_3$ $\sin$ i$_3$ = 14 M$_{\rm Jup}$ orbiting \object{HW~Vir} in 12.7 yrs. The outer one is a brown dwarf or low-mass main-sequence star of M$_4$ $\sin$ i$_4$ = 30--120 M$_{\rm Jup}$ on a 55 yrs orbit. The uncertainty of the orbital period of the latter is large ($\pm$ 15 yrs) because the observational time base is still too short. 
\citetads{2012A&A...543A.138B} find that such a system would be stable over more than 10$^7$ yrs, in spite of the sizeable interaction. Since no process other than the light-travel time variations had to be invoked, the planetary hypothesis of the eclipse-time variations has been revived.

The Beuermann team announced the detection of third bodies 
to the sdB binaries 
\object{NSVS14256825} and \object{HS~0705+6700} from cyclic variations in their measured orbital period. The third object in \object{NSVS14256825} is a giant planet with a mass of roughly 12 M$_{\rm Jup}$, whereas, in the case of \object{HS~0705+6700}, it is a brown dwarf of 31 M$_{Jup}$ if the orbit is coplanar with the binary \citepads{2009ApJ...695L.163Q}. An extended series of eclipse timings is consistent with the presences of a third body to the \object{HS~0705+6700} system and hints at the possible presence of a fourth \citepads{2015arXiv150204366P}.

For \object{NSVS14256825}, the existence of a fourth body was proposed \citepads{2013ApJ...766...11A}. 
However, the orbital stability analysis by \citetads{2013MNRAS.431.2150W} 
suggested that proposed orbits for the two planets are extremely unstable on time-scales of less than a thousand years, regardless of the mutual inclination between the planetary orbits. 
\citetads{2014MNRAS.438..307H} reanalyzed the photometric data  
and find no evidence for the existence of a second planet. The one-companion model is shown to be poorly constrained by the existing data set, because various models result in substantially different orbits despite similar statistical significance. The baseline of timing data needs to be extended to nail down the orbit of the third body.

Finally, \object{NY Vir} is a pulsating sdB binary harboring a circumbinary planet \citepads{2012ApJ...745L..23Q}. 
The issue of stability
arose in this case also when \citetads{2014MNRAS.445.2331L} 
presented a new model based on historical eclipse times combined with their long-term CCD data and found that the periodic variations are  most likely caused  by a pair of light-travel-time effects due to the presence of two planets of M$_3$ $\sin$ i$_3$ = 2.8 M$_{\rm Jup}$, and M$_4$ $\sin$ i$_4$ = 4.5$_{\rm Jup}$, respectively. However, their dynamical analysis suggests that long-term stability requires the outer companion to orbit on a moderately eccentric orbit. Further monitoring is required to corroborate this issue. 
The most recent discovery is a low mass circumbinary companion to the pulsating 
sdB \& dM system \object{2M1938+4603} \citepads[][see also Sect. \ref{sect:eclipse_puls}]{2015A&A...577A.146B}. 

However, the travel time variation could be due to other non-circumbinary mechanisms. The Applegate mechanism \citepads{1992ApJ...385..621A}, which is caused by gravitational coupling of the orbit to changes of oblateness of a magnetically active star, is often considered, but the luminosity of the convective secondaries is mostly found to be insufficient to drive the Applegate mechanism \citepads[e.g. for \object{HS~0705+6700,}][]{2015arXiv150204366P}.

The lesson learned from the \object{HW~Vir} case is that one has to be patient and never stop observing.
Despite of the problems described, evidence is accumulating that, indeed, sdB binaries of the HW~Vir type are orbited by additional substellar companions that may have survived the giant-phase evolution of their host binary.

\subsection{Planet formation: first or second generation}\label{sect:planet_form}

Substellar circumbinary objects have also been discovered in other post common envelope binary (PCEB) systems, hosting white dwarfs instead of sdB stars \citepads[e.g.][]{2013A&A...555A.133B}. 

Among the ten well studied systems all but AA~Dor show period changes calling for the presence of circumbinary planets \citepads{2013A&A...549A..95Z}. 
This frequency (90\%) of substellar-mass circumbinary companions to PCEB systems is surprisingly high when compared to that of close main-sequence binaries, for which \citetads{2012Natur.481..475W}  
found from the {\it Kepler} sample that more than ~1\% of such binaries have giant planets on nearly coplanar orbits.

In order to test the hypothesis that the PCEB companions could indeed be of first generation, \citetads{2013A&A...549A..95Z} 
compared binary population models with observational and theoretical results for the formation of circumbinary giant planets and concluded that only 10\% of the PCEBs could have first generation giant planets. 

The circumbinary planets may not have existed before the common-envelope phase, but may have formed from the material of the common envelope \citepads{2014MNRAS.444.1698B}.

Recent simulations of common envelope evolution showed that some fraction ($\approx$10\%) of the ejected material may remain bound to the PCEB \citepads{2014A&A...563A..61S} and is likely to form a circumbinary disk, sufficiently massive for gravitational instabilities to occur and form giant planets. However, the process has to be sufficiently fast to explain systems as young as  NN~Ser (1 Myr), a white dwarf binary with two circumbinary massive planets similar to HW~Vir \citepads{2013A&A...555A.133B}. A detailed discussion can be found in \citetads{2013A&A...549A..95Z}.   

Second-generation planets might form more efficiently if remnants of lower mass planets survived the common envelope phase and provided seeds to accrete material quickly to form giant planets \citepads[hybrid first- and second-generation scenario,][]{2013A&A...549A..95Z,2014A&A...563A..61S}.

\citetads{2014A&A...563A..61S} 
and \citetads{2015AN....336..458S} 
modelled the case of NN Ser, 
 and find that the current data cannot be explained by pure first generation models, whereas the second generation scenario 
naturally explains the observed masses of the two circumbinary planets.
\citetads{2014MNRAS.444.1698B} 
carried out similar calculations, compared them to twelve PCEB systems that host planets and concluded from angular momentum considerations that they are more likely to be of first generation. However, \citetads{2014A&A...563A..61S} 
 point out that there are probably two populations,
where some systems form primordially with the progenitor star and some are of second generation.

\citetads{2013A&A...549A..95Z} suggested that the eclipse timing variations may not be caused by the existence of third and fourth bodies at all, but by the secondary being magnetically active. They suggest that this could be tested because the period variations should not occur in close WD binaries with a second WD component as the secondary is not likely to be active.
Hence the formation of circumbinary planets to PCEBs remains an open question. 

\clearpage

\newpage
\section{Pulsating hot subdwarf stars}\label{sect:asteroseismology}

\begin{figure}
\begin{center}
\includegraphics[width=0.9\textwidth]{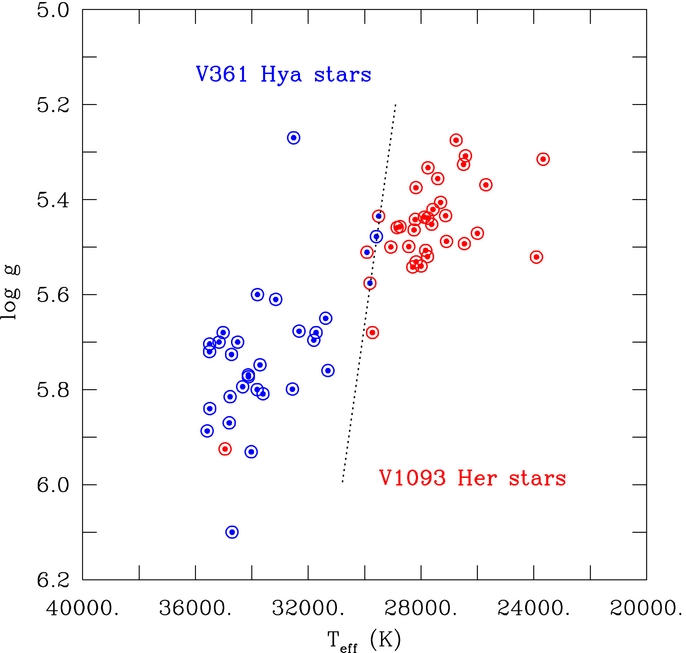}
                \end{center}
\caption{Distribution of pulsating sdB stars in the effective temperature--surface gravity plane. The locations of 28 short-period p-mode pulsators of the V361 Hya type are indicated in blue, while those of 30 long-period g-mode variables of the V1093~Her type are shown in red. Three hybrid pulsators, showing simultaneously both p-modes and g-modes, are shown in red and blue, lying at the common boundary between the two distinct domains, defined approximately by the dotted line. LS IV$-$14$^\circ$116 finds itself (the red symbol in the V361 Hya domain) totally out of place.
From \citetads{2011ApJ...734...59G}; copyright ApJ; reproduced with permission.}
\label{fig:kiel_pulsators}
\end{figure}

\begin{figure}
\begin{center}
\includegraphics[width=0.6\textwidth]{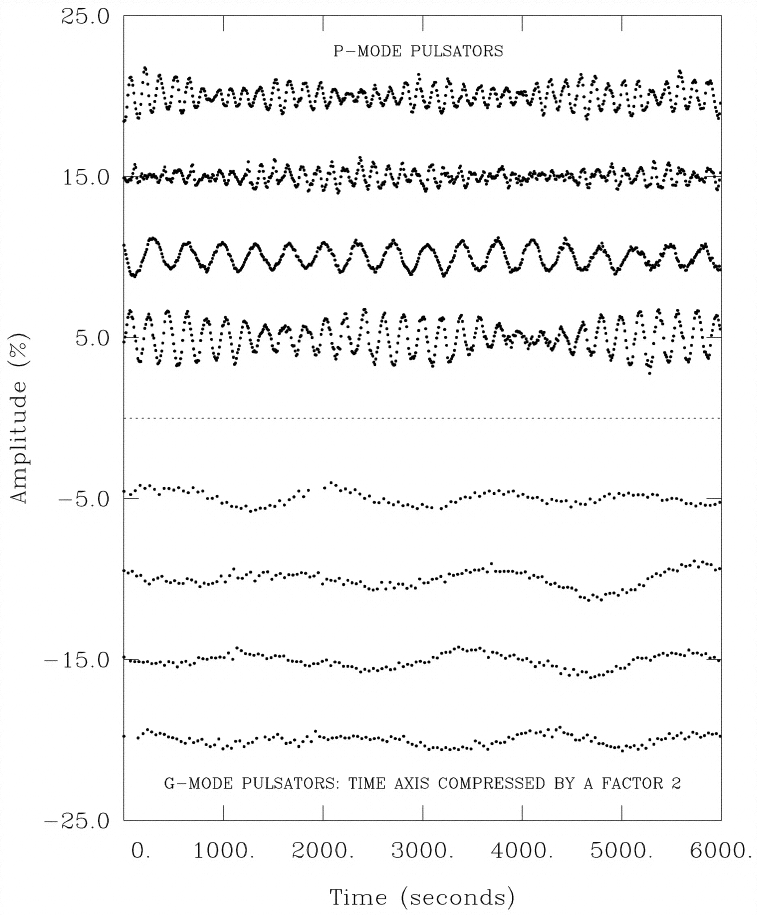}
                \end{center}
\caption{
Representative light curves of four short-period (top) and four long-period (bottom) pulsating sdB stars. The short-period p-mode pulsators (V361 Hya  stars) are, from top to bottom, PG 1047+003, PG 0014+067, Feige 48, and KPD 2109+4401. These curves are ``white-light'' light curves expressed in terms of percentage of residual amplitude relative to the mean brightness of the star. The curves have been shifted by arbitrary amounts in the vertical direction away from the zero point for visualization purposes. The long-period g-mode pulsators (V1093~Her stars) have been observed through different bandpasses. The pulsators are, from top to bottom, PG 1716+426 (R), PG 0850+170 (R), PG 1338+481 (B), and PG 1739+489 (V). Note that the time axis refers to the top half of the figure; the light curves in the bottom half have been compressed by a factor of 2. From \citetads{2003ApJ...597..518F}; copyright ApJ; reproduced with permission.}
\label{fig:p_g_pulsators}
\end{figure}

Multi-periodic light variations of low amplitudes (few milli-mag)
and periods of a few minutes were discovered in sdB stars \citepads[now termed V361\,Hya
  stars, ][]{1997MNRAS.285..640K} 
at almost the
same time at which they were predicted by theory to be caused by non-radial 
pulsations \citepads{1996ApJ...471L.103C}. 
 The driving mechanism was readily identified
as being due to an iron opacity bump \citepads{1997ApJ...483L.123C}.
The V361\,Hya stars are found amongst the hotter sdB
stars with 28,000 $< T_{\rm eff} <$ 35,000~K and 5.2 $< \log g <$ 6.1 (see Fig. \ref{fig:kiel_pulsators}). 
The periods suggest that the stars are $p$-mode
pulsators, offering the opportunity of using oscillations to probe the
interior of sdB stars. 
Asteroseismology of sdB stars received another boost when pulsations
with periods of 45\,min--2\,hr were discovered \citepads[][ now termed V1093~Her stars]{2003ApJ...583L..31G}. Typical light curves are shown in Fig. \ref{fig:p_g_pulsators}. 
The much longer period
pulsations found in these stars indicate that they are gravity modes. The stars
are typically cooler than the $p$-mode pulsators  (see Fig. \ref{fig:kiel_pulsators}). The pulsations in both groups are driven by an
opacity bump due to iron ionization 
\citepads[and other iron-group elements, see][]{1997ApJ...483L.123C,2003ApJ...597..518F,2008ASPC..392..231F}.
However, iron must be enhanced by diffusion 
processes in the sub-photospheric layers in order to drive the pulsations. 

\begin{figure*}
\centering
  \includegraphics[width=0.45\textwidth]{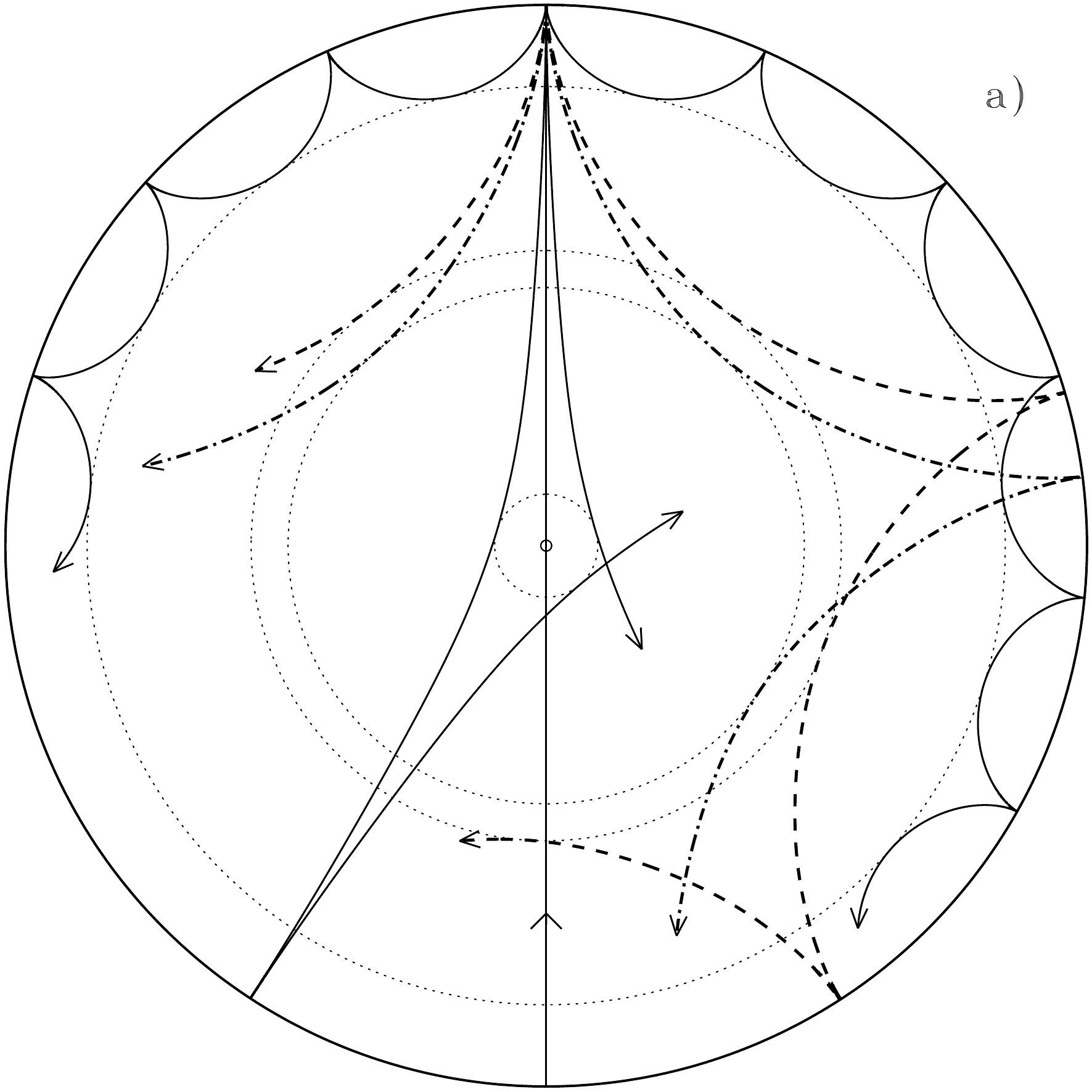}
  \includegraphics[width=0.45\textwidth]{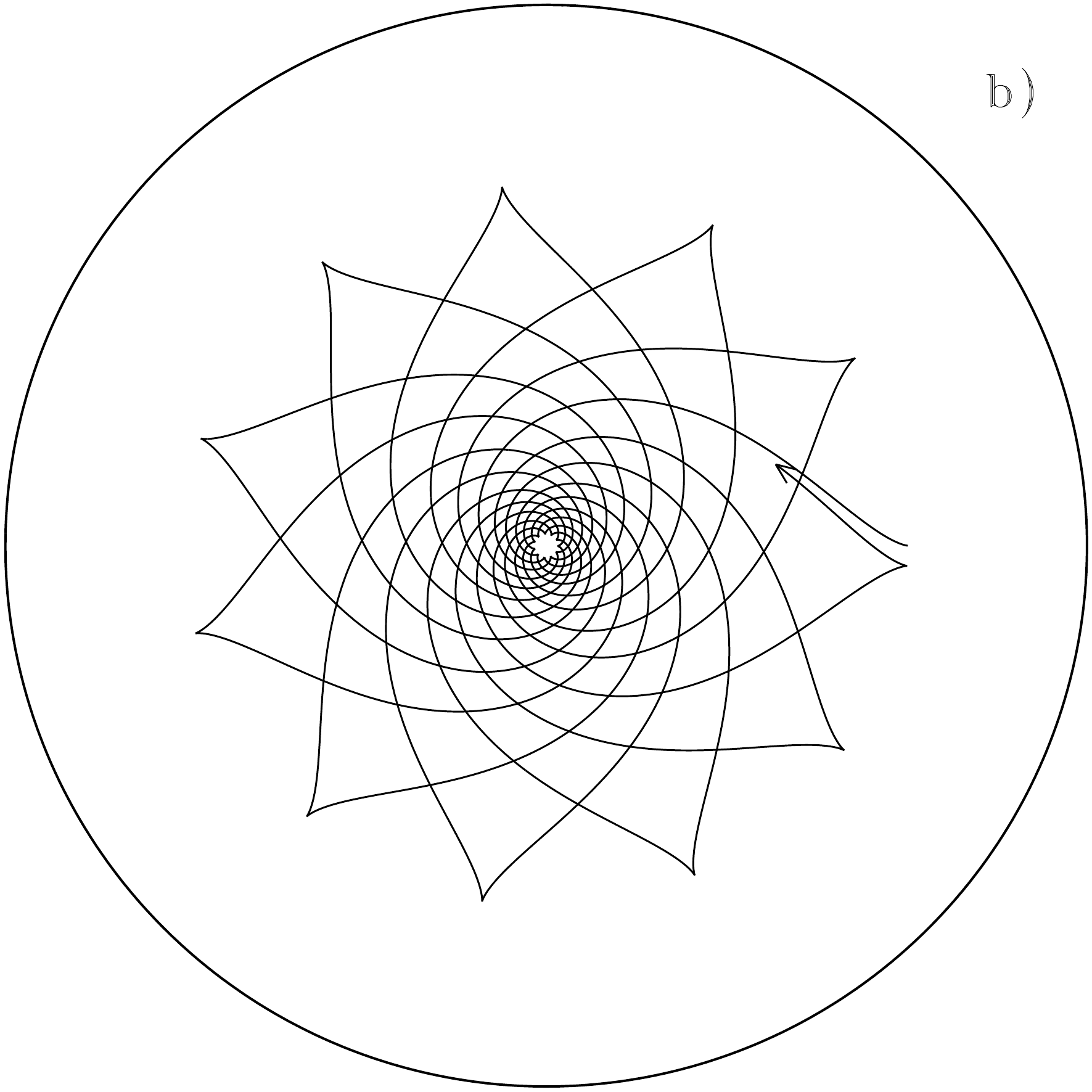}
\caption{Illustration of the propagation of sound and gravity waves in a cross-section of the
solar interior. The acoustic ray paths (panel a)
are {{bent}} by the increase in sound speed
with depth until they reach the inner turning point
(indicated by the dotted circles) where they undergo total internal
refraction.
At the surface the acoustic waves are reflected by the rapid decrease
in density.
Rays show corresponding modes with frequency $3000 \mu Hz$ and degrees
(in order of increasing penetration depth) $l = 75$, $25$, $20$ and $2$;
the line passing through the center schematically illustrates the behavior
of a radial mode.
The gravity-mode ray path (panel b) corresponds to a mode of frequency
$190 \mu Hz$ and degree 5. From \citetads{2007A&ARv..14..217C}; copyright A\&ARv; reproduced with permission.
}
\label{fig:propagation_demo}       
\end{figure*}

Of great importance for the development of asteroseismology are the 
so-called hybrid pulsators 
which show both short
period $p$-mode pulsations as well as long-period $g$-mode pulsations
\citepads[e.g.][]{2006A&A...445L..31S} 
and which lie at the temperature boundary ($\approx$28\,000~K) between both classes of pulsating stars (see Fig. \ref{fig:kiel_pulsators}).
Acoustic waves (p-modes) propagate in the outer regions of the star, whereas 
g-modes do in the deep interior (see Fig. \ref{fig:propagation_demo} for an illustration). Hence, the internal structure of the star can be probed if both types of pulsations are detected in a star.

Dedicated surveys that targeted sdB stars in the predicted instability strip were partially successful {{at finding}} new sdB pulsators. For example a survey at the Nordic Optical Telescope 
\citepads{2010A&A...513A...6O} monitored more than 300 sdBs predicted to lie in the instability strip but discovered only twenty new short-period pulsators
which means that only about 10\% of the stars in the V361~Hya instability strip are actually pulsating with amplitudes of a few mmag, detectable from ground\footnote{\citetads{2010A&A...513A...6O} also provide limits on the pulsation amplitudes for 285 objects with no obvious variations at the expected time scales.}.

Searches for long-period
pulsations require a lot more observing time and photometric precision. New discoveries have, for instance, been reported from ground-based studies, e.g. 
two pulsators of V1093~Her type \citepads[GALEX J0321+4747 \& GALEX J2349+3844,][]{2012ASPC..452..121K}. 
The companion to the sdB in GALEX J0321+4747 is a low mass dwarf, whereas in the case of GALEX J2349+3844 it is probably a white dwarf. 
Another important discovery from the MUCHFUSS project was \object{FBS 0117+396}, a hybrid pulsator in an sdB+dM system \citepads[][ see Sect. \ref{sect:synchro_puls}]{2013A&A...559A..35O}. 
The {\it Kepler} mission provided a much better option to search for long-period sdB pulsators and, indeed, discovered more than a dozen of them \citepads{2010MNRAS.409.1470O,	
2011MNRAS.414.2860O}.

Pulsation modes can be described by three quantized numbers, $n$, $\ell$, and $m$, where n is the number of radial nodes between
center and surface, $\ell$ that of { surface} nodes { perpendicular to the pulsation axis}, and m
the number of surface nodes passing through the pulsation axis.
The radial fundamental mode is assigned 0,0,0 and negative $n$ denote gravity (g) and positive $n$ {{pressure}} (p) modes.

Ground-based observations did not reach sufficient precision and resolution to pin down those numbers unambiguously, leaving the theorists with the so-called forward modelling technique of quite incomplete frequency spectra  \citepads[e.g.][]{2007CoAst.150..241C,2008A&A...489..377C}. 
The breakthrough came with the {\it Kepler} satellite mission that achieved just that goal. It is not exaggerating to say that {\it Kepler} data have revolutionized asteroseismology of pulsating stars in general and of pulsating sdB stars in particular. 

However, before exploring the {\it Kepler} era, we shall first 
 discuss fundamental issues and techniques such as the 
 location of the instability strips for $p$-mode and
$g$-mode pulsators, respectively (Sect. \ref{sect:strip}).
The {\it Kepler} satellite provided broadband photometry of unprecedented precision and frequency resolution, whereas ground-based telescopes offer complementary techniques such as multi-band photometry and time-series spectroscopy, which will be discussed in Sect. \ref{sect:multi_spectro}. We shall turn to the sdB  pulsators in the {\it Kepler} field (Sect. \ref{sect:kepler_puls}), review the perspective to probe the stellar interior and determine stellar ages (Sect. \ref{sect:internal_age}), visit the pulsating sdBs in eclipsing binaries (Sect. \ref{sect:eclipse_puls}), revisit the rotation characteristics derived from multiplet splittings and discuss the impact on synchronization time scales
(Sect. \ref{sect:synchro_puls}). New classes of pulsating hot subdwarf star have also been discovered, e.g. in the globular cluster $\omega$ Cen (Sect. \ref{sect:cluster_puls}), as well as the unique pulsating He-sdB star LS~IV-14$^\circ$116 (Sect. \ref{sect:lsiv_puls}). Finally we take a glance at the ongoing {\it Kepler K2} mission.

\subsection{The instability strips}\label{sect:strip}

The location of the instability strip in the 
T$_{\rm eff}$--$\log$ g-plane is of foremost interest. For the V361~Hya pulsators the prediction from models provide an excellent match to the 
observations \citepads[see][]{2007CoAst.150..241C}, 
as the excited modes have periods very similar to those observed in these stars. 

However, the predicted
instability strip is wider than observed and pulsators and 
non-pulsators coexist in the same region of the T$_{\rm eff}$--$\log$ g- diagram, i.e. only one out of ten stars pulsates. Because ground-based observations may have missed low-amplitude pulsators, this might be just an observational selection effect. High-precision {\it Kepler} photometry, however, has confirmed the scarcity of p-mode pulstors\footnote{\citetads{2011MNRAS.414.2860O} surveyed 32
sdB pulsator candidates hotter than 28 000 K and found only one pulsator of V361 Hya type, a transient one and one hybrid pulsator} indicating that the {{V361 Hya}} instability strip is by no means pure \citepads{2011MNRAS.414.2860O}.

The long-period, low-amplitude oscillations of the V1093~Her stars are much harder to detect from the ground. {\it Kepler} photometry has revealed that almost all 
sdB stars which have effective temperatures in the predicted instability strip in the {\it Kepler} field are indeed pulsating. 

The theoretical blue edge of the V1093~Her stars posed a problem to modelers. Early models predicted the blue edge at about 5\,000~K cooler than observed \citepads[see][]{2007CoAst.150..241C}. 
However, \citetads{2006MNRAS.372L..48J, 2007MNRAS.378..379J} found 
that adding Ni and using opacities from the Opacity Project 
(rather than OPAL) shifts the $g$-mode
instability strip by 5\,000 K to the blue. 
Artificial abundance enhancements of Fe and Ni were introduced in the pulsation driving layers.
Therefore, a more detailed investigation of the driving mechanism is required 
accounting for Ni and other iron-group elements with abundance 
stratifications shaped by diffusion. 
Self-consistent calculations that account for an enrichment of iron group elements in the driving regions resolved the issue \citepads{2014A&A...569A.123B}.
In fact, \citetads{2014A&A...569A.123B} found that the abundance enhancements of Fe and Ni were actually previously underestimated. 
The new models predict the instability strip of observed g-mode pulsators very well for masses close to 0.47 M$_\odot$, which solved the blue edge problem of the sdB g-mode instability strip.  

\subsection{Multi-color light curves and time-series spectroscopy}\label{sect:multi_spectro}

Monochromatic light curves are typically used to study the pulsations of hot subdwarf stars, with the {\it Kepler} space mission being the flagship tool that provided enormous progress in the field (see Sect. \ref{sect:kepler_puls}).
{{Often observations are done in white light; that is, no filters are employed, in order to improve throughput and, hence, the signal-to-noise ratio.}}  However, multi-color photometry provides another important tool to study the properties of pulsating stars.
The three-channel imager ULTRACAM \citepads{2007MNRAS.378..825D}
has provided 
 excellent light curves in three filters \citepads[e.g.][]{2004MNRAS.352..699J,2005MNRAS.362...66J,2006MNRAS.367.1317A,
2013A&A...559A..35O}. 
An illustrative example is the light curve of the HW Vir binary NY~Vir \citepads{2007A&A...471..605V}, which shows the pulsations in addition to the reflection effect and the eclipses (see Fig. \ref{fig:ny_vir}). Also the four-channel imager BUSCA at Calar Alto observatory \citepads{1999SPIE.3649..109R}  has provided multi-color observation of pulsating sdB stars \citepads[e.g.][]{2003A&A...401..289F}. 
The degree {{index $\ell$}} can be inferred from multi-color lightcurves and spectrophotometry by making use of the frequency dependence of the amplitude of an oscillation and its phase, and synthesizing the brightness variation expected from temperature, radius, and surface gravity perturbations across the stellar disk from model atmospheres. The optical UV band is of particular importance because of the much higher photospheric opacities in this wavelength range compared to filters transmitting light redder than the Balmer edge. 
\citetads{2005ApJS..161..456R} developed the theoretical framework and showed that non-adiabatic effects are significant. Temperature and radius changes turned out to dominate the brightness variations, while surface gravity perturbations play a minor role only.

\subsubsection{Mode identification from multi-color and spectro-photometry}\label{sect:multi_phot}

\citetads{2006ApJS..165..551T} 
reviewed the state-of-the-art
from a homogeneous analysis of multi-color data sets of six V361 Hya pulsators available at that time using the models of \citetads{2005ApJS..161..456R}. 
Their results suggested that a majority of the modes must have l-values of 0, 1, and 2 and predicted that $\ell=4$ modes in rapidly pulsating B subdwarfs have a better visibility than $\ell=3$ ones. \citetads{2008ASPC..392..297C} 
used high-precision UBV photometry of Balloon090100001 to show that its dominant pulsation mode is a radial one, while eight other modes have $\ell=1$ or 2. 
The rapid pulsator HS\,2201+261 (=V391 Peg) is a high priority target to asteroseismology because it hosts a Jupiter-mass planet \citepads{2007Natur.449..189S}.
Subsequent WHT/ULTRACAM photometry allowed the dominant mode to be identified as
a radial one and the second-highest amplitude mode as an $\ell=1$ mode \citepads{2010AN....331.1034S}. 

\begin{figure*}
\begin{minipage}{0.46\textwidth}
\includegraphics[width=\textwidth]{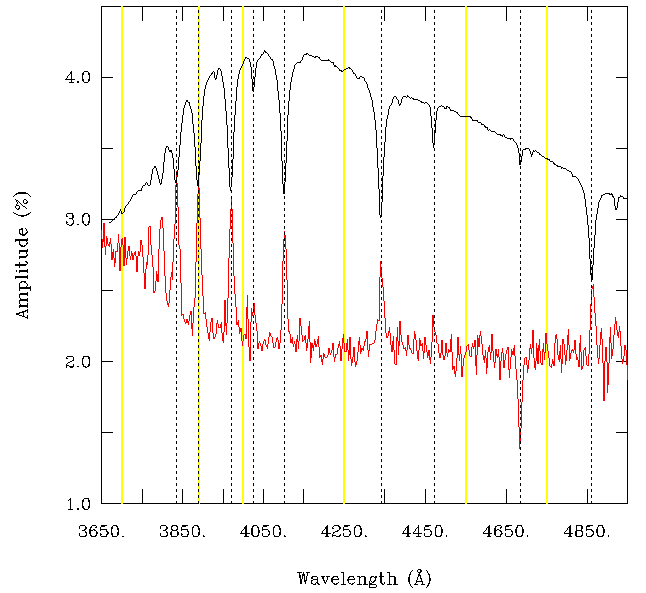}
\end{minipage}
\hspace{1cm}
\begin{minipage}{0.46\textwidth}
\includegraphics[width=\textwidth]{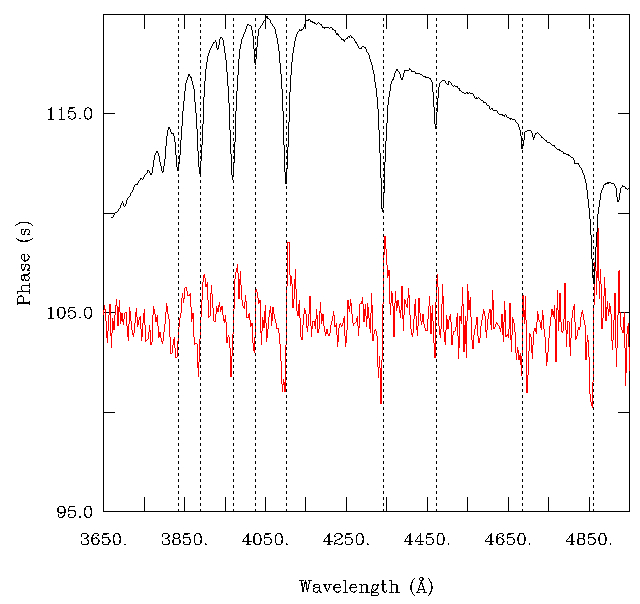}
\end{minipage}
\caption{{\it Left panel:} Observed amplitude of flux variations of EC\,20338$-$1925 as a function of wavelength for the dominant pulsation mode $f_1$ (red). The averaged spectrum (black) is plotted on an arbitrary y-axis scale and also the central wavelengths of the Balmer and prominent He lines are marked by dotted vertical lines. 
{\it Right panel:} The same as the left panel, but for the observed phase.
Increasing amplitudes of the flux variations correspond to increasing opacities e.g. in the Balmer line cores and the Paschen continuum.
From \citetads{2010A&A...522A..48R}; copyright A\&A; reproduced with permission.}\label{fig:spectro_puls}
\end{figure*}

Spectrophotometric studies of EC 20338$-$1925 and EC 01541$-$1409 using ESO VLT/FORS were presented by \citetads{2010A&A...522A..48R} and \citetads{2014A&A...563A..79R}. 
Again the dominant modes turned out to be radial modes, while the much lower amplitude mode is an $\ell=2$ pulsation (see Fig. \ref{fig:spectro_puls}).

\subsubsection{Mode-identification from time-series spectroscopy}\label{sect:puls_spectro} 

\begin{figure}
\begin{center}
\includegraphics[width=\textwidth]{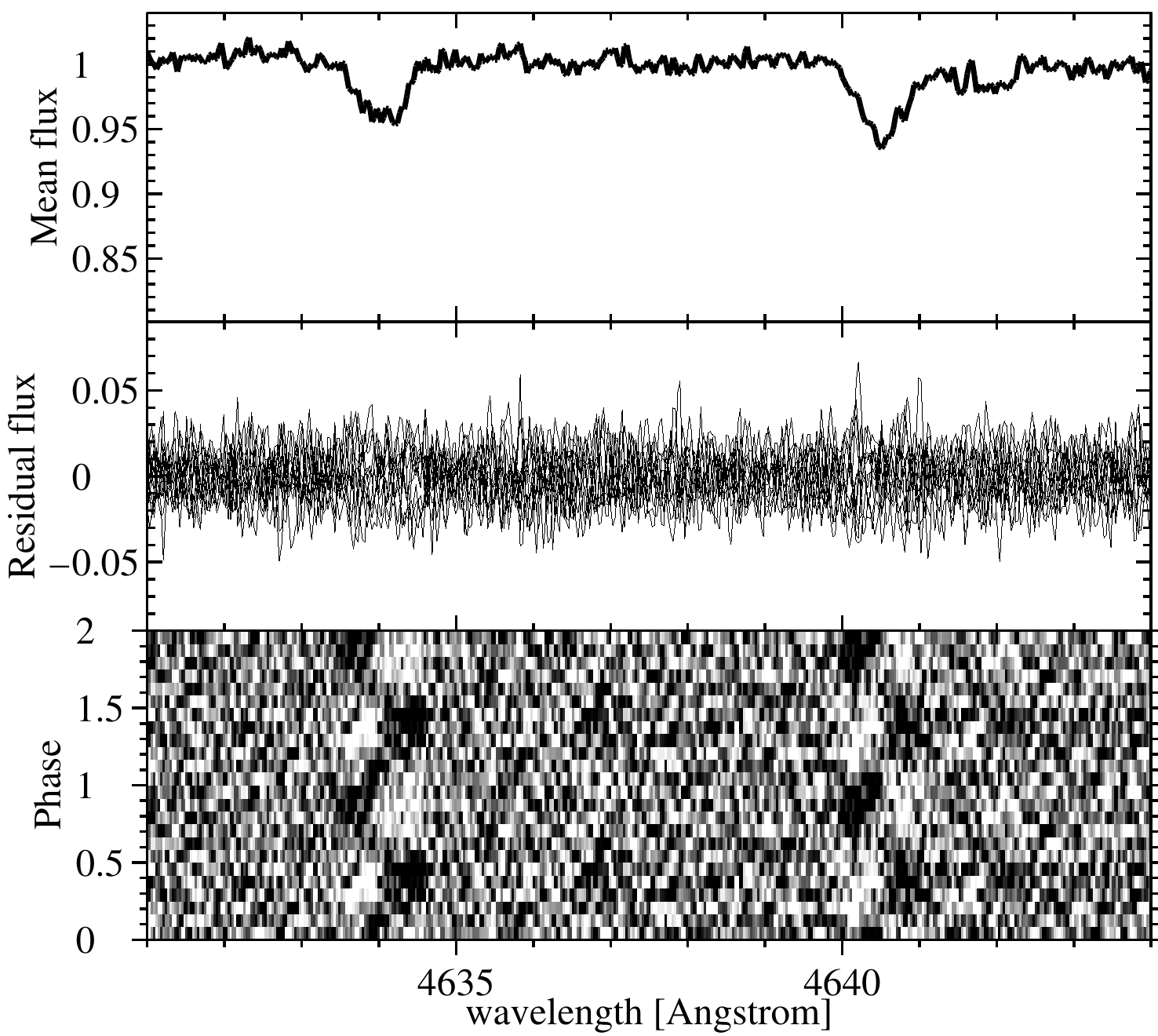}
                \end{center}
\caption{Line-profile variability in the strongest N {\sc iii} lines (4634.2 and 4640.6 \AA) in the spectrum of the pulsating sdB star QQ~Vir. The top panel
shows the mean spectrum and the middle one twelve phase-binned spectra, after subtraction of the mean spectrum. These residual spectra are displayed as a function of pulsational phase in the grey-scale representation at the
bottom. From \citetads{2010Ap&SS.329..163T}; copyright Ap\&SS; reproduced with permission.}
\label{fig:lpv_pg1325}
\end{figure}

Time-series spectroscopy allow stellar surface motions to be traced. This method was pioneered for sdB stars by 
\citetads[e.g.][]{2000ApJ...537L..53O,2005A&A...440..667O} 
who were successful in detecting  
many modes in velocity that were also seen in the light curves.
Line profile variations caused by pulsations can be monitored \citepads[see e.g. Fig. \ref{fig:lpv_pg1325};][]{2010Ap&SS.329..163T}  
and allow the identification of pulsation modes \citepads[e.g.][]{2007A&A...471..605V,2010Ap&SS.329..163T,2008CoAst.157..112T,2008MNRAS.385..255B}. 

\begin{figure}
\begin{center}
\includegraphics[width=0.8\textwidth]{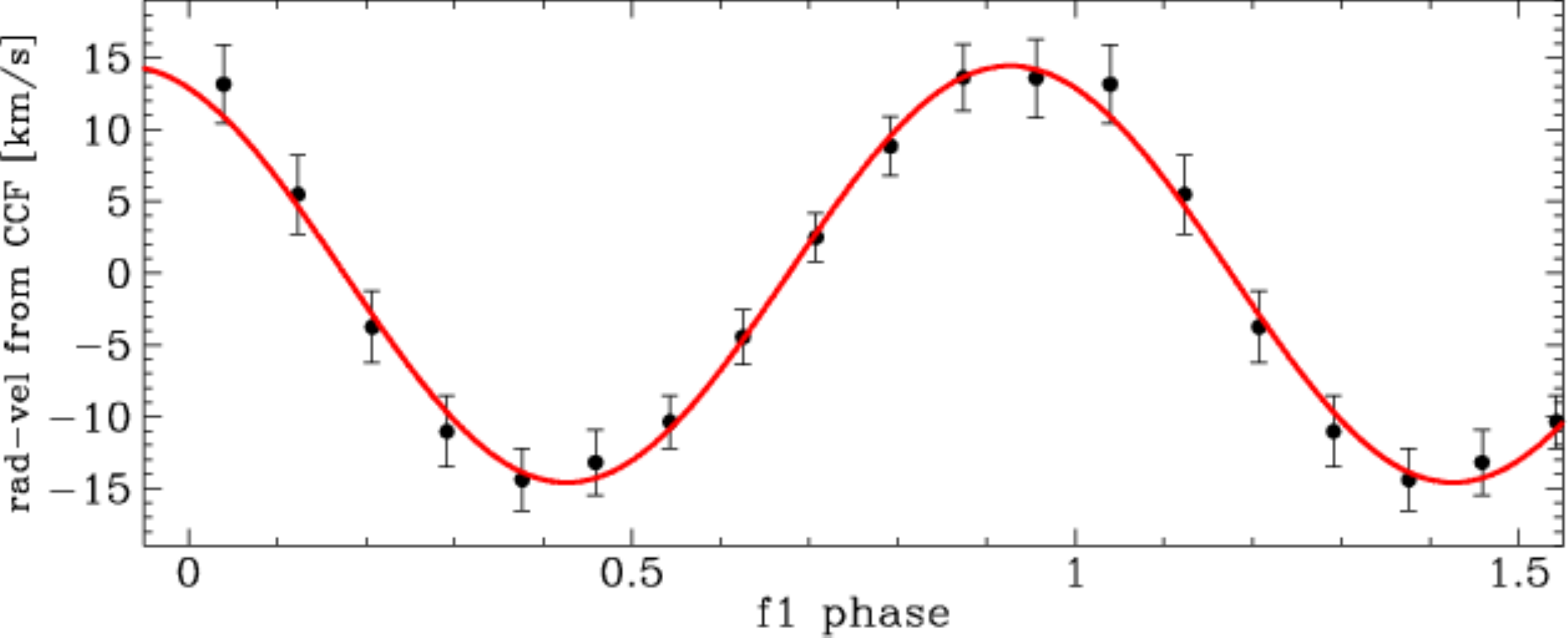}
                \end{center}
\caption{Main pulsation mode radial-velocity variation of Balloon 090100001 as obtained from the cross-correlation analysis. 
From \citetads{2008A&A...492..815T}; copyright A\&A; reproduced with permission.}
\label{fig:lpv_balloon}
\end{figure}

\begin{figure}
\begin{center}
\includegraphics[width=0.8\textwidth]{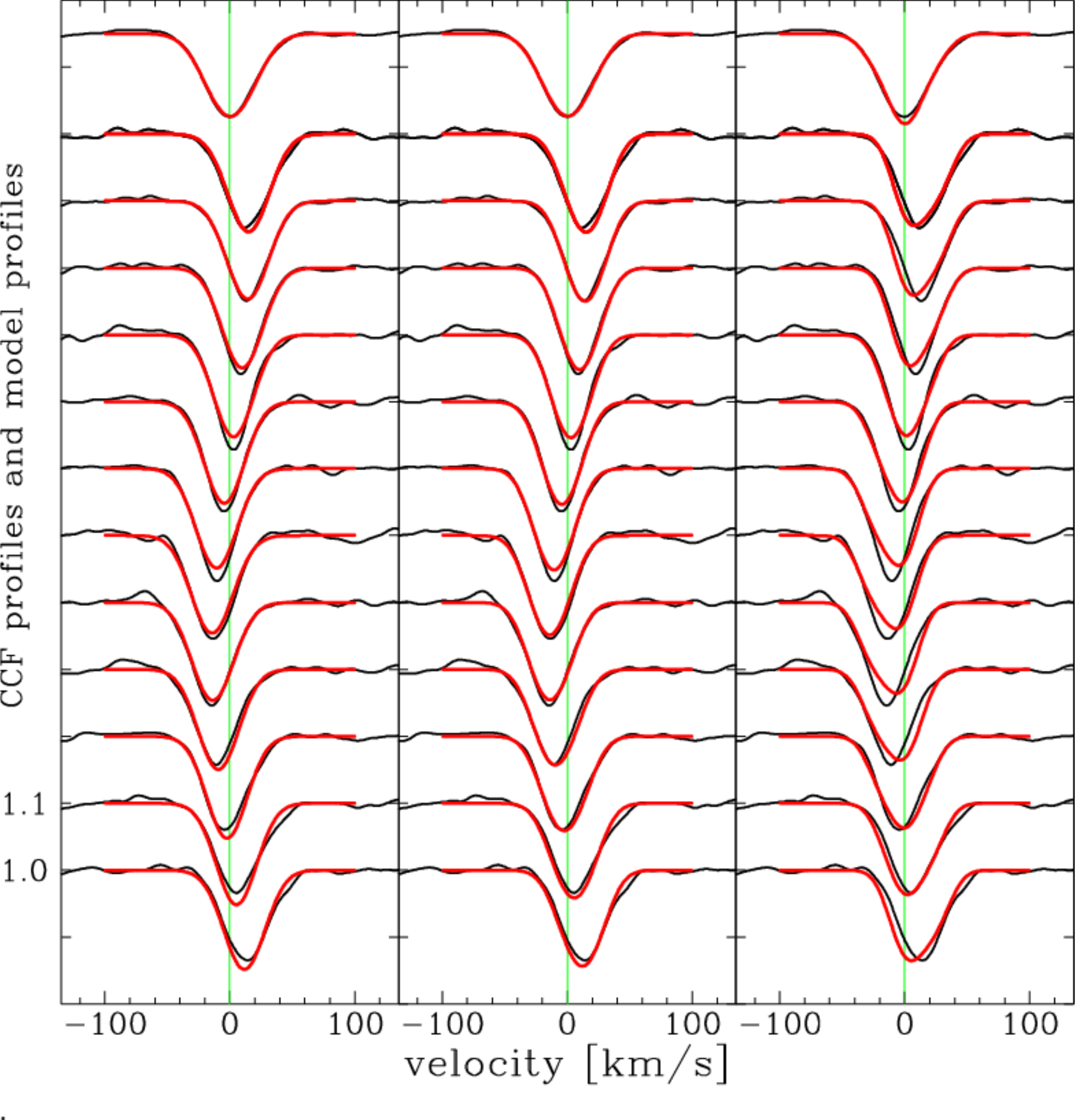}
                \end{center}
\caption{Observed cross-correlation functions (CCFs), offset as a function of pulsation phase by 0.1 continuum units (phase is increasing upwards), with the mean of all CCFs at the top. Overplotted  are model predictions (in red): {\it Left panel:} Model profile variations for a radial pulsation mode: $\ell =0$, $i=70^\circ $, surface velocity amplitude 22 km s$^{-1}$. {\it Middle panel:} Profile variations expected for a non-radial pulsation with $\ell =1$, m=-1, $i=50^\circ $, surface velocity amplitude 35 km s$^{-1}$. {\it Right panel:} profile variations expected for a non-radial pulsation with $\ell =2$, m=-1, $i=40^\circ $, surface velocity amplitude 38 km s$^{-1}$, which is the best-fit $\ell =2$ mode. Both the radial and $\ell =1$ mode fit better than any $\ell =2$ mode. From \citetads{2008A&A...492..815T}; copyright A\&A; reproduced with permission.}
\label{fig:rvc_balloon}
\end{figure}

\begin{figure}
\begin{center}
\includegraphics[width=0.6\textwidth]{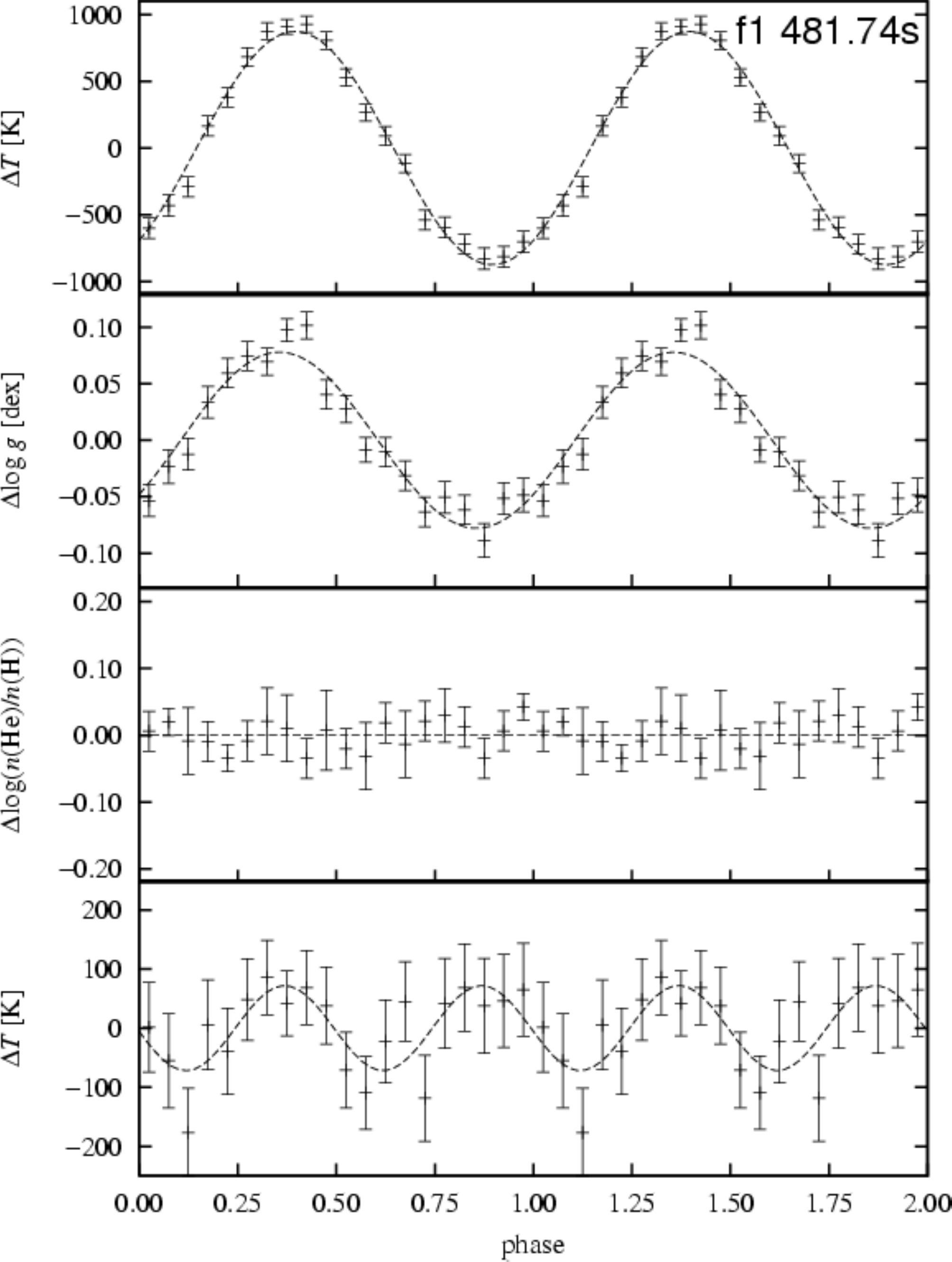}
                \end{center}
\caption{The variations of the atmospheric parameters of PG\,1605+072 with pulsation phase derived from a quantitative spectral analysis of optical spectra (statistical error bars and sine fit are also shown). Upper panels: temperature and surface gravity variations. Lower panels: He/H abundance and the temperature residuals (with a sine fit for the first harmonic).
From \citetads{2007A&A...473..219T}; copyright A\&A; reproduced with permission.}
\label{fig:teff_g_var}
\end{figure}

A beautiful example is the bright high-amplitude pulsating subdwarf B star Balloon 090100001, for which \citetads{2008A&A...492..815T} 
presented an analysis of time-resolved high-resolution spectroscopy using 
56 narrow absorption lines to compute cross-correlation functions that for each individual pulsation phase represent the average line-profile shape. 
The line profile variations with phase are displayed in Fig. \ref{fig:lpv_balloon} and the radial velocity curve in Fig. \ref{fig:rvc_balloon}. The pulsation amplitude of the main mode decreased from 19 km\,s$^{-1}$ in the first epoch of observations (2004) to 14.5 km\,s$^{-1}$ about two years later. The main pulsation mode is identified as $\ell=1$ in accordance with the results of the light curve analysis by \citetads{2008ASPC..392..297C}. 
In addition to radial velocity variations, radial and non-radial pulsations
lead to changing physical conditions (temperature and density variations) when 
parts of the stellar surface expand or contract. This is witnessed by variations {{in}} the shape of the line profile in the disk-integrate spectrum. Therefore, time-series spectroscopy can also be used to deduce temperature and gravity variations of pulsating stars, which will then allow mode identifications from fits of observed line shape variations to synthetic profiles calculated from appropriate model atmospheres \citepads[e.g.][]{2007A&A...473..219T}.
Despite the stars' faintness, such analyses have recently been carried out 
for V361 Hya 
stars \citepads[e.g.][]{2004A&A...419..685T,2007A&A...473..219T,2008A&A...492..815T,2009A&A...505..239V,2010Ap&SS.329..167O}. 
As an illustrative example the temperature and gravity variations from the main pulsation mode in PG\,1605+072 are shown in Fig. \ref{fig:teff_g_var} \citepads{2007A&A...473..219T}. 
A cleaning procedure for phase binned spectra has been devised by
\citetads{2007A&A...473..219T} 
which allowed T$_{\rm eff}$ and $\log$ g
variations as small as 100~K and 0.01~dex, respectively, to be measured.

In most cases, the strongest mode 
is a radial one. Often the power spectrum is dominated by a single
high-amplitude mode which renders the analysis of fainter modes difficult. 

In summary, both multi-color lightcurves and time-series spectroscopy demonstrate that rapidly pulsating sdB stars, indeed, follow  the expected amplitude hierarchy and the dominant pulsations correspond to radial modes. 

\subsection{The {\it Kepler} legacy of pulsating sdB stars}\label{sect:kepler_puls}

\begin{figure}
\begin{center}
\includegraphics[width=0.95\textwidth]{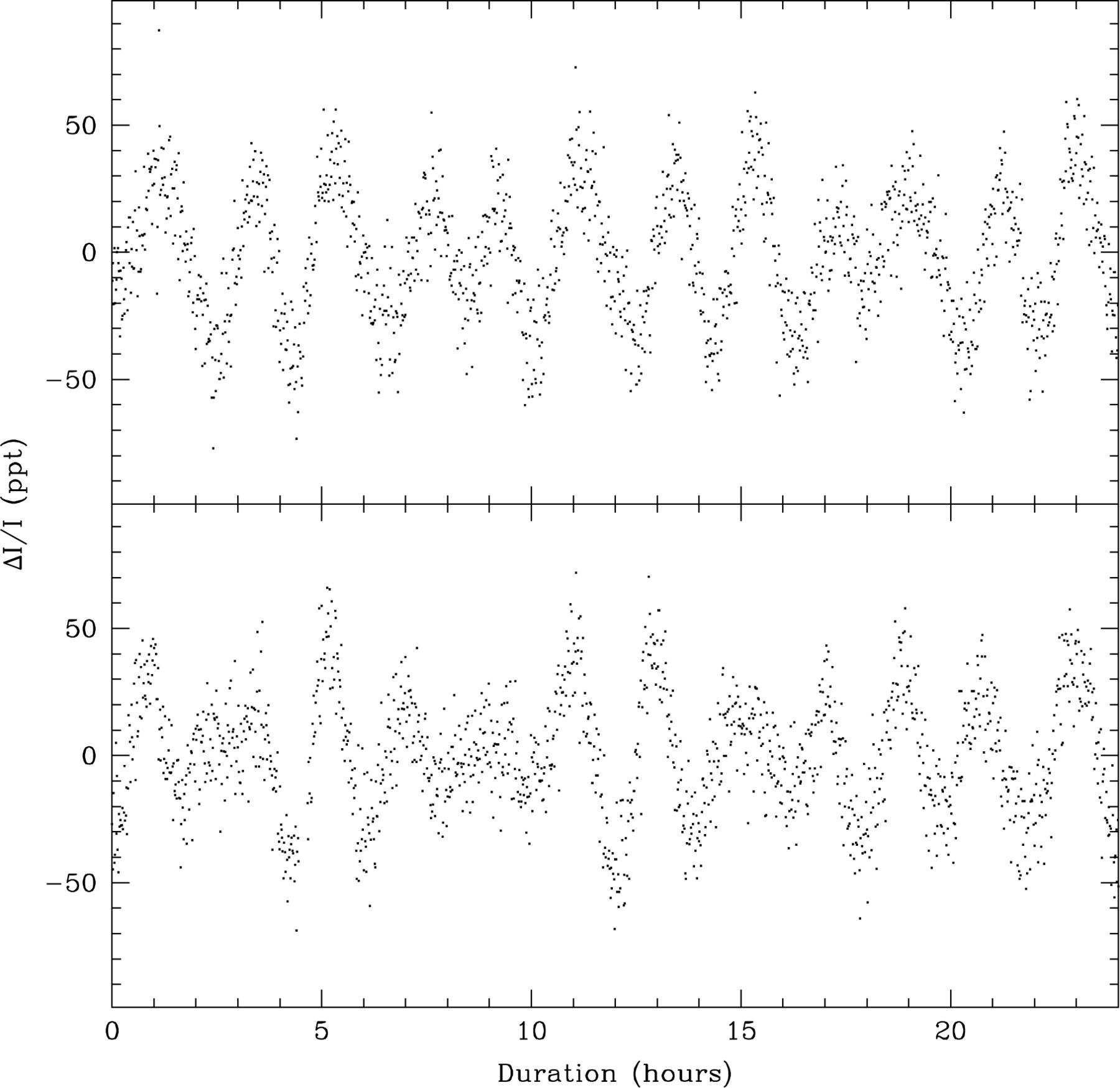}
                \end{center}
\caption{Sample {\it Kepler} light curve of KIC\,10670103, showing 24 h of data near the beginning (top) and end (bottom) of the 33 months long run with an impressive duty cycle of 93.8\%. 
From \citetads{2014MNRAS.440.3809R}; copyright MNRAS; reproduced with permission.}
\label{fig:kepler_lc}
\end{figure}

Ground-based observations of pulsating stars suffer from many aliasing problems and the limited photometric accuracy and duration. Space missions such as {\it CoRoT} \citepads{2009A&A...506..411A} and {\it Kepler} \citepads{2010ApJ...713L..79K} largely overcome those limitations by providing long-term, uninterrupted, high-precision photometry. {\it Kepler} used two sampling rates (30 or 1\,minute, respectively), and achieved very high precision \citepads[some tens of ppm, typically, see Fig. \ref{fig:kepler_lc} for an example,][]{2014MNRAS.440.3809R}, 
 interrupted only every quarter of a year for reorientation of the spacecraft and monthly interruptions for data downlink during four years of operation. 

A survey of the {\it Kepler} field for pulsating sdB stars by \citetads{2010MNRAS.409.1470O,	
2011MNRAS.414.2860O}
resulted in 13 g-mode V1093~Her pulsators and two multi-periodic V361 Hya {{pulsators}}\footnote{Four He-sdOB stars were also found in the {\it Kepler} field, none of which pulsates but two show irregular light variations \citepads{2010MNRAS.409.1470O,2013MNRAS.429.3207J}.}. The number of known sdB pulsators increased to 18 when another three V1093~Her pulsators were found in the open cluster NGC 6791 \citepads{2011ApJ...740L..47P,2012MNRAS.427.1245R}.

The results from the analyses of {\it Kepler} light curves had an enormous 
impact on our understanding of pulsating sdB stars, the long-period V1093~Her pulsators in particular. 
Already the analysis of data from the first year of the {\it Kepler} mission revealed 
very rich frequency spectra of most sdB pulsators. Their analysis led to important discoveries, especially the 
g-mode periods turned out to be evenly spaced  \citepads[e.g.][]{2011MNRAS.414.2885R} 
and frequency multiplets became apparent \citepads{2012AcA....62..179B,2012A&A...544A...1T,2012AcA....62..343B,2012MNRAS.422.1343P}.
Both patterns allowed pulsational and rotational periods to be assigned.
Recently, the first analyses of the full {\it Kepler} data sets revealed stunning results, which we shall discuss in a bit more detail highlighting the detection of trapped modes, which pave the way to probe the internal structure and age of the stars (see Sect. \ref{sect:internal_age}), slow rotation and stochastic variations discovered for the first time.
 

\subsubsection{Mode identification of g-mode pulsations}\label{sect:g_mode_ident}

\begin{figure}
\begin{center}
\includegraphics[width=0.95\textwidth]{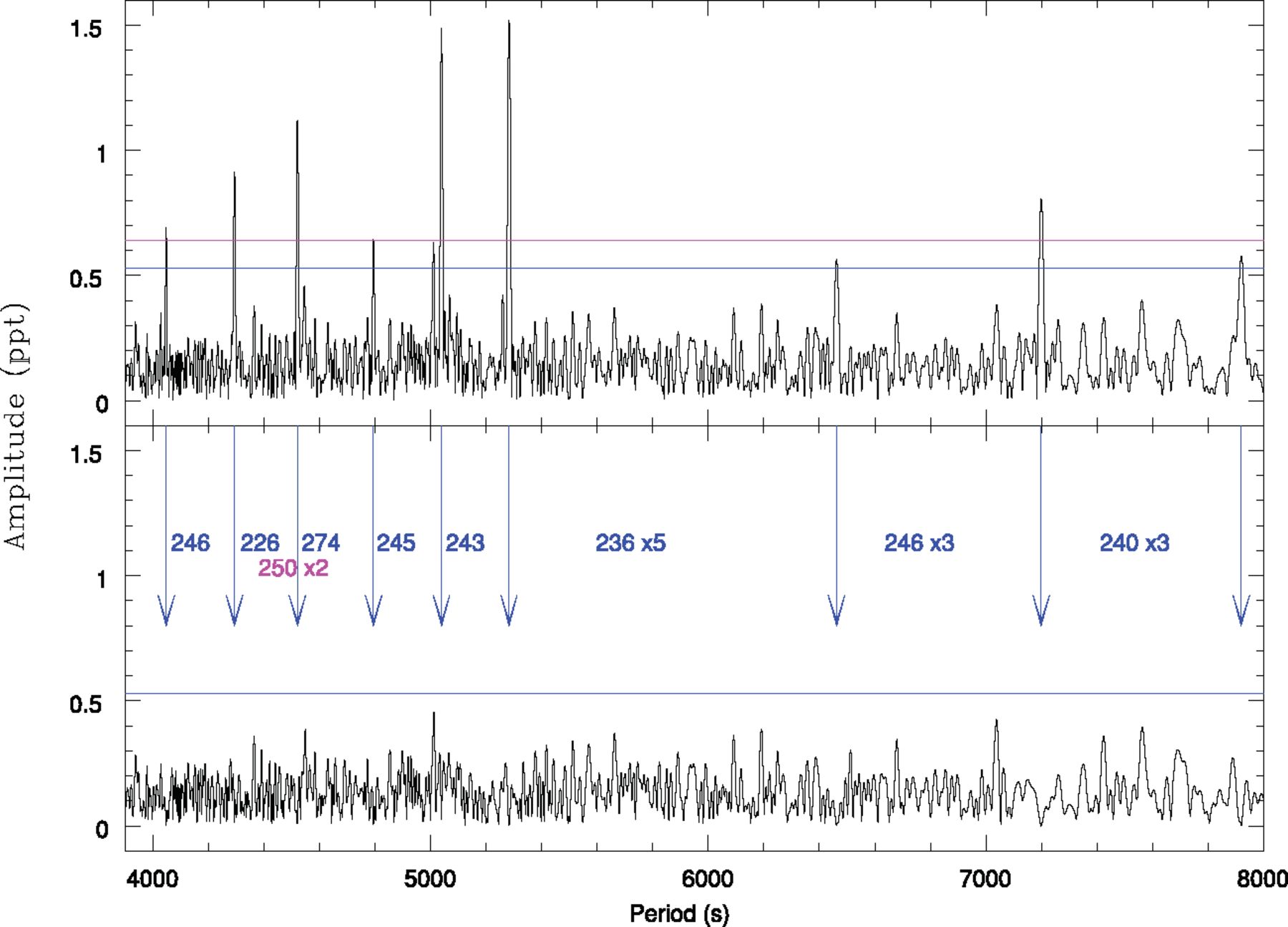}
                \end{center}
\caption{Temporal spectra of NGC\,6791-B3. The top panel shows the original Fourier transform (FT) with the 4$\sigma$ detection threshold and the false alarm probability = 99.9\% limit. The bottom panel shows the pre-whitened FT. The fitted periods are indicated by arrows and the period spacings between them are given as integer numbers. Note the nearly constant period spacing as predicted by theory for the asymptotic limit. From \citetads{2012MNRAS.427.1245R}, copyright MNRAS with permission.}\label{fig:asymptotic}
\end{figure}

Even period spacings as predicted for the asymptotic limit have been found in the {\it Kepler} light curves of many sdB pulsators indicating high radial order modes. An example is NGC\,6971-B3 (see Fig. \ref{fig:asymptotic}), for which \citepads{2012MNRAS.427.1245R} derived period spacings near 245 seconds as appropriate for $\ell=1$ modes.

\begin{figure}
\begin{center}
\includegraphics[width=0.95\textwidth]{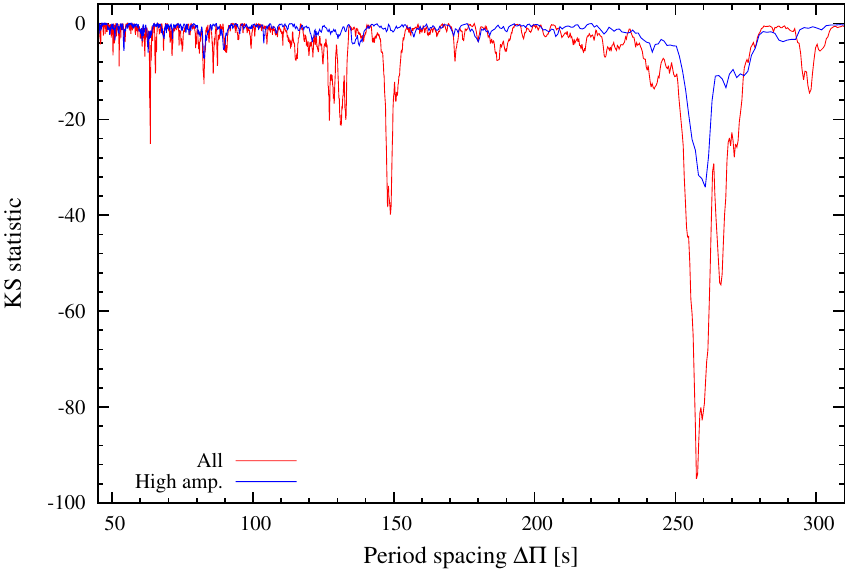}
                \end{center}
\caption{The g-mode pulsator KIC\,10553698:
Kolmogorov-Smirnov test statistic for the full frequency list (red) and high-amplitude (A$>$ 1000 ppm) modes (red) respectively. From \citetads{2014A&A...569A..15O}, 
copyright A\&A, reproduced with permission.}\label{fig:ks_test}
\end{figure}

For most of the {\it Kepler} sdB pulsators the mode spectrum is much more complex than that of NGC 6791-B3 with
more than hundred oscillation frequencies detected.
A Kolmogorov-Smirnov test is then applied to identify the  most  frequently  observed  spacing  in  a  dataset. \citetads{2014A&A...569A..15O} 
studied the light curve of KIC 10553698A 
(see Fig. \ref{fig:ks_test}).
Two  minima are seen
around $\Delta$P=260 s and $\Delta$P=150s which can be readily identified as high-order (n=10--35) modes with $\ell=1$ or $\ell=2$, respectively, because the  
$\Delta$P relation between the two peaks matches the value of $\frac{1}{\sqrt{3}}$ expected 
from the asymptotic approximation. When only the high-amplitude modes ($>$ 1000 ppm) are considered the peak at $\Delta$P=150s disappears, indicating that they are $\ell=1$ modes. 

\paragraph{KIC 10553698A: trapped modes} 

\begin{figure}
\begin{center}
\includegraphics[width=0.95\textwidth]{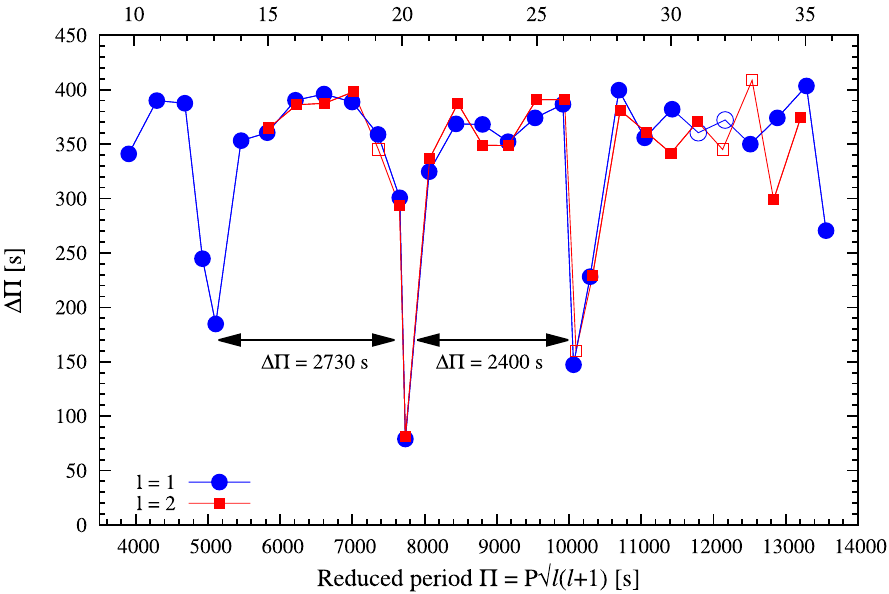}
                \end{center}
\caption{The g-mode pulsator KIC\,10553698: Period difference ($\Delta Pi$) between consecutive modes of the $\ell=1$ and $\ell=2$ sequences, after converting to reduced periods. The asymptotic order of the modes, n, is indicated on the upper axis.
From \citetads{2014A&A...569A..15O}, 
copyright A\&A, reproduced with permission.}\label{fig:spacings}
\end{figure}

By sequencing the observed periods, \citetads{2014A&A...569A..15O} noted that six modes were missing from the sequences, indicating that they are trapped modes.  
The trapping signature can be demonstrated in a diagram, where the period difference between consecutive  modes  are  plotted  against reduced period \citepads[$\frac{P}{\sqrt{l(l+1)}}$, see e.g. Fig. 3 of][]{2002ApJS..139..487C}.  For KIC 10553698A drop-offs in both the $\ell=1$ and $\ell=2$ sequences were found by \citetads{2014A&A...569A..15O} to occur for radial order n=20,26,27 and for the $\ell=1$ sequence at n=13 (see Fig. \ref{fig:spacings}). These are g-modes trapped by transition zones such as the boundary of the convective core and the transition layers of changing chemical composition inside the star (conf. Sect.
 \ref{sect:internal_age}).  

\subsubsection{Rotation of pulsating sdB stars}\label{sect:puls_rotation}

\begin{figure}
\begin{center}
\includegraphics[width=0.95\textwidth]{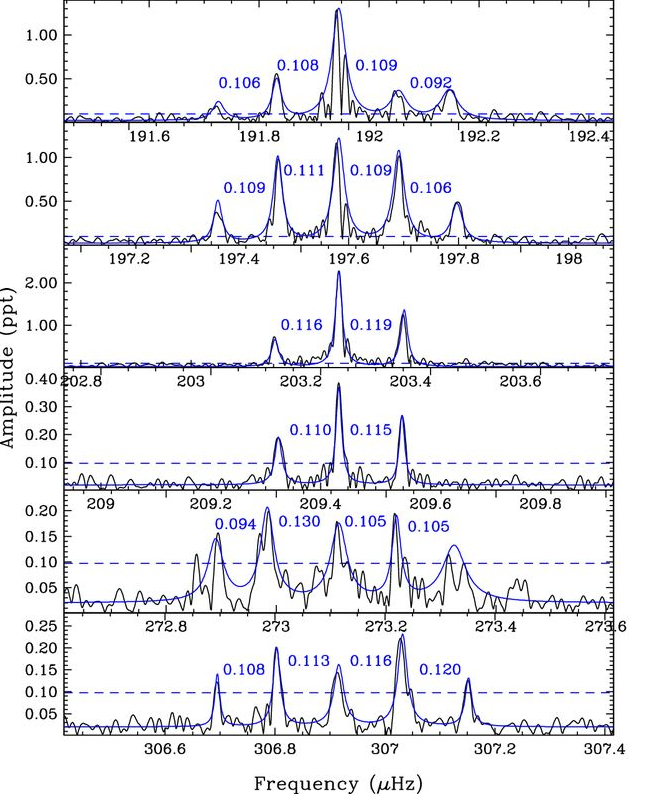}
                \end{center}
\caption{KIC 10670103: Fourier transform spectrum: $\ell=2$ multiplets with splittings of roughly 0.11$\mu$Hz. From \citetads{2014MNRAS.440.3809R}; copyright A\&A; reproduced with permission.}
\label{fig:multiplets}
\end{figure}

From equidistant splitting of multiplets it became possible to determine the rotation of many {\it Kepler} sdB pulsators. A particularly striking example is KIC 10670103. \citetads{2014MNRAS.440.3809R} detected as many as 278 pulsation frequencies in the 2.75 years of {\it Kepler} data with periods ranging 0.4 to 11.8 hours and amplitudes between 0.1 and 14 ppt. Splitting into frequency multiplets is obvious (see Fig. \ref{fig:multiplets}), which translates into a rotation period of 88 $\pm$ 8 days. Such a slow rotation was unexpected but has also been seen in   
other {\it Kepler} pulsators, with rotation periods between $\approx$ 7 and $\approx$100 days \citepads[see Table 2 of][]{2014MNRAS.440.3809R},
 irrespective of whether the star is single or in a binary.  Surprisingly, no mode splitting could be found from the {\it Kepler} lightcurve of KIC 8302197,  which implies that either the star is rotating very slowly with a period exceeding 1000 days or the rotation axis is seen pole-on \citepads{2015A&A...573A..52B}.

\subsubsection{KIC\,2991276: Stochastic variations of a p-mode pulsator}

\begin{figure*}
\begin{center}
\includegraphics[width=0.95\textwidth]{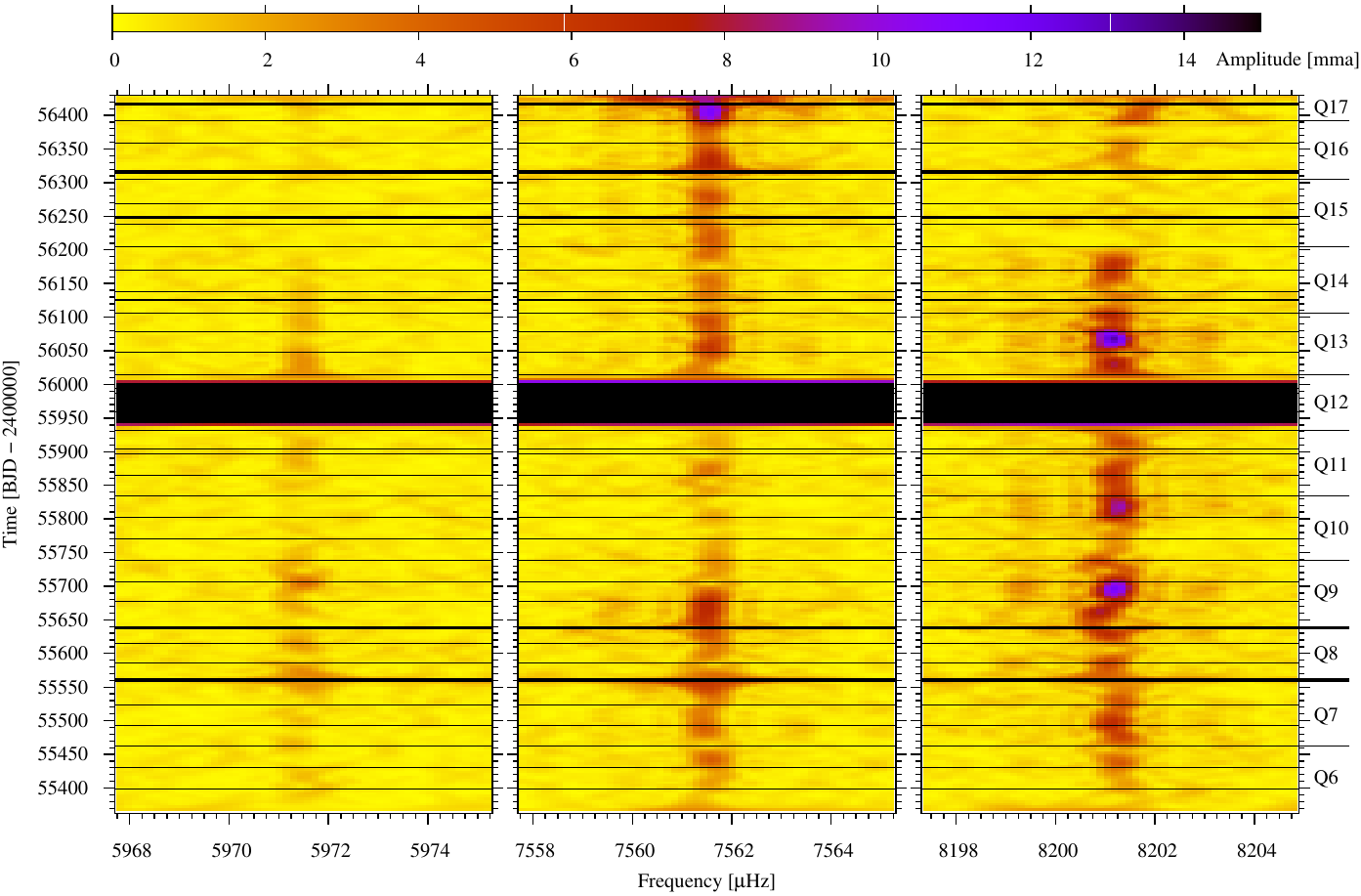}
                \end{center}
\caption{Stochastic variations of the strongest pulsations in KIC\,2991276.
Sliding Fourier transforms are computed for 20-day chunks of data. Amplitudes of up to 1.4\% are reached in individual months, but vary substantially in amplitude as well as phase on timescales of about a month and occasionally disappear completely. The horizontal lines reflect data gaps.  
From \citetads{2014A&A...564L..14O}; copyright A\&A; reproduced with permission.}
\label{fig:stochastic}
\end{figure*}

It is remarkable that some short-period pulsators (V361-Hya stars) show 
coherent pulsations that are stable in phase over many years. The most important case is HS\,2201+261 \citepads{2007Natur.449..189S}, 
which showed periodic phase variations (P=3.2 years) caused by an orbiting planet (see Sect. \ref{sect:planets}). On the other hand, amplitude variations seam to be quite common among V361-Hya stars, and recently, even evidence for
stochastic pulsations in the sdB star KIC 2991276 was presented
by \citetads{2014A&A...564L..14O}. 
Their analysis of the {\it Kepler} light curve which spans no less than 1051.5\,days revealed that the pulsations in KIC\,2991276 lose coherence on timescales of $\approx$60 days or more and are therefore stochastic in nature.
 Such stochastic oscillations are normal for solar-like pulsations
 and have already been suggested  
 for the high-amplitude sdB pulsator PG~1605+072 \citepads{2005ApJ...622.1068P}, 
 but KIC\,2991276 is the first sdB pulsator for which stochasticity has been established beyond doubt (see Fig. \ref{fig:stochastic}). Because the oscillations are driven by the Z-bump, it is not obvious what causes the stochastic behavior. External influence by an unseen close compact companion would be an option. However, the star does not show radial velocity variations and is therefore likely a single star unless a potential  companion orbits on a highly inclined orbit.
 \citetads{2014A&A...564L..14O}, therefore, favor   
thermohaline convection in the driving zone which may temporarily stall the driving by changing the chemical composition. 
 
Evidence is growing that KIC\,2991276 may not be a unique case but several sdB pulsators observed with the {\it Kepler} 
satellite show some degree of stochasticity ({\O}stensen, 7$^{th}$ meeting on hot subdwarfs and related objects\footnote{\url{http://www-astro.physics.ox.ac.uk/~aelg/SDOB7/booklet.pdf}}), which may be related to the question of why oscillation were found only for about 10\% of rapidly pulsating sdB stars in the instability strip  \citepads{2010A&A...513A...6O}.  

\subsubsection{Pulsating sdB stars in binaries}

{\it Kepler} data have also revealed that four pulsating sdB stars are actually in binaries,
including one  eclipsing  binary { \citepads[KIC\,9472174=2M1938+4603,][]{2010MNRAS.408L..51O}}, showing light variations due to the reflection effect in addition to oscillations.  
The three non-eclipsing sdB+dM binaries, host g-mode sdB pulsators which 
have rather long orbital  periods  of  0.443\,d, 0.395\,d, and 0.399\,d \citepads{2010MNRAS.409.1487K,2010MNRAS.409.1509K,2011ApJ...740L..47P}, respectively, difficult to discover from ground. Radial velocity monitoring of the {\it Kepler} sdB pulsators has confirmed binarity and led to the discovery of three additional ones. Hence, at least eight of the 18 pulsating sdB stars in the {\it Kepler} field are binaries { \citepads{2014ASPC..481..287T,2016A&A...585A..66B}.} 
Three of them have been 
used to investigate tidal synchronization timescales by comparing rotation periods 
measured from rotational splittings of pulsation frequency multiplets to the binaries' orbital periods \citepads[][see Sect. \ref{sect:synchro_puls}]{2012MNRAS.422.1343P}.

\subsection{Internal structure and stellar age from asteroseismology}\label{sect:internal_age}

One of the first great achievements from asteroseismology of {\it CoRoT} and {\it Kepler} light curves was the determination of the internal structure
of the stellar core and the age of sdB pulsators. 
 A general review {{of}} stellar age determinations via asteroseismology is given by \citetads{2015AN....336..477A}. 
Because the g-modes are sensitive to the deep interior of the star, the structure of the helium-burning core, e.g. its size and composition, can be studied.  The mixed convective core is bound by the chemical transition from He/C/O to He composition.  

The first successful case study targeted \object{KPD 0629$-$0016}, which was discovered by 
\citetads{2007MNRAS.377.1605K} 
to be  a slowly pulsating sdB star.  \citetads{2010A&A...516L...6C} 
detected many g-mode pulsations in its {\it CoRoT} light curve, carried out the first detailed
asteroseismic analysis, and found the total mass of M$_{\rm tot}$=0.471
$\pm$0.002 M$_\odot$ to be in excellent agreement with the canonical mass. The
age of the star with respect to the ZAEHB was derived to be 42.6$\pm$1.0 Myr. 
In two subsequent studies, \citetads{2010ApJ...718L..97V} and \citetads{2011A&A...530A...3C} used the same technique to 
analyze {\it Kepler} light curves of \object{KPD 1943+4058} and \object{KIC 2697388}, respectively, which are also g-mode pulsators. 
\object{KPD 1943+4058} was found to be slightly more massive (M$_{\rm tot}$=0.4{ 96}
$\pm$0.002 M$_\odot$) and younger ({ 18.4$\pm$0.1} Myrs) than \object{KPD 0629$-$0016}.
Their hydrogen-rich envelopes turned out to be thicker than for the hotter short-period sdB pulsators, which is expected from models. For \object{KIC 2697388} no unique set of parameters could be found, but two families of solution were identified which matched the observation equally well. Nevertheless an upper limit for the age of 55 Myr (since the zero-age EHB) was derived. 

All three studies concluded that the mixed convective cores are more massive than expected from evolutionary tracks.
In the case of \object{KPD 1943+4058}, for instance, more than half (57\%) of the total mass of the star is in the convective core, while only 25\% of its helium nuclear fuel was burned. Therefore, it was concluded that extra mixing has to occur early-on in the evolution of the helium cores of sdB stars, caused, for example, by core overshooting, semi-convection and/or differential rotation.
In order to remedy this discrepancy, \citetads{2015MNRAS.452..123C} 
carried out an extensive investigation of the pulsation properties of core helium burning stars considering four different modes of mixing (no overshooting, standard-overshoot, semi-convection, and a new maximum-overshoot scheme) as well as varying the input physics (stellar composition and the He-burning reaction rates). The latter as well as 
three of the mixing schemes failed to increase the mixed-convective core to the size required to match the observations.
The only successful models invoked a maximal-overshoot scheme proposed by \citetads{2015MNRAS.452..123C}, which, however, is an ad-hoc scheme to ensure a maximum size of the core but lacks a physical explanation. \citetads{2015ApJ...806..178S}
added atomic diffusion processes and confirmed this result by calculations with the MESA code. Because sufficiently large convective core masses resulted only for extreme overshooting, \citetads{2015ApJ...806..178S} called for realistic 3D hydrodynamical modelling of the convective core boundary \citepads{2013ApJ...769....1V}.

\subsection{Stellar mass and radius: Asteroseismology vs. eclipsing binaries}\label{sect:eclipse_puls}

Mass and radius are the most fundamental parameters of a star, which can be derived by studying double-lined, eclipsing binaries by making use of Kepler's 3rd law. Asteroseismology provides an important alternative, already demonstrated convincingly for about a dozen of sdB pulsators \citepads{2012A&A...539A..12F,2014ASPC..481..229V}. 
Hence it is obvious 
that pulsating stars  
in eclipsing binary systems are important benchmarks to understand structure and evolution of hot subdwarf stars because they allow   
cross-checking of results from asteroseismology with dynamical estimates
\citepads{2014arXiv1404.7501H}\footnote{A general review of asteroseismology of eclipsing binary stars in the {\it Kepler} era \citepads{2014arXiv1404.7501H} %
includes sdB binaries.}. In 2009 only one such system, \object{NY~Vir} was known, which { \citetads{2008A&A...489..377C}} termed a ``Rosetta stone'' for the field. The asteroseismic modeling 
combined with a full orbital solution allowed mass and radius of the sdB star to be determined to an unprecedented precision 
\citepads{2007A&A...471..605V,2013A&A...553A..97V}.

NY~Vir remains the benchmark because it is a bright and well studied eclipsing sdB+dM system, which hosts a V361~Hya pulsator. 
Interest in this system increased when two circumbinary planets were discovered \citepads{2012ApJ...745L..23Q,2014MNRAS.445.2331L} 
 making NY~Vir one of the 18 known systems to host a circumbinary planet \citepads{2015MNRAS.446.1283C,2015MNRAS.448L..16B}.
Hence, the light variations of NY~Vir are regularly monitored \citepads[e.g.][]{2014MNRAS.445.4247K}, 
 also by making use of robotic telescopes \citepads[see e.g.][]{2014MNRAS.440.1490C}. 

As already pointed out, siblings to NY~Vir have been discovered recently by the MUCHFUSS survey, and by other studies. The {\it Kepler} satellite also discovered pulsating sdB stars in eclipsing binaries, which have their light curves measured to very high quality and length. Arguably the most important discovery is 
\object{2M1938+4603} = \object{KIC\,9472174} 
which is even brighter than NY~Vir \citepads{2010MNRAS.408L..51O}.  

An unusually large number of oscillation frequencies were identified in the {\it Kepler} light curve of \object{2M1938+4603} at amplitudes too low to be detectable from the ground.
Fifty five pulsation frequencies between  
$50-4500\mu$Hz, were attributed to both p-mode and g-mode pulsations making  \object{2M1938+4603} a hybrid pulsator \citepads{2010MNRAS.408L..51O}.
Combined with the radial velocity semi-amplitude (see Sect. \ref{sect:binaries}) and spectroscopic gravity, a mass of 
$M=0.48\pm0.03$ M$_\odot$ was derived, making the star a prime target for an asteroseismology analysis.

\subsection{Rotation and tidal synchronization}\label{sect:synchro_puls}

For a pulsating star the rotation rate can be derived from the characteristic even splitting of the oscillation frequency multiplets (if $\ell \ge$ 1). From ground-based light curves the explicit identification is difficult. However, the {\it Kepler} mission  
provided light curves for several sdB pulsators that allowed rotational splittings to be identified and rotation rates derived.

From ground-based observations the $p$-mode pulsations of NY~Vir and Feige~48 
have successfully been modelled and the rotation profile determined. 
\citetads{2008A&A...489..377C} found that the rotation of the sdB star in the HW~Vir binary NY~Vir (orbital period P=0.101~d)
is synchronized to its orbit  
and that the sdB rotates as a solid body for at least the outer half of the star. 
The second case is Feige 48, a non-eclipsing sdB+dM binary 
\citepads{2014ASPC..481...91L} of relatively long orbital period, for which \citetads{2008A&A...483..875V} 
found tidally locked rotation and solid body rotation of the outer part of the 
star. However, this conclusion was called into question when \citetads{2014ASPC..481...91L} 
revised the orbital period downwards to P=8.25~h. 

Synchronization time scales are predicted to strongly depend on the orbital period. Therefore it is important to study binaries of different orbital periods and ages.
We might expect that the systems of shortest orbital periods are more likely to have reached synchronization (see Sect. \ref{sect:bin_synchro}).

{\it Kepler} light curves of sdB pulsators are very important in this respect. 2M1938+4603 has a short orbital period of about 1/8\,d. The analysis of the rich pulsation spectrum of its primary derived from the {\it Kepler} light curve led  \citetads{2010MNRAS.408L..51O} 
to conclude that the sdB primary is rotating with a period close to the orbital one demonstrating that the sdB's rotation is indeed synchronized. However, for binaries with longer periods, this was found not to be the case. 
\citetads{2011ApJ...740L..47P}, for example,  
were able to infer a rotation period of 9.63\,d for the sdB binary NGC~6791/B4. Because its orbital period is much shorter (0.399\,d) this  
demonstrates that the rotation of the primary is not
tidally locked to the orbit. The same is true 
 for both of the g-mode pulsators of \citetads{2010MNRAS.409.1509K}, which have orbital periods similar to that of  
 NGC 6791/B4 but longer than that of 2M1938+4603.

From these results \citetads{2013A&A...559A..35O} concluded that tidal synchronization of the primary 
in sdB+dM binaries is efficient at periods of 1/8 d but not at 2/5 d. 
A new sdB pulsator, \object{FBS 0117+396}, was found in an 1/4 d orbit with a dM companion. Two short (337 and 379s), as well as eight long periods (45min to 2.5h) were discovered, making \object{FBS 0117+396} a hybrid pulsator in an sdB+dM binary.

Because of its intermediate orbital period, \citetads{2013A&A...559A..35O} suggested that \object{FBS 0117+396}, would be important to investigate at which point the tidal forces become sufficiently strong to enforce synchronization in sdB+dM systems. However, the rotational period splittings have not yet been determined.

\subsubsection{Radial differential rotation}\label{sect:differential_rotation}

Recently, radial differential rotation has been detected in red giants and HB clump stars  as well as in a white dwarf binary from {\it Kepler} data. 
In red giants the rotational splitting of mixed modes showed that in several cases the stellar cores rotate at least five times faster than the stars' envelopes \citepads[e.g. ][]{2012Natur.481...55B}. In the white dwarf binary SDSS J1136+0409, the core rotates at a period of P=2.49 $\pm$ 0.53~h considerably faster than the orbital 
period \citepads[P$\approx$6.9h, ][]{2015MNRAS.451.1701H}, indicating that the core rotation is not tidally locked to the orbit.
\citetads{2012A&A...548A..10M} studied a large sample of red giants and conclude from the observed rotational splittings that the mean core rotation significantly slows down during the last stages of the red giant branch, more quickly than expected. \citetads{2015AN....336..477A} concludes: 
``Core-to-envelope rotation rates during the red-giant stage are far lower than theoretical predictions, pointing towards the need to include new physical ingredients that allow strong and efficient coupling between the core and the envelope in the models of low-mass stars in the evolutionary phase prior to core helium burning.''

In the light of these discoveries, it would be very interesting to search for radial differential rotation  in sdB pulsators that show $p$- and $g$-modes at the same time.
The number of such hybrid pulsators keeps growing thanks not only to the {\it Kepler} mission but also through ground-based discoveries \citepads[e.g.][]{2011MNRAS.413.2838B}. Because $p$- and $g$-mode pulsations probe different regions inside the star, the combination of both allows  
the rotation of the interior with depth to be traced. A first result was presented by  \citetads{2015ApJ...805...94F} who analyzed the {\it Kepler} light curve of  the hybrid-pulsator \object{KIC 3527751} and found that the rotation period derived from the $p$-mode splitting 
(P=15.3 $\pm$ 0.7 d), which trace the outer envelope, is smaller than that from the splitting of the g-modes (P=42.6 $\pm$ 3.4 d) which probe deep layers of the star. They conclude that the core rotates more slowly than the outer regions of the star.
If confirmed, this finding would be the first discovery of radial differential rotation of an sdB and in fact very surprising, because it is the opposite to what is expected from the results for the rotation of RGB stars and the white dwarf binary discussed above.

\subsection{A new class of sdO pulsators in globular clusters}\label{sect:cluster_puls}

Also among the sdO stars, a rapid pulsator  was discovered  \citepads{2006MNRAS.371.1497W}, which exhibits very rapid p-mode pulsations with periods of
60 to 120 s. 
\citetads{ 2008A&A...486L..39F} 
readily explained the pulsation as driven by an iron group opacity bump.
However, despite a lot of observational effort, searches for pulsating subluminous O stars met with little success { \citepads[e.g.][]{2007MNRAS.379.1123R,2014ASPC..481..153J}}. The only new discovery is { the reclassification of }\object{EO Ceti} { (= PB 8783), the second sdB star found to pulsate \citepads{1997MNRAS.285..645K}, which turns out to be of spectral type sdO ( $\approx$50,000K or hotter)}, with an F-type companion \citepads{2012ASPC..452..233O,2014ASPC..481..115V}. { However, irregular yet unexplained variations in light or radial velocity were found for some He-sdO stars \citepads{2014ASPC..481..161G,2015A&A...577A..26G}.} 

\citetads{2011ApJ...737L..27R} %
carried out a search for sdO pulsators in the globular cluster $\omega$ Cen similar to the field star \object{SDSS~J160043.6+074802.9}
\citepads[{ V499\,Ser,}][]{2006MNRAS.371.1497W}, 
which is a very hot, helium-enhanced star \citepads[68,500 K, log (He/H)=
0.64, ][]{2011ApJ...733..100L}. 
 
The five pulsators that were discovered by \citetads{2011ApJ...737L..27R}
appear to form a homogeneous class in terms
of pulsation properties and atmospheric parameters. Like for V499\,Ser 
the pulsation periods are short (80 to 125 s), but $\omega$ Cen stars are considerably cooler (48,000 to 54,000 K) { than the field sdO} and helium-poor rather than helium-rich. Therefore, 
 \citetads{2014IAUS..301..289R} argue that these stars do not exist among the field sdO population, because a dedicated monitoring of 36
sdO stars in the field \citepads{2014ASPC..481..153J} 
 failed to uncover a single pulsator. 
{  \citetads{2016arXiv160205470R} 
 identified the same $\kappa$ mechanism that excites the p-mode instability in the sdB variables to drive the rapid pulsations in the $\omega$ Cen variables. 
 However, the models predict excited pulsation at somewhat higher temperatures and shorter periods than observed. This discrepancy may be resolved, when more heavy element opacities were included in the calculations. }  
 
 Hence, globular clusters seem to be a more promising place to search for pulsating stars among O subdwarfs.
Indeed, six pulsating subdwarfs were found among the blue hook stars in the massive globular cluster NGC~2808 \citepads{2013ApJ...777L..22B}. 
Their pulsation periods range from 85 to 149 s and UV amplitudes of 2.0 to 6.8\%, similar to that of the $\omega$ Cen stars. However, their atmospheric parameters require a proper determination. They span a rather wide range of UV color which might indicate a much wider spread in temperature than for the $\omega$ Cen stars. { Hence, besides the binary fraction and the atmospheric parameters the pulsational properties of field and cluster population differ markedly. However, if its effective temperature and helium content could be better constrained, \object{EO Ceti} (see above) may turn out to be the first field counterpart to the $\omega$ Cen variables.}

\subsection{The enigmatic intermediate He-sdB LS IV $-$14$^\circ$116 (V0366 Aqr)}\label{sect:lsiv_puls}

LS IV $-$14$^\circ$116 { (V0366 Aqr)} is the only intermediate He-sdB showing multi-periodic light variations \citepads{2005A&A...437L..51A}, 
very likely caused by non-radial g-mode pulsations \citepads{2011ApJ...734...59G} 
because of the long periods (1953 s for the dominant mode and 2620 -- 55084 s for others). 
Its effective temperature and gravity are inconsistent with domains known to be unstable to g-mode oscillations, that is LS IV$-$14$^\circ$116 is out of place in the
T$_{\rm eff}$ -- log g -- diagram (see Fig. \ref{fig:kiel_pulsators}, according to which {\emph short}-period p-mode pulsations were expected). Recently,
\citetads{2015MNRAS.446.1889J} detected radial velocity variations of $>$5 km\,s$^{-1}$  caused by the dominant oscillation mode.

\citetads{2011ApJ...741L...3M} and \citetads{2013EPJWC..4304004M} 
suggested that LS IV $-$14$^\circ$116 might represent the first known case of $\epsilon$-driven pulsation. \citetads{2012MNRAS.423.3031N} suggested that the He-sdB stars have not yet settled onto the helium main-sequence and are still evolving towards it. 
Driving by the $\epsilon$-mechanism would be acting on helium-burning shells which occur sequentially when the star transits from the tip of the red-giant branch to the EHB. 
Recently, \citetads{2015A&A...576A..65R} derived a significantly higher surface gravity for 
LS IV $-$14$^\circ$116, which places the star on the EHB; that is, in the core helium-burning phase, during which the $\epsilon$ mechanism is unlikely to work.
The evolution of late flashers may reach high gravities as well for low mass progenitors just before settling on the EHB (Miller-Bertolami, priv. comm.). 
Hence, the evolutionary status of LS IV $-$14$^\circ$116 and the driving of its pulsations remain issues to be solved.

\subsection{The {\it K2} Mission, {\it Kepler} goes on}


The two-wheel mission of the {\it Kepler} space telescope ({\it K2}) has already generated important results. \citetads{2014MNRAS.442L..61J} studied the Kepler {\it K2} light curve of the pulsating subdwarf B star EQ { Psc}. Besides the rich g-mode pulsation spectrum, light variation of 2\% amplitude were discovered, probably caused by the reflection effect. This indicates that EQ PSc  is a binary with a cool companion orbiting the sdB in 19.2 h, the longest period found for a sdB+dM binary up to now.
{ \citetads{2016MNRAS.458.1417R} discovered another new g-mopde pulsator, \object{PG 1142-037}, from the first {\it K2} campaign (14 periodicities between 0.9 and 2.5 hr with amplitudes below 0.35 ppt). In addition variations caused by ellipsoidal deformation and Doppler boosting reveal that the sdB has a compact companion (most likely a white dwarf) in a 13h orbit. Despite the close orbit no rotationally split pulsation multiplets were found, indicating that the rotation period is longer than 45 days and, therefore, \object{PG 1142-037} is the first case for non-synchronized rotation in an ellipsoidal variable.}

\clearpage

\newpage
\section{Low mass white dwarfs: The close relatives of sdB stars}\label{sect:lmwd_elm}

\begin{figure}
\begin{center}
\includegraphics[width=0.75\textwidth]{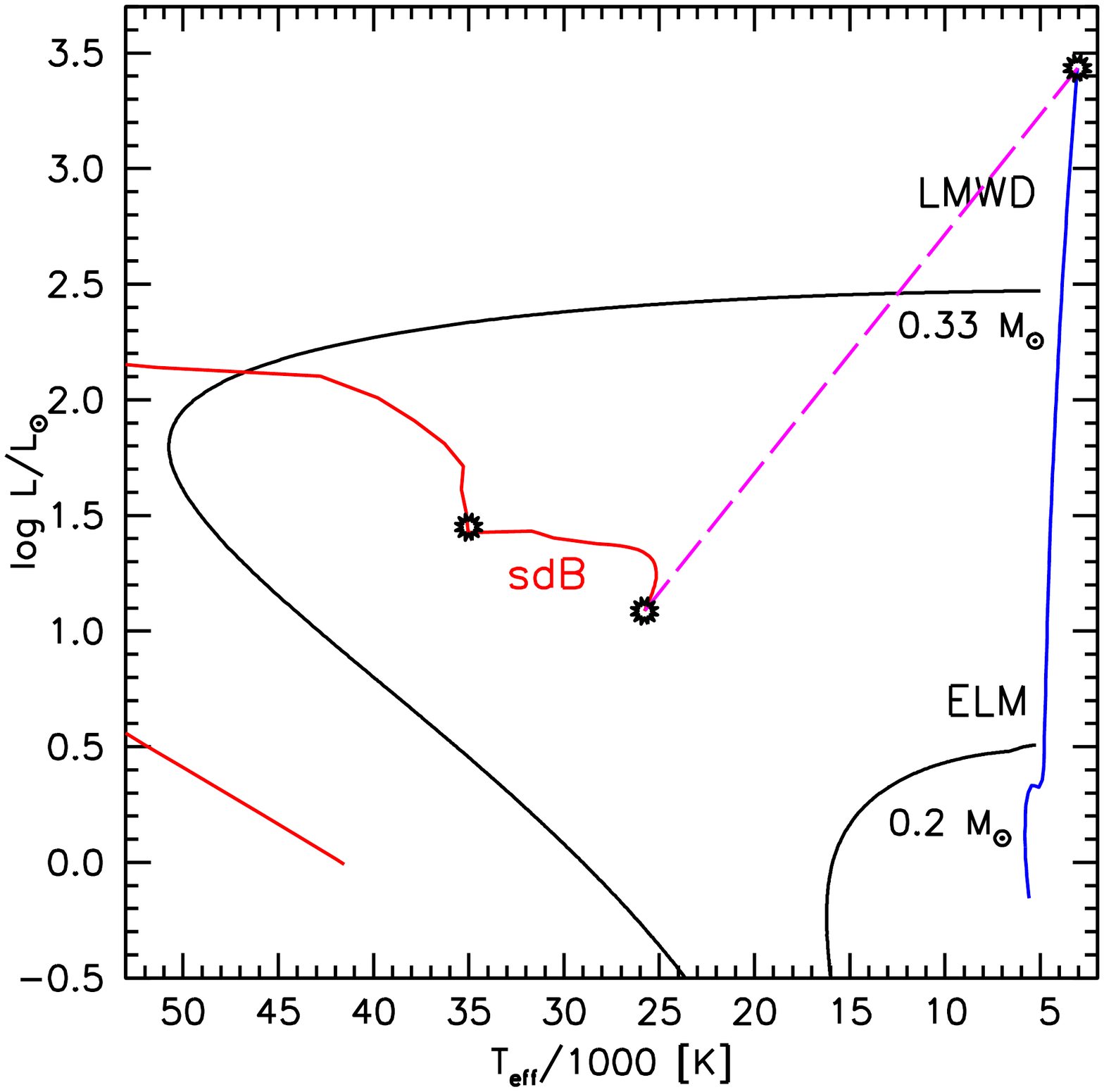}
                \end{center}
\caption{Sketch of the evolution of an 1 M$_\odot$ star (blue) from the zero age main sequence to the tip of the red giant branch, stripped at a core mass of $\approx$ 0.2M$_\odot$ to form an extremely low mass (ELM) white dwarf; stripped at $\approx$ 0.33 M$_\odot$ to form a low mass white dwarf (LMWD), and at the onset of the helium flash to form an sdB star (dashed line). The beginning and the end of the core helium burning of the sdB evolution (red line) is also marked. 
}
\label{fig:rgb_stripping_sketch}
\end{figure}

The formation of sdB stars through mass transfer in close binaries requires some fine tuning, because mass transfer must start close to the tip of the red giant branch, that is the core must have grown close to the canonical mass to ignite helium burning when the red giant progenitor fills its Roche lobe. Hence binaries must exist which result from envelope stripping of a red giant star well before core helium burning ignites (see Fig. \ref{fig:rgb_stripping_sketch}). The remnant of such an event would be a helium white dwarf. As in the case of the sdB stars, the companion is expected {to} be a either a normal dwarf star or a white dwarf. 

\begin{figure*}
\begin{center}
\includegraphics[width=0.99\textwidth]{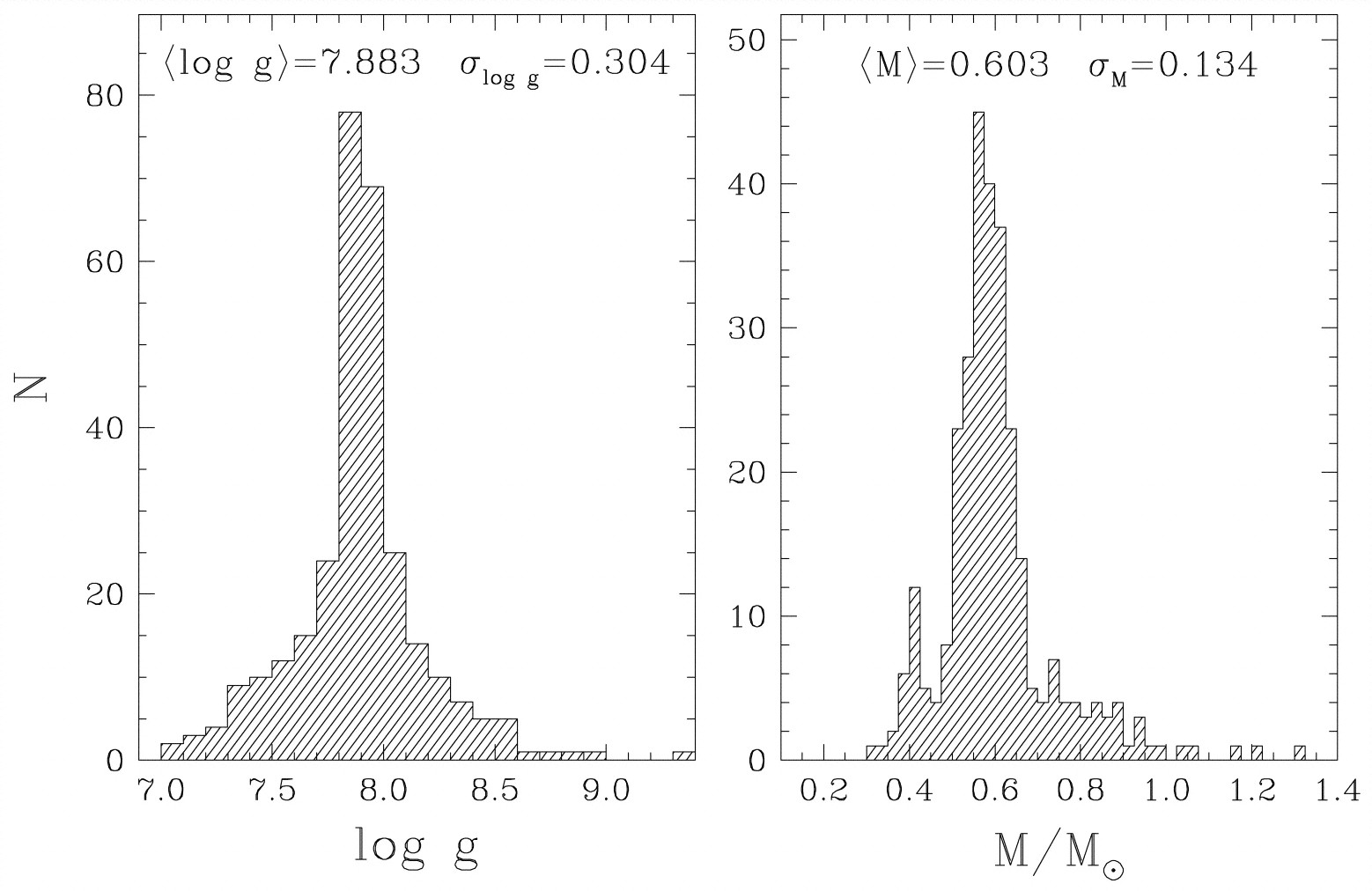}
                \end{center}
\caption{Gravity and mass distribution of DA white dwarfs. Note the presence of a low-mass population at M $\approx$0.4 M$_\odot$. From \citetads{2005ApJS..156...47L}; copyright ApJ; reproduced with permission.}
\label{fig:mass_distribution_lbh}
\end{figure*}

The mass distribution of DA white dwarfs revealed the existence of low mass white dwarfs \citepads{1992ApJ...394..228B} 
with masses below the canonical mass for helium ignition (0.47 M$_\odot$) which were, therefore, considered as helium core white dwarfs.
Later studies of larger samples \citepads[e.g. ][ see Fig. \ref{fig:mass_distribution_lbh}]{2005ApJS..156...47L} confirmed the existence of a population of white dwarfs with a mean mass of 0.4~M$_\odot$. 
Since the evolutionary time for a single star to evolve into a helium-core white dwarf would exceed the age of the Universe by far, \citetads{1992ApJ...394..228B} concluded that they must be in close binaries, which was, indeed, confirmed {{in}}
 many cases by observations \citepads[][and others]{1995MNRAS.275..828M}  thereafter. 

\subsection{Low mass white dwarfs (0.3 to 0.5 M$_\odot$) }\label{sec:lmwd}

About 10\% of the white dwarfs have masses below the 0.5 M$_\odot$ limit and are often referred to as helium white dwarfs.
However, it is premature to assume that white dwarfs with masses below the canonical mass for core helium burning are of helium composition, because intermediate mass ($\geq$ 2.3 M$_\odot$) red giants may ignite helium in non-degenerate conditions 
at core masses as low as 0.33 M$_\odot$ \citepads{2002MNRAS.336..449H,2008A&A...490..243H,2009JPhCS.172a2011P}. 
Hence, a bona-fide helium-core object has to be less massive than 0.33 M$_\odot$. We shall address such stars in 
see Sect. \ref{sect:elm}.

\subsubsection{Low mass white dwarf in single degenerate systems: WD+dM}

Because of their similar absolute visual magnitudes, dM stars can be detected as companions to hot white dwarfs. Their spectral energy distribution is unique, with the white dwarf dominating the blue, the dM the red part of the spectrum. Hence, they are easy to detect and have, indeed, 
been discovered in large numbers \citepads{2013MNRAS.433.3398R}, and 
a significant fraction of them are hosting white dwarfs of low mass  \citepads[$<$0.5 M$_\odot$,][]{2011MNRAS.413.1121R}. 
Compact WD+dM binaries may be the product of common-envelope evolution and, therefore, we might expect that the properties of those binaries are similar to that of the sdB+dM binaries. 
Many of the WD \& dM systems are radial-velocity variable and orbits have be determined for several {{dozen}} with periods ranging from 0.08 to almost 10 days \citepads{2011A&A...536A..43N,2012MNRAS.419..806R}. 

\citetads{2015A&A...576A..44K} 
compared the period distributions of the WD+dM and sdB+dM systems and find the former to be much wider than the latter. Hot subdwarfs with dM companions have very short periods (less than $0.3$\,days). However, this difference is likely {{to be}} due to observational bias, because the WD+dM systems are identified spectroscopically, while most sdB+dM systems were identified photometrically via the reflection effect, the amplitude of which decreases with increasing period, making it difficult to find long period systems. 
 
\citetads{2011MNRAS.413.1121R} found that 1\% to 20\% of the low mass white dwarfs in their sample are not radial-velocity variable on time scales of weeks or longer. Hence they must be wide binaries not formed by common envelope evolution, similar to    
IP~Eri, a K0 (sub)giant + He-WD system \citepads{1995A&A...299L..29V},  with an eccentric orbit ($e$=0.25) and a long period of P=1071d \citepads{2014A&A...567A..30M}. It is interesting to note that the orbit of IP~Eri  is very similar to those of the long-period sdB \& F/G/K binaries discussed in Sect. \ref{sect:double_bin_rv}. 

For an extensive discussion on the origin of low-mass white dwarfs see  
\citetads{2011MNRAS.413.1121R}, who also discussed several evolutionary scenarios that may lead to wide binaries and single low mass white dwarfs, including triple mergers, similar to that put forward by \citetads{2011ApJ...733L..42C} 
to explain wide sdB\&F/G/K binaries\footnote{More exotic scenarios for the formation of He-WDs have also been proposed that do not involve binary evolution, e.g. stellar collisions in the dense cores
of globular clusters \citepads{2008ApJ...683.1006K}
or tidal stripping of a red giant star by a supermassive black hole \citepads{2014ApJ...788...99B}.  
}. 

\subsubsection{Low mass white dwarfs in double degenerate systems}

The SPY survey of DA white dwarfs has found
39 double degenerates among 679 observed white dwarfs which corresponds to a binary frequency of 5.7\% \citepads{2009A&A...505..441K}. 
The binary frequency for low mass white dwarfs is much higher \citepads[$>$70\%, ][]{2011ApJ...730...67B}, 
even larger than that of sdB stars. The number of double degenerates has been boosted recently, by the discovery of dozens of extremely low mass white dwarfs, for which the binary frequency is close to 100\% (see Sect. \ref{sect:elm}).
 
A spectacular recent addition  is WD~1242$-$105 in the solar neighborhood, at a distance of only 39~pc. It is a double-lined spectroscopic binary consisting of two white dwarfs of 0.56 M$_\odot$ and a 0.39 M$_\odot$, respectively, in an 0.1188 day orbit \citepads{2015AJ....149..176D}.

\subsubsection{Single low mass white dwarfs}

However, not all low-mass
WDs are found in binary systems \citepads{2000MNRAS.319..305M,2007ASPC..372..387N}.
\citetads{2011ApJ...730...67B} estimate that the fraction of apparently single 
low-mass white dwarfs is less than 30\%. The origin of these stars remains a puzzle just like the origin of  
the singleton sdB stars. Similar scenarios as for the formation of  sdB singles have been advocated \citepads{2009A&A...493.1081J,2011MNRAS.413.1121R}. 

\subsection{Thermally bloated hot white dwarfs, proto-helium white dwarfs or extremely low mass white dwarfs}\label{sect:elm}

Even the largest sample of white dwarfs \citepads{2015MNRAS.446.4078K} 
did not find white dwarfs with masses below 0.3 M$_\odot$. 
Only recently several dozen of these extremely low-mass (ELM) white dwarfs have been found. White dwarfs are usually thought to have surface gravities exceeding $\log$ g = 7 and correspondingly show strongly Stark broadened spectral lines. The newly discovered white dwarfs of very low mass ($\approx$0.15 to 0.3 M$_\odot$) have lower gravities ranging {{from}} $\log$ g = 4.5 to 7. Therefore, their colors and spectra are quite different from classical white dwarfs. Actually, very-low mass white dwarfs have been found by a survey that {{was}} primarily aimed at finding hypervelocity stars  
of late B-type \citepads{2006ApJ...647..303B} 
by exploring a sparsely populated region in a two-color diagram \citepads{2010ApJ...723.1072B,2011ApJ...727....3K,2012ApJ...751..141K,2013ApJ...769...66B}. 
The low gravities indicate that the stars are bloated, that is they have not yet reached the cooling sequence. When such a star is formed by detaching from its binary Roche lobe 
it retains a rather thick envelope ($\approx$ 0.01 M$_\odot$) which sustains hydrogen burning resulting in transition times to the cooling sequence up to several billion years 
\citepads[depending on the stellar mass,][]{2014A&A...571L...3I}. 
These very-low mass objects were termed ``helium core white dwarf progenitors'' \citepads{2003A&A...411L.477H}, 
``proto-helium white dwarfs'' \citepads{2014A&A...571L...3I},``thermally bloated, hot white dwarf" \citepads{2011ApJ...728..139C}, 
``pre-He-WD'' \citepads{2013Natur.498..463M}, 
or ``Extremely low mass white dwarfs'' \citepads{2010ApJ...723.1072B}, 
which we regard as synonymous.

Interest in low mass white dwarfs arose when such objects were discovered as  companions to millisecond pulsars.
\citetads{1996ApJ...467L..89V} 
showed that the companion of a neutron star in the millisecond pulsar \object{PSR J1012+5307} is a low-mass helium-core object, for which \citetads{1998A&A...339..123D} determined a mass of M=0.19$\pm$0.02 M$_\odot$ and a cooling age of 6$\pm$1 {Gyr}. 
\citetads{2014A&A...571L...3I} list another eight milli-second pulsars in short period binaries with a low-mass ($<$0.21 M$_\odot$) proto-helium white dwarf.
Further discoveries unrelated to pulsar binaries followed \citepads{2003A&A...411L.477H,2004ApJ...606L.147L,2006ApJ...643L.123K,2006BaltA..15...61O,2007ApJ...660.1451K,2010A&A...516L...7K}, but the 
 number of such stars remained small until 
a dedicated search was initiated using the SDSS data base \citepads{2010ApJ...723.1072B,2011ApJ...727....3K,2012ApJ...751..141K,2013ApJ...769...66B}. 
Because these stars are expected to result from close binary evolution, they were monitored for radial-velocity variations with a very high success rate. The majority of them is, indeed,
found to have orbital periods shorter than one day, the companions being white dwarfs as well.

\paragraph{An ELM white dwarf in a triple system}

\citetads{2014Natur.505..520R}
reported the discovery of PSR J0337+1715, a triple system of a neutron star orbited by {\it two} white dwarfs.
PSR J0337+1715 contains a 1.438 M$_\odot$ radio millisecond pulsar and two low mass WDs. The inner WD companion is an ELM with 0.197 M$_\odot$ on a 1.63~d circular orbit. The system is highly hierarchical because the outer white dwarf is distant orbiting in 327 days  \citepads{2014Natur.505..520R}. 
The system must have survived {\it three} phases of mass transfer and, in addition, has managed to survive a supernova explosion. \citetads{2014ApJ...781L..13T} presented a model to explain the evolution of this fascinating triple and find 
the two white dwarf companions to match the theoretical expectations from the mass -- orbital period relation of WDs \citepads[e.g.][]{1999A&A...350..928T}.   


\citetads{2014ApJ...783L..23K} discovered the optical counterpart of the radio source and identified it with the inner white dwarf companion. 
Their spectroscopic analysis resulted in  
the mass and radius of the inner WD which are fully consistent with predictions for a young ELM WD. Its high effective temperature (15800K) implies that the inner white dwarf is relatively young and formed last, as expected from models \citepads{2014ApJ...781L..13T}.  

\paragraph{Eclipsing ELM white dwarfs}

Eclipsing binaries are of great importance, because their masses and radii can be tightly constrained.
\citetads{2016MNRAS.458..845H} 
 found an eclipsing binary consisting of two low mass white dwarfs of 0.38  and 0.23 M$_\odot$, respectively.
Now six eclipsing systems are known, with CSS 41177 being the only double-lined eclipsing WD system known \citepads{2014MNRAS.438.3399B,2015MNRAS.448..601B}, for which mass and radius could be determined to unprecedented precision. 

\subsubsection{Confusion with other types of star}\label{sect:mimick}

\begin{figure}
\begin{center}
\includegraphics[width=0.6\textwidth]{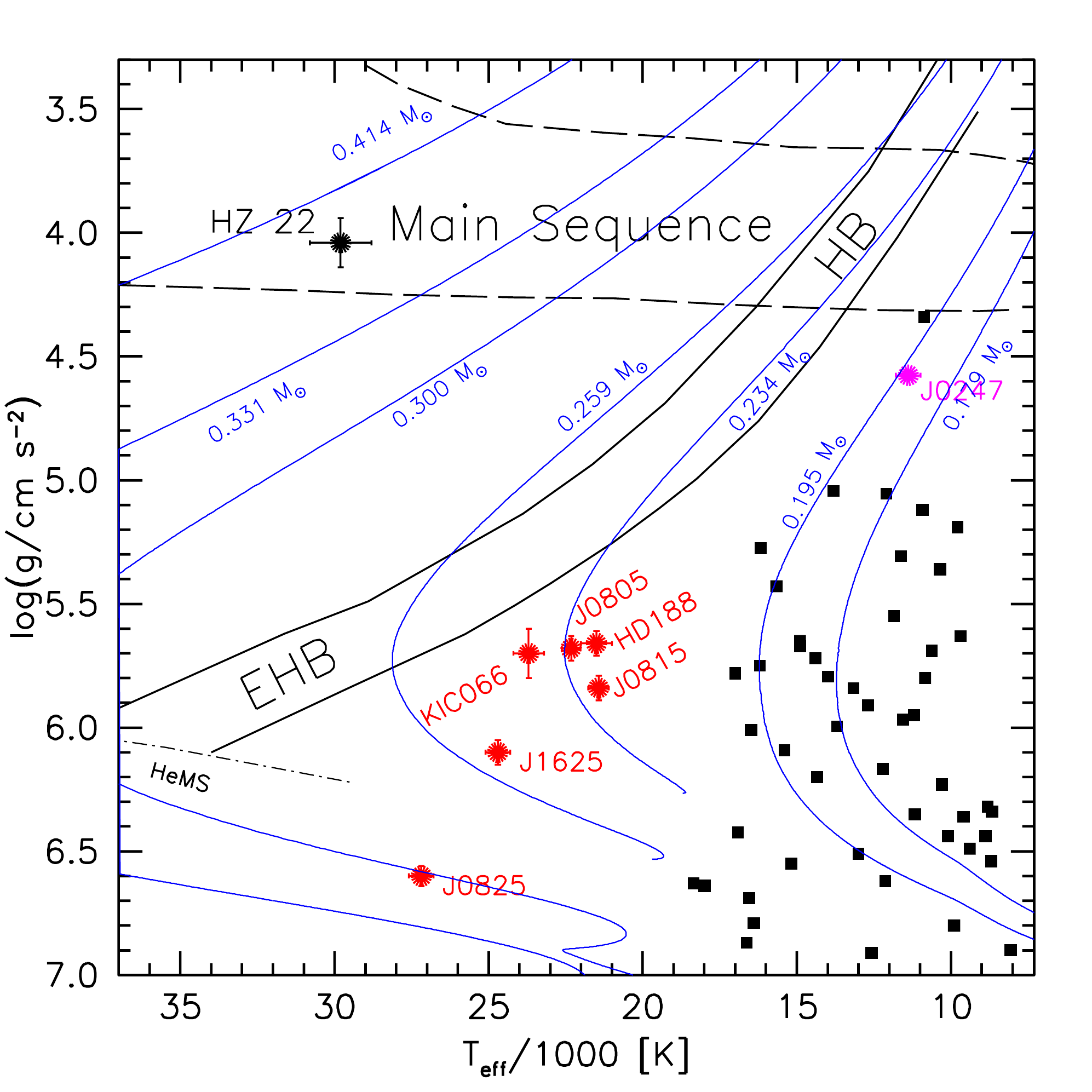}
                \end{center}
\caption{Distribution of proto-helium white dwarfs in the T$_{\rm eff}$ -- $\log$ g diagram from the list compiled by \citetads{2013A&A...557A..19A}. 
Tracks for the evolution of proto-helium white dwarfs are from 
\citetads{1998A&A...339..123D} and labelled {{with}} their mass. 
The main sequence band is shown (dashed lines) as well as the HB and EHB band. 
Note that the tracks cross both bands leading to confusion for objects in the overlapping regions.
The values for WASP~J0247$-$25B are from \citetads{2013Natur.498..463M}, 
for HZ~22 from \citetads{1997ApJ...491..172S},  
 for GALEX J0805$-$1058 (G0805) from \citetads{2015MNRAS.450.3514K}, 
 for SDSS~J1625+3632 (J1625), SDSS~J0815+2309 (J0815) and SDSS J0825+1152 (J0825) are from \citetads{2014ApJ...794...35G}. 
Other abbreviations: KIC066=KIC 06614501, HD188=HD188112}
\label{fig:hrd_elm}
\end{figure}

The transition track of a proto-helium white dwarf from the red giant branch to the cooling sequence leads through a wide area of the (T$_{\rm eff}, \log~g)$ plane and across the main sequence and the (extreme) horizontal branch, and therefore, proto-helium white dwarfs may be confused with normal stars. In fact a few such cases are known. The most extreme case, HZ~22, even mimics an early type main-sequence star, while the sdB stars HD~188112, KIC 06614501, SDSS J0815+2309,
SDSS~J1625+3632 and GALEX J0805$-$1058 mimic sdB stars, albeit at a somewhat higher gravity (see Fig. \ref{fig:hrd_elm}). 

\paragraph{HZ~22 (UX CVn) - imitating a massive B star} 
HZ~22  \citepads[UX CVn,][]{1972ApJ...174...27Y,1973A&A....23....1G}, 
is star No. 22 in the very first survey for faint blue stars at high Galactic latitudes \citepads{1947ApJ...105...85H}. 
The object was discussed at the {\it First conference on faint blues stars} \citepads{1965fbs..conf.....L} because its spectrum mimics that of an early-type main-sequence star, which 
would place the star very far from the Galactic plane, where no young star had been found at that time. 
Light and radial-velocity variation showed that it must be a single-lined binary. The orbital period turned out to be about half a day 
 \citepads{1972ApJ...174...27Y}, which is too short to accommodate a massive B star. Hence HZ~22 must be a subluminous B star \citepads{1969AcA....19..165S}.
 The strong ellipsoidal light variations indicate that the companion must be a white dwarf \citepads{1982PASP...94..815Y}.     
Using evolutionary models \citetads{1973A&A....23..281T} found that HZ~22 is a helium-core object of 0.3--0.4 M$_\odot$. \citetads{1978A&A....70..451S} carried out similar calculations and derived a mass of 0.39 M$_\odot$ confirming the identification of HZ~22 as a proto-helium core white dwarf. 
The star remained unique, when \citetads{1974ApJS...28..157G} completed their seminal work on faint blue stars. 
\citetads{2002ARep...46..127S} revisited the star and derived atmospheric parameters and metal abundances from a detailed quantitative spectral analysis
of medium-resolution (2.6\AA) optical spectra obtained at the 6m-telescope of the Special Astrophysical Observatory. The helium, carbon and iron deficiency
are consistent with an old evolved star, i.e. support the interpretation as a proto-helium core white dwarf. In the light of {these} results \citetads{2002ARep...46..656S} reconsider the evolutionary status of the system and confirm the previous interpretation of the system. 
\citetads{2002ARep...46..127S} noted that their spectrum of HZ~22 taken in 1999 differed from that published by \citetads{1973A&A....23....1G}, in particular with respect to the He~{\sc ii} 4686\AA\ line which is present in their spectrum but absent in the early one. This led to the conjecture that the star has increased its temperature by 2000K within 40 years and, hence, must be in phase of rapid evolution.
This conjecture, however, needs confirmation, because the of the respective uncertainties of the temperature determinations.    
Although HZ~22 remains a unique and fascinating object, it has not gotten the attention in recent years that it deserves.

\paragraph{Proto-helium white dwarfs mimicking sdB stars.}
\citetads{2003A&A...411L.477H} found the sdB star HD~188112 to be of somewhat higher gravity than a core-helium burning EHB star from a quantitative spectral analysis. Its Hipparcos parallax  combined with the spectroscopic gravity allowed the mass to be determined to be ~0.23 M$_\odot$, consistent with evolutionary model predictions for stripped RGB stars by \citetads{1998A&A...339..123D} \citepads[see also][]{2016A&A...585A.115L}. 
 The star turned out to be a short period binary with a white dwarf companion of at least 0.7 M$_\odot$. 
Two very similar sdB binaries were discovered recently in the {\it Kepler} field \citepads[KIC\,6614501, ][]{2012MNRAS.424.1752S}  
 and in the {\it GALEX} survey \citepads[GALEX\,J080510.90$-$105834.00, ][]{2015MNRAS.450.3514K}, respectively. 
A star of similar temperature but somewhat higher gravity was found by \citetads{2011ApJ...727....3K}.

\subsubsection{Evolution of extremely low mass white dwarfs}

Thermal instabilities in the hydrogen burning shell are expected to occur during the post-RGB evolution \citepads{2013arXiv1303.6652G}, 
but theoretical models \citepads{1999A&A...350...89D,2000MNRAS.316...84S,2004ApJ...616.1124N,2007MNRAS.382..779P,2013arXiv1303.6652G,2013A&A...557A..19A,2014A&A...571L...3I} disagree about the thickness of the hydrogen envelope and 
the range of helium core masses for which shell flashes occur. 
Because hydrogen is being burnt efficiently during a flash, the evolutionary lifetime is shortened considerably, if such a flash occurs. 
Many of these differences may be due to differences in the treatment of the physical processes within the donor star and the secular evolution of
the binary itself \citepads{2004ApJ...616.1124N}.  
The least massive stars are predicted not to suffer from flashes and therefore, quiescent hydrogen burning guarantees a long evolutionary time scale (up to a few Gyrs). \citetads{2013A&A...557A..19A} 
showed that gravitational settling has an important effect on the occurrence and strength of flashes which reduce the hydrogen envelope. 

\subsection{Abundance pattern of ELM white dwarfs}

Few ELM white dwarfs 
are sufficiently bright for detailed quantitative abundance analyses.
Nevertheless, \citetads{2014ApJ...794...35G} 
were able to determine abundances (or upper limits thereof)
of He, Ca, and Mg in their sample of 61 stars. Calcium lines were found in all
stars with $\log$ g $<$ 6.0.

\begin{figure*}[]
\begin{center}
\includegraphics[width=0.85\textwidth]{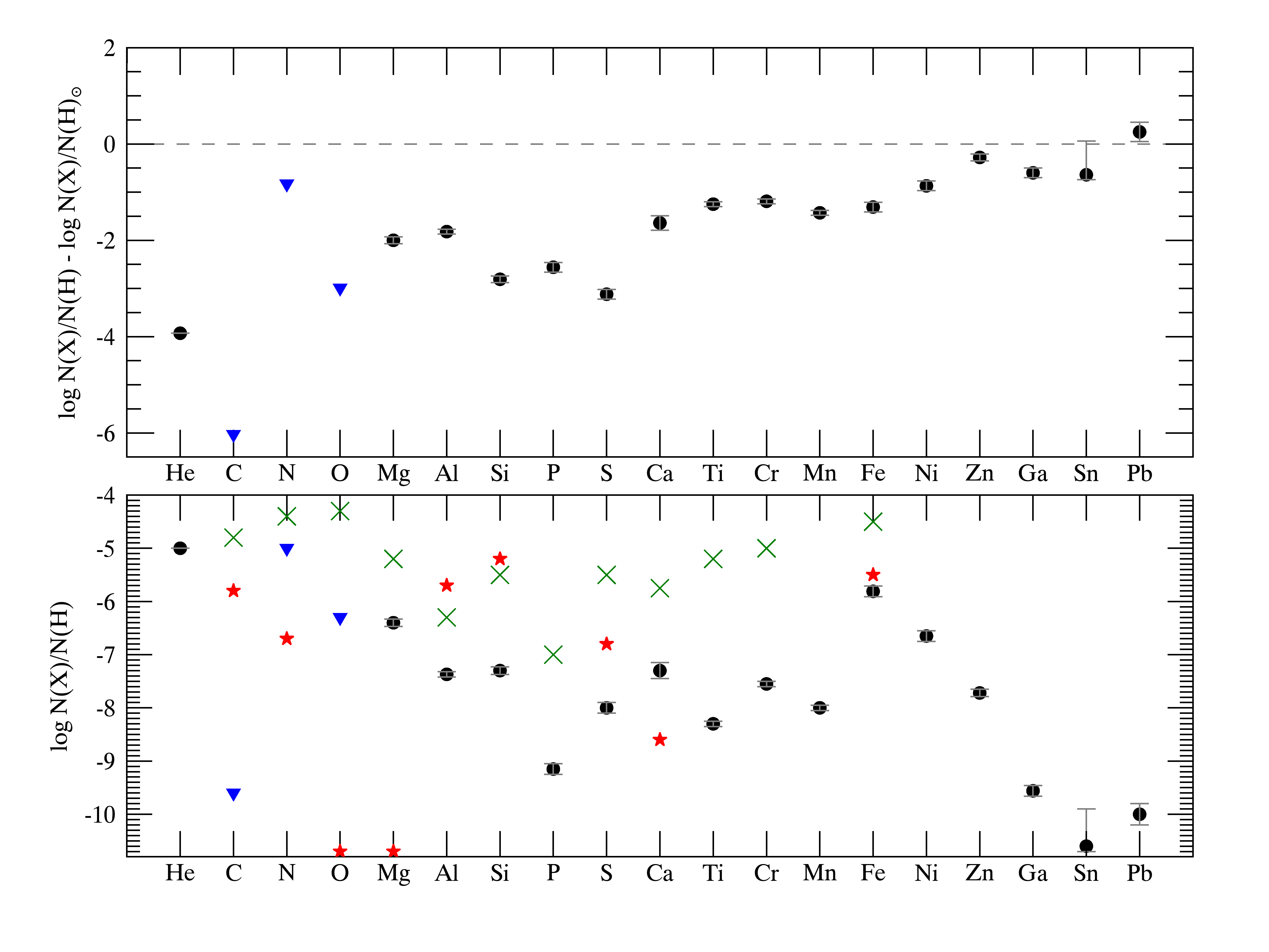}
\caption{Summary of the determined chemical composition of HD~188112 \citepads{2016A&A...585A.115L}. The top panel shows the abundances relative to the solar ones \citepads{2009ARA&A..47..481A} 
and the bottom panel the absolute ones. Downward triangles indicate upper limits determined for C, N, and O. The asterisks on the bottom panel indicate abundances predicted by radiative levitation (P. Chayer, private communication). No radiative support is expected for oxygen and magnesium {{(asterisks at the bottom of the panel)}}. The green crosses indicate the average abundances for sdB stars. The diffusion model fails to reproduce the observed abundances of oxygen and Mg, dramatically, while it matches those of silicon and iron. From \citetads{2016A&A...585A.115L}; copyright A\&A; reproduced with permission.} 
\label{fig:abu_hd188}
\end{center}
\end{figure*}

Detailed abundance studies are available for four objects, only.
\citetads{2016A&A...585A.115L} analyzed very high resolution ultraviolet spectra of HD~188112 taken with the Hubble Space telescope's STIS spectrograph and derived abundances of 14 metals and upper limits for C, N, and O, which turn out to be all subsolar (with the exception of lead that is solar, see upper panel of Fig. \ref{fig:abu_hd188}). A comparison with predictions from diffusion models was disappointing. While the Fe abundance is quite close to the prediction, the derived abundances of all other elements were either well below the predicted ones (C, Al, Si, S, Fe) or much higher
(Ca by 1.5 dex, for Mg no radiative support at all is expected). 

GALEX J1717+6757 \citepads[discovered by][]{2011ApJ...737L..16V} 
has a surface gravity of log g = 5.67, almost identical to that of HD 188112, but is significantly cooler (T$_{\rm eff}$ = 14,900 K) and of lower mass (0.19~M$_\odot$). 
\citetads{2014MNRAS.444.1674H}
derived abundances for nine metals, all of them being higher (by roughly 0.5 to 2.0 dex) than in HD 188112. Unlike for HD~188112,
radiative levitation for N, O, Si, P, and
Fe gives predictions in line with the star's abundance pattern \citepads{2014MNRAS.444.1674H}.  

Two {\it metal-rich} ELM white dwarfs are also known: the companion of a massive mil-
lisecond pulsar  PSR J1816+4510
\citepads[T$_{\rm eff}$ =16,000 K, log g=4.9,][]{2013ApJ...765..158K}, 
and SDSS J0745+1949
(T$_{\rm eff}$ =8,380 K, $\log$ g=6.2), a tidally distorted star \citepads{2014ApJ...781..104G}. 
The analyses of their optical spectra allowed for
the abundance determination
of Mg, Ca, Ti, Cr, and Fe, which turned out to be close to solar in the
case of J0745, while He, Mg, Ca, Si and Fe are about 10 times
solar in J1816. 

It seems that the diverse abundance pattern of the four stars can neither be explained by diffusion models nor by mixing during repeated shell flashes.

\subsection{Pulsations of ELM white dwarfs}

Photometric monitoring of ELM white dwarfs led to the discovery of pulsations. The first object,  
SDSS~J1840+6423 \citepads{2012ApJ...750L..28H} 
is  in a 4.59 hr orbit with another white dwarf 
and shows multiperiodic light variations  with  a  dominant  period  of  about
4698s.  Four additional  pulsating ELM white dwarfs
were found soon thereafter \citepads{2013MNRAS.436.3573H,2013ApJ...765..102H}, and 
a sixth member of this new class of pulsating white dwarfs has been reported recently \citepads{2015ASPC..493..217B}. 
All stars have effective temperatures between 7800K and 9500K, gravities between log~g= 6.0 and 6.8. 
The light 
curves are multiperiodic, with a handful of periodicities between 1300 s to 7000s.
The detected periodicities of the light curves are caused by
g-mode pulsations \citepads{2012ApJ...750L..28H}, thus, opening a new opportunity and challenge for asteroseismology. 

These observations triggered several investigations to model pulsations of ELM white dwarfs and to identify the driving mechanism  
\citepads[e.g.][]{2010ApJ...718..441S,2012A&A...547A..96C,2013ApJ...762...57V,2015ASPC..493..221C}, 
which led to the conclusion that the pulsating ELM white dwarfs form an extension of the ZZ~Ceti instability strip \citepads{2013ApJ...762...57V} albeit at lower temperature. 

\subsection{ELM white dwarf and sdB\&WD binaries in comparison}\label{sect:elm_sdb_comp}

At the time of writing the ELM survey has identified {88 ELM white dwarfs \citepads{2016ApJ...818..155B}, of which one dozen appear to be non-variable.
From  a ''clean'' subsample\footnote{The subsample contains 
only binaries with radial velocity semi-amplitudes larger than 75~km\,s$^{-1}$.
for which
the catalog should be 95\% complete.} of 65 ELM WD binaries \citepads{2016ApJ...818..155B} found that the orbital periods have a lognormal
distribution with a median period of 5.4 hours, which is consistent with predictions from binary population synthesis models \citepads{1998MNRAS.296.1019H}. The distribution of companion mass is best described by a normal
distribution with a mean of 0.76 M$_\odot$. Typically the systems' total masses of about one solar mass, with a mass ratio of $\approx$1:4.}

\begin{figure*}
\begin{center}
\includegraphics[width=0.6\textwidth]{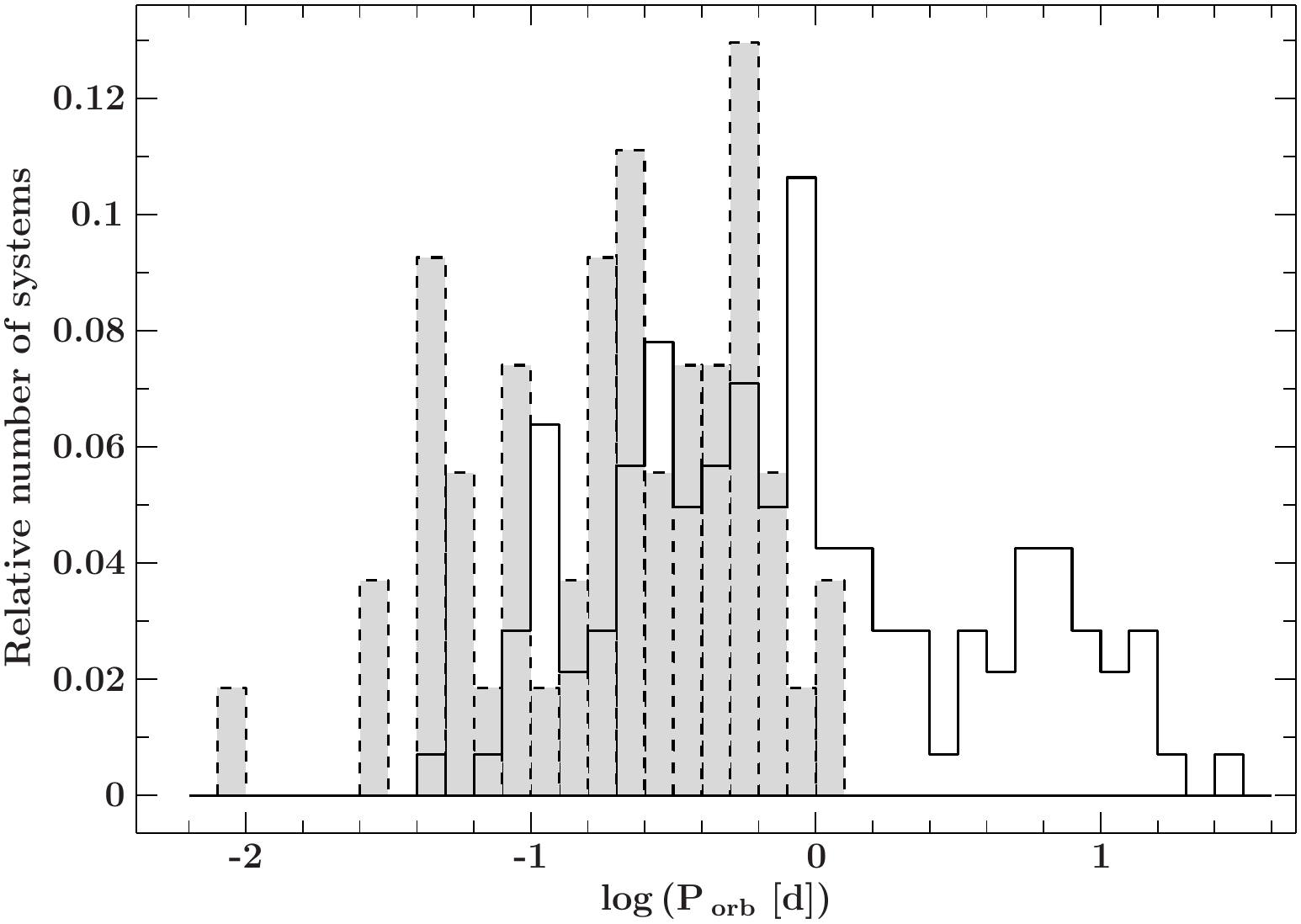}
                \end{center}
\caption{Comparison of orbital periods of sdB binaries with confirmed WD companions to the known ELM-WD binaries with orbital solutions (grey shaded area) taken from \citetads{2014ApJ...794...35G}.  From \citetads{2015A&A...576A..44K}, copyright A\&A with permission.}
\label{fig:elm_period}
\end{figure*}

\begin{figure}
\begin{center}
\includegraphics[width=0.6\textwidth]{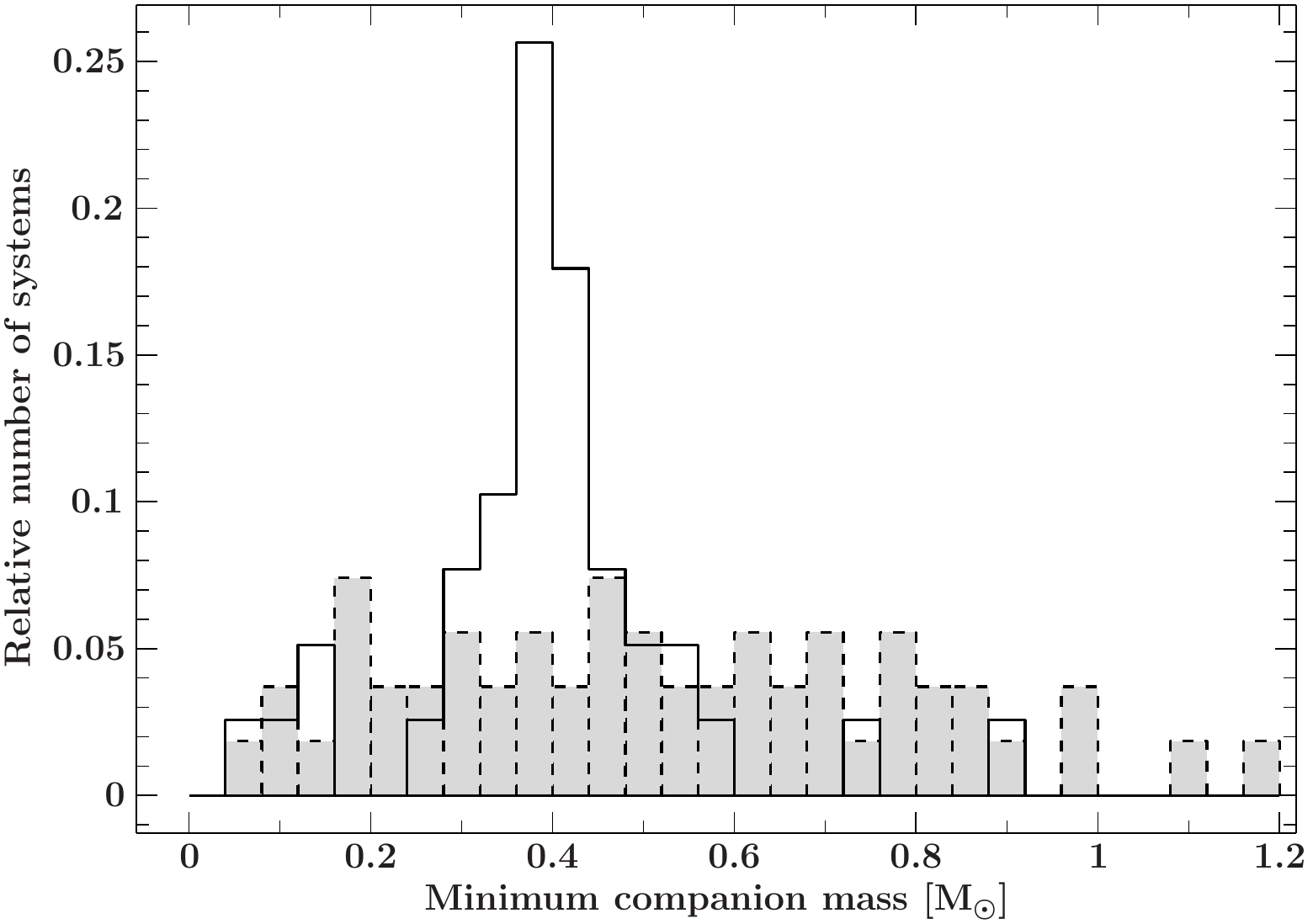}
                \end{center}
\caption{Comparison of minimum companion masses of sdB binaries to the ELM-WD binaries (grey shaded area) taken from \citetads{2014ApJ...794...35G}. 
From \citetads{2015A&A...576A..44K}; copyright A\& A; reproduced with permission.}
\label{fig:elm_mass}
\end{figure}

\citetads{2015A&A...576A..44K} 
compared the distributions of orbital periods and minimum companion masses  of 55 ELM white dwarf binaries 
from \citetads{2014ApJ...794...35G} to their sdB\&WD sample (Figs.~\ref{fig:elm_period} and \ref{fig:elm_mass}). 
No ELM white dwarf with an orbital period significantly longer than one day has been found yet, which, however, is likely to be a selection effect. 
For orbital periods between a quarter of a day and one day both distributions are very similar. However, for the shortest periods ($<0.25$\,days) ELM white dwarfs are more numerous compared to the sdB+WD systems (Fig.~\ref{fig:elm_period}), that is ELM white dwarfs are formed preferentially with shorter periods than sdB\&WD systems, which may be expected, because the progenitor of an ELM white dwarf 
may fill its Roche lobe early-on during its red giant evolution.

ELM-WD companions cover a wider range of minimum masses than that of sdB\&WD (Fig.~\ref{fig:elm_mass}); that is, masses extend to low as well as to high masses in comparison to the sdB companions. There is no apparent preferred mass which compares with the peak at $0.4$\,M$_\odot$ for the WD companions to sdB stars. 

The white dwarf nature of the companion to most ELM-WDs and many sdB stars is inferred from the mass function which {{only allows a lower limit to be derived}}. 
Additional information, such as ellipsoidal light variations or the lack of a reflection effect and of X-ray emission \citepads{2011A&A...536A..69M,2011ApJ...727....3K,2014MNRAS.438L..26K}, help to identify the companion as a white dwarf. 
However, more massive companions may exist. The existence of neutron {{star}} companions to ELM white dwarfs is obvious from the systems hosting milli-second pulsars, but no pulsar companion has as yet been detected to accompany a hot subdwarf star, although such systems are predicted by theory \citepads{2003ApJ...597.1036P}. 
However, the pulsar signal can only be detected if the orientation of the beam is appropriate. Hence, other ELM and sdB binaries may host neutron stars as well but do not appear as radio sources. 
In order to estimate the neutron star fraction in the ELM survey sample \citetads{2014ApJ...797L..32A} 
developed a statistical model to infer the companion mass distribution. Applying it to 55 ELM white dwarfs they find the distribution to be consistent with no neutron star companion in the sample and derive an upper limit to the neutron star fraction of 16\% (within 1$\sigma$ limit).

\begin{figure}
\begin{center}
\includegraphics[width=0.9\textwidth]{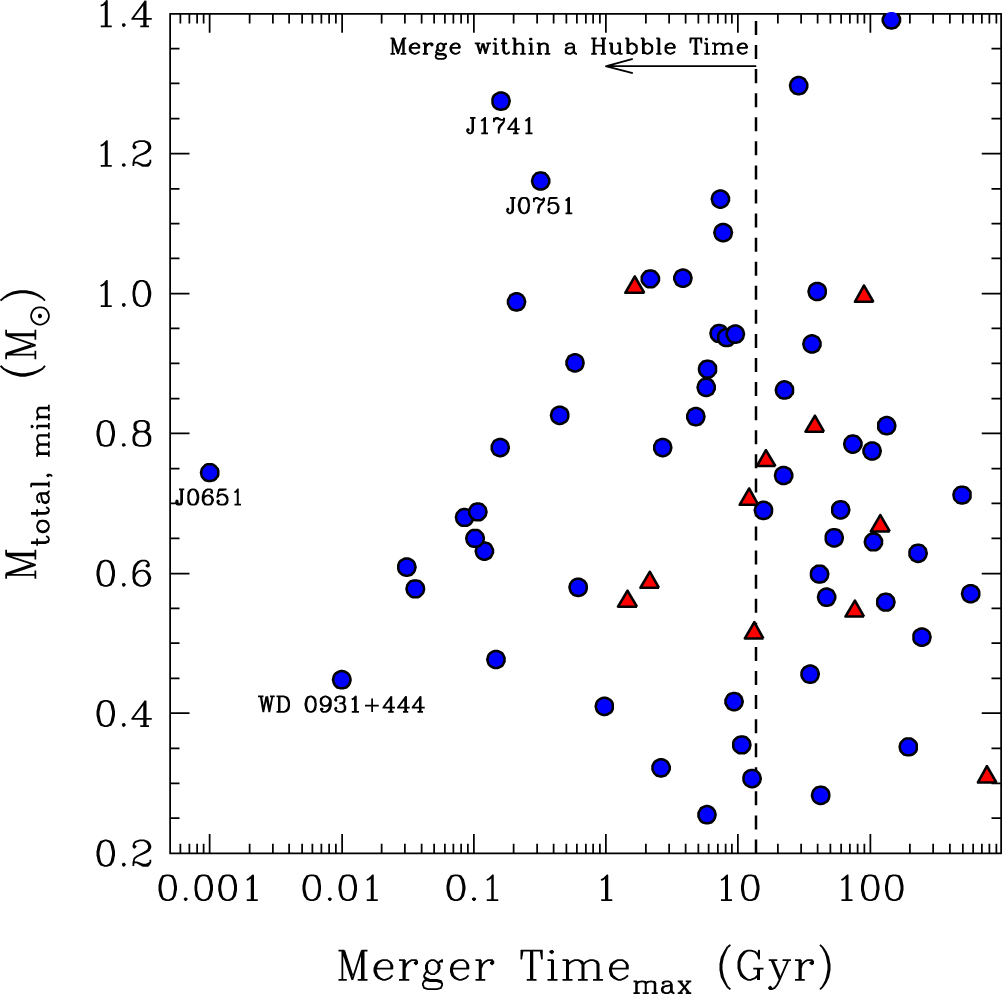}
                \end{center}
\caption{Minimum total mass as a function of the maximum merger time for the entire ELM Survey sample (blue circles and red triangles).
From \citetads{2015ApJ...812..167G}; copyright ApJ; reproduced with permission.}
\label{fig:elm_merger}
\end{figure}

\begin{figure}
\begin{center}
\includegraphics[width=0.9\textwidth]{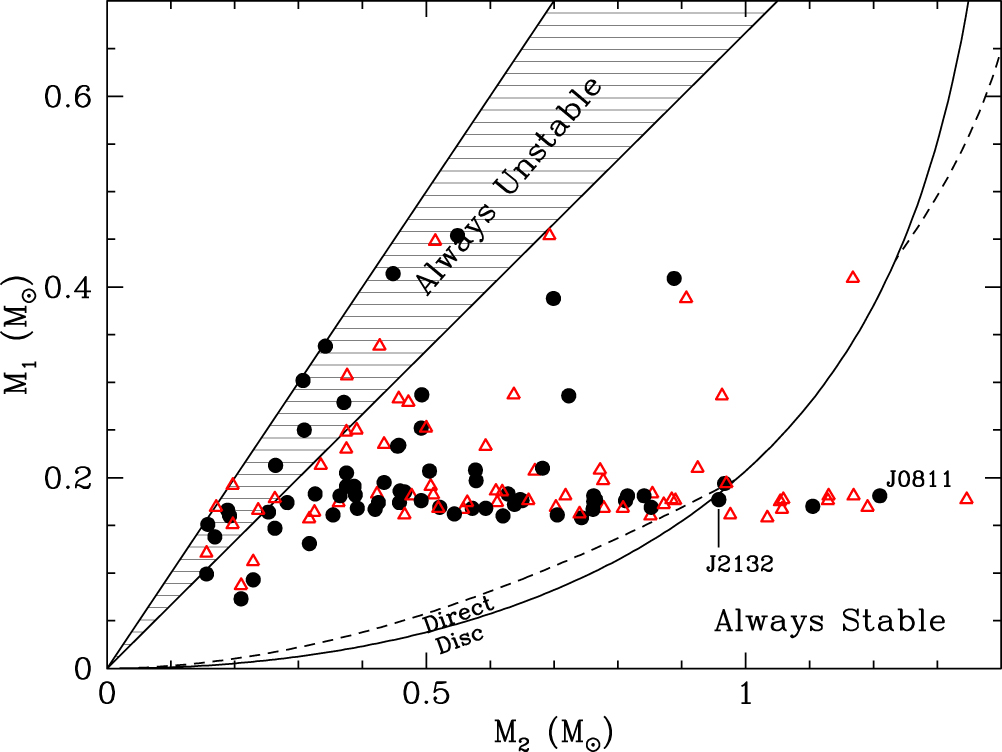}
                \end{center}
\caption{Plot of M$_1$ vs. M$_2$ for the entire ELM-WD sample. The companion masses are shown for inclinations of 90$^\circ$ (black dots)  and 60$^\circ$ (red triangles). The majority of the ELM-WD binaries lie in the region between the regions of stable and unstable mass transfer.
From \citetads{2015ApJ...812..167G}; copyright ApJ; reproduced with permission.}
\label{fig:elm_stability}
\end{figure}

The orbital periods of the ELM-WD binaries are short and, therefore, many are expected to start transferring mass within a Hubble time (see Fig. \ref{fig:elm_merger}).
The nature of the mass transfer phase depends on the mass ratio q of the two components. If mass transfer is stable a disk will form around the more massive white dwarf resulting in an AM CVn system. 
If the mass is accreted directly, the orbit may be destabilized and a merger occurs \citetads{2004MNRAS.350..113M}. The stability criteria for different binaries are shown in Fig. 
\ref{fig:elm_stability} 
and compared to the observed ELM-WD mass distribution. 
Four ELM-WD systems lie in the stable mass transfer regime and, therefore, should end up as AM CVn systems.
Most ELM-WD binaries, however, lie between the stable and unstable regions and the final merger products remain uncertain.

\subsection{Low-mass proto-helium white dwarfs with main-sequence companions}

We also expect extremely low-mass helium-core objects with main-sequence companions to exist, which result when the more massive star starts to fill its Roche lobe when it is still a subgiant or very early on the first giant branch. This may result in either a stable RLOF or in a common envelope ejection event.

Amongst the {\it Kepler} Objects of Interest(KOI) two early-type stars, KOI-74 (spectral type A, orbital period P=5.2 d) and KOI-81 (late B-type, P=23.9 d), show transits of planet-size objects. However, the minima of their light curves were shallower during the transit than during occultation of the companion \citepads{2010ApJ...713L.150R}. 
 Hence the companion must be hotter than its host star. \citetads{2010ApJ...713L.150R} and \citetads{2010ApJ...715...51V} concluded that the companions must be low mass white dwarfs. Doppler boosting present in the {\it Kepler} light curve allowed \citetads{2010ApJ...715...51V} 
to determine the companion mass to KOI-74 to be 0.22 M$_\odot$.  
 Using HST/COS spectra, \citetads{2015ApJ...806..155M} 
  were able to reconstruct the UV spectrum of the companion to KOI-81 and to measure the companion's radial velocity curve resulting in a mass of M=0.194$\pm$0.02 M$_\odot$ and a radius of 0.0911$\pm$0.0025 R$_\odot$. The UV Spectrum resembled that of the sdB star CD$-$64$^\circ$481, and the effective temperature was constrained to $>$ 19,000K.
 Hence the star is very similar to the helium core subdwarf HD~188112.
   
Further discoveries of similar objects of a low mass white dwarf with an A-dwarf secondary from {\it Kepler} light curves followed  \citepads[KIC10657664, ][]{2011ApJ...728..139C}, and \citepads[KIC 9164561, KIC 10727668, ][]{2015ApJ...803...82R}
with periods between 1.3 and 5.2 days and masses determined from Doppler boosting. In addition a F-type secondary was found in KOI-1224
\citepads{2012ApJ...748..115B}. 
Recently, \citetads{2015ApJ...815...26F} added another four {\it Kepler} systems consisting of A type stars hosting extremely low mass white dwarf companions with orbital periods between 1.17 and 3.82 days and companion masses from 0.19 to 0.22 M$_\odot$

Such eclipsing binaries have also been discovered from ground-based (SuperWASP) lightcurves \citepads{2011MNRAS.418.1156M,2014MNRAS.437.1681M}. The host stars are of spectral type A and orbital periods range from 0.7 to 2.2 days. 
The class was named ``EL CVn'' stars after the prototype\footnote{It is interesting to note that 
Regulus A (a bright B7 star) has a low mass white dwarf companion
in a 40 day orbit
\citepads{2008ApJ...682L.117G}. Making use of the white dwarf period - mass relation \citetads{2009ApJ...698..666R} showed that the white dwarf companion must be of low mass (0.28 $\pm$0.05~M$_\odot$). Hence, the inner binary of Regulus qualifies as a EL CVn system.}.


Perhaps not surprisingly, the low mass helium-core objects in KOI-71, KOI-81, and 1SWASP~J024743.37$-$251549.2 (J0247-25B)
were found to be pulsating \citepads{2010ApJ...713L.150R,2013Natur.498..463M}. 
The oscillations of J0247-25B are mixed non-radial modes \citepads{2013Natur.498..463M}, that is they 
 behave like g-modes near the stellar core and like p-modes in the outer layers. This paves the way to probe its interior structure and to measure its internal rotation profile \citepads{2015AN....336..477A}. The pulsations are probably driven by the  $\kappa$-mechanism, i.e. by the change in opacity in the second partial ionization zone of helium.
Hence, \citetads{2013Natur.498..463M} conjecture that other ELM white dwarfs with effective temperatures similar to J0247-25B may also show pulsations. Indeed
\citetads{2014MNRAS.444..208M} found a second candidate pulsator (WASP 1628+10B, T$_{\rm eff} \approx 9200$K) as a companion to an A2V star which itself is a $\delta$ Scuti pulsator.

\citetads{2013MNRAS.435..885J} explored the boundaries of the instabilities of low- to high-order radial oscillations in ELM WDs and found them to depend strongly on chemical composition and
radial order number. For J0247-25B they conclude that the envelope must be hydrogen deficient (0.2 $<$ X $<$0.3), in agreement with expectations from evolution models, but still lacks observational confirmation.

\clearpage

\newpage
\section{Kinematics, population membership and a unique hyper-velocity star}\label{sect:kinematic}

Hot subdwarf stars are found in all stellar populations. Globular cluster hot subdwarfs are population II stars, while the members of open clusters belong to the disk population. Most of the hot subdwarfs in the field belong to an old disk population {\citepads{1997A&A...317..689T,1997A&A...327..577D,2004A&A...414..181A,2015MNRAS.450.3514K}.} 
 Finally, hot subdwarfs have been discovered in the Galactic bulge \citepads{2005ApJ...633L..29B}. 
 However, some high-velocity hot subdwarf stars may have been ejected from their place of birth \citepads{2011A&A...527A.137T}, that is they may be run-away stars. An extreme such case is US~708, a hyper-velocity star (HVS), that travels so fast that it will escape from the Galaxy \citepads{2005A&A...444L..61H}. 
 
\subsection{Population membership}

For normal stars chemical tagging allows the identification of stellar population membership. However, this is not possible for hot subdwarf stars,
because atmospheric diffusion has altered the abundance pattern and washed out any information on the original metallicity. Therefore, we are left with the Galactic kinematics of the hot subdwarf stars to 
assign an individual hot subdwarf to a stellar population.
To do this the full 6D phase space information needs to be available. Distances are usually derived from the atmospheric parameters by assuming a canonical sdB mass. Proper motions can be drawn from large astrometric catalogs but typically have substantial uncertainties. 
Although straightforward the radial velocity measurement requires a considerable effort. Because the frequency of radial-velocity variables amongst the sdB stars is very large (about 50\% with periods less than 30 days), multi-epoch observations are needed to rule out RV variability or derive the systemic velocity from the RV curve of a close binary system.

Systemic radial velocities are now available for 142 sdB binaries  
\citepads{2015A&A...576A..44K} and have recently been used to determine their population membership from kinematic diagnostic diagrams such as the {$v_\phi$  -- $v_r$} and the $e$--$J_z$ diagrams (Eichie et al., in prep.; see Fig. \ref{fig:sdb_kin_groups}) {where $v_\phi$ is the velocity component in the direction of Galactic rotation, $v_r$ the component in Galactic radial direction\footnote{{$v_\phi$  and $v_r$ often referred to as $V$ and $U$, but may be confused with the cartesian velocities $v_x$  and $v_y$ \citepads{1987AJ.....93..864J}, see also \citetads{2015A&A...576A..65R}.}},
 $J_z$ the component of the angular momentum of the star's Galactic orbit perpendicular to the Galactic disk and $e$ the eccentricity of the 
Galactic orbit.}


Accordingly, the vast majority of binaries belongs to the 
thin disk, with a few thick disk stars and a single halo object, SDSS~J002323-002953. 
Most of the stars in the sample are nearby, and therefore it is not surprising that they belong to the thin disk population.

However, the MUCHFUSS survey, which includes much fainter and, therefore, distant hot subdwarfs, has identified a couple of high-velocity stars that may belong to the halo population\footnote{Halo and thick disk stars have also been identified amongst the ELM white dwarfs \citepads{2015ApJ...812..167G} and the EL CVn binaries \citepads{2014MNRAS.437.1681M} 
discussed in Sect. \ref{sect:lmwd_elm}.}. 

\subsubsection{High velocity hot subdwarfs stars: run-away stars or genuine  halo stars}\label{sect:sd_hvs}

\begin{figure}
\begin{center}
\includegraphics[width=0.5\textwidth]{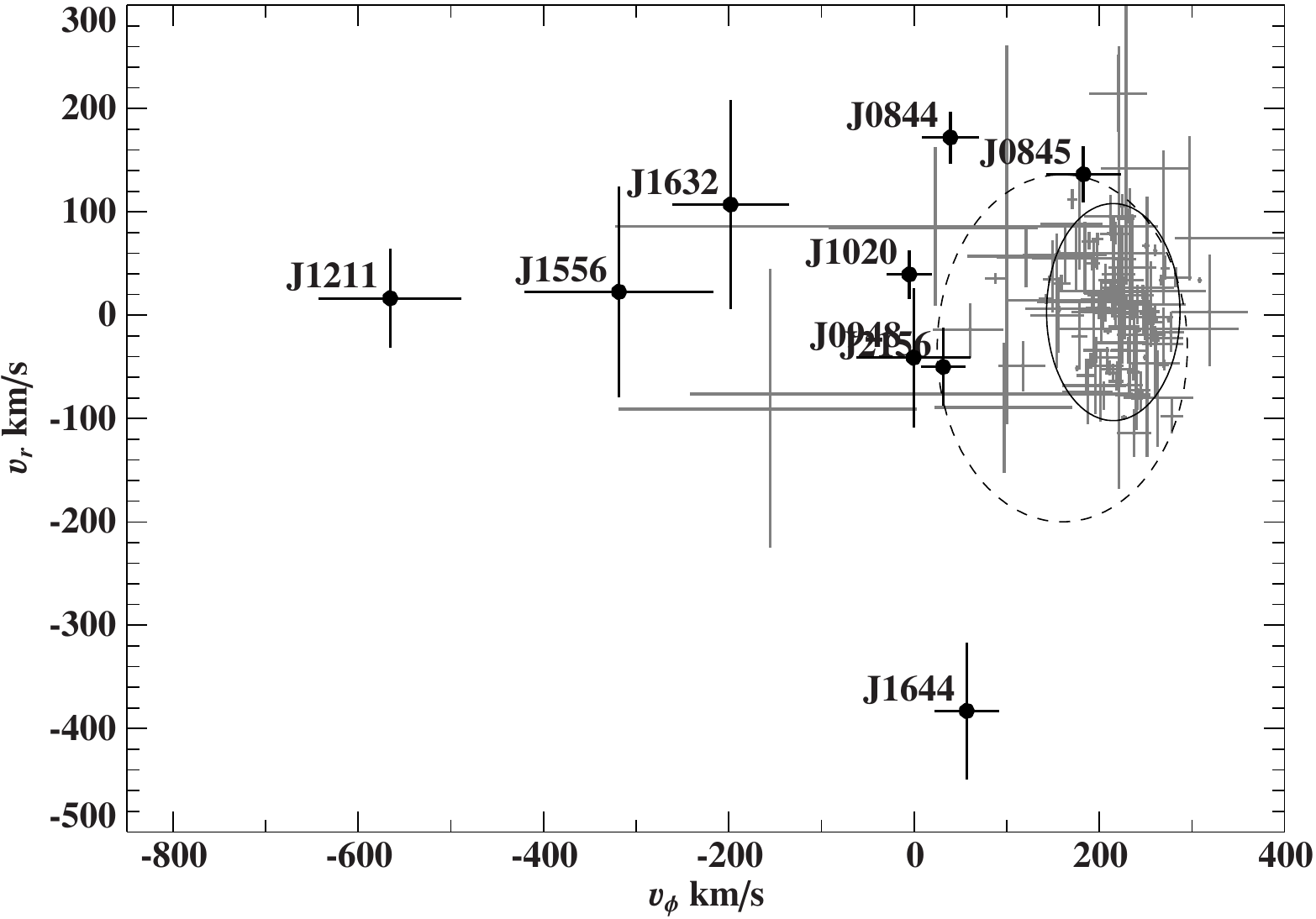}
\includegraphics[width=0.5\textwidth]{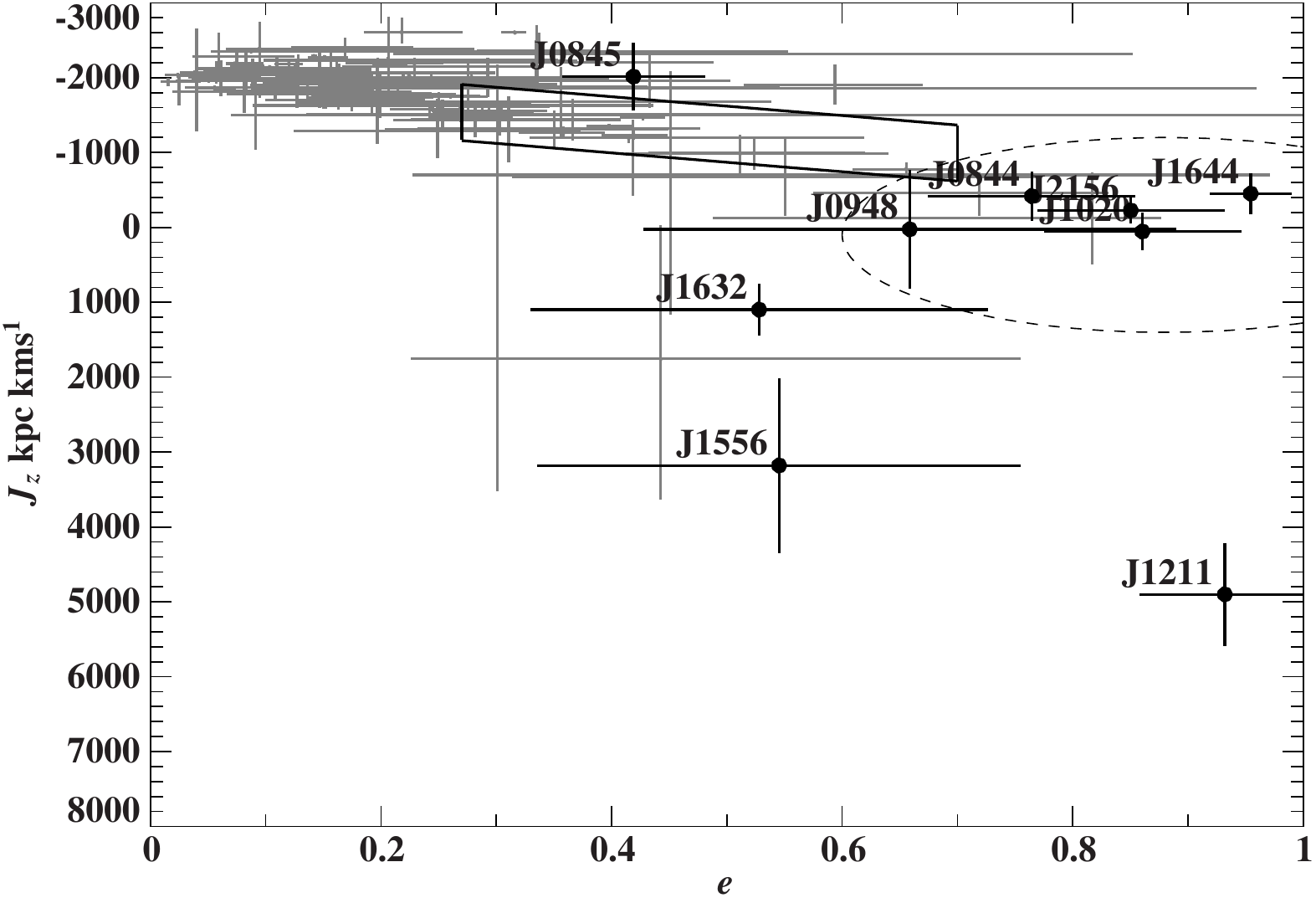}
\end{center}
\caption{{v$_\phi$  -- v$_r$ diagram (upper panel) and e-J$_z$ diagram (lower panel) for nine high-velocity sdB stars \citepads{2011A&A...527A.137T} labelled by their abbreviated name \citetads[see][]{2011A&A...527A.137T}. The sample of binary sdBs from 
\citetads{2015A&A...576A..44K} and analyzed from PPXML proper motions by Eichie et al. (in prep., grey error bars) 
 serves as a comparison. The solid ellipses render the 3$\sigma$-thin (solid) and thick disk contours (dashed) in the $v_\phi$  -- $v_r$ diagram (upper panel), while the solid box in the $e$-J$_z$ diagram (lower panel) marks the thick disk region as specified by \citetads{2006A&A...447..173P}. 
The values for J1211$+$1437 are taken from Nemeth et al. (2016) using model II of \citetads{2013A&A...549A.137I}. 
 Accordingly, eight stars would be assigned halo membership (J0845 is considered a thick disk star). \citetads{2011A&A...527A.137T} identify two groups, one (G1) lying inside the dashed ellipse in the $e$-J$_z$ diagram (lower panel), while those with positive J$_z$ form their group G2. 
From Ziegerer (priv. comm.)}}
\label{fig:sdb_kin_groups}
\end{figure}

\citetads{2011A&A...527A.137T} 
analyzed the kinematics of ten high-velocity hot subdwarfs, starting from their own proper motion measurements. Using spectroscopic distances and radial velocities 
Galactic trajectories were calculated \citepads[see][for details]{2013A&A...549A.137I}. 
 Accordingly, nine sdB stars were found to belong to the halo and one to the thick disk.  Two distinctive kinematic groups emerged from their analysis (see Fig. \ref{fig:sdb_kin_groups}): the normal halo subdwarfs (G1) with low Galactic rotation and the extreme halo subdwarfs (G2) on highly-eccentric retrograde orbits. \citetads{2011A&A...527A.137T} also considered that the objects are actually run-away disk stars ejected from their place of birth in the Galactic plane. A close inspection of their Galactic trajectories showed that 
the members of class G2 stars would come from the outer disk excluding an origin in the Galactic center.
The members of  the G1 class, however, would originate from the inner Galactic disk or bulge, including the Galactic center, and might have been ejected by {{a super-massive black-hole}} slingshot mechanism \citepads{1988Natur.331..687H}. 
{Some sdB stars of the sample of \citetads{2011A&A...527A.137T} that are approaching the Earth, J1644$+$4523 in particular, have to be bound to the Galaxy and therefore might provide constraints on the mass of the dark matter halo.}

\subsubsection {J1211$+$1437 -- an extremely fast halo hot subdwarf star in a wide binary system}

\begin{figure*}
\begin{center}
\includegraphics[width=0.99\textwidth]{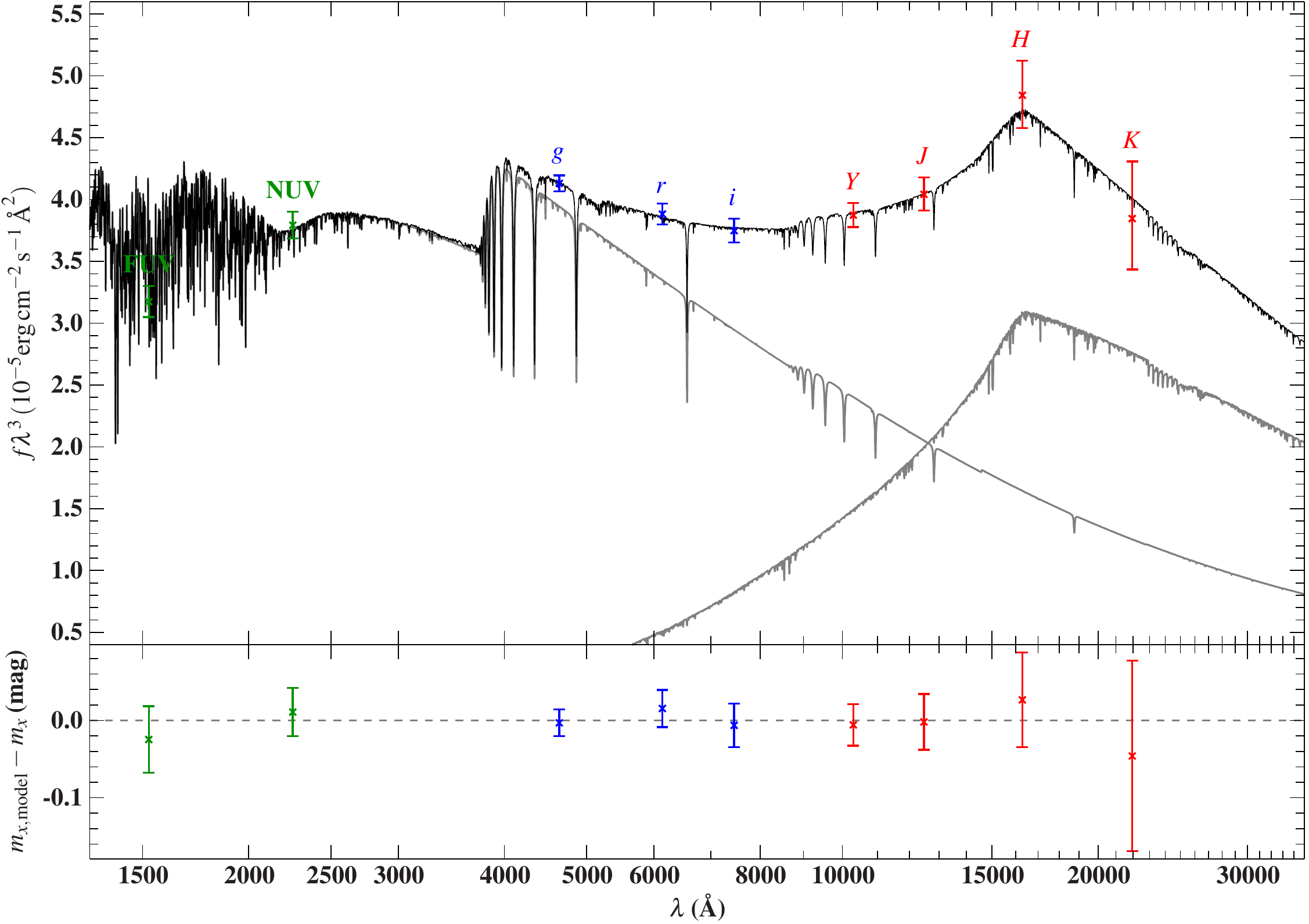}
\end{center}
\caption{Comparison of synthetic and observed photometry of the fast sdB+K binary J1211$+$1437 \citepads{2016ApJ...821L..13N}. Top panel: the spectral energy distribution. The colored data points are fluxes, which are converted from observed magnitudes, and the solid gray line is the composite (sdB+K) model. The individual contributions are plotted in light gray. Bottom panel: the residuals show the differences between synthetic and observed magnitudes. The photometric systems have the following color code: GALEX (green), SDSS (blue), UKIDSS (red). This is a modified version of  Fig. 2 of\citetads{2016ApJ...821L..13N}, from A. Irrgang (priv. comm.)
}
\label{fig:sed_j1211}
\end{figure*}

{The most interesting star in the sample, however, turned out to be J1211$+$1437, because recent spectroscopic follow-up \citepads{2016ApJ...821L..13N} 
revealed a metal-weak K-type main-sequence companion in a wide orbit (see Fig. \ref{fig:sed_j1211} for the spectral energy distribution).
Whether this binary with a Galactic rest frame velocity of 570 km\,s$^{-1}$ is bound or unbound to the Galaxy depends on the Galactic mass model preferred. 
Its Galactic kinematic rules out an origin in the Galactic center, but essentially all other acceleration mechanisms discussed for HVS and
runaway stars can be excluded as well. The binary is too fragile to survive  dynamical interaction in a dense stellar population or the
 kick of a core-collapse supernova. Hence, \citetads{2016ApJ...821L..13N} 
 concluded that J1211$+$1437 is either a bound extreme halo object
or was accreted from the a former satellite galaxy torn to shreds by the tidal forces of the Milkyway \citepads{2009ApJ...691L..63A}. 
However, kinematically J1211$+$143 could not be associated to any stellar stream in the halo that might represent the debris 
of a destroyed satellite galaxy.
}
 
\subsection{The unique hyper-velocity star US~708 - a remnant donor of a SN Ia explosion} 

 \citetads{2005A&A...444L..61H} 
discovered that the sdO star \object{US~708} shows an exceptionally high radial velocity of +708$\pm$15\,km\,s$^{-1}$. Accordingly, its Galactic
rest frame velocity ($>$757~km\,s$^{-1}$) 
exceeds the local Galactic escape velocity. At the time of discovery US~708 was only the second hyper-velocity star known to be unbound to the Galaxy\footnote{An HVS survey has increased the number of known HVSs to about two dozen
\citepads{2014ApJ...787...89B}. A comprehensive review is presented by \citetads{2015ARA&A..53...15B}.}. 
A quantitative non-LTE model atmosphere analysis of optical spectra obtained 
with the Keck~I telescope by \citetads{2005A&A...444L..61H} 
showed that US~708 is a normal \emph{helium-rich} sdO 
at a distance of 19~kpc. 

\begin{figure}
\begin{center}
\includegraphics[width=0.9\textwidth]{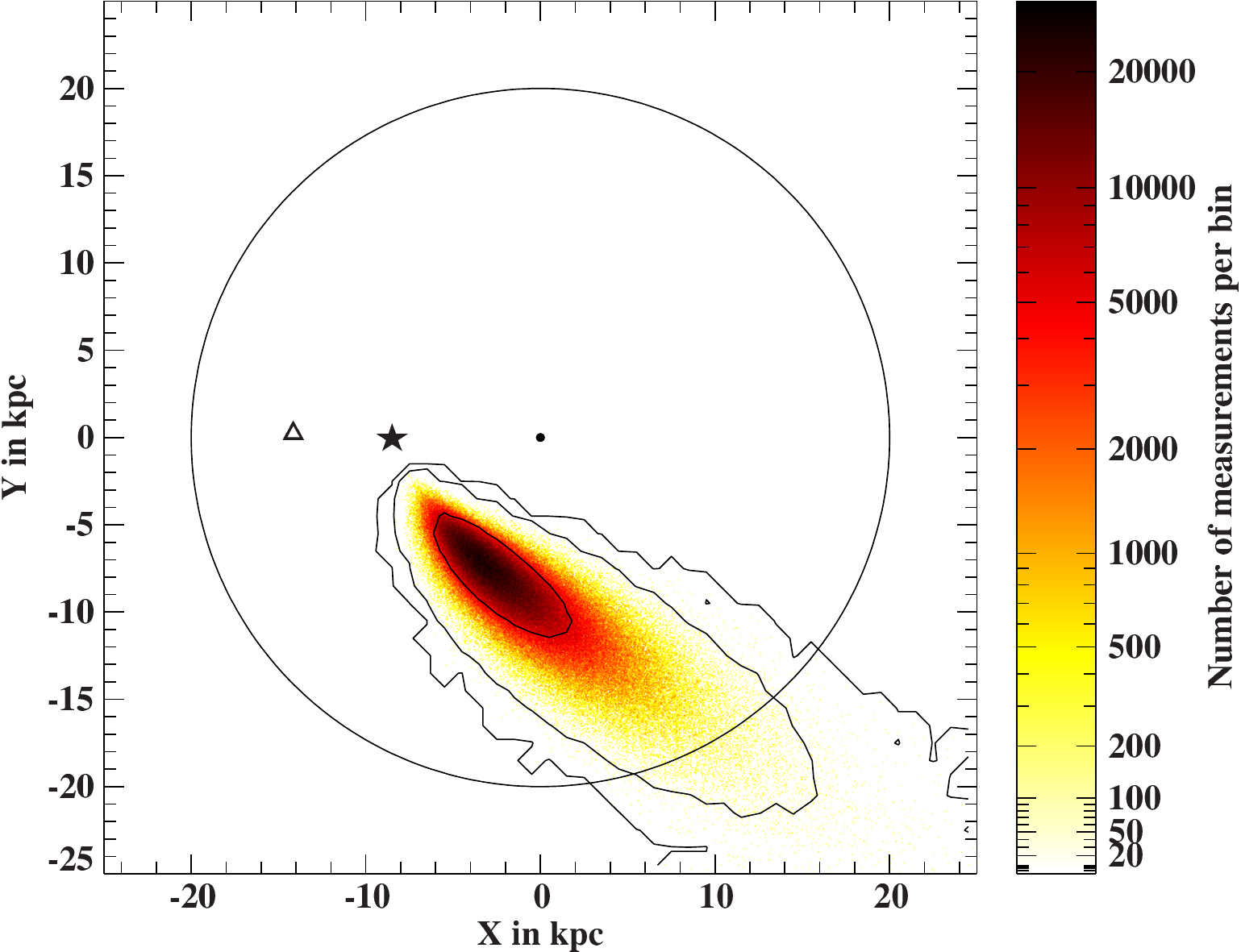}
\end{center}
\caption{Origin of US\,708. Monte Carlo simulation of the past trajectory of US\,708. The color-coded bins mark the positions, where the star crossed the Galactic disc, which is shown pole-on. The contours correspond to the $1\sigma$, $3\sigma$, and $5\sigma$ confidence limits. The position of the Galactic center is marked by the black dot, the position of the Sun as asterisk. The current position of US\,708 is marked by a triangle. From \citetads{2015Sci...347.1126G}; copyright Science; reproduced with permission.}
\label{fig:us708_origin}
\end{figure}

It has been suggested by \cite{1988Natur.331..687H} 
that such
hyper-velocity
stars can be formed by the tidal disruption of a binary through 
interaction with
a super-massive black hole. It is plausible
that US~708 might have originated from the Galactic center, because this is the only place in the Galaxy known to host a super-massive black hole,  
and the time of flight from the Galactic center to its present position is  
sufficiently short (36~Myrs, Fig. \ref{fig:us708_origin}). Whether or not the star originated in the Galactic center can
only be answered from very high precision proper-motion measurements.

\citetads{2015Sci...347.1126G} revisited US~708 by obtaining new spectra at better resolution at the Keck telescope to improve the radial velocity measurement. In addition, they {{were able}} to determine the star's proper motion by combining position measurements from astrometric photographic plates with modern CCD measurements (SDSS, PANSTARRS). By calculating Galactic trajectories in the Galactic potential the place of origin was found in the Galactic disk, but far from the Galactic center, which was excluded at a 5$\sigma$ confidence level.

\begin{figure}
\begin{center}
\includegraphics[width=0.9\textwidth]{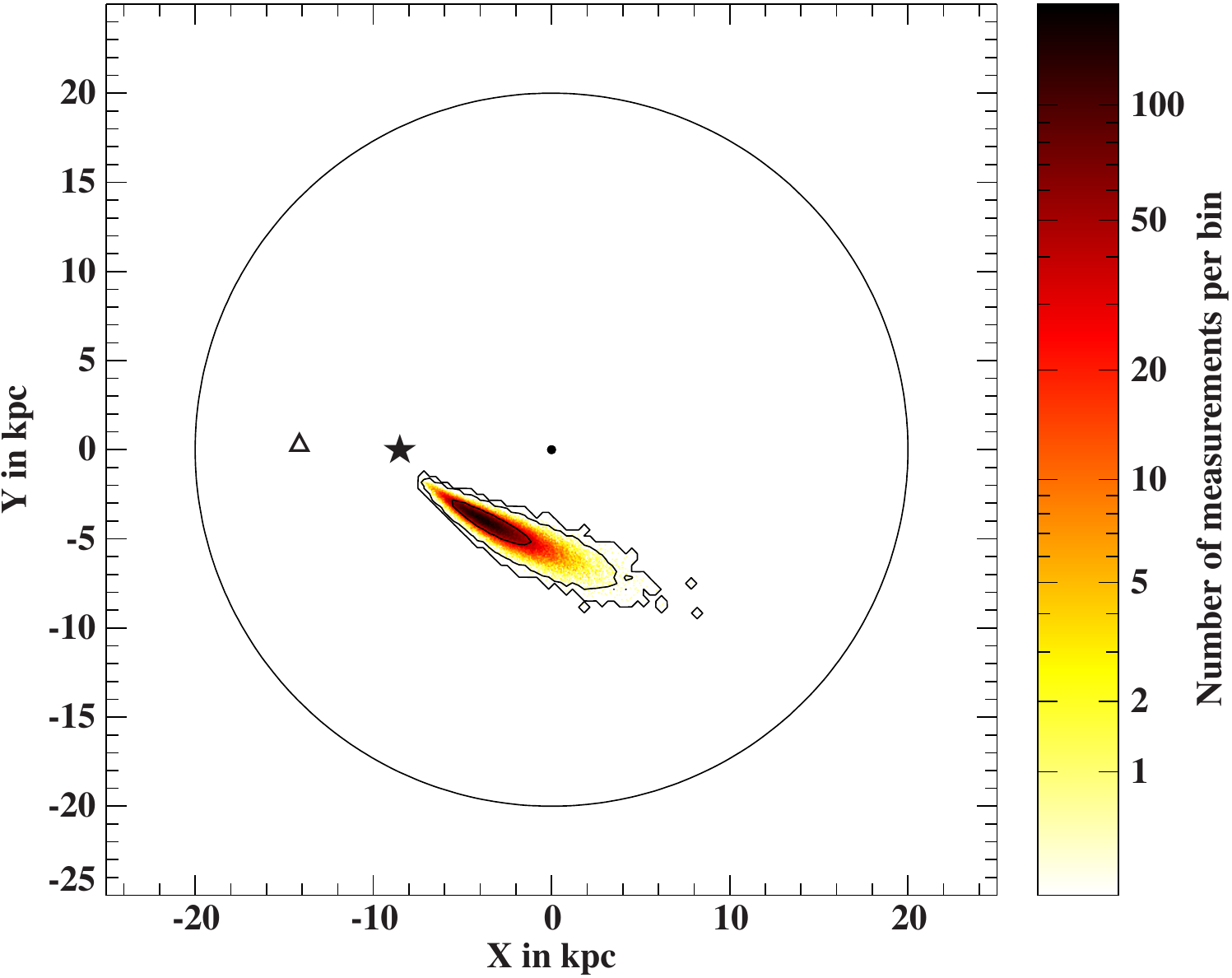}
\end{center}
\caption{Same as Fig.~\ref{fig:us708_origin} but based on the {\it HST} proper motion of US\,708 \citepads{2015ApJ...804...49B}. From Ziegerer (priv comm.).}
\label{fig:us708_origin_hst}
\end{figure}

The proper motion motion has also been determined recently from {\it HST} astrometry \citepads{2015ApJ...804...49B}. The conclusion is the same as from the ground-based measurements of \citetads{2015Sci...347.1126G}, 
i.e. the Galactic center is excluded as the place of origin (see Fig. \ref{fig:us708_origin_hst}). Accordingly, the Galactic rest-frame velocity of US~708 is 1082 $\pm$ 19 km\,s$^{-1}$
and the ejection velocity from the disk 894 $\pm$ 13 km\,s$^{-1}$ (Ziegerer, priv. comm.), which makes US 708 the fasted HVS star known.

\subsubsection{A link to type Ia supernovae}

\begin{figure}
\begin{center}
\includegraphics[angle=270,width=0.9\textwidth]{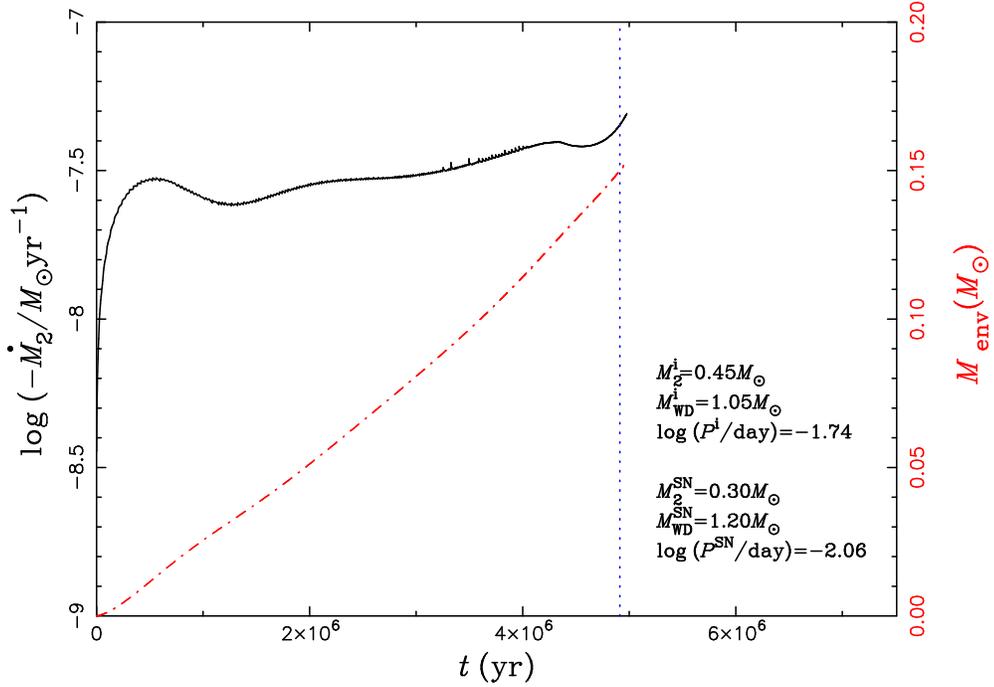}
\end{center}
\caption{a) Evolution of the progenitor after the He star fills its Roche lobe: Mass-transfer rate (full drawn) and the mass of the WD envelope (dashed-dotted) are varying with time. The dotted vertical line indicates the position where the double-detonation may happen (the mass of the He shell increases to $\sim0.15\,M_{\odot}$). The initial binary parameters and the parameters at the moment of the SN explosion are also given. 
From \citetads{2015Sci...347.1126G}; copyright Science; reproduced with permission. Combine with the next 2 Figures into one Figure. a,b,c.}
\label{fig:us708_progenitor_evol1}
\end{figure}

\begin{figure}
\begin{center}
\includegraphics[angle=270,width=0.9\textwidth]{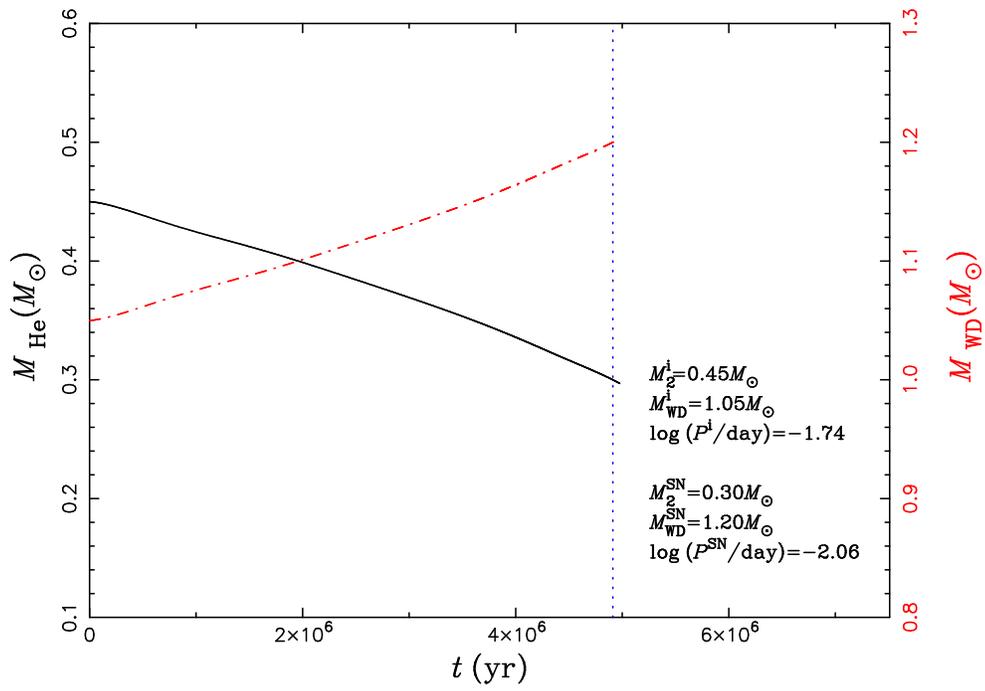}
\end{center}
\caption{b) Evolution of the progenitor after the He star fills its Roche lobe:
Same as a) but for the mass of the He-star (solid line) and of the WD (dash-dotted line). 
From \citetads{2015Sci...347.1126G}; copyright Science; reproduced with permission.}
\label{fig:us708_progenitor_evol2}
\end{figure}

\begin{figure}
\begin{center}
\includegraphics[width=0.9\textwidth]{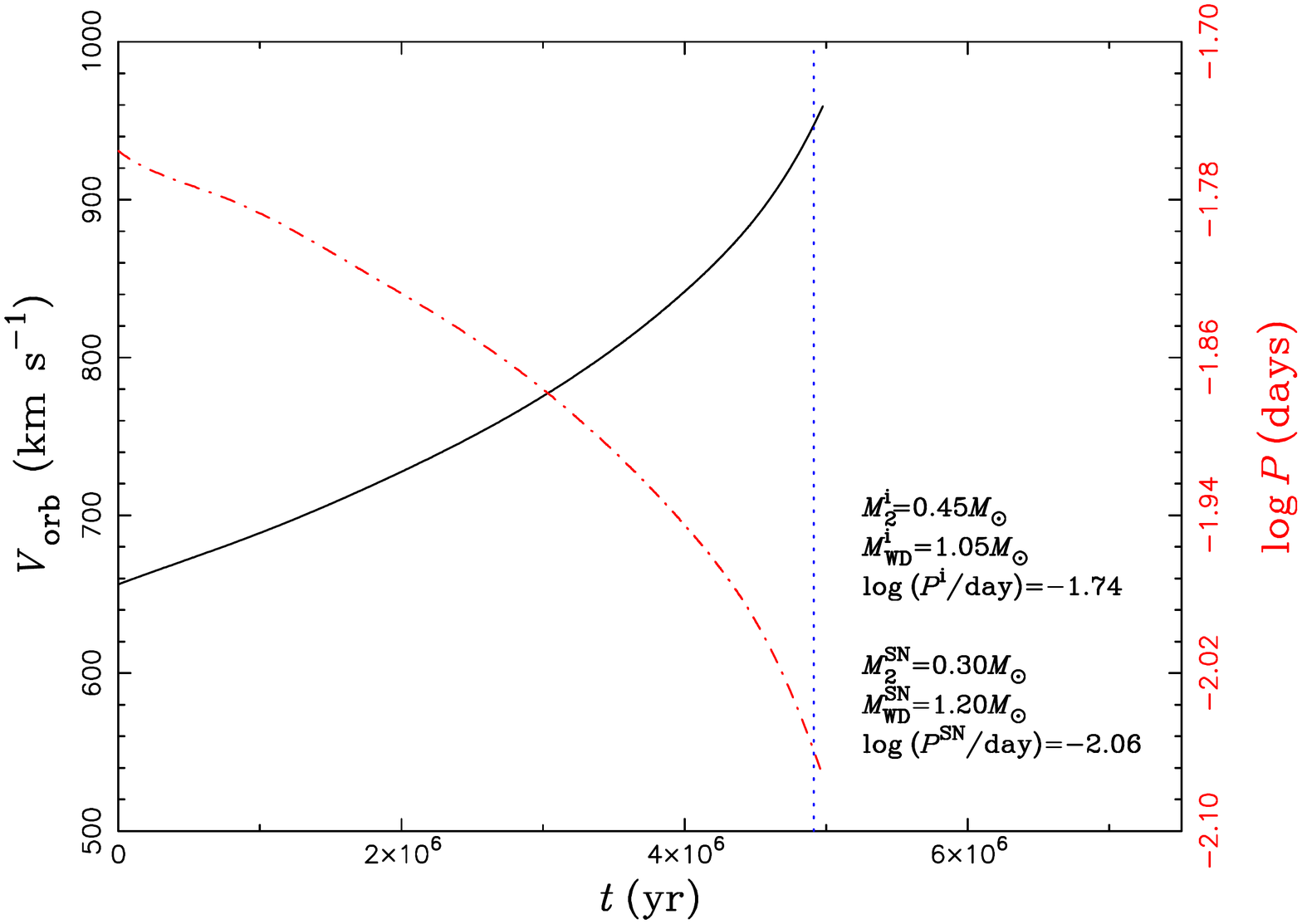}
\end{center}
\caption{c) Evolution of the progenitor after the He star fills its Roche lobe:
Same as a.), but for the Radial velocity semiamplitude (solid line) and the orbital period (dash-dotted line) of the binary. 
From \citetads{2015Sci...347.1126G}; copyright Science; reproduced with permission.}
\label{fig:us708_progenitor_evol3}
\end{figure}

\citetads{2009A&A...493.1081J}
suggested a link 
to type Ia supernovae by considering single low-mass (hyper-velocity) white dwarfs and He-sdO stars as the donor remnants to the exploding white dwarfs.
 \citetads{2010Ap&SS.329....3J,2011MNRAS.410..984J} investigated the case of the hypervelocity sdO US 708. 
\citetads{2015Sci...347.1126G} investigated the evolution of a progenitor system consisting of an sdB star and a massive C/O white dwarf. The evolution is depicted in Figs. 
\ref{fig:us708_progenitor_evol1} to \ref{fig:us708_progenitor_evol3}.
Accordingly, the donor was released at the moment of explosion and travelled away at approximately the orbital velocity. In order to gain a space velocity as high as that of US~708
the progenitor binary must have been extra-ordinarily tight and the white dwarf companion rather massive. \citetads{2015Sci...347.1126G} assumed that the system consisted of a compact helium star of $\sim0.3\,M_{\rm \odot}$ and a massive carbon-oxygen WD ($1.0-1.2\,M_{\rm \odot}$) with an orbital period of about $10\,{\rm min}$\footnote{Such tight binaries do exist as witnessed by  the eclipsing He-WD+CO-WD binary SDSS\,J065133+284423 \citepads[orbital period of only $12\,{\rm min}$,][]{2011ApJ...737L..23B}.} at the time of explosion. 

The rotation of the components in such ultra-short binaries are assumed to be tidally locked to the orbit. Hence, we expect the surviving remnant to be rapidly rotating. 
\citetads{2015Sci...347.1126G} derived the rotational velocity from their Keck spectra, which turned out to be
v$_{\rm rot} \sin$ i = 115$\pm$8 km\,s$^{-1}$ 
much higher than  observed for single He-sdO stars \citepads{2009JPhCS.172a2015H}, but smaller than the expected rotation velocity 
at the time of ejection ($\approx$350 km s$^{-1}$). Accounting for the evolution of the sdO star, which led to an expansion of the star \citepads{2013ApJ...773...49P}, 
and assuming conservation of angular momentum, \citetads{2015Sci...347.1126G} estimate $\approx$120 km s$^{-1}$ in perfect agreement with the observed rotation rate.
This lends strong support to the SN Ia donor remnant scenario. 

Another imprint of a thermonuclear explosion of the original white dwarf primary would be the debris deposited on the surface of the surviving donor  
\citepads{2013ApJ...774...37L} 
which could possibly be traced by UV spectroscopy. 
The scenario also predicts a low mass {for} the He-sdO star because of the transfer of a significant fraction of the helium envelope prior to the explosion and additional few percent by ablation during the explosion \citepads{2013ApJ...774...37L}.

Up to now, US~708 remained unique; the only known hyper-velocity hot subdwarf star. This, however, is expected to change, when the {\it Gaia} measurements become available. 

 \clearpage

\newpage

\section{Conclusion}\label{conclusion}

This review updates and extends that of \citetads{2009ARA&A..47..211H} 
and includes a section about the new classes of extremely low mass (ELM) white dwarfs, which are the stripped helium cores of red giant stars similar to many sdB stars in binaries. The latter, however, managed to ignite core helium burning and therefore populate the extreme horizontal branch, while the ELM white dwarfs evolve towards the fully degenerate configuration, eventually experiencing hydrogen shell flashes. 
Many subluminous B stars reside in binaries accompanied by either white dwarfs on close orbits (periods of 0.1 to 30 days), main sequence stars of M-type 
on even closer orbits, or F/G/K type on wide orbits (periods of 700-1200 days).
Similarly, almost all ELM white dwarfs are found in close binaries with another white dwarf or with an A- or F-type main sequence star (''EL CVn'').
Since the orbital separations in the short period sdB binaries are only a few solar radii, much smaller than the size of the red-giant progenitor, their orbit must have shrunk considerably with respect to the original one. Therefore, the progenitor system must have undergone a common-envelope phase, during which the companion was engulfed in the red giant's envelope leading to a spiral-in due to friction and eventually to the ejection of the envelope.    
     
Progress in the field has been enormous over the last seven years. Many of the new discoveries, however, {{cannot be easily explained}} and pose new questions:

\begin{itemize}

\item Quantitative spectral analyses of high-resolution optical and ultraviolet spectra have established a coarse abundance pattern of sdB stars. On average the abundances of the lighter elements (helium to sulfur) are subsolar, while those of heavier elements (Ca to Pb) are enriched with respect to the Sun reaching factors larger than one hundred, in a few cases up to ten thousand (e.g. for Pb). Notable exceptions are nitrogen and iron which  are close to solar with remarkably small star-to-star scatter. On the other hand, very large star-to-star scatter is obvious for carbon and silicon. There is no doubt, that these patterns result from atmospheric diffusion processes. Diffusion theory has advanced considerably and included turbulence, but is still limited by the lack of atomic data for many heavy elements.    

\item Many of the compact companions in close binaries are white dwarfs. 
With respect to our understanding of binary evolution, it would be important to find out how many of them are of helium composition, hence, originate from the first giant branch, or of C/O composition, hence from the second giant branch.
Neutron-star or black-hole companions to sdB stars are predicted but have not yet been discovered, while ELM white dwarfs have been found as partners to milli-second pulsars already. Binary sdB stars with massive white dwarf companions are candidate supernova Ia progenitors both in the double degenerate scenario (e.g. KPD 1930+2752) and the single degenerate scenario as helium
donors for double detonation supernovae (e.g. CD$-$30$^\circ$11223). {For a long time the candidate SN Ia progenitor HD~49798  was the only known X-ray source among hot subdwarf stars. Modern X-ray satellites have now identified a handful of X-ray sources among the hottest sdO stars, including two non-binaries.}

\item The wide orbits of sdB binaries with main-sequence companions of spectral type F/G/K result from stable Roche lobe overflow. Only recently, their periods have been determined to lie between 700 and 1200 days. Because the sdB star will evolve directly into a white dwarf, we expect a population of F/G/K stars with unseen white dwarf companions to exist \citepads{2016arXiv160401613P} with similar orbital periods. 
Such systems will be easily detectable by {\it Gaia} as astrometric binaries.

\item Substellar companions to sdB stars have also been found. For HW~Vir systems the companion mass distribution extends from 0.2 M$_\odot$ to below the stellar mass limit; that is, into the brown dwarf regime.
A giant planet companion to the p-mode pulsator V391 Peg marked the first discovery of a planet that survived the red giant evolution of its host star.
Evidence for Earth-size planets to two pulsating sdB stars has been reported  and circumbinary giant planets or brown dwarfs have been found around HW~Vir systems by the eclipse timing technique. The high incidence of circumbinary substellar objects favors a second generation nature; that is, the planets are formed from remaining common-envelope material. However, none of the candidates have been confirmed by an independent method. Therefore, it remains unclear whether those substellar companions are real.

\item Thanks to {\it Kepler} light curves, asteroseismology advanced to determine the 
rotation rates of a handful of pulsating sdB stars from the splittings of frequency multiplets.
The stars are found to rotate much more slowly (tens of days) than expected.
What causes efficient angular momentum transport and loss, probably early-on on the red giant branch? Does radial differential rotation occur? 
Asteroseismology of gravity-mode pulsators allowed the internal structure of some sdB stars to be probed. The size of the convective cores are larger than {the models imply} indicating that convection is more efficient than  {assumed by today's standard physical models}.
  
\item Almost half of the sdB population is apparently single, which may be explained in the context of binary evolution, if they result from binary mergers of helium white dwarfs, or {{result}} from the disruption of a companion; that is, massive enough to initiate common-envelope ejection but of too low mass to survive. However, an sdB star may form in isolation by internal processes that allow the helium core of a red giant to grow beyond the canonical size and cause delayed helium flashes, when the star has already departed from red giant branch. 
Could rapid core rotation be the clue?

\item What is the origin of the helium subdwarf O stars?
The binary frequency of He-sdO stars is much lower than that of the sdB stars, indicating that they arise from a different evolutionary path than the latter. The merger of two helium white dwarfs is the most popular scenario for the origin of He-sdO, rivaled by the late hot flasher scenarios. The merger scenario may explain the observed carbon and nitrogen abundance pattern at least qualitatively, whereas the very late flasher may be able to match the nitrogen-rich stars. However, the predicted location of the He-sdOs in the T$_{\rm eff}$ - $\log$ g diagram
does not match the spectroscopic one. It is, though, notoriously difficult to determine the surface gravity of He-sdO stars from optical spectra. Independent
gravity estimates, would be needed. The origin of erratic radial velocity variations and strong magnetic fields also awaits an explanation.   

\item The small class of intermediate He-sdB stars contains chemically very peculiar members, in particular the halo star LS~IV $-$14$^\circ$ 116. Large enrichment of heavy elements such as zirconium, yttrium, and lead are common to 
those stars. In addition, LS~IV $-$14$^\circ$ 116 is a unique pulsator.  
What causes those pulsations? Are the intermediate He-sdBs objects in transition; that is, evolving onto the extreme horizontal branch as suggested by \citetads{2012ASPC..452...41J}? 

\item The hot subdwarf population in globular clusters appears to be different from that of the field. Counterparts to the blue hook stars have not yet been found in the field population, whereas the extremely helium rich sdO stars, which are frequent in the field, seem to be lacking in globular clusters. Similarly, the counterparts to the pulsating sdO stars in $\omega$ Cen have not yet been found in the field population.  

\item The discovery of the extremely low-mass white dwarfs filled a gap in the mosaic of binary stellar evolution. As for the hot subdwarfs the binary frequency is high, actually even higher than that of the sdB stars. Companions are either white dwarfs, neutron stars (pulsars) or main sequence stars (''EL CVn'' stars). Multi-periodic oscillations were also discovered and shown to be related to the ZZ Ceti strip of pulsating white dwarfs. 
 
\end{itemize} 

{{In order to tackle the newly revealed shortcomings improvement to models of the stellar interior, envelope, and atmosphere as well as additional observations and facilities are needed. Atomic data for heavy elements are required to improve diffusion calculations, which are crucial for our understanding of pulsation driving, radiative levitation, atmospheric abundance pattern and stratification of O/B subdwarfs as well as ELM white dwarfs. The lack of atomic data (opacities and line broadening) still limits the accuracy for atmospheric parameter determination and chemical abundances.}}

{{
The treatment of convection in stellar interiors remains a challenge despite of decade-long efforts. Asteroseismology of hot subdwarfs adds new constraints to improve interior models, because it allows the size of the convective core to be constrained. The}} {\it MESA} {{code provides new options to tackle the problems.
Asteroseismology of red giants and hot subdwarf stars revealed that additional physical mechanisms are required to understand their evolution, in particular the angular momentum transport, tidal synchronization and the origin of turbulence remain to be clarified. In order to understand the evolution of close binaries the physics of the common envelope ejection as well as of stellar mergers need to be understood. The hot subdwarf stars and ELM white dwarf will provide ideal laboratories to this end.}}

{On the observational side the} {{Kepler}} {mission is going on and we can expect surprises from the {\it K2} campaign both for asteroseismology and binary research. The massive ground-based photometric surveys, such as the Palomar Transient facility, will be upgraded 
\citepads{2014SPIE.9147E..79S} and new 
projects will start culminating in the {\it LSST} \citepads{2009arXiv0912.0201L}. They will increase the inventory of binary hot subdwarfs, low mass white dwarfs an EL CVn systems significantly and may even lead to the detection of optical transients caused by white dwarf mergers to form hot subdwarf stars.
New spectroscopic surveys such as 
{\it LAMOST} already discovered and characterized hot subdwarf stars \citepads{2016ApJ...818..202L} and will go on to do so.} 

The masses of hot subwarf stars, both single or in binaries, are the key to understand their evolution. The tools available are the analysis of eclipsing binary orbits and asteroseismology. Therefore, NY Vir, a pulsating sdB star in an eclipsing binary was considered the ''Rosetta stone'' by {\citetads{2008A&A...489..377C}}
because it allows both techniques to be applied and cross-checked and was unique at the time of writing. 
In the mean time a few similar systems have been found. Combining dynamical and seismic analyses of eclipsing binaries with pulsating components will put stringent constraints on evolutionary models \citepads[see] [for an example]{2014MNRAS.442..616T}. 

The ability to derive masses is also impeded by the poor knowledge of the stars' distances. In most cases spectroscopic distances are used, which rest on an adopted subdwarf mass and the atmospheric parameters determined by quantitative analyses. The largest contribution to the error budget usually comes from the surface gravity.  
This will soon be remedied by the  
{\it Gaia} astrometric mission \citepads{2001A&A...369..339P}, 
which will provide parallaxes to 10 $\mu$as precision in the most favorable cases. Parallax measurements for the near-by hot subdwarfs will 
provide a benchmark to test atmospheric models (surface gravity) and/or the mass distribution of binary as well as single hot subdwarf stars. 
Already the first data release, expected in 2016, will provide parallaxes and proper motions for many of the 2.5 million stars from the Tycho-2 catalog \citepads{2015A&A...574A.115M} 
to sub-mas precision. 
This catalog contains about 180 hot subdwarfs of all types and will provide a first benchmark, although not yet full-fledged. {{{\it Gaia} will be very efficient to study astrometric binaries with periods from a few weeks to years, which makes it complementary to many radial velocity studies which are biased towards short periods ($<$30 days).}} 

{\it Gaia} will provide proper motions for a huge number of hot subdwarf stars that will allow the different stellar populations to be disentangled. The halo members will be particularly interesting, because they allow a meaningful comparison with
the population of globular cluster hot subdwarfs. 

{The {\it Gaia} survey will be complemented by large spectroscopic surveys using multi-object spectrographs like {\it 4MOST} \citepads{2014SPIE.9147E..0MD} and {\it WEAVE} \citepads{2012SPIE.8446E..0PD}. They will allow large samples of faint blue stars to be studied in detail, including new types of hot subluminous stars.}    
The fastest hot subdwarfs will allow  the mass of the dark matter halo to be constrained further and other unbound hyper-velocity stars similar to the sdO hyper-velocity star US~708 
will be discovered. 
 
\clearpage

{ACKNOWLEDEMENT.} I thank the editor, Dr. Jeff Mangum, for the invitation and his patience with the slow progress of my work. I thank the organizers of the ''7$^{th}$ meeting on hot subdwarf stars and related objects'', Tony Lynas-Gray and Philipp Podsiadlowski, for offering such 
a lively environment to discuss with many colleagues. {Financial support for publication costs by the Deutsche Forschungsgemeinschaft (grant He1356/45-2) is gratefully acknowledged.}
I {thank} Andreas Irrgang, Peter N{\'e}meth, and Eva Ziegerer for providing me with their figures prior to publication, and to Stephan Geier, Simon Jeffery, and Thomas Kupfer for providing updated versions of their figures and John Telting for his original figures.
My sincere thanks go to Warren Brown, Stefan Dreizler, Stephan Geier, Andreas Irrgang, Simon Jeffery, Thomas Kupfer, Marilyn Latour, Tony Lynas-Gray, Sandro Merreghetti, Marcello Miller-Bertolami,  Sabine Moehler, Peter N{\'e}meth, Roy {\O}stensen, Nicole Reindl, Veronika Schaffenroth, and Markus Schindewolf for their most helpful comments, corrections and suggestions.

\appendix
\section{Atmospheric parameters and abundances of He-sdO stars}

The atmospheric parameters and helium, carbon, and nitrogen abundances of He-sdO stars from \citetads{phd_hirsch2009} are listed in table \ref{tbl:spyresult} as plotted in Figs. \ref{fig:spy}, \ref{fig:spy_hqs_galex_pg}, \ref{fig:spyabun} and \ref{fig:spytefflogg}, and 
\ref{fig:merger_hirsch}.
 
 
\begin{table*}[t]
\caption{The atmospheric parameters,nitrogen and carbon abundance of He-sdO stars (extracted from table 11.3 of \citetads{phd_hirsch2009}. 
Abundances are given by mass fractions.}
\label{tbl:spyresult}
\vspace*{5mm}
\begin{tabular}{rllllll}
\hline\hline
Nr. &Star		&\multicolumn{1}{c}{\teff}&\multicolumn{1}{c}{\logg}&\multicolumn{1}{c}{\massy}&\multicolumn{1}{c}{\massn}&\multicolumn{1}{c}{\massc}\\
\hline
1 & HE 1256$-$2738	& 39571	 & 5.66		& $-$0.094 & $-$2.44 $\pm$ 0.09	  & $-$1.50	  $\pm$ 0.18  \\[0.0mm]
2 & [CW83]0904$-$02	& 46170	 & 5.64		& $-$0.014 & $-$3.60 $\pm$ 0.16	  & $-$1.56	  $\pm$ 0.19  \\[0.0mm]
3 & HE 1446$-$1058	& 45240	 & 5.67		& $-$0.013 & $-$2.62 $\pm$ 0.11	  & $-$1.60	  $\pm$ 0.08  \\[0.0mm]
4 & HE 0958$-$1151	& 44229	 & 5.39		& $-$0.011 & $-$3.24 $\pm$ 0.04	  & $-$1.64       $\pm$ 0.01  \\[0.0mm]
5 & HE 0016$-$3212	& 39186	 & 5.13		& $-$0.014 & $-$2.29 $\pm$ 0.07	  & $-$1.64	  $\pm$ 0.06  \\[0.0mm]
6 & WD 2020$-$253	& 44105	 & 5.51		& $-$0.011 & $-$2.68 $\pm$ 0.06	  & $-$1.68	  $\pm$ 0.11  \\[0.0mm]
7 & HE 1142$-$2311	& 51154	 & 5.38		& $-$0.016 & $<-$4.44		  & $-$1.70 	              \\[0.0mm]
8 & HE 1203$-$1048	& 44806	 & 5.62		& $-$0.009 & $<-$4.46		  & $-$1.71	  $\pm$ 0.06  \\[0.0mm]
9 & HE 1251$+$0159	& 45637	 & 5.63		& $-$0.010 & $<-$4.46		  & $-$1.72	  $\pm$ 0.16  \\[0.0mm]
10 & HE 2203$-$2210	& 47049	 & 5.60		& $-$0.009 & $-$3.45 $\pm$ 0.02	  & $-$1.72	  $\pm$ 0.03  \\[0.0mm]
11 & HE 0952$+$0227	& 44266	 & 5.45		& $-$0.009 & $-$3.08 $\pm$ 0.04	  & $-$1.75	  $\pm$ 0.04  \\[0.0mm]
12 & [CW83]0832$-$01	& 43953	 & 5.66		& $-$0.008 & $-$2.77 $\pm$ 0.17	  & $-$1.79	  $\pm$ 0.20  \\[0.0mm]
13 & HE 0414$-$5429	& 43970	 & 5.52		& $-$0.005 & $-$2.95 $\pm$ 0.47    & $-$2.04	  $\pm$ 0.18  \\[0.0mm]
14 & HE 1136$-$1641	& 43957	 & 5.58		& $-$0.006 & $-$2.62 $\pm$ 0.10	  & $-$2.05	  $\pm$ 0.02  \\[0.0mm]
15 & HE 0914$-$0341	& 45496	 & 5.72		& $-$0.004 & $-$3.20 $\pm$ 0.34	  & $-$2.26       $\pm$ 0.03  \\[0.0mm]
16 & HE 2347$-$4130	& 45040	 & 5.78		& $-$0.002 & $-$3.52 $\pm$ 0.12   & $-$2.40	  $\pm$ 0.01  \\[0.0mm]
17 & HE 1258$+$0113	& 39169	 & 5.66		& $-$0.237 & $-$2.55 $\pm$ 0.23	  & $<-$4.17                  \\[0.0mm]
18 & HE 1238$-$1745	& 38743	 & 5.48		& $-$0.272 & $-$2.76 $\pm$ 0.17	  & $<-$4.14	  	      \\[0.0mm]
19 & CD $-$24 9052	& 41700	 & 5.49		& $-$0.195 & $-$2.39 $\pm$ 0.08	  & $<-$4.21                  \\[0.0mm]
20 & HE 1135$-$1134	& 40358	 & 5.38		& $-$0.107 & $-$2.53 $\pm$ 0.24	  & $<-$4.31                  \\[0.0mm]
21 & HE 1310$-$2733	& 40225	 & 5.45		& $-$0.105 & $-$2.28 $\pm$ 0.06	  & $<-$4.31                  \\[0.0mm]
22 & HE 1136$-$2504	& 41212	 & 5.65		& $-$0.097 & $-$2.31 $\pm$ 0.20   & $<-$4.32                  \\[0.0mm]
23 & HE 1511$-$1103	& 41090	 & 5.46		& $-$0.008 & $-$2.29 $\pm$ 0.09	  & $-$4.33	  $\pm$ 0.03  \\[0.0mm]
24 & WD 2204$+$071	& 40553  & 5.52	        & $-$0.078 & $-$2.23 $\pm$ 0.34	  & $<-$4.36                  \\[0.0mm]
25 & HD 127493	& 42484	 & 5.60		& $-$0.027 & $-$2.50 $\pm$ 0.12	  & $<-$4.46	  	 	      \\[0.0mm]
26 & CD $-$31 4800	& 43080	 & 5.87		& $-$0.002 & $-$2.50 $\pm$ 0.17	  & $-$4.49	  $\pm$ 0.20  \\[0.0mm]
27 & HE 0342$-$1702	& 41082	 & 5.59		& $-$0.004 & $-$2.56 $\pm$ 0.06	  & $<-$4.51	  	      \\[0.0mm]
28 & WD 2258$+$155	& 40619	 & 5.71		& $-$0.001 & $-$2.76 $\pm$ 0.06	  & $<-$4.52	  	      \\[0.0mm]
29 & HE 0031$-$5607	& 39367	 & 5.58		& $-$0.002 & $-$2.55 $\pm$ 0.04	  & $<-$4.52                  \\[0.0mm]
30 & WD 0447$+$176	& 40545	 & 5.54		& $-$0.001 & $-$2.71 $\pm$ 0.01	  & $<-$4.52                  \\[0.0mm]
31 & HE 1316$-$1834	& 41170  & 5.30	        & $-$0.003 & $-$2.20 $\pm$ 0.34   & $<-$4.52                  \\[0.0mm]
33 & HE 0155$-$3710	& 40521	 & 5.61		& $-$0.002 & $-$2.81 $\pm$ 0.01	  & $<-$4.52                  \\[0.0mm]
33 & HE 0001$-$2443	& 39840	 & 5.69	        & $-$0.002 & $-$2.47 $\pm$ 0.34   & $<-$4.52 \\
\hline                                          
\end{tabular}                                     
\end{table*}                                      
\clearpage                                     

\bibliography{heber}
\bibliographystyle{apj}




\end{document}